\documentclass{article}
\usepackage{times}
\usepackage[pdftex]{graphicx}
\usepackage{amssymb,amsfonts,amsmath,amsthm}
\usepackage{float}
\usepackage{dsfont}
\usepackage{xcolor}
\usepackage{wrapfig}

\usepackage{amsfonts}
\usepackage{amssymb}
\usepackage[dvips]{epsfig}
\usepackage{delarray}
\usepackage{hyperref}
\usepackage{url}
\usepackage{anysize}
\usepackage{multirow}
\usepackage{amsmath}
\usepackage{graphicx}
\usepackage[OT1]{fontenc}
\usepackage{amsthm,amsmath}%,natbib}
\usepackage{hyperref}
\usepackage{hypernat}
\usepackage{color}
\usepackage[dvips]{epsfig}
\usepackage{delarray}
\usepackage{hyperref}
\usepackage{url}
\usepackage{colortbl}
\usepackage{anysize}
\usepackage{multirow}
\usepackage{amsmath}
\usepackage{amsthm}
\usepackage{amsmath, amssymb}
\usepackage[ruled,noend,linesnumbered]{algorithm2e}
\usepackage{rotating}

\numberwithin{equation}{section}

\newtheorem{theorem}{Theorem}

\newtheorem{lemma}{Lemma}

\newtheorem{ass}{Assumption}

\newcommand\half {\frac{1}{2}}

\newcommand\EE {\mathbb E}

\newcommand\RR {\mathbb R}
\newcommand\PP {\mathbb P}
\newcommand\QQ {\mathbb Q}

\def\bone{\mathbf{1}}

\def\tC{\tilde{C}}

\def\tV{\tilde{V}}

\def\halpha{\hat{\alpha}}

\def\hbeta{\hat{\beta}}

\def\hD{\hat{\Delta}}

\def\bT{\bar{T}}
\def\bbeta{\bar{\beta}}

\def\tkappa{\tilde{\kappa}}

\def\bOmega{\bar{\Omega}}

\def\tkappa{\tilde{\kappa}}

\def\hkappa{\hat{\kappa}}

\usepackage{color}

\begin{document}

\title{\textbf{Simulation of Implied Volatility Surfaces via Tangent L\'{e}vy Models}}
\author{R.~Carmona, Y.~Ma and S.~Nadtochiy}
\date{Current version: Mar 24, 2015}

\maketitle

\begin{abstract}
In this paper, we implement and test two types of market-based models for European-type options, based on the tangent L\'evy models proposed in \cite{carmona2010tangentlevy} and \cite{carmona2011tangentlevy}. As a result, we obtain a method for generating Monte Carlo samples of future paths of implied volatility surfaces. These paths and the surfaces themselves are free of arbitrage, and are constructed in a way that is consistent with the past and present values of implied volatility.
We use a real market data to estimate the parameters of these models and conduct an empirical study, to compare the performance of market-based models with the performance of classical stochastic volatility models. We choose the problem of minimal-variance portfolio choice as a measure of model performance and compare the two tangent L\'evy models to SABR model. Our study demonstrates that the tangent L\'{e}vy models do a much better job at finding a portfolio with smallest variance, their predictions for the variance are more reliable, and the portfolio weights are more stable. To the best of our knowledge, this is the first example of empirical analysis that provides a convincing evidence of the outperformance of the market-based models for European options using real market data.
\end{abstract}

\section{Introduction\label{se:intro}}

The existence of liquid markets for equity and volatility derivatives, as well as a well-developed over-the-counter market for exotic derivatives, generates a need for a modeling framework that is consistent across time and across financial instruments. Within this framework, once a model is chosen so that it matches both the present prices of liquid instruments and their past dynamics, it is expected to produce more realistic results for the problems of pricing and hedging of exotic instruments. In addition, such models can be used to quantify the risk embedded in portfolios of derivative contracts. Needless to say, evaluating and managing the risk of such portfolios is crucial for proper functioning of the financial markets: recall, for example, that VIX index, itself, is a portfolio of European options written on S\&P 500.

In this paper we investigate an arbitrage-free modeling framework for multiple European-type options written on the same underlying, which is consistent across time and products. In particular, this framework allows to resolve one of the nagging challenges of \emph{quant} groups supporting equity trading: i.e. how to generate realistic Monte Carlo scenarios of implied volatility surfaces which are consistent with present and historical observations? As mentioned above, such models can be used to address the problems of pricing, hedging and risk management. Herein, we implement several such models using real market data and \emph{conduct a numerical experiment which demonstrates clearly the advantages of this modeling approach}.

The attempts to model the dynamics of implied volatility surface directly can be dated back as early as the ``sticky smile model" and the ``sticky delta model" (also known as ``floating smile model") (see Section~6.4 of \cite{Rebonato2004volatility} for the definitions).  As an improvement of the two models, Cont et al. later proposed a multi-factor model of implied volatility surface in \cite{Cont2002dynamics} and \cite{cont2002stochastic}, where they applied a Karhunen-Lo\`{e}ve decomposition on the daily variations of implied volatilities. It turns out that the first three eigenvectors could explain most of the daily variance, and a mean-reverting factor model based on the three eigenvectors is then constructed for future implied volatility surface.  
The major issue with these early attempts is that the proposed models for the dynamics of implied volatility are either too restrictive, not allowing to match the historical evolution of implied volatility, or too loose, so that they may contain \emph{arbitrage opportunities}. While the importance of the first issue for any time-series analysis is very clear, the second one deserves a separate discussion. Indeed, what do we mean by arbitrage opportunities in a model for implied volatility and why do we need to avoid it? There are two types of arbitrage opportunities we refer to: \emph{static} and \emph{dynamic}. A given implied volatility surface contains static arbitrage if it is impossible to obtain such a surface in any arbitrage-free model for the underlying. The fact that not every surface can be an arbitrage-free implied volatility simply follows from the well-known static no-arbitrage restrictions on the shape of a call price surface: e.g. monotonicity and convexity in strikes, etc (cf. \cite{Cousot2007conditions} and \cite{derman1998tree}). Notice that a violation of any of these conditions leads to an obvious arbitrage opportunity which is very easy to implement, hence, it is natural to assume that every implied volatility surface is free of static arbitrage.  This, in turn, implies that any realistic simulation algorithm for future implied volatility surfaces has to produce surfaces that are arbitrage-free: otherwise, the algorithm generates outcomes that are simply impossible. The static no-arbitrage conditions are rather difficult to state explicitly, in terms of the implied volatility surface itself (without mapping it to a call or put price surface first). Nevertheless, it is not hard to deduce from the existing necessary conditions (cf. \cite{lee2005implied}) that the set of arbitrage-free implied volatility surfaces forms a ``thin" set in the space of all (regular enough) functions of two variables. Hence, it is a non-trivial task to construct a modeling framework that excludes static arbitrage in the implied volatility surface. The dynamic arbitrage adds to this problem, and it refers to a restriction on the evolution (i.e. the time increments) of implied volatility surface, rather than its values at a fixed moment in time. This restriction follows from the same arbitrage considerations for option prices. However, its associated arbitrage strategies are not as straightforward as in the case of static arbitrage. In addition, the simulated implied volatility surfaces that contain only dynamic arbitrage are, typically, very close to the ones that are arbitrage-free, when the time horizon is small (it is related to the fact that dynamic arbitrage only changes the drift term of the implied volatility, which is much smaller than the diffusion term, for small times). This is why, eliminating the dynamic arbitrage in a model for implied volatility surface is often viewed as a ``second priority" for risk management. Nevertheless, we believe that a good model should exclude both types of arbitrage, in order to produce realistic dynamics of implied volatility surface (for risk management) and eliminate the possible arbitrage opportunities (for pricing).

We have already mentioned that it is not a trivial task to construct a model of implied volatility that excludes arbitrage opportunities. In fact, when trying to model the surface directly, the first challenge that one faces is: how to describe the space possible implied volatility surfaces? Note that, as discussed above, the existing characterizations of arbitrage-free implied volatility surfaces are rather implicit. In addition, if the resulting space is not an open subset of any linear space (which it is not), what kind of mathematical tools can be used to describe evolution in space? Recall, for example, that all statistical models of time-series are defined on linear spaces (or those that can be easily mapped in to a linear space). Hence, it appears natural to map the space of possible implied volatility surfaces to an open set in a linear space, and then proceed with the construction of arbitrage free models. Such mapping became known as a \emph{code-book} mapping, and it turns out that it can be constructed by means of the so-called \emph{tangent models} (cf. \cite{carmona2009localvol}, \cite{carmona2010tangentlevy}, \cite{carmona2011tangentlevy}). The concept of a tangent model is very close to the method of \emph{calibrating} a model for underlying to the target derivatives' prices prices (in the present case, European options calls). Consider a family of arbitrage-free models for the underlying, $\mathcal{M}(\theta)$, parameterized by $\theta$, taking values in a ``convenient" set $\Theta$ (an open set of a linear space). For any given surface of option prices (or, equivalently, any given implied volatility surface), we can try to calibrate a model for this family to a given surface of option prices (or, equivalently, to a given implied volatility surface). In, other words, we attempt to find $\theta\in\Theta$ such that: $C^{\theta}(T,K) = C(T,K)$, for all given maturities $T$ and strikes $K$, where $C(T,K)$ is the given call price, and $C^{\theta}(T,K)$ is the call price produced by the model $\mathcal{M}(\theta)$. If the above calibration problem has a unique solution, we obtain a one-to-one correspondence between the call price surfaces and the models in a chosen family: $\theta \leftrightarrow C^{\theta}$. For every call price surface $C=C^{\theta}$, the associated (calibrated) model $\mathcal{M}(\theta)$ is called a tangent model.\footnote{It is important to remember that any such model serves \emph{only as a static description} of option prices, and it does not describe their dynamics!} Notice that $C^{\theta}$ is always arbitrage-free, hence, we obtain the desired code-book mapping $C=C^{\theta}\mapsto \theta$. Now, the problem of static arbitrage has been resolved, and one simply needs to prescribe the distribution of a stochastic process $\left(\theta_t\right)$, taking values in a convenient set $\Theta$, in order to obtain a model for the dynamics of call prices $\left(C_t=C^{\theta_t}\right)$, and, in turn, the dynamics of implied volatility surface. Finally, one needs to characterize all possible dynamics of $\left(\theta_t\right)$ that produce no dynamic arbitrage in the associated call prices $\left(C^{\theta_t}\right)$. An interested reader is referred to \cite{carmona2011tangentlevy}, for a more detailed description of this general algorithm, and, for example, to \cite{carmona2009localvol}, \cite{carmona2010tangentlevy}, \cite{kallsen2010HJM}, \cite{Zhao2010dtl}, \cite{emmanuel2014discrete}, \cite{richter2014discretetime}, for the analysis of specific choices of the families of models $\left\{\mathcal{M}(\theta)\right\}$.

The idea of modeling prices of derivative contracts directly dates back to the work of Heath, Jarrow and Morton \cite{heath1992bond}, who analyzed the dynamic of bond prices along with the short interest rate. Such models have become known as the \emph{market-based models} (or simply \emph{market models}), as opposed to the classical \emph{spot models}, since the former are designed to capture the evolution of the entire market, including the liquid derivatives. This approach has been extended to more general mathematical settings, as well as to other derivatives' markets. The list of relevant works includes \cite{derman1998tree}, \cite{schonbucher1999market}, \cite{Schonbucher.credit}, \cite{schweizer2008term}, \cite{schweizer2008arbitrage}, \cite{damir2012bridge}, in addition to those mentioned in the previous paragraph. Even though the notions of code-book and tangent models never appear in these papers, almost all of them follow the algorithm outlined in the previous paragraph (and described in more detail in \cite{carmona2011tangentlevy}), in order to construct a market-based model.

Even though various code-books for implied volatility surface (or, equivalently, for call price surface) have been proposed and the corresponding arbitrage-free dynamics have been characterized, it was not until very recently that some of these models were implemented numerically. As is shown in the rest of the paper, the lack of such results is not a surprise given the complexity of the models. So far, the numerical implementations are mostly based on \emph{tangent L\'{e}vy models} proposed in \cite{carmona2010tangentlevy} and \cite{carmona2011tangentlevy}: as the name suggests, this corresponds to a code-book which is constructed using non-homogeneous L\'evy (or, additive) models as the tangent models. 
Karlsson \cite{karlsson2011consistent} implements a class of tangent L\'{e}vy models with absolutely continuous L\'evy densities and no continuous martingale component. Zhao \cite{Zhao2010dtl} and Leclercq \cite{emmanuel2014discrete}, on contrary, implemented the tangent L\'{e}vy models whose L\'{e}vy measure is purely atomic in the space variable. As opposed to \cite{Zhao2010dtl}, the work of Leclercq \cite{emmanuel2014discrete} allows for tangent models with continuous martingale component and includes options with multiple maturities, but it does require that the L\'evy density possess certain symmetry, which may limit the ability of the model to capture the skew of the implied smile. All of the works \cite{karlsson2011consistent}, \cite{Zhao2010dtl}, \cite{emmanuel2014discrete} estimate the parameters of the model from real market data. In addition, \cite{emmanuel2014discrete} conducts a numerical experiment comparing the performance of a market-based model to a classical spot model. The actual results of this experiment, however, do not provide a convincing evidence in favor of the market-based approach. We believe that the latter is simply due the choice of experiment and to the deficiency of the theory, and we intend to demonstrate it in the present work.

The purpose of this paper is to propose implementation methods for two classes of tangent L\'{e}vy models -- with continuous and discrete L\'evy measures. These methods provide practical algorithms for simulating future arbitrage-free implied volatility surfaces, which are consistent with both present and past observations. Our first method is similar to the one used in \cite{karlsson2011consistent}, but with a different ``dynamic fitting" part, and the second method is in the spirit of \cite{emmanuel2014discrete}, although we avoid the assumption of symmetry of a L\'evy measure made in \cite{emmanuel2014discrete}. However, the most important original contribution of this paper is the \emph{numerical experiment which uses real market data to demonstrate clearly the advantages of market-based models for implied volatility (or, option prices)}, as compared to the classical spot models. To the best of our knowledge, this is the first convincing empirical analysis that justifies the use of market-based approach for modeling implied volatility surface.

The rest of the paper is organized as follows. Section 2 starts by reviewing the work on tangent L\'{e}vy models with continuous L\'evy density and continuous martingale component, developed in \cite{carmona2011tangentlevy}. Then, we introduce the implementation approach for this models, which is based on double exponential jump processes, hence the name ``Double Exponential Tangent L\'{e}vy Models". Section 3 introduces the implementation method for tangent L\'{e}vy models with discrete L\'evy density, called ``Discrete Tangent L\'{e}vy Models". The two approaches are then tested against a popular classical model in a portfolio optimization problem in Section 4. Section 5 concludes the paper by highlighting the main contributions and the future work. Appendices A--C contain technical proofs and derivations, Appendix D contains all tables and graphs.

\section{Double exponential tangent L\'{e}vy models}\label{sec:DETLM}

\subsection{Model setup and consistency conditions}\label{sec:nosym}

In this subsection, we review and update the results of \cite{carmona2010tangentlevy}, which serve as a foundation for the analysis in subsequent sections. 
Herein, we assume that the interest and dividend rates for the underlying asset are zero. In the implementation that follows, we discount the market data accordingly, to comply with this assumption. As in \cite{carmona2011tangentlevy}, we denote by $(S_t)_{t\ge 0}$ a stochastic process representing the underlying price, and assume that the true dynamics of $S$ under the pricing measure $\QQ$ are given by:
\begin{equation}
\label{fo:nosym:Sdynamics}
S_t = S_0 + \int_0^t \int_{\RR} S_{u-}(e^x-1)[M(dx,du)-K_u(x)dxdu].
\end{equation}
Here, $M$ is a general integer-valued random measure (not necessarily a Poisson measure!), whose compensator is $K_{u,\omega}(x)dxdu$, where $(K_{t})_{t\ge 0}$ is a predictable stochastic process taking values in the function space $\mathcal{B}_0$, defined in \eqref{fo:review:B0}.

For any fixed time $t\ge 0$ and a given value of $S_t$, a stochastic process $(\tilde{S}_{T})_{T\ge t}$ is said to be \emph{tangent} to the true model $(S_t)_{t\geq 0}$ if the time-$t$ prices of all European call options written on $S$ can be obtained by pretending the future risk-neutral evolution of the index value is instead given by $(\tilde{S}_T)_{T\ge t}$ from $t$ on. Throughout this section, for any fixed $t\geq0$, we assume that the tangent processes $\tilde{S}$ is in the form
\begin{equation}
\label{fo:nosym:jumpunderlying}
\tilde{S}_T = S_t + \int_t^T \int_{\RR} \tilde{S}_{u-}(e^x-1)\left[N_t(dx,du)-\kappa_t(u,x)dx du\right],
\end{equation}
for $T\in[0,\bar{T}]$, where $N_t(dx,du)$ is a Poisson random measure associated with the jumps of $\log \tilde{S}$ whose compensator is given by a deterministic measure $\kappa_t(u,x)dxdu$.
Notice that the law of $\tilde{S}$ is uniquely determined by $(S_t,\kappa_t)$. Let $C_t^{S_t,\kappa_t}(T,x)$ denote the option prices generated by $(\tilde{S}_u)_{u \geq t}$, i.e
\begin{equation}\label{fo:nosym:Cdef}
C_t^{S_t,\kappa_t}(T,x) := \EE \left[(\tilde{S}_T - e^x)^+ | \tilde{S}_t = S_t \right],\qquad \forall T\geq t,\,\, x\in\RR.
\end{equation}
The concept of a \emph{tangent model}, then, requires that, for each fixed $t \in [0, \bar{T})$,
\begin{equation}
C_t^{S_t,\kappa_t}(T,x) = \EE \left[(S_T-e^x)^+ |\mathcal {F}_t \right], \qquad \forall T\geq t,\,\,\forall x\in\RR. \label{fo:nosym:C}
\end{equation}
Thus, at each time $t$, we obtain the \emph{code-book} for call prices, given by $(S_t, \kappa_t)$.
Of course, the value of the code-book may be different at a different time $t$. Hence, we consider the dynamic tangent L\'{e}vy models characterized by a pair of stochastic processes $(S_t, \kappa_t)_{t\in[0,\bar{T}]}$ that satisfies \eqref{fo:nosym:C}. Here, $S$ is a positive martingale with dynamics given by \eqref{fo:nosym:Sdynamics}; $\kappa$ is progressively measurable positive stochastic process taking values in $\mathcal{B}$ (cf. \eqref{fo:review:B}). The dynamics of $S_t$ and $\kappa_t$ are given by
\begin{equation}\label{fo:nosym:dyn}
\left\{
\begin{array}{c}
{ S_t = S_0 + \int_0^t \int_{\RR}S_{u-}(e^x-1)[M(dx,du)-K_u(x)dxdu],}\\
{}\\
{ \kappa_t(T,x) = \kappa_0(T,x) + \int_0^{t}\alpha_u(T,x) du + \sum_{n=1}^m \int_0^{t} \beta^{n}_u(T,x) dB^n_u,}\\
\end{array}
\right.
\end{equation}
where
$(\alpha_t)_{t\in[0,\bar{T}]}$ is a progressively measurable integrable stochastic process with values in $\mathcal{B}$, and, for each $n\in\{1,\cdots,m\}$, $(\beta^n_t)_{t\in [0,\bar{T}]}$ is a progressively measurable square integrable stochastic process taking values in $\mathcal{H}$ (cf. \eqref{fo:review:H}).

Notice that \eqref{fo:nosym:dyn} defines the dynamics of the code-book $(S_t,\kappa_t)_{t\in[0,\bar{T}]}$, but it does not ensure that it does, indeed, produce tangent models at each time $t$: in other words, there is no guarantee that \eqref{fo:nosym:C} holds. Thus, additional ``consistency" conditions have to be enforced to obtain models which are, indeed, tangent to the true underlying process. As shown in \cite{carmona2010tangentlevy}, this consistency is, in fact, equivalent to the fact that call prices generated by these tangent models are free of dynamic arbitrage. In order to present the main consistency result, we state the following regularity assumptions on $\beta$.

\begin{ass}
For each $n\leq m$,
almost surely, for almost every $t\in[0,\bT]$, we have:

\begin{description}

\item[RA1] $\label{eq.betaAltAssumption} \sup_{T\in[t,\bT]}\int_{-1}^1 \left|\beta^n_t(T,x)\right|dx <\infty$

\item[RA2]  For every $T\in[t,\bT]$, the function $\beta^n_t(T,\,\cdot\,)$ is absolutely
continuous on $\RR\setminus\left\{0\right\}$.

\item[RA3] For any $T\in[t,\bT]$, $\int_{\RR}\left(e^x-1\right)\beta^n_t(T,x)=0$.

\end{description}

\end{ass}

Finally, we introduce some extra notation and formulate the consistency result, which is a simple corollary of Theorem 12 in \cite{carmona2010tangentlevy}.
\begin{equation}\label{eq.Psi.betabar}
\bbeta^n_t(T,x) :=\int_{t\wedge T}^T \beta_t^n(u,x) du.
\end{equation}

\begin{theorem}\label{theo:nosym:conditions}
(Carmona-Nadtochiy 2012) Assume that $\left(S_t\right)_{t\in[0,\bT]}$ is a true martingale, $\beta$ satisfies the above regularity assumptions RA1-RA4, and $\kappa_t(T,x)\geq0$, almost surely for all $t\in[0,\bT)$ and almost all $(T,x)\in[t,\bT]\times\RR$.
Then the processes $\left(S_t,\kappa_t\right)_{t\in[0,\bT]}$ satisfying (\ref{fo:nosym:dyn}) are \emph{consistent}, in the sense that \eqref{fo:nosym:C} holds, if and only if the following conditions hold almost surely for almost every
$x\in\RR$ and $t\in[0,\bT)$, and all $T\in(t,\bT]$:

\begin{enumerate}
\item \textbf{Drift restriction:}
\begin{align} \label{fo:implementation:driftcondition}
\alpha_t(T,x) = & - \sum_{n=1}^m \bigg\{\int_\RR \bbeta^n_t(T,y) \beta^n_t(T,x-y) dy \nonumber \\
& -  \bbeta^n_t(T,x) \cdot \int_\RR \beta^n_t(T,z)dz  - \beta^n_t(T,x) \cdot \int_\RR \bbeta^n_t(T,z)dz\bigg\}. 
\end{align}

\item \textbf{Compensator specification:} $K_t(x) = \kappa_t(t,x)$.
\end{enumerate}
\end{theorem}

Theorem~\ref{theo:nosym:conditions}, along with equations \eqref{fo:nosym:dyn} provide a general method for constructing a market-based model for call prices (i.e. an arbitrage-free dyanimc model for implied volatility surface). Indeed, choosing $(\beta^1_t\ldots,\beta^m)_{t\in[0,\bar{T}]}$, we use the drift restriction in Theorem~\ref{theo:nosym:conditions} and the second equation in \eqref{fo:nosym:dyn} to generate the paths of $(\kappa_t)_{t\in[0,\bar{T}]}$. Finally, to generate the paths of $(S_t)_{t\in[0,\bar{T}]}$, one can use the compensator specification in Theorem~\ref{theo:nosym:conditions} and the first equation in \eqref{fo:nosym:dyn}, after representing the random measure $M$ through its compensator $K$ and a Poisson random measure $N$ (as shown in \cite{carmona2010tangentlevy}). However, in the present paper we avoid simulating $(S_t)_{t\in[0,\bar{T}]}$ at all, by simply noticing that
$$
\frac{1}{S_t}C_t^{S_t,\kappa_t}(T,x+\log S_t) = \EE \left[(\tilde{S}_T/S_t - e^x)^+ | \tilde{S}_t = S_t \right]
= \EE \left[(\tilde{S}_T - e^x)^+ | \tilde{S}_t = 1 \right]
= C_t^{1,\kappa_t}(T,x),
$$
$$
\frac{1}{S_t}C_t^{S_t,bs}(T,x+\log S_t;\sigma) = C_t^{1,bs}(T,x;\sigma),
$$
where $C_t^{S_t,bs}(T,x)$ is the Black-Scholes price at time $t$ of a call option with maturity $T$ and strike $e^x$ given that the level of underlying is at $S_t$ and the volatility is $\sigma$. At any time $t$, regardless of the value of $S_t$, if we find the level of $\sigma$ that makes the right hand sides of the two equations above coincide, then the option prices in the left hand sides have to coincide as well. This means that we can obtain the implied volatility of $C_t^{S_t,\kappa_t}$, in the maturity and log-moneyness variables, by computing the corresponding implied volatility of $C_t^{1,\kappa_t}$, for which we do not need to generate $S_t$.

\subsection{Implied volatility simulation with tangent L\'{e}vy models\label{sec:implementation}}

We first introduce the general framework of the simulation procedure. Our procedure has two stages, \emph{estimation} and \emph{simulation}. The estimation stage, where the additive density of the tangent process as well as its dynamics are fitted to market data, is performed in two steps: 
\begin{itemize}
\item \emph{Static fitting}. In static fitting, the additive density $\kappa_t$ for each day $t$ is obtained by least squares optimization which minimizes the squared difference between model prices and actual market prices. Notice that for any given day $t$, $\kappa_t$ is fixed and there is no dynamics involved, which explains the term `static'.

\item \emph{Dynamic fitting}. In dynamic fitting, we recover the dynamics of the time series $(\kappa_t)$. In view of the drift restriction in Theorem~\ref{theo:nosym:conditions}, this boils down to determining the volatility terms $\{\beta^n\}_{n = 1}^m$. This is done by applying the Principle Components Analysis to the time series of $(\kappa_t)_t$.
\end{itemize}
Once the estimation is completed, we generate the future paths of $(\kappa_t)$ using Euler scheme Monte Carlo applied to the second equation in \eqref{fo:nosym:dyn}. From the simulated additive densities, we compute call prices $C_t^{1,\kappa_t}$ and, then, implied volatilities by inverting the Black-Scholes formula. 

Within the general framework, the simulation stage is generic, but the static part of the estimation stage can be quite different depending on the specific subclass of tangent L\'{e}vy densities $\kappa(u,x)$ that we fit to option price at any given time. In this section, we implement the procedure with the L\'evy densities arising from the double exponential L\'{e}vy models proposed by Kou in \cite{Kou2002model}. The small number of parameters in double exponential models and the availability of an analytical pricing formula for call options make the resulting family of tangent L\'{e}vy models fairly easy to calibrate.

\subsection{Market data} \label{sec:implementation:data}

We use \emph{SPX} (S\&P 500) call option prices provided by OptionMetrics, an option database containing historical prices of options and their underlying instruments. Throughout the paper, we use the option data from two time periods: Jan. 2007 - Aug. 2008 and Jan. 2011 - Dec. 2012. Table~\ref{tb:implementation:SPX} gives a quick summary of the two periods.  We cut off the first period at August 2008 to reduce the impact of the financial crisis. 

On each day of a period, we only keep the options with time to maturity less than one year, whose best closing bid price and best closing offer price are both available, and take the average of the two prices as the option price. To ensure the validity of all prices, the contracts with zero open interest are excluded. As a result, there are roughly 10 to 80 call contracts with valid prices available for each maturity. The log-moneyness (more precisely, the put log-moneyess, defined as $\log(K/S_t)$) of these call options ranges roughly from -0.3 to 0.1, varying for different $t$ and $T$.  Our calibration also requires dividend and interest rate data available on OptionMetrics and the homepage of U.S. Department of Treasury, respectively. This dividend yield is recovered from option prices via put-call parity with the method proposed in \cite{AitSahalia1997nonparametricestimation}. On day $t$, we denote the dividend yield by $q_t$, and the risk-free rate between $t$ and $T$ by $r_{t,T}$.
To simplify our implementation, we perform a simple transformation on the market data so that we can assume that the interest and dividend rates are both zero from now on:
\begin{align} \label{fo:dtl:transform}
C^{mkt}_t(T,x) & = e^{q_t(T-t)} \bar{C}_t^{mkt}(T,\bar{x}), \quad \text{with} \quad x = \bar{x} - (r_{t,T}-q_t)(T-t),
\end{align}
where $\bar{C}_t^{mkt}(T,\bar{x})$ is the market price of a call option with maturity $T$ and strike $e^{\bar{x}}$.
The adjusted call prices $C^{mkt}_t(T,x)$, corresponding to maturity $T$ and strike $e^x$, are then consistent with the assumption of zero interest and dividend rates (i.e. they do not contain arbitrage under thee assumptions). IN a similar way, we define the adjusted bid and ask prices, $C^{mkt,b}_t$ and $C^{mkt,a}_t$.

In this Section and Section~\ref{sec:dtL}, we will perform the calibration of the tangent L\'{e}vy models on the time span from Jan. 3, 2007 to Dec. 31, 2007, denoted by $[t_0,\bT]$. In Section~\ref{sec:comparison}, data from both periods will be used to test the performance of the tangent L\'{e}vy models.

\subsection{Static fitting}
\label{sec:implementation:static}

Before we proceed with the static fitting, let us first have a quick review of the double exponential model. In such a model, the logarithm of underlying follows a pure jump L\'evy process whose jump sizes have a double exponential distribution. More specifically, assuming no diffusion term, the dynamics of the underlying are given by
\begin{equation} \label{fo:implementation:kou}
d\hat{S}_t = \mu\, \hat{S}_{t-} dt + \hat{S}_{t-}\,  d \left(\sum_{i=1}^{N_t}(\exp(Y_i) - 1)\right),
\end{equation}
where $\mu$ is the drift term, $N_t$ is a Poisson process with rate $\lambda$, $\{Y_i\}$ is a sequence of i.i.d. random variables with asymmetric double exponential distribution, independent of $N_t$. The density of an asymmetric double exponential distribution is given by
\begin{equation}
f_Y(y) = p \cdot \lambda_1 e^{-\lambda_1 y} \bone_{y\geq 0} + q \cdot \lambda_2 e^{\lambda_2 y} \bone_{y < 0},
\end{equation}
where $p,q \geq 0, \ p+q = 1$ represent the probabilities of positive and negative jumps, and $\lambda_1 > 1, \lambda_2 > 0$ are the parameters of the two exponential distributions. In other words, a double exponential model is a martingale model for the underlying whose logarithm is a pure jump L\'evy process, with the L\'{e}vy density
\begin{equation}\label{fo:implementation:kou_eta}
\eta(x) = \lambda ( p \cdot \lambda_1 e^{-\lambda_1 x} \bone_{x\geq 0} + q \cdot \lambda_2 e^{\lambda_2 x} \bone_{x < 0} ).
\end{equation}

One of the advantages of double exponential models is the availability of analytical pricing formulas for European call options, which could greatly simplify the calibration. \cite{Kou2002model} gives the pricing formula for double exponential models with a diffusion term. A minor modification of the derivation in \cite{Kou2002model} gives us the pricing formula in absence of the diffusion term, as shown in the lemma below (its proof is given in Appendix B).

\begin{lemma}\label{cor:implementation:koupricing}
Under the assumptions of zero interest and dividend rates, assume, in addition, that the underlying process $S$ follows a double exponential process with L\'{e}vy density given by \eqref{fo:implementation:kou_eta}, under the risk-neutral probability measure. Then, the price of a European call option with strike $K$ and maturity $T$ is given by
\begin{align} \label{fo:implementation:Kouprice}
C_t^{\lambda, \lambda_1, \lambda_2, p}(T,\log K) & = S_t\Psi \left( - \lambda \zeta, \lambda^*, p^*, \lambda_1^*, \lambda_2^*; \log\left(K/S_t\right), T-t \right) \nonumber \\
& \qquad \qquad - K \Psi \left( - \lambda \zeta, \lambda, p, \lambda_1, \lambda_2; \log\left(K/S_t\right), T-t \right),
\end{align}
where
\begin{align*}
& p^* = \frac{p}{1+\zeta} \cdot \frac{\lambda_1}{\lambda_1 - 1}, \quad \lambda_1^* = \lambda_1 - 1, \quad \lambda_2^* = \lambda_2 + 1,\\
& \quad \lambda^* = \lambda(\zeta+1), \quad \zeta = \frac{p\lambda_1}{\lambda_1-1} + \frac{q\lambda_2}{\lambda_2 + 1} - 1,
\end{align*}
and the function $\Psi$ is given by:
\begin{align}\label{fo:implementation:psi}
& \qquad \Psi(\mu, \lambda, p, \lambda_1, \lambda_2; a, T) \nonumber \\
& = \pi_0 \bone_{a - \mu T \leq 0} + \sum_{n=1}^\infty \pi_n \sum_{k=1}^n P_{n,k} \left[ \sum_{i=0}^{k-1} \frac{(\lambda_1 (a-\mu T))^i}{i!} e^{-\lambda_1 (a-\mu T)} \bone_{a - \mu T \geq 0} + \bone_{a - \mu T<0} \right] \nonumber \\
& \qquad + \sum_{n=1}^\infty \pi_n \sum_{k=1}^n Q_{n,k} \left( 1 - \sum_{i=0}^{k-1} \frac{(-\lambda_2 (a - \mu T))^i}{i!} e^{\lambda_2 (a-\mu T)} \right) \bone_{a - \mu T < 0},
\end{align}
with
\begin{equation*}
\pi_n = \frac{e^{-\lambda T}(\lambda T)^n}{n!}
 \end{equation*}
and
\begin{align*}
P_{n,k} & = \sum_{i = k}^{n-1}
\binom{n-k-1}{i-k} \binom{n}{i} \left(\frac{\lambda_1}{\lambda_1+\lambda_2}\right)^{i-k} \left(\frac{\lambda_2}{\lambda_1+\lambda_2}\right)^{n-i} p^i q^{n-i}, \quad 1 \leq k \leq n-1,\\
Q_{n,k} & = \sum_{i = k}^{n-1}
\binom{n-k-1}{i-k} \binom{n}{i} \left(\frac{\lambda_1}{\lambda_1+\lambda_2}\right)^{n-i} \left(\frac{\lambda_2}{\lambda_1+\lambda_2}\right)^{i-k} p^{n-i} q^i, \quad 1 \leq k \leq n-1,\\
P_{n,n} & = p^n, \quad Q_{n,n} = q^n.
\end{align*}
\end{lemma}

For each $T_l$, with $l=1,\ldots,L$, we would like to find the set of parameters $\{\lambda, \lambda_1, \lambda_2, p\}$ that minimizes the difference between the market and the model prices. 
For practical reasons, we will work with time values instead of options prices. The market time value and the model time value are calculated as follows
\begin{align*}
V_t^{mkt, j}(T_l) & = C_t^{mkt}(T_l,e^{x_j}) - (S_t - e^{x^j})^+,\\
V_t^{\lambda, \lambda_1, \lambda_2, p, j}(T_l) & = C_t^{\lambda, \lambda_1, \lambda_2, p}(T_l,e^{x_j}) - (S_t - e^{x^j})^+. 
\end{align*}
There are two reasons for working with time values. Firstly, the time values go to zero for very large and very small log-moneyness, which allows us to truncate the $x$-space with negligible numerical errors. Secondly, time values and option prices are often of different magnitudes, especially for in the money options, with option prices much greater than time values, hence, working with time values is likely to result in smaller numerical errors. 
For fixed time $t$ and fixed maturity $T_l$, the optimization problem can be written as
\begin{align}\label{fo:implementation:opt_ori}
\min_{\lambda > 0, \lambda_1>0, \lambda_2>0, p \in(0,1)} \quad & \sum_{j=1}^N \omega_j |V_t^{\lambda, \lambda_1, \lambda_2, p, j}(T_l) - V_t^{mkt,j}(T_l)|^2,
\end{align}
where $\omega_j = \left|C_t^{bid}(T_l,e^{x_j}) - C_t^{ask}(T_l,e^{x_j})\right|^{-2}$ are the weights we put on different options to take into account the difference in liquidity (measured by bid-ask spread).
For every fixed maturity $T_l$, the solution of the above optimization problem, $(\lambda^{l}, \lambda^{l}_1, \lambda^{l}_2, p^l)$, yields the L\'evy density $\eta_t(T_l,x)$ via \eqref{fo:implementation:kou_eta}.
Then, we search for a function $\kappa_t(\cdot,\cdot)$, such that
\begin{equation}\label{eq.kappa.fit.eta}
\eta_t(T_l,x) = \frac{1}{T_l-t}\int_t^{T_l}\kappa_t(u,x)du,
\end{equation}
for every maturity $T_l$ and all $x\in\RR$. The resulting tangent model on day $t$ is defined as a martingale model for the underlying whose logarithm is a pure jump additive (non-homogeneous L\'evy) process, with the L\'evy density $\kappa_t(\cdot,\cdot)$. It is easy to see that the call prices produced by this model, for every maturity $T_l$ and strike $e^{x_j}$, coincide with the prices produced by the double exponential model, $C_t^{\lambda^l, \lambda^l_1, \lambda^l_2, p^l}(T_l,e^{x_j})$.
Thus, for a given $t$, the problem of static fitting is essentially a series of optimization problems \eqref{fo:implementation:opt_ori}, over all maturities $T_l$, along with the fitting problem \eqref{eq.kappa.fit.eta}.

At the first glance, the optimization in \eqref{fo:implementation:opt_ori} seems to have four parameters. However, the following constraints will reduce the number of parameters to two in our calibration:
\begin{itemize}
\item To improve the stability of small-jumps intensity over time, we would like the L\'evy density $\eta(T_l,x)$ to be continuous in $x$. The continuity at $x = 0$ requires
\begin{equation} \label{fo:implementation:etacont}
p \cdot \lambda_1 = (1-p) \cdot \lambda_2 \Leftrightarrow \lambda_2 = \frac{p}{1-p} \lambda_1.
\end{equation}

\item In view of the results in Section~\ref{sec:nosym}, we have to impose the symmetry condition RA3 on $\beta^n$'s. A simple application of It\^o's lemma shows that, for the symmetry condition RA3 to hold, it suffices to choose every $\kappa_t$, so that
$$
\int_\RR (e^x-1) \kappa_t(T,x) dx
$$
is a deterministic function of $T-t$, for all times $0\leq t<T\leq \bar{T}$. To achieve this, in view of \eqref{eq.kappa.fit.eta}, we need to choose every $\eta_t(T_l,\cdot)$ so that the symmetry index
\begin{equation} \label{fo:implementation:symmetry}
\Xi(T-t) := \int_\RR (e^x-1) \eta_t(T,x) dx = \lambda\left( \frac{p}{\lambda_1 - 1} - \frac{1-p}{\lambda_2 + 1} \right)
\end{equation}
is a deterministic function of $T-t$.
This yields:
\begin{equation}
p = \frac{-(1+\Xi(T-t)/\lambda)(\lambda_1 -1)}{\Xi(T-t)/\lambda(\lambda_1 -1)^2 - 2(\lambda_1-1) - 1},
\end{equation}
where $\Xi$ is a fixed (estimated a priori) function.
\end{itemize}

With the two constraints, our calibration takes only two variables: $\lambda$ and $\lambda_1$. The condition $p \in (0,1)$ transforms to the following condition on $\lambda_1$:
\begin{equation} \label{fo:implementation:intlambda_1}
\lambda_1 \in \left\{
\begin{array}{c}
{(1,\infty)}, \quad \text{if} \quad \Xi(T-t) \leq 0,\\
{\left(1, 1+\frac{1}{\Xi(T-t)}\right), \quad \text{if} \quad \Xi(T-t) > 0.}\\
\end{array}
\right.
\end{equation}
As a result, the optimization problem \eqref{fo:implementation:opt_ori} can be re-written as
\begin{align}\label{fo:implementation:opt}
\min_{\lambda > 0, \lambda_1 \in I_{\lambda_1}} \quad & \sum_{j=1}^N \omega_j |V_t^{\lambda, \lambda_1, j}(T_l) - V_t^{mkt,j}(T_l)|^2,
\end{align}
where $I_{\lambda_1}$ is the interval defined in \eqref{fo:implementation:intlambda_1}.
The symmetry index function $\Xi(\tau)$, for all $\tau \in \RR_+$, can be obtained on the first calibration day $t=0$, solving a three-variable optimization problem,
\begin{align}\label{fo:implementation:optinit}
\min_{\lambda > 0, \lambda_1 > 1, p \in(0,1)} \quad & \sum_{j=1}^N \omega_j |V_{0}^{\lambda, \lambda_1, p, j}(T_l) - V_{0}^{mkt,j}(T_l)|^2,
\end{align}
and setting
\begin{equation}\label{fo:implementation:symmetryinit} 
\Xi(T_l) = \lambda\left( \frac{p}{\lambda_1 - 1} - \frac{1-p}{\lambda_2 + 1} \right),
\end{equation}
for every maturity $T_l$, and, finally, interpolating linearly between every $T_{l-1}$ and $T_l$.
We summarize the calibration procedure for $\left\{\eta_t(T_l,\cdot) \right\}$ in the following algorithm:

\vspace{5mm}
\begin{algorithm}[H]
\label{algo:implementation:calibration}
\caption{Algorithm for calibrating $\left\{\eta_t(T_l,\cdot) \right\}$}
Preprocess the market data according to \eqref{fo:dtl:transform}\;
For $t=0$, run the three-variable optimization \eqref{fo:implementation:optinit}, without the symmetry condition, for all maturities, and compute $\Xi(\cdot)$ by \eqref{fo:implementation:symmetryinit} and linear interpolation\;
For the subsequent days $t \in (0,\bT]$, run the two-variable optimization \eqref{fo:implementation:opt}, with already estimated $\Xi$, to obtain the time series of L\'{e}vy densities $(\eta_t)_{t \in [0,\bT]}$.
\end{algorithm}
\vspace{5mm}

\begin{figure}[htp]
  \begin{center}
       \includegraphics[width=0.5\columnwidth]{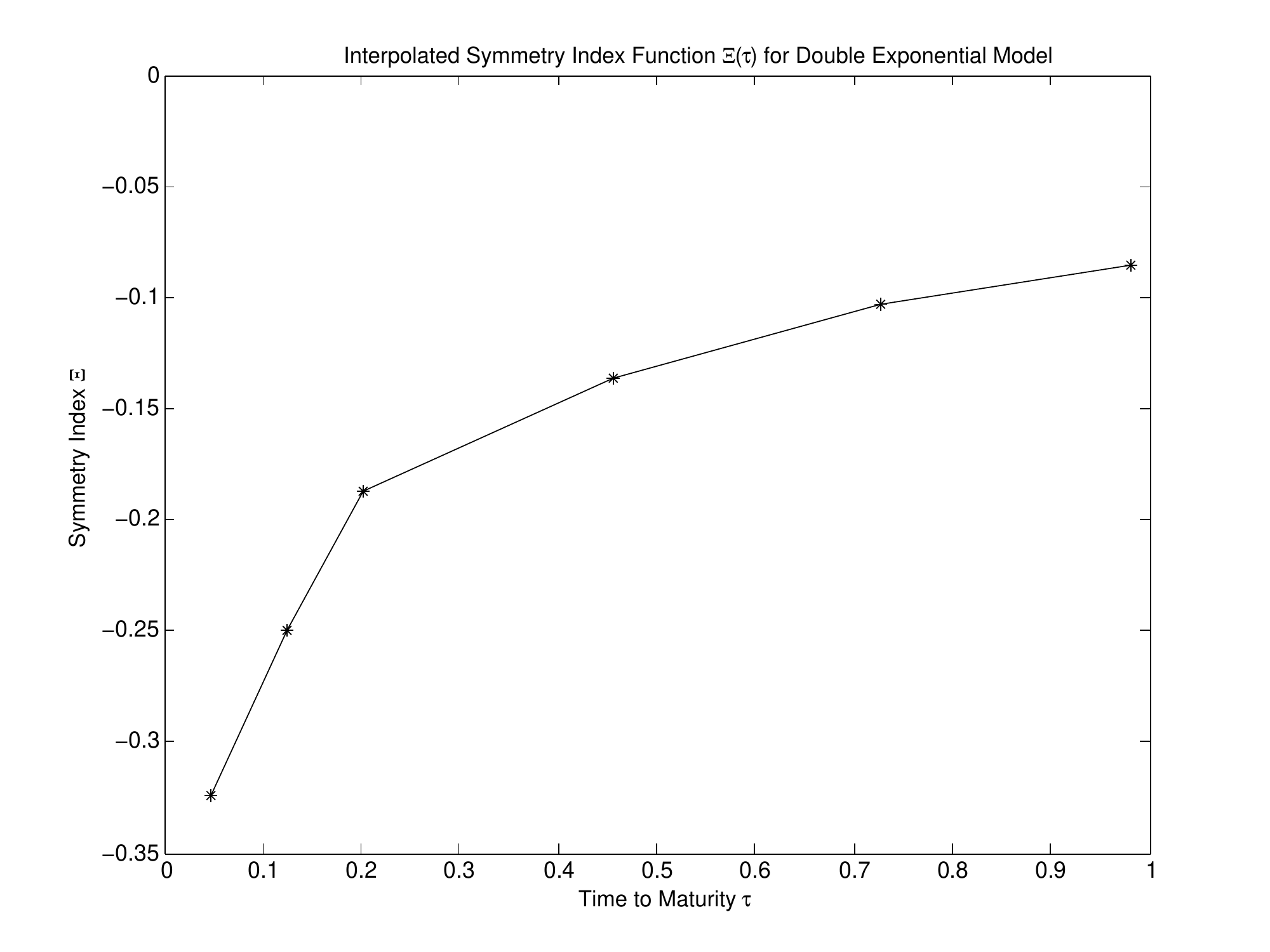}
   \end{center}
   \vspace{-10pt}
   \caption{Symmetry index $\Xi$ as a function of time to maturity with double exponential model}
   \label{fg:implementation:symmetry}
\end{figure}

\begin{figure}[htp]
   \begin{center}
       \begin{tabular}{cc}
       \includegraphics[width=0.45\columnwidth]{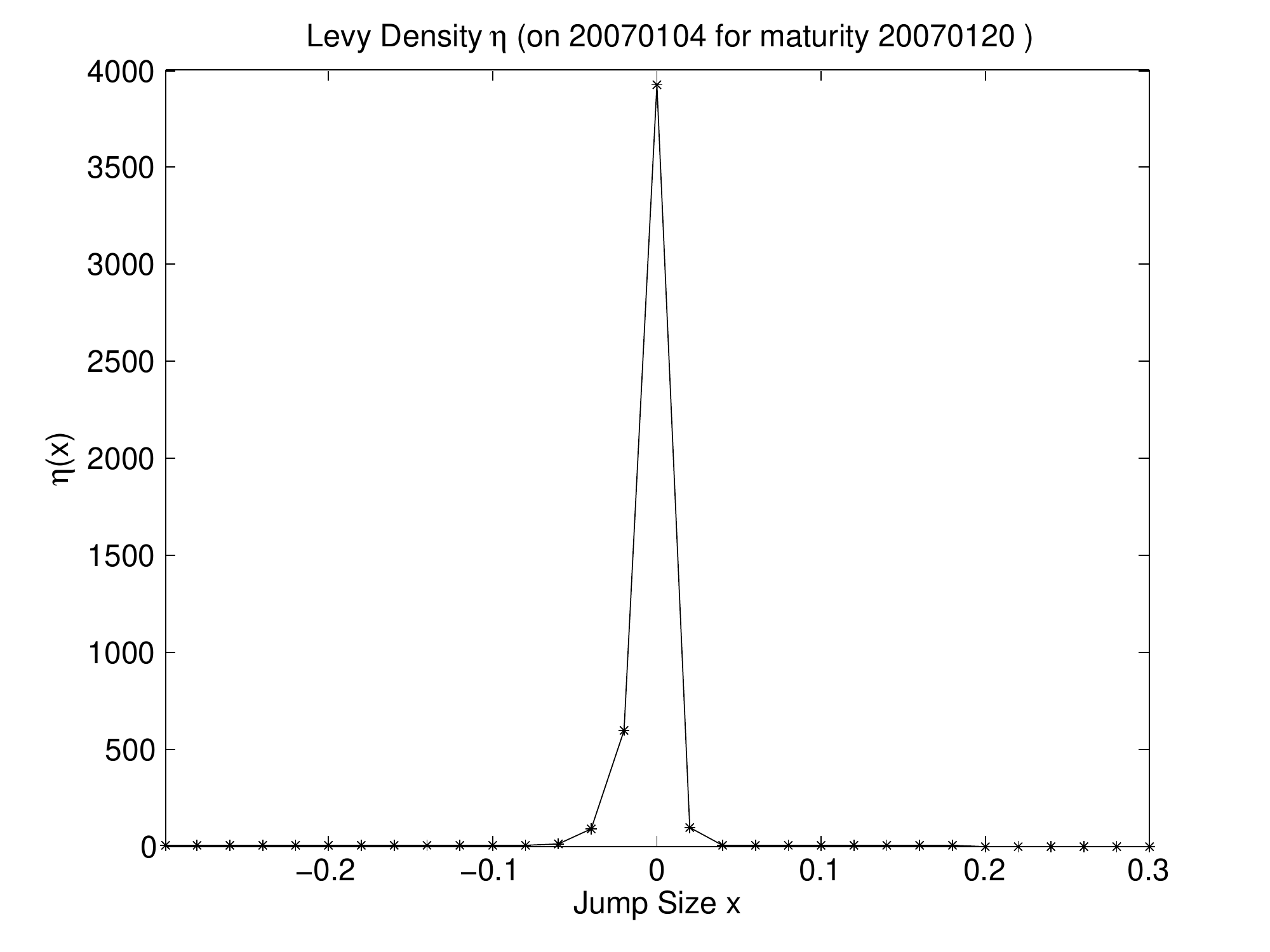}
       &
       \includegraphics[width=0.45\columnwidth]{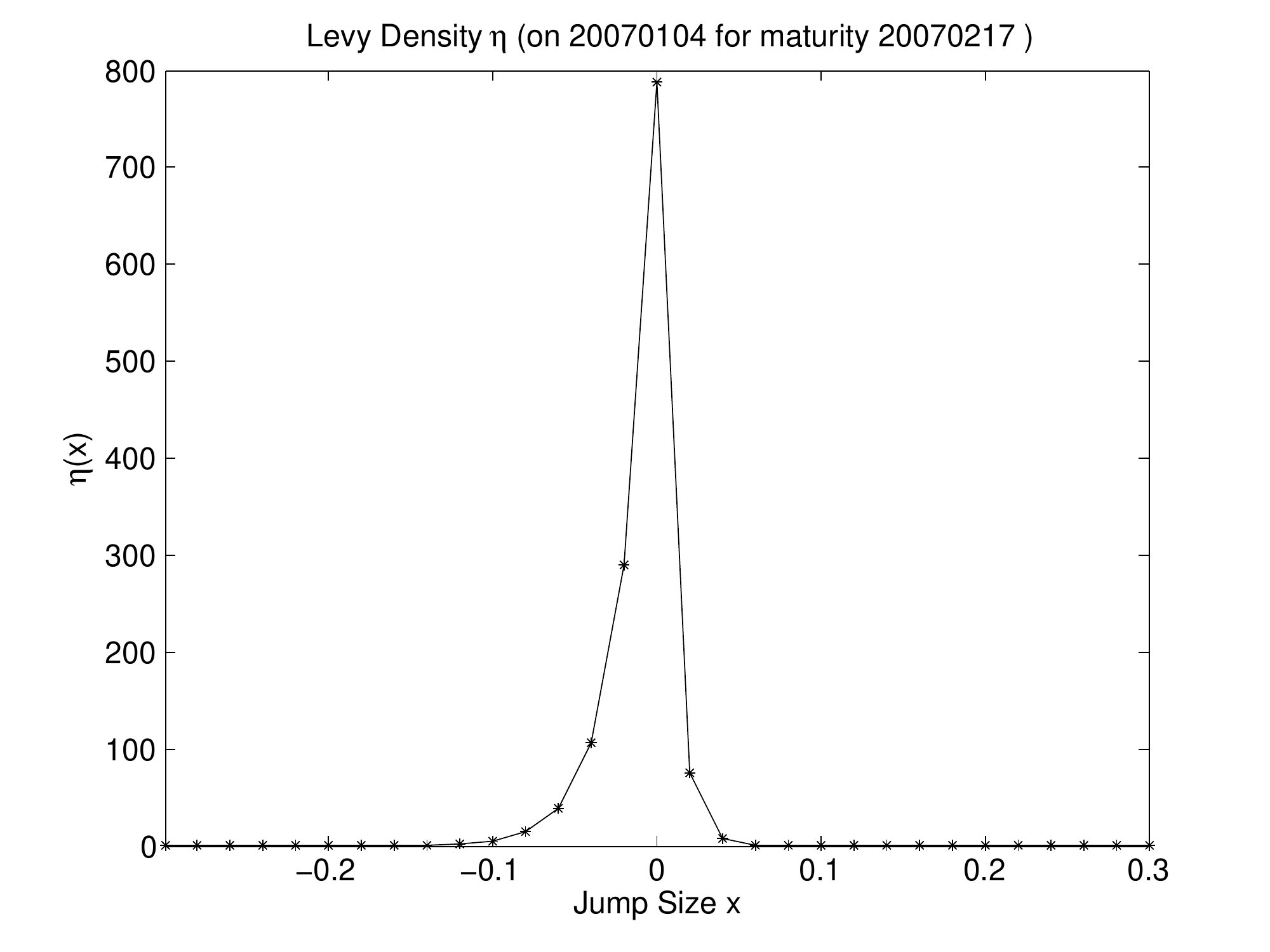}
       \\
      \fontsize{7}{12}\selectfont (a) Maturity Jan.20,2007
     &
      \fontsize{7}{12}\selectfont (b) Maturity Feb.17,2007
      \\
      \includegraphics[width=0.45\columnwidth]{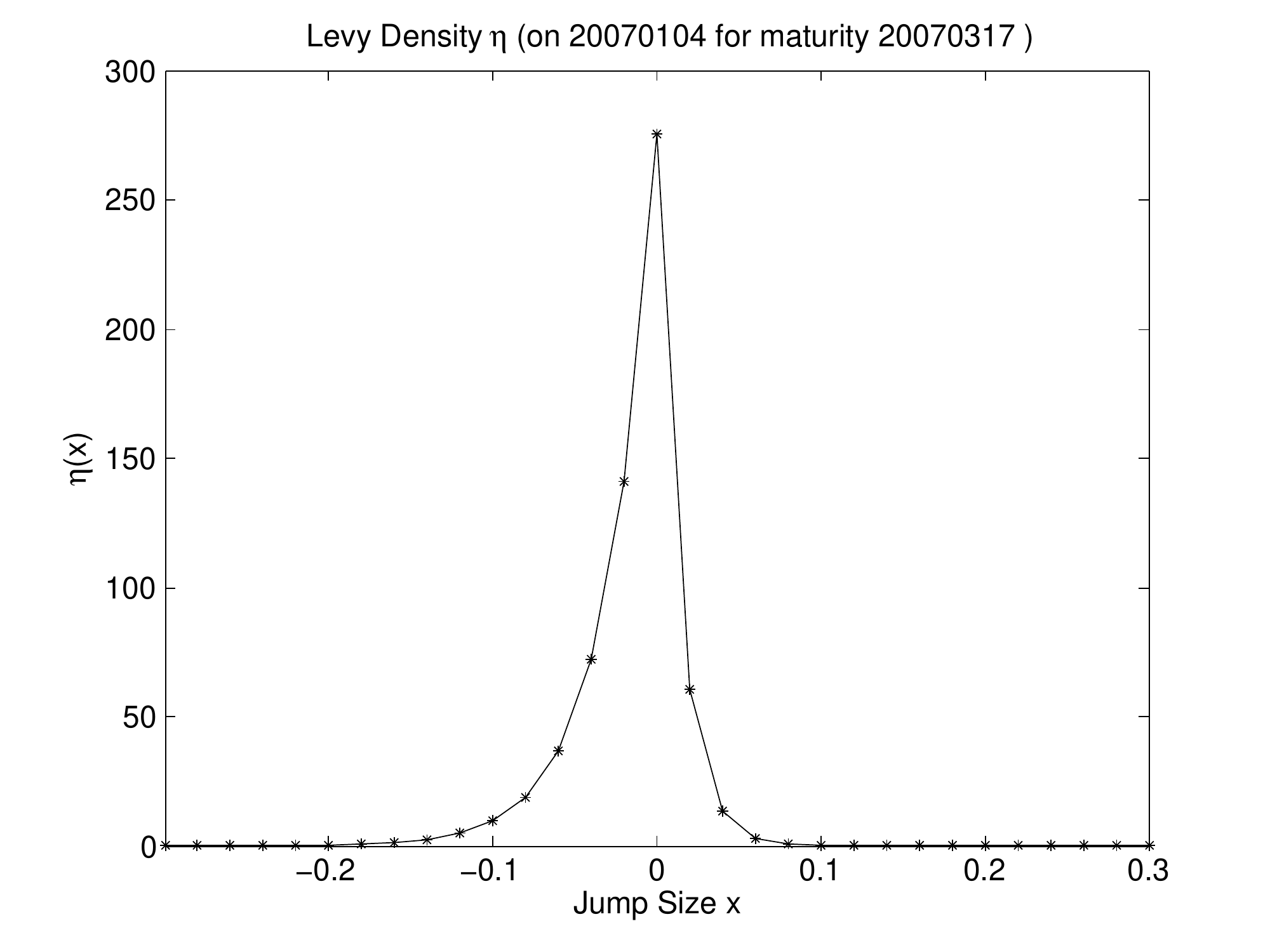}
       &
       \includegraphics[width=0.45\columnwidth]{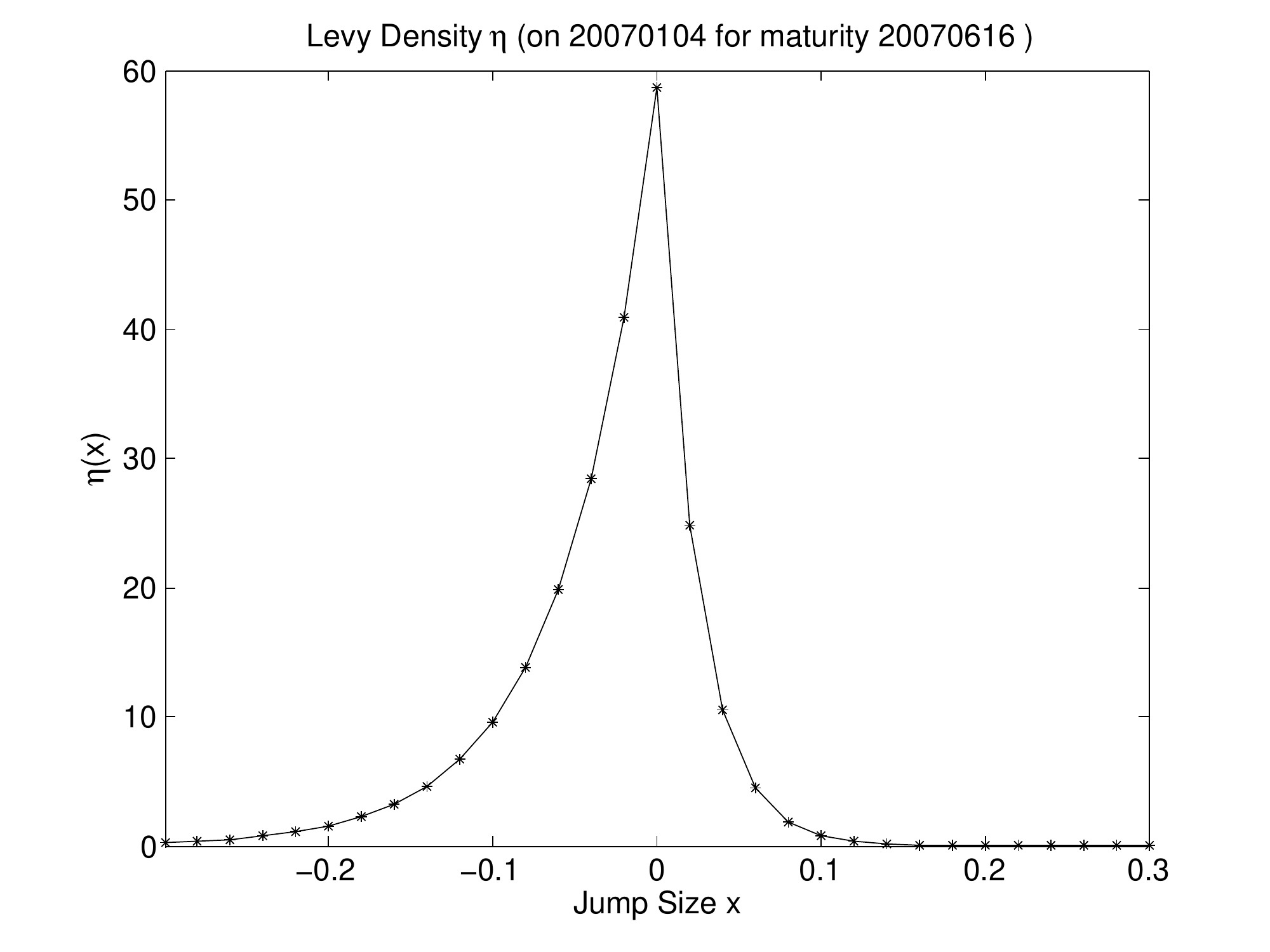}
       \\
      \fontsize{7}{12}\selectfont (c) Maturity Mar.17,2007
     &
      \fontsize{7}{12}\selectfont (d) Maturity Jun.16,2007
      \\
      \includegraphics[width=0.45\columnwidth]{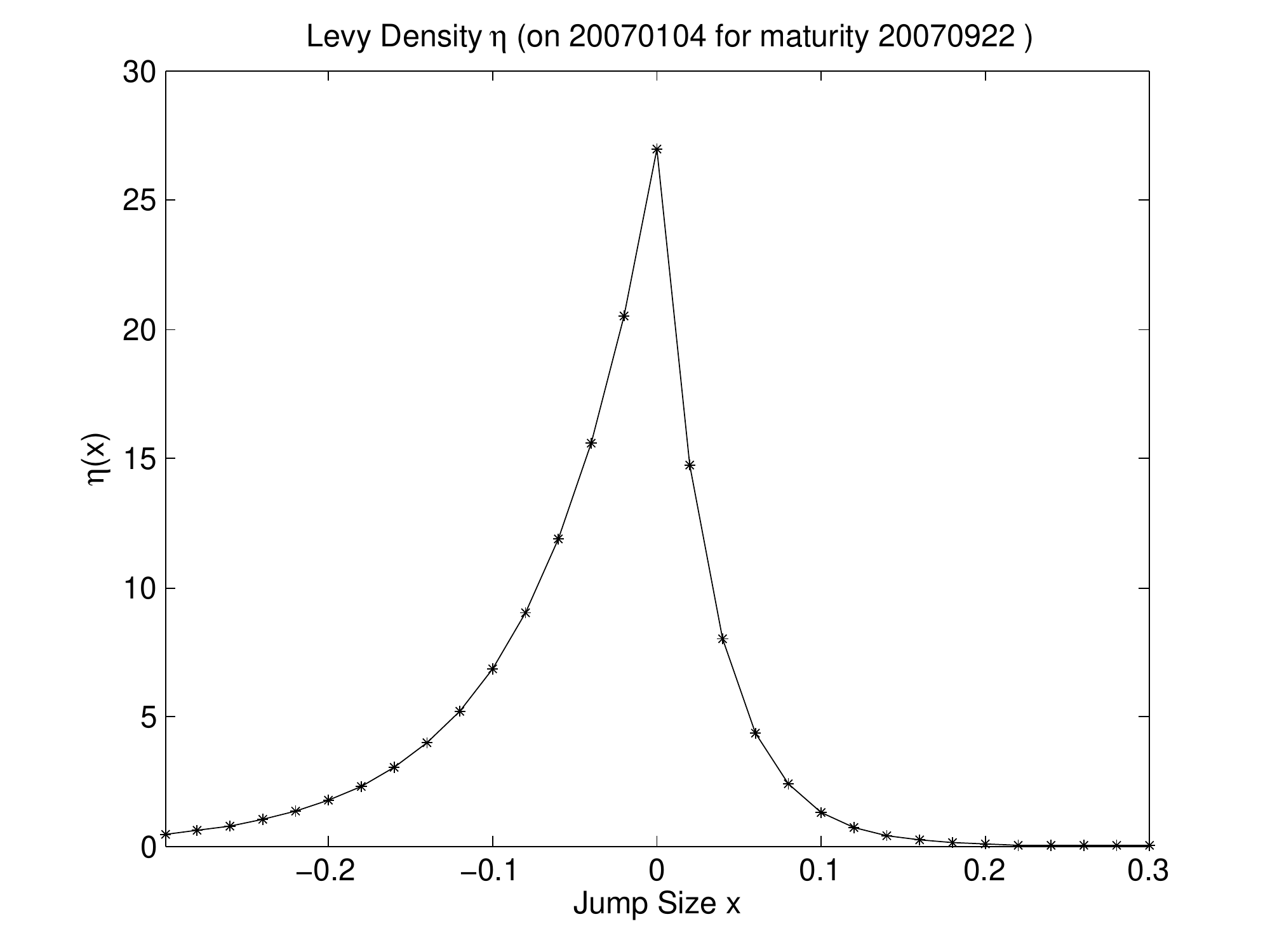}
       &
       \includegraphics[width=0.45\columnwidth]{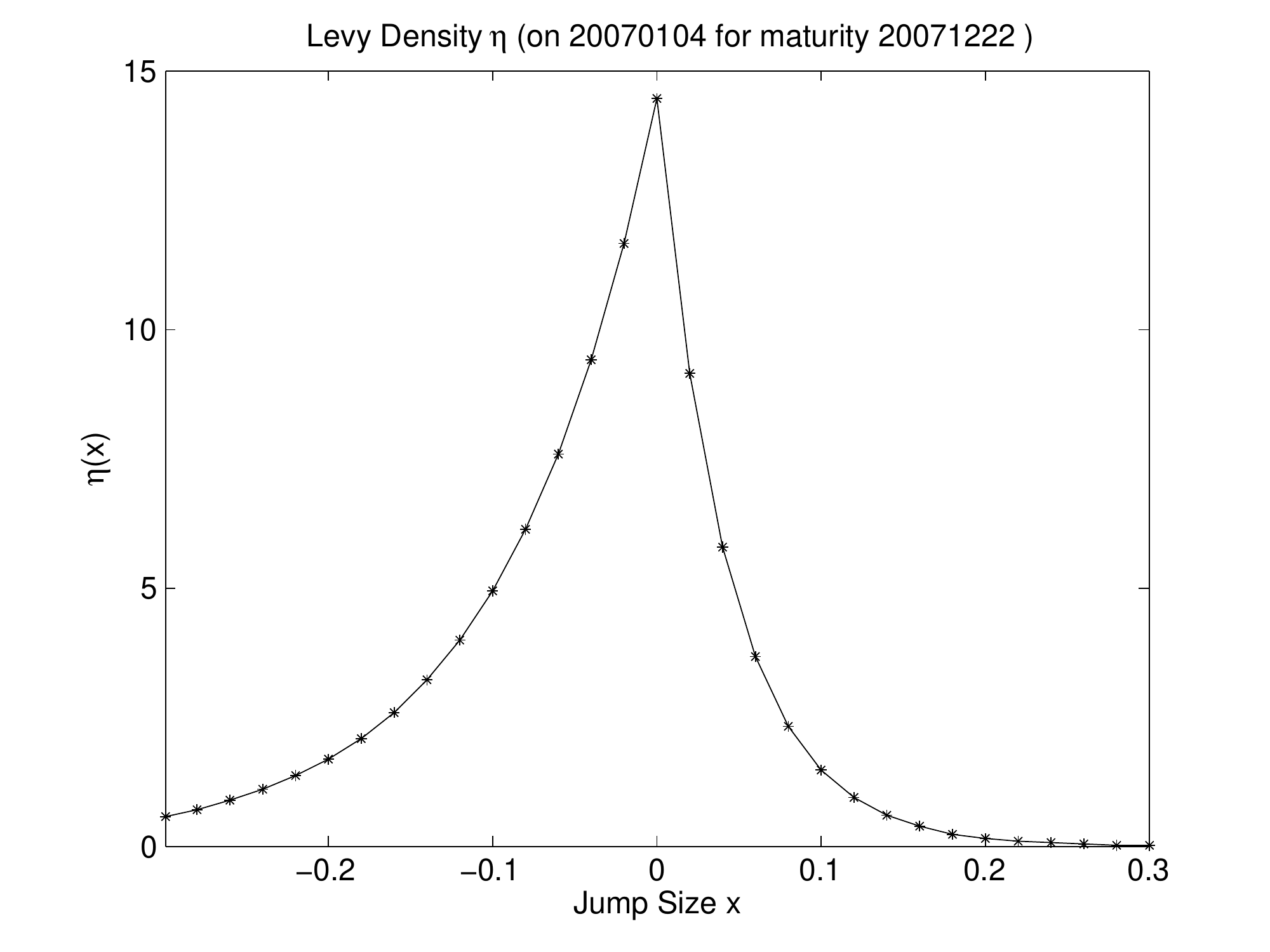}
       \\
      \fontsize{7}{12}\selectfont (e) Maturity Sep.22,2007
     &
      \fontsize{7}{12}\selectfont (f) Maturity Dec.22,2007
\\
       \end{tabular}
   \end{center}
   \vspace{-10pt}
   \caption{Calibrated densities $\eta$ for DETL model on the second day, Jan. 4, 2007}
   \label{fg:implementation:eta_day2}
\end{figure}

\begin{figure}[htp]
   \begin{center}
       \begin{tabular}{cc}
       \includegraphics[width=0.45\columnwidth]{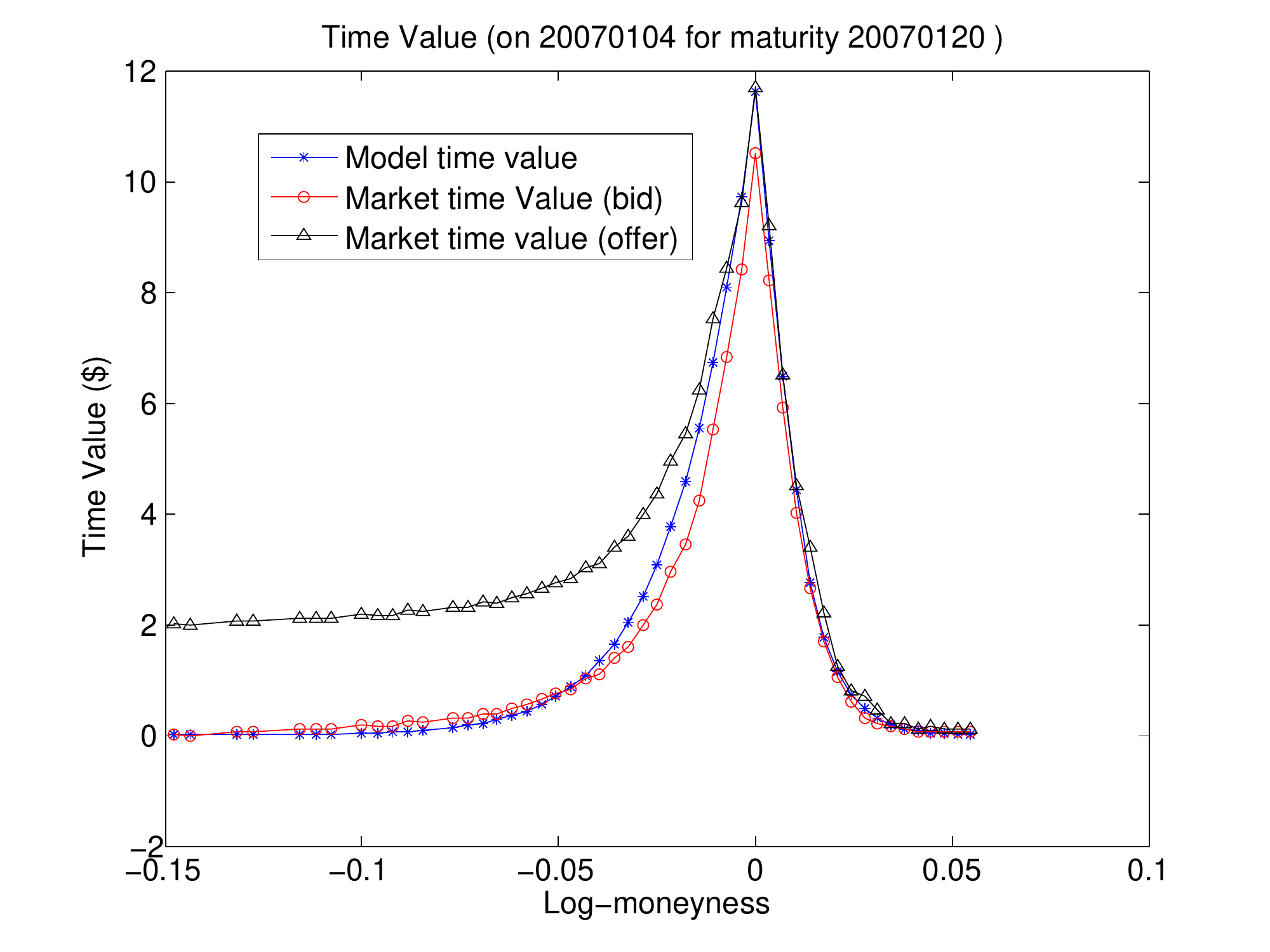}
       &
       \includegraphics[width=0.45\columnwidth]{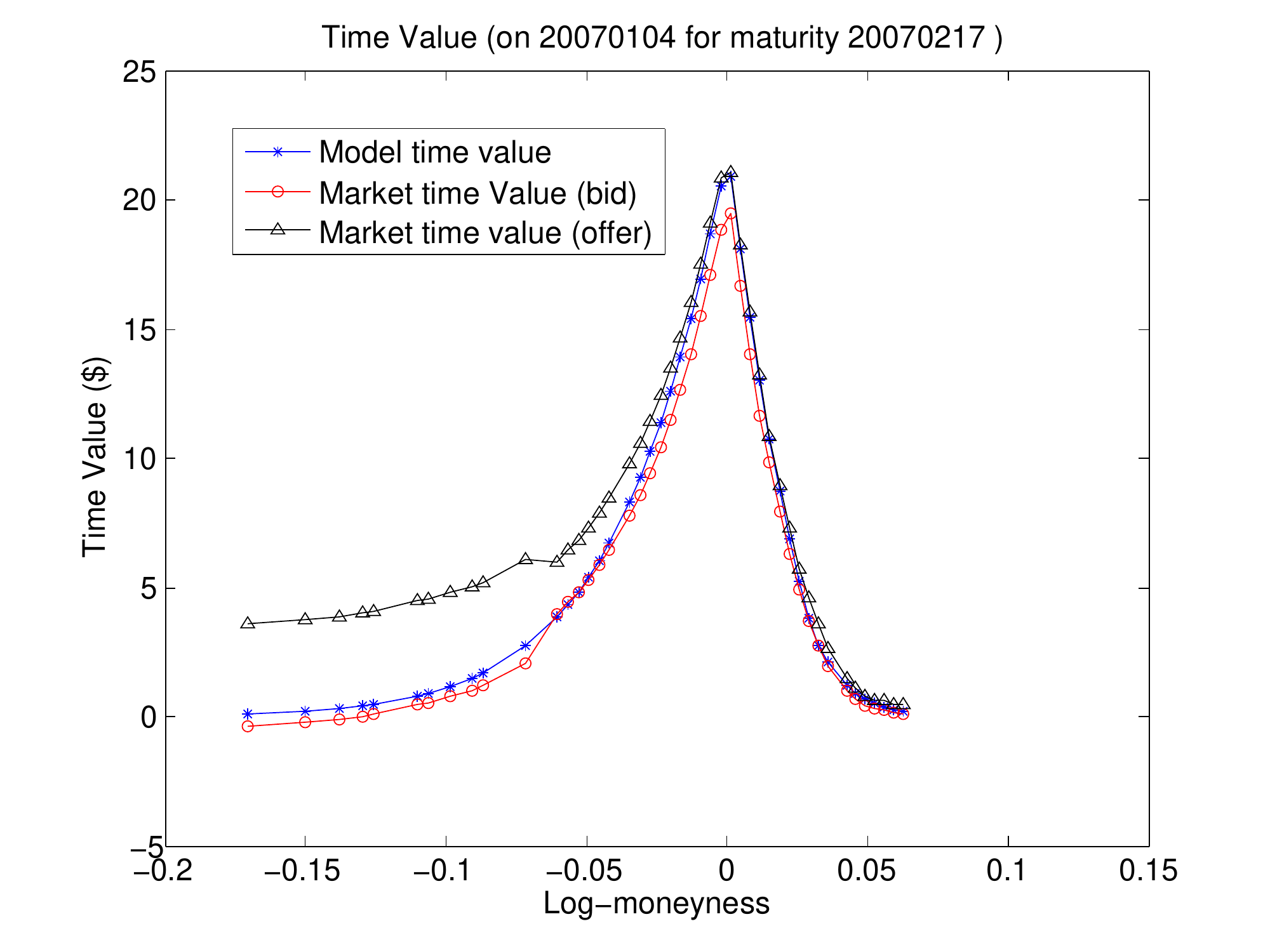}
       \\
      \fontsize{7}{12}\selectfont (a) Maturity Jan.20,2007
     &
      \fontsize{7}{12}\selectfont (b) Maturity Feb.17,2007
      \\
      \includegraphics[width=0.45\columnwidth]{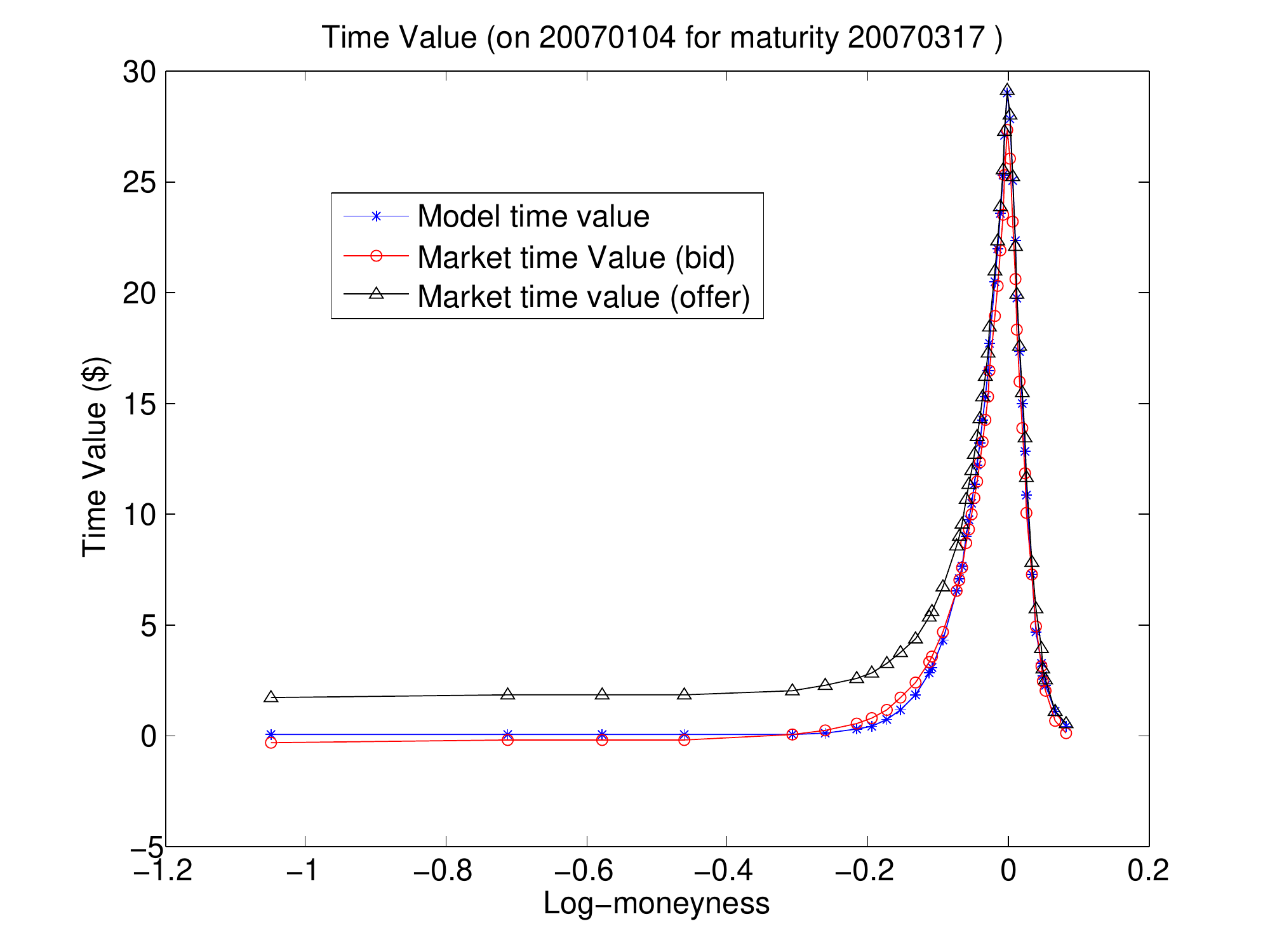}
       &
       \includegraphics[width=0.45\columnwidth]{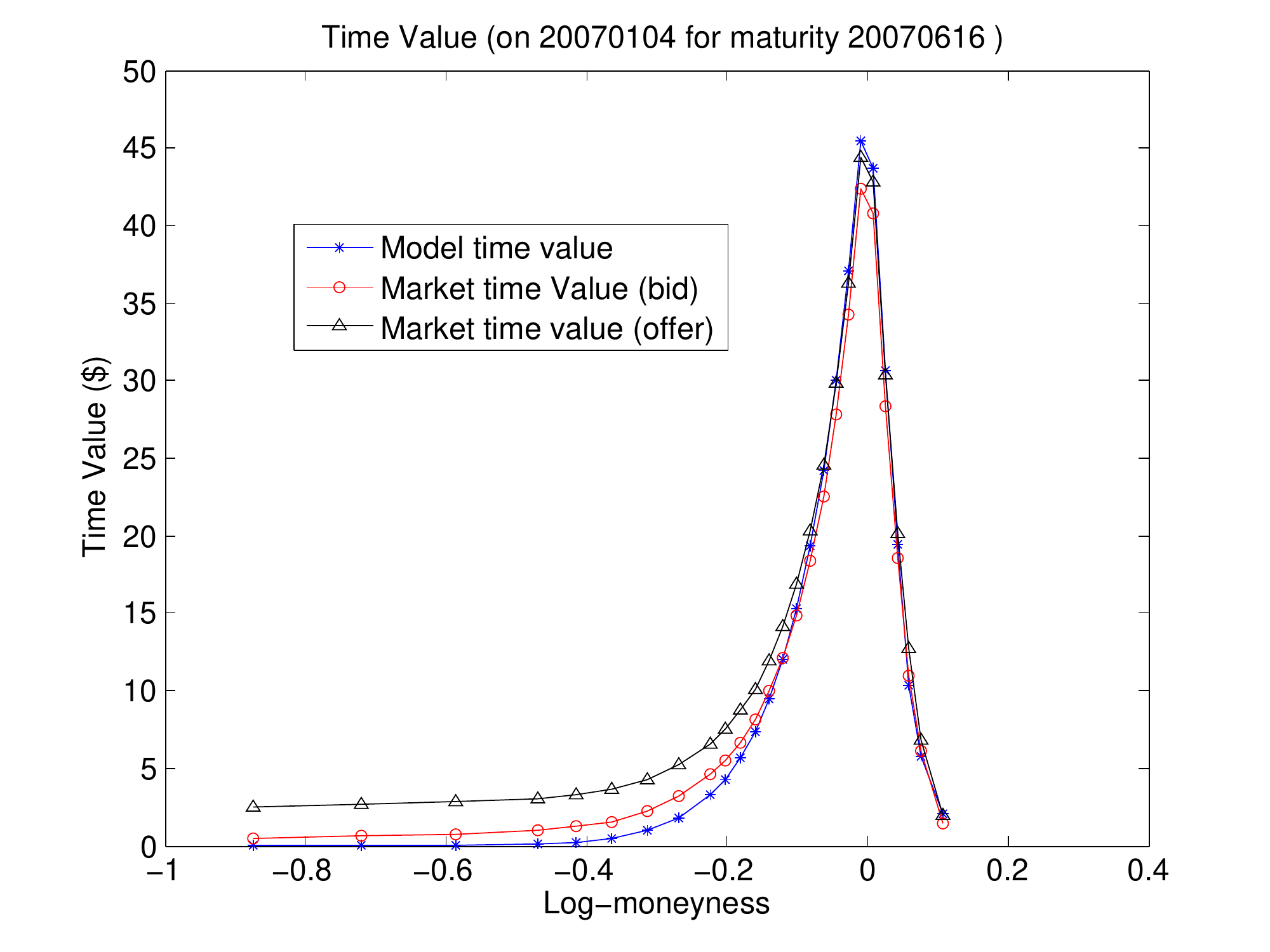}
       \\
      \fontsize{7}{12}\selectfont (c) Maturity Mar.17,2007
     &
      \fontsize{7}{12}\selectfont (d) Maturity Jun.16,2007
      \\
       \includegraphics[width=0.45\columnwidth]{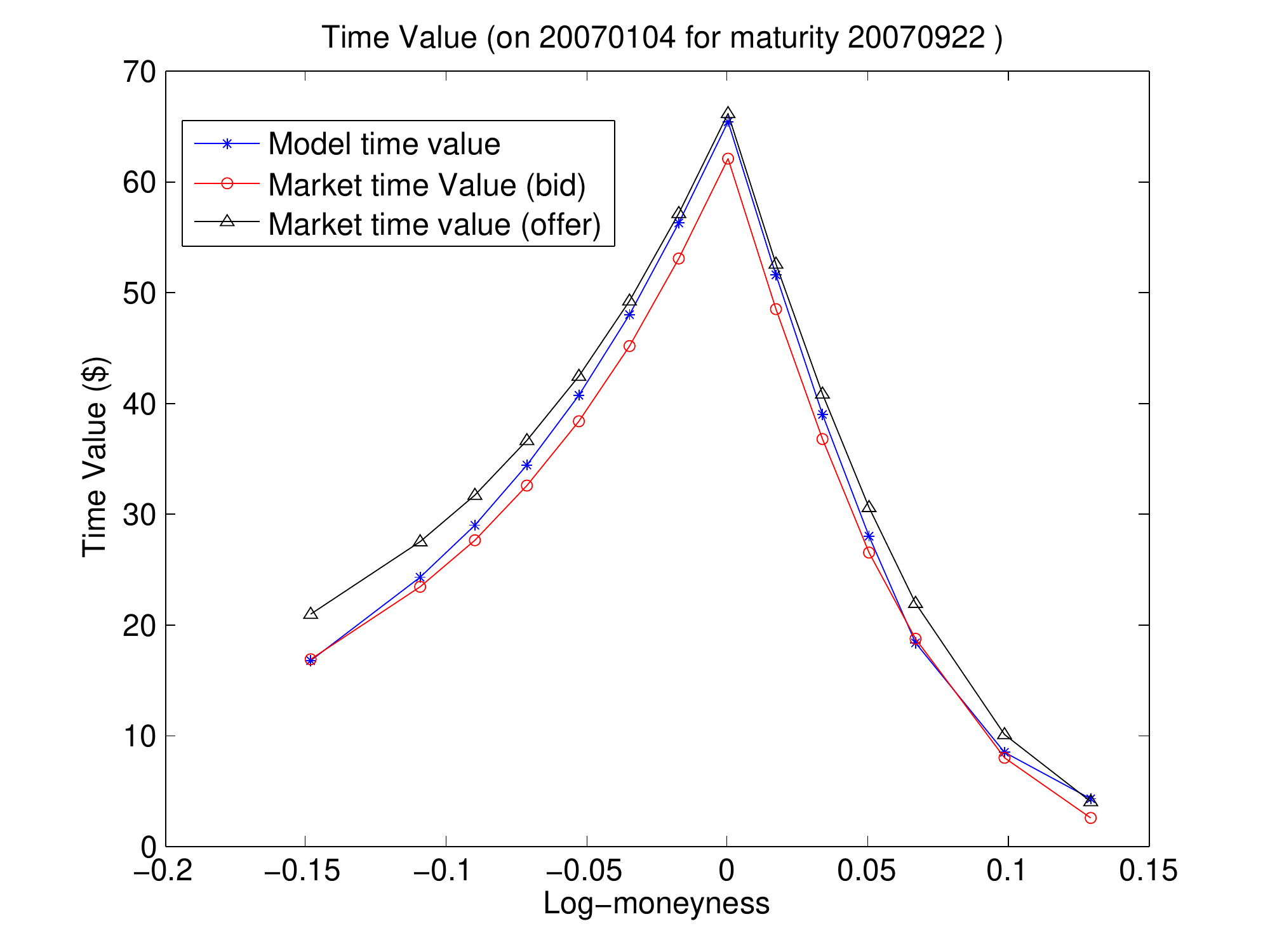}
       &
       \includegraphics[width=0.45\columnwidth]{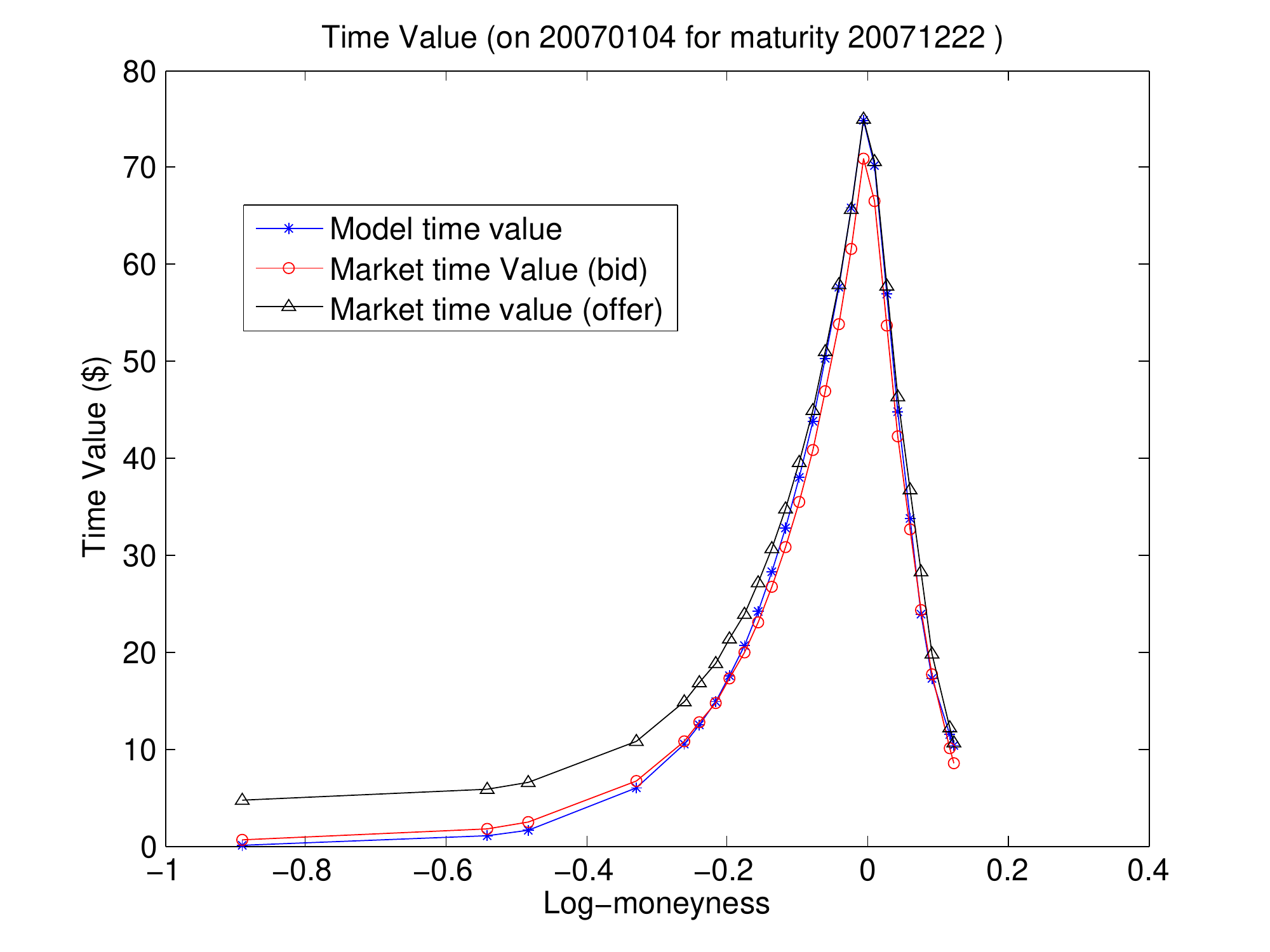}
       \\
      \fontsize{7}{12}\selectfont (e) Maturity Sep.22,2007
     &
      \fontsize{7}{12}\selectfont (f) Maturity Dec.22,2007
\\
       \end{tabular}
   \end{center}
   \vspace{-10pt}
   \caption{Calibrated time values for DETL model on the second day, Jan. 4, 2007}
   \label{fg:implementation:timevalue_day2}
\end{figure}

Below are the calibration results. The L\'{e}vy densities $\eta$ on Jan. 3, 2007 -- the first day of calibration -- is obtained by the three-variable optimization \eqref{fo:implementation:optinit}. From the calibrated parameters, we compute the symmetry index $\Xi$ via \eqref{fo:implementation:symmetryinit}, which is shown in Figure~\ref{fg:implementation:symmetry}. With the symmetry index $\Xi$, we run the two-variable optimization \eqref{fo:implementation:opt} on the following day, Jan. 4, 2007, and obtain the L\'{e}vy densities $\eta$ shown in Figure~\ref{fg:implementation:eta_day2}. The corresponding time values are shown in Figure~\ref{fg:implementation:timevalue_day2}.
We can see that the calibration results are quite precise in the sense that the time value falls between the bid and the ask values most of the time. As for the calibrated L\'{e}vy densities $\eta$, its values tend to decrease as the time to maturity increases (cf. Figure~\ref{fg:implementation:eta_day2}). The magnitude of $\Xi$ ( which measures the ``asymetry" of the L\'evy measure) is decreasing with maturity as well. Both results are in line with empirical findings on jump intensities and volatility skews.

Next, for every day $t$, we need to find $\kappa_t$ that satisfies \eqref{eq.kappa.fit.eta}.
Notice that, if $\eta_t(T,x)$ is differentiable in $T$, we obtain: 
\begin{equation} \label{fo:implementation:etakappa2}
\eta_t(T,x)+(T-t)\frac{\partial\eta_t(T,x)}{\partial T} = \kappa_t(T,x),
\end{equation}
for each $x \in \RR$. The relationship \eqref{fo:implementation:etakappa2} can be used to back out the additive densities $(\kappa_t)_{t\in[0,\bT]}$ from the calibrated L\'{e}vy densities $(\eta_t)_{t\in[0,\bT]}$. However, the calibrated densities $\eta_t(T,\cdot)$ are only defined for $T=T_l$, hence, we need to interpolate them across maturities.
An analysis of the calibrated L\'{e}vy densities shows that $\eta_t(T,x)$ generally exhibits one of the following two patterns as a function of $T$.   
\begin{itemize}
\item For small jump sizes $x$, $\eta_t(T,x)$ decreases rapidly as $T$ increases. To ensure that the recovered $\kappa$ is non-negative, we used a combination of exponential function and power function
\begin{equation}
\eta_t(T,x) = c_1 (T-t)^{c_2} + c_3(T-t) \exp(-c_4 (T-t)) + c_5
\end{equation}
to fit $\eta$, for any fixed $x$. The corresponding L\'evy density $\kappa$ can then be computed as
\begin{equation}
\kappa_t(T,x) = c_1(c_2+1)(T-t)^{c_2} + \exp(-c_4(T-t))(2c_3(T-t)-c_3c_4(T-t)^2) + c_5.
\end{equation}

\item For large jump sizes $x$, $\eta_t(T,x)$ increases as $T$ increases. The function we used to fit this scenario is a simple polynomial function
\begin{equation}
\eta_t(T,x) = c_1 (T-t)^4 + c_2 (T-t)^3 + c_3(T-t)^2 + c_4(T-t) + c_5. 
\end{equation}
Then, $\kappa$ is computed as
\begin{equation}
\kappa_t(T,x) = 5c_1(T-t)^4 + 4c_2(T-t)^3 + 3c_3(T-t)^2 + 2c_4(T-t) + c_5.
\end{equation}
\end{itemize}

\begin{figure}[htp]
   \begin{center}
       \begin{tabular}{cc}
       \includegraphics[width=0.45\columnwidth]{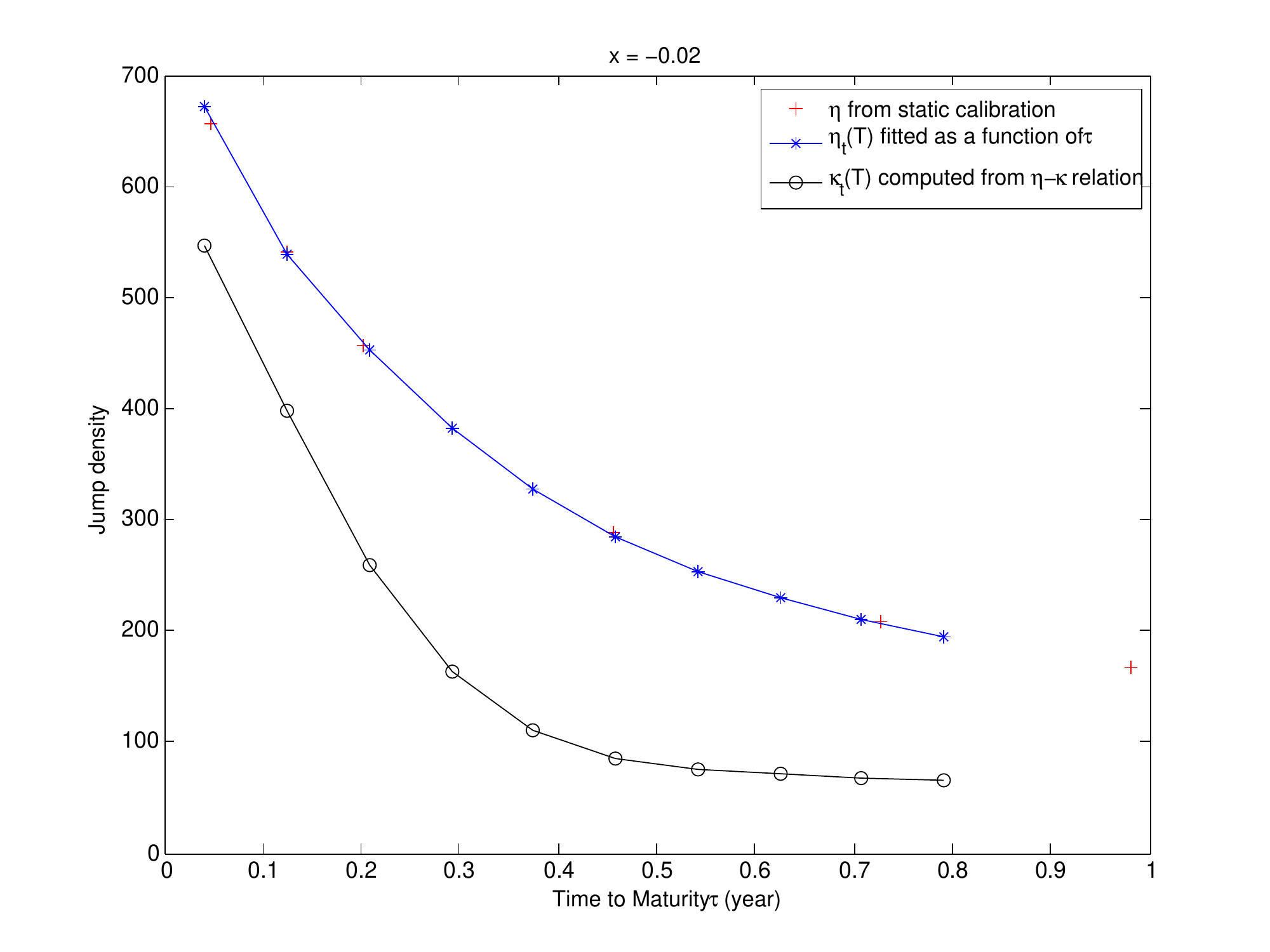}
       &
       \includegraphics[width=0.45\columnwidth]{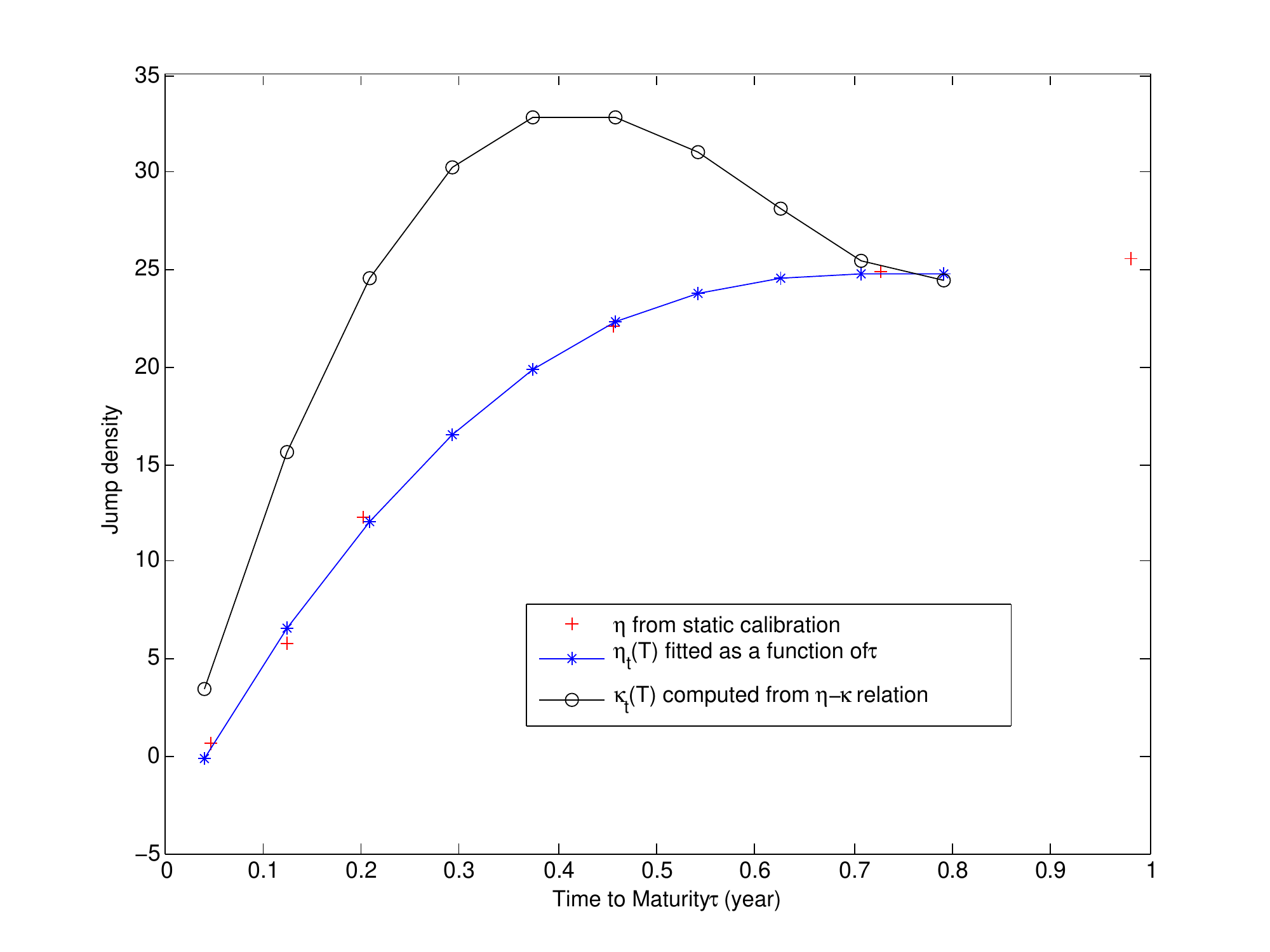}
       \\
      \fontsize{7}{12}\selectfont (a) $\eta$ and $\kappa$ for small jumps 
     &
      \fontsize{7}{12}\selectfont (b) $\eta$ and $\kappa$ for large jumps
      \\
       \includegraphics[width=0.45\columnwidth]{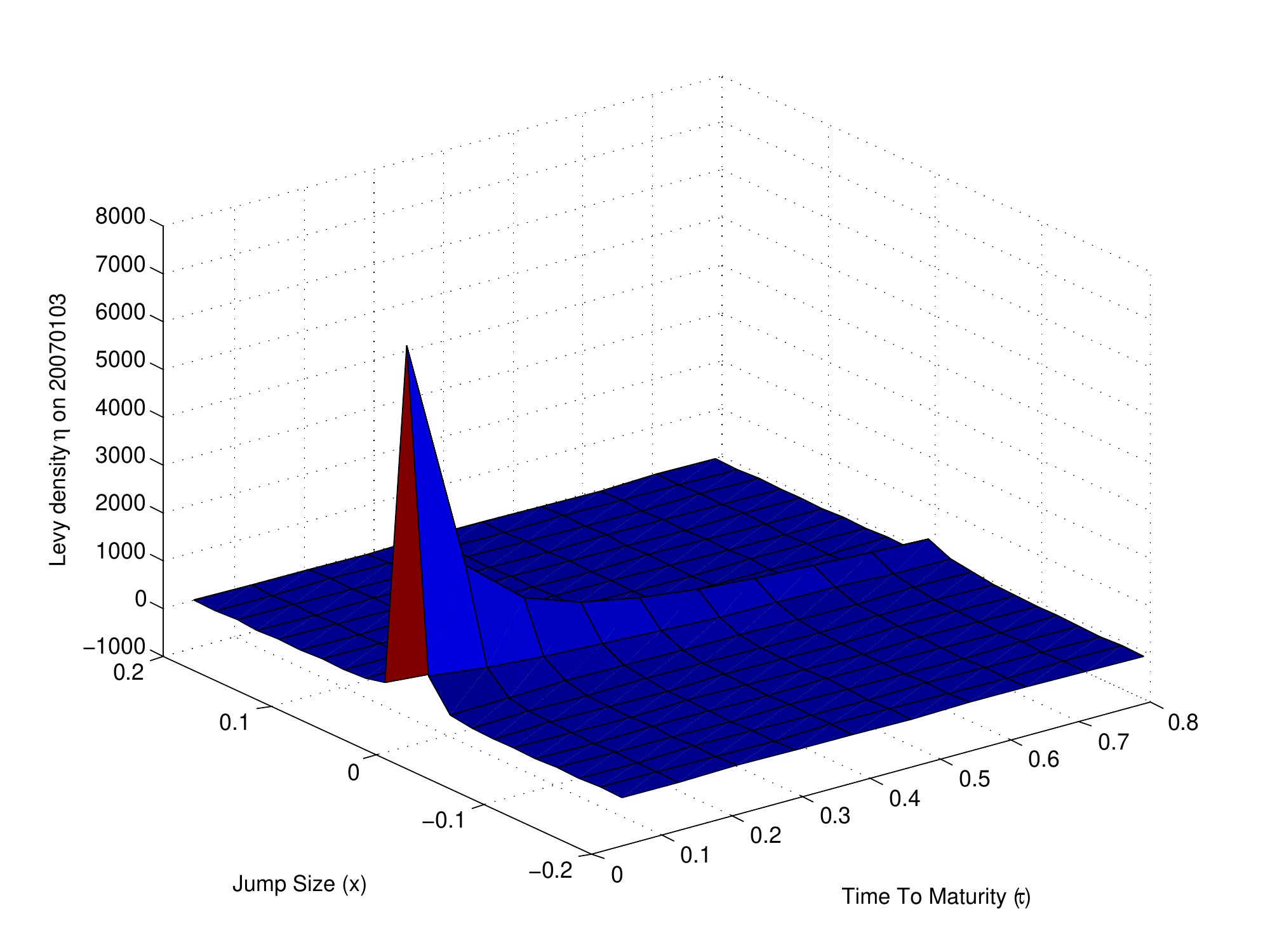}
       &
       \includegraphics[width=0.45\columnwidth]{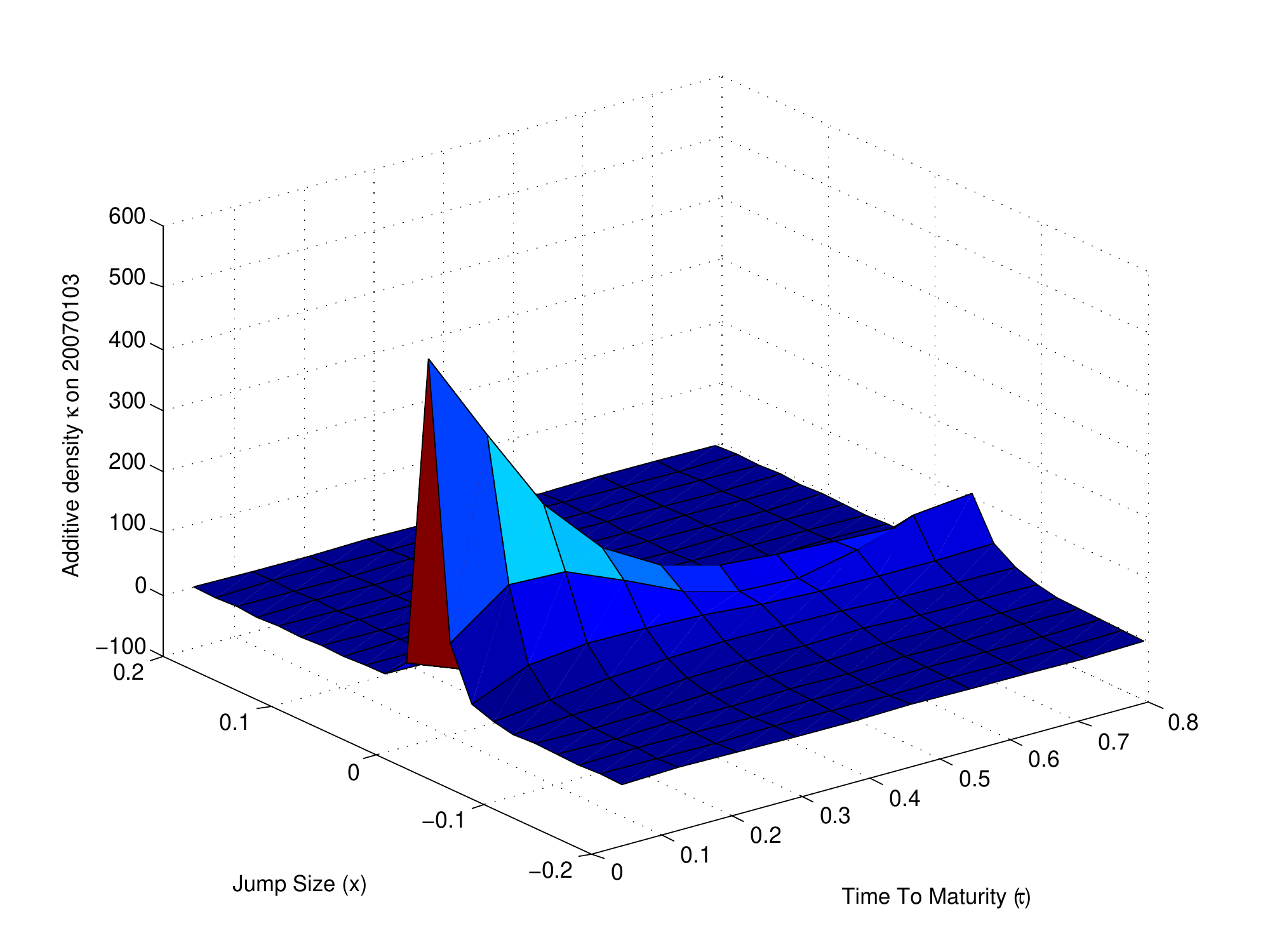}
       \\
      \fontsize{7}{12}\selectfont (c) L\'{e}vy density $\eta$   
     &
      \fontsize{7}{12}\selectfont (d) Additive density $\kappa$ reconstructed from $\eta$ in (c)      
      \\
       \end{tabular}
   \end{center}
   \vspace{-10pt}
   \caption{Calculating L\'evy density $\kappa$ from $\eta$}
   \label{fg:implementation:etatokappa}
\end{figure}

An illustration of the two scenarios together with an example of the reconstructed $\kappa$ is shown in Figure~\ref{fg:implementation:etatokappa}.

\subsection{Dynamic fitting}
\label{sec:implementation:dynamic}

Recall that, in view of \eqref{fo:nosym:dyn}, the L\'evy density $\kappa$ has the following dynamics: 
\begin{equation} \label{fo:implementation:kappadynamics}
d\kappa_t(T,x) = \alpha_t(T,x)dt + \sum_{n=1}^m \beta^n_t(T,x) dB_t^n. 
\end{equation}   
In the dynamic fitting, we need to assume that the time increments of $\kappa$ are stationary, which is only natural if we work with the time to maturity $\tau = T - t$ instead of the maturity $T$. Namely, we define $\hkappa_t(\tau,x) = \kappa_t(t+\tau,x)$ and its dynamics
\begin{equation}\label{fo:implementation:tkappadynamics} 
d\hkappa_t(\tau,x) = \halpha_t(\tau,x) dt +  \sum_{n=1}^m \hbeta^n_t(\tau,x) dB_t^n.
\end{equation}   
A simple application of It\^o's formula shows that
\begin{equation}
\halpha_t(\tau,x) = \alpha_t(t+\tau,x) + \frac{\partial \kappa_t(t+\tau, x)}{\partial T}\,\,\,\,\,\,\, \text{and}\,\,\,\,\,\,\,\hbeta^n_t(\tau,x) = \beta^n_t(t+\tau,x).
\end{equation}
To simulate future implied volatility surfaces, all we need are the diffusion terms $\hbeta^n$'s, because the drift term $\halpha$ can be computed from $\hbeta^n$'s. We assume that $\hbeta^n_t(\tau,x)$'s are deterministic and constant as functions of $t$, for any $(\tau,x)$ (from a finite family of points). Then, every increment $\Delta\hkappa_t = \hkappa_{t}-\hkappa_{t-1}$ is a sum of a Gaussian random vector, corresponding to the diffusion part, and a vector that corresponds to the drift term (we view every surface as a vector whose entries correspond to different values of $(\tau,x)$). Notice that the distribution of the Gaussian component is completely determined by its covariance matrix, hence, we will aim to choose $\hbeta^n$'s to match the estimated covariance matrix. Assuming that the drift term is bounded, it is easy to notice that the standard estimate of the covariance of $\Delta \hkappa_t$ also provides a consistent estimate of the covariance of the aforementioned Gaussian vector, asymptotically, as the length of the time increments converges to zero. In the actual computations, we use daily increments -- these are small compared to the time span of the entire sample, which is one year.
%Based on \eqref{fo:implementation:tkappadynamics}, $\hbeta^n$'s represent the diffusion (noise) part in $\hkappa$'s daily variation. It's not difficult to see, by changing the coordinates, that multiple choices of $\hbeta^n$'s may lead to the same dynamics of $\hkappa$. To reduce the computational load in the stage of simulation, it is better to choose the coordinates so that the variance of the noise term can be expressed in as few terms as possible. 
To fit $\hbeta^n$'s to the estimated covariance matrix, it is natural to use the Principal Component Analysis (PCA), which finds the directions that explain most of the variance in the increments $\Delta\hkappa_t$. 
%Namely, we assume that $\{\Delta\hkappa\}_t$ is stationary, and that $\hbeta^n_t$'s are deterministic and constant as functions of $t$ -- then they can be deduced from the covariance matrix of $\{\Delta\kappa\}_t$ -- and this can be done via PCA.
However, the PCA can not be applied directly because the number of points on the surface is close to the sample size, which is 251: for each $t$, we have call prices for 10 maturities and 21 jump sizes, which gives us 210 points on the $\hkappa$ surface after static fitting. To reduce the number of points, we pick every other maturity and the 7 jump sizes whose intensities are larger than others across time $t$. This gives us $5*7 = 35$ points on the reduced surface of $\{\Delta \hkappa_t\}_{t\in[0,\bT]}$. 

\begin{figure}[htp]
   \begin{center}
       \begin{tabular}{cc}
       \includegraphics[width=0.45\columnwidth]{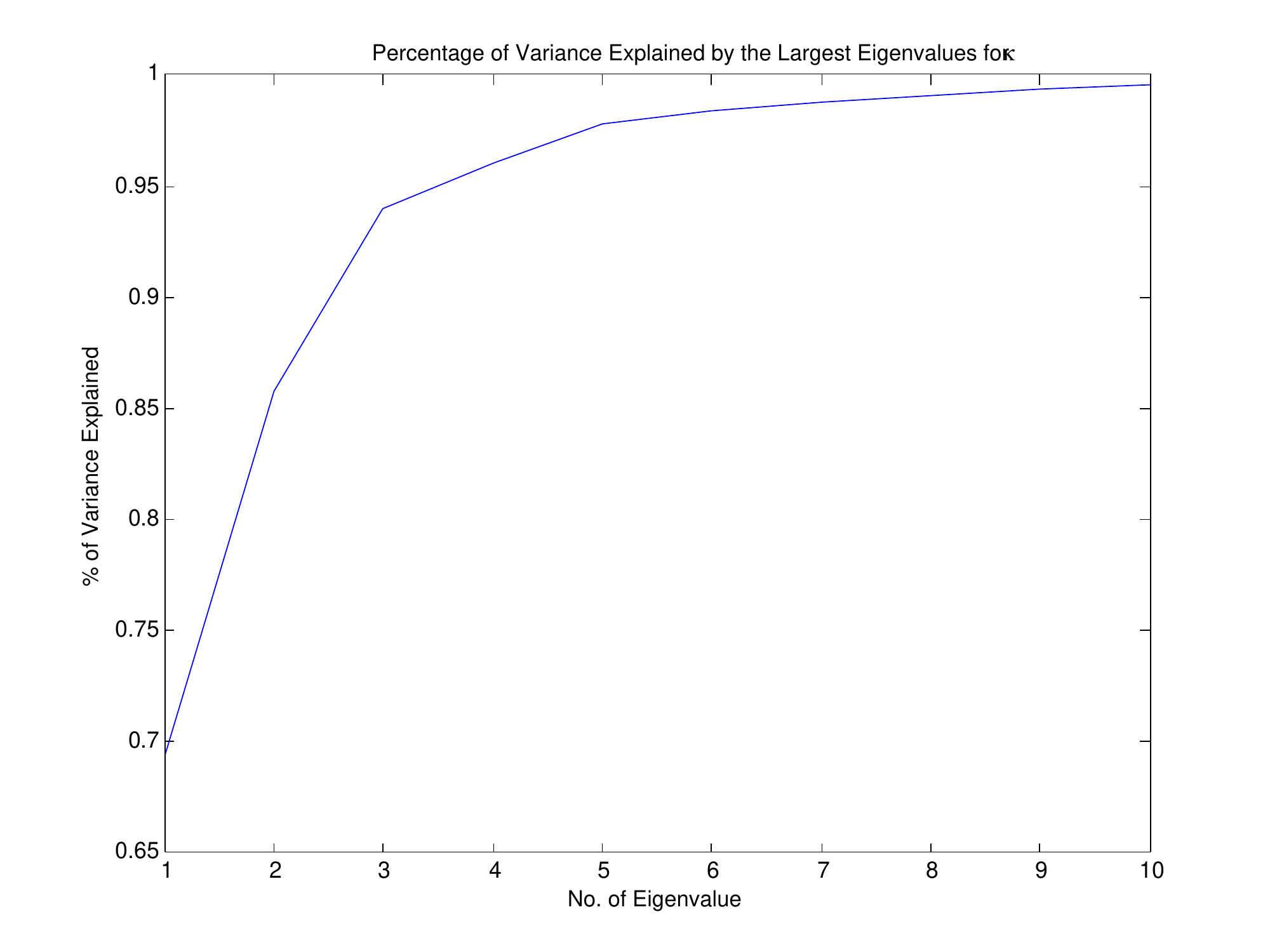}
       &
       \includegraphics[width=0.45\columnwidth]{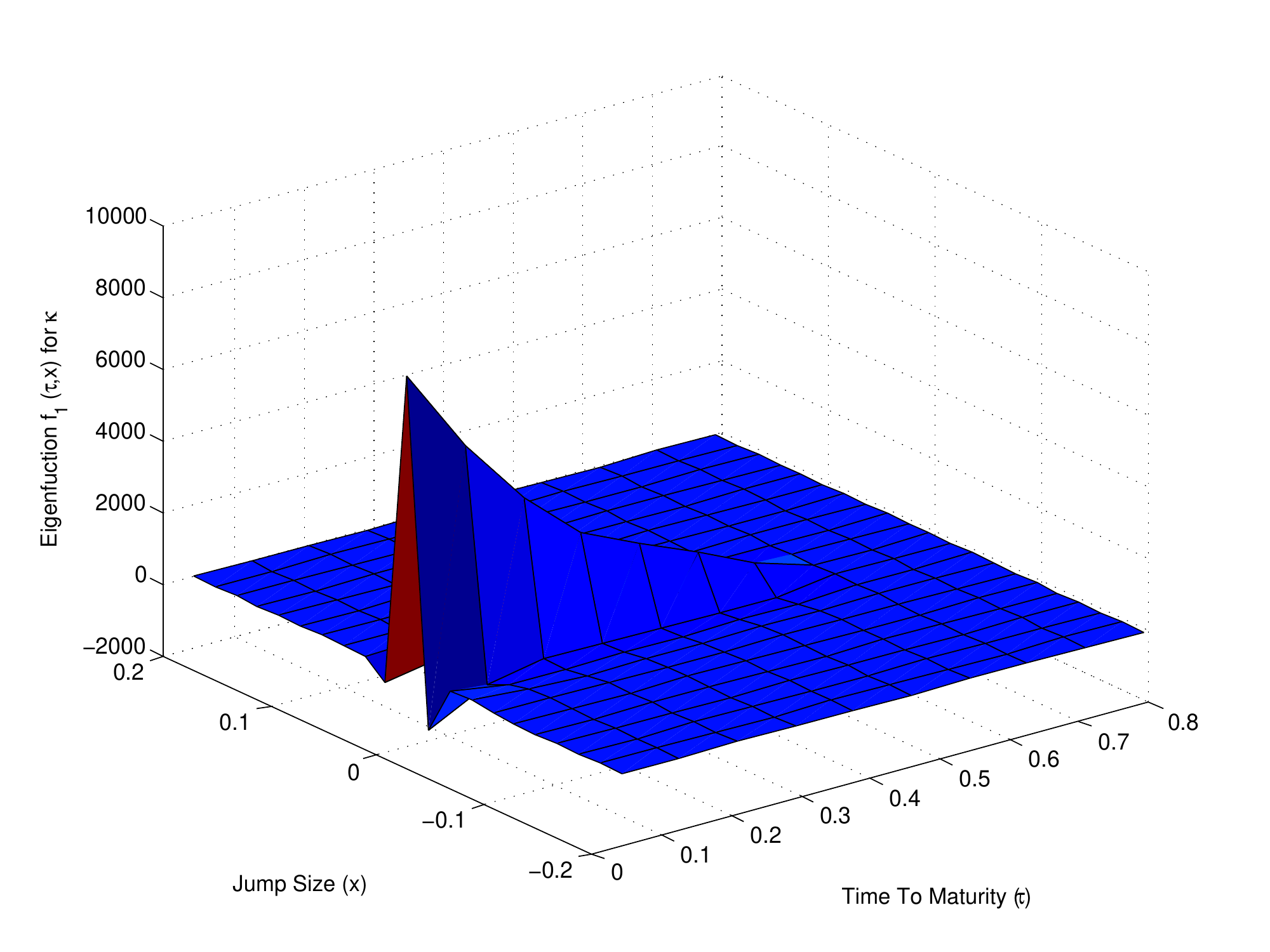}
       \\
      \fontsize{7}{12}\selectfont (a) Percentage of variance explained by the eigenmodes
     &
      \fontsize{7}{12}\selectfont (b) The first eigenmode scaled by $\sqrt{\lambda_1}$
      \\
       \includegraphics[width=0.45\columnwidth]{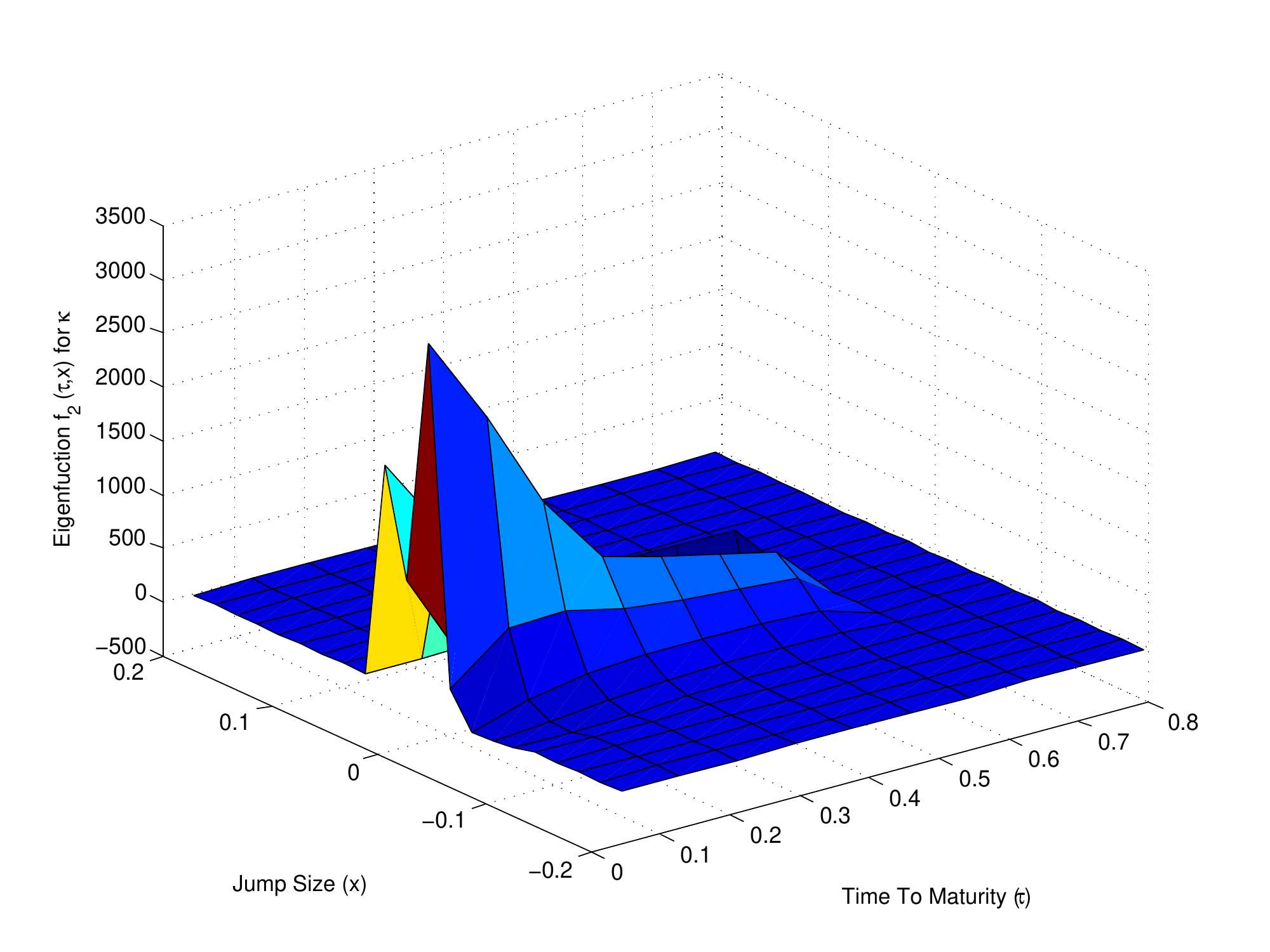}
       &
       \includegraphics[width=0.45\columnwidth]{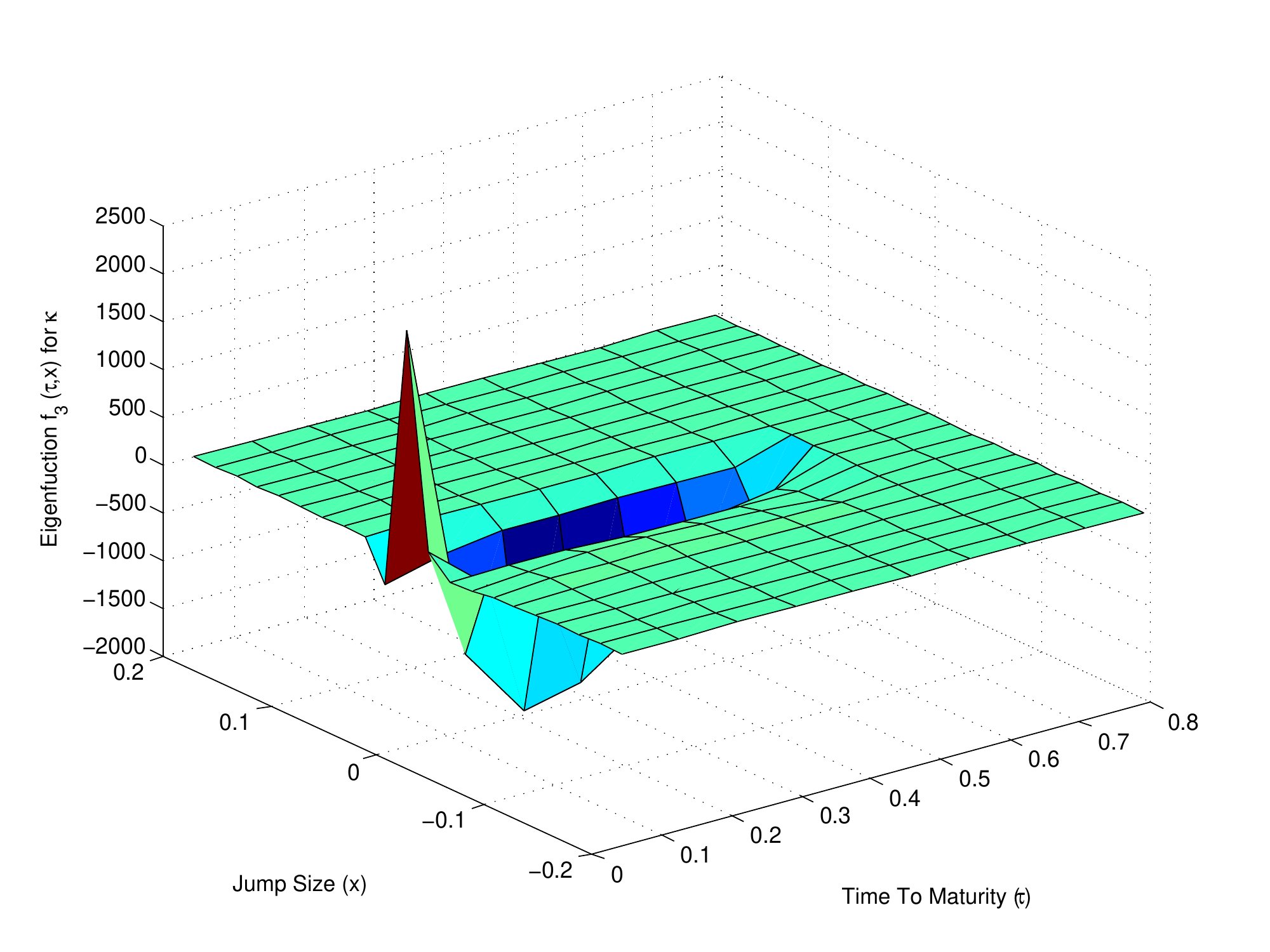}
       \\
      \fontsize{7}{12}\selectfont (c) The second eigenmode scaled by $\sqrt{\lambda_2}$
     &
      \fontsize{7}{12}\selectfont (d) The third eigenmode  scaled by $\sqrt{\lambda_3}$    
      \\
       \end{tabular}
   \end{center}
   \vspace{-10pt}
   \caption{Eigenvalues and eigenmodes of $\Delta\hkappa$ for DETL model}
   \label{fg:implementation:eigen}
\end{figure}

\begin{figure}[htp]
  \begin{center}
       \includegraphics[width=0.5\columnwidth]{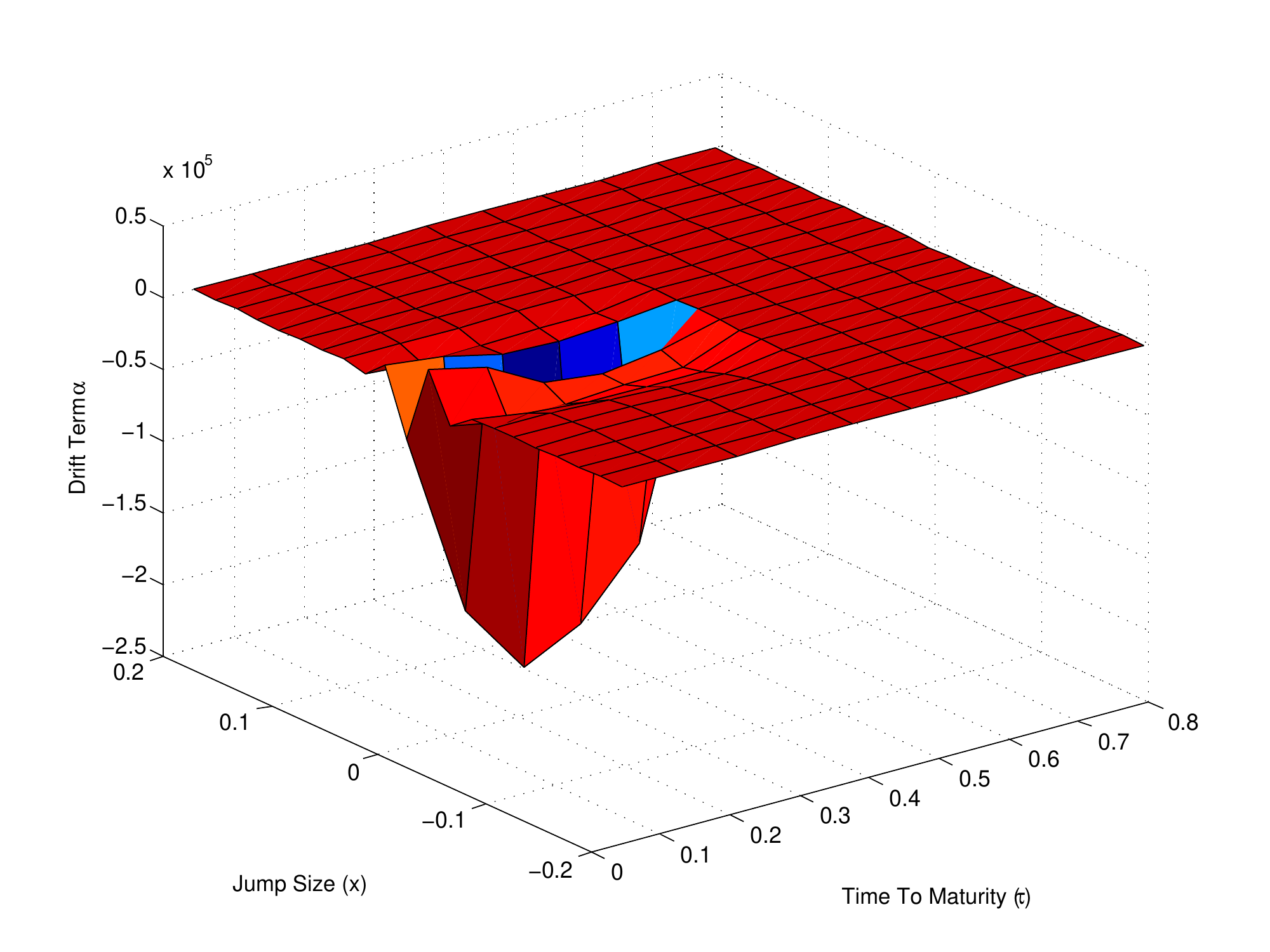}
   \end{center}
   \vspace{-10pt}
   \caption{The drift term $\alpha$ for DETL model}
   \label{fg:implementation:drift}
\end{figure}

Applying PCA on the reduced surface, we see that the first three eigenmodes $\{f^n(\tau,x)\}_{n=1}^3$ explain over $93\%$ of the daily variance of $\hkappa$, as shown in Figure~\ref{fg:implementation:eigen}. To extend the values of the eigenmodes to other points (i.e. other jump sizes and maturities), we simply perform a linear interpolation.
The first three eigenmodes have very unique characteristics. The first eigenmode takes the most prominent feature of $\hkappa$ - the densities are concentrated around small jumps at very short time to maturity. This eigenmode can be understood as a combination of the ``level" factor and the ``slope" factor (appearing in a typical PCA result for yield curve dynamics) along both the maturity and the jump size directions. 
%The first eigenmode sets the general shape of $\hkappa$ (and also $\kappa$). 
The second eigenmode shows the curvature along the jump size direction, and the third eigenmode shows the curvature along the time to maturity direction. 
As the eigenmodes $\{f^n(\tau,x)\}_{n=1}^3$ are normalized, to obtain the diffusion terms $\hbeta^n$'s, we need to multiple the eigenmodes by the loading factors:
\begin{equation}
\hbeta^n_t(\tau,x) = \sqrt{\lambda_n} \cdot f^n(\tau,x), \quad n = 1,2,3.
\end{equation}
Once we have $\hbeta^n$'s, we change the variables to pass to $\beta^n$'s and calculate the drift term $\alpha$ according to \eqref{fo:implementation:driftcondition}. Figure~\ref{fg:implementation:drift} shows the drift term $\alpha$ computed according to \eqref{fo:implementation:driftcondition}. 
Notice that $\halpha$ can then be computed as
\begin{equation}
\halpha_t(\tau,x) = \alpha_t(t+\tau,x) + \frac{\partial \kappa_t(t+\tau, x)}{\partial T},
\end{equation}
where we have no problem with evaluating the partial derivative, as, in the static fitting stage, $\kappa_t$ was interpolated across maturities.

 \subsection{Monte Carlo simulation of implied volatility surfaces}
\label{sec:implementation:simulation}

Once all the terms in the right hand side of \eqref{fo:implementation:tkappadynamics} are estimated, we can, for example, apply and explicit Euler scheme to simulate the future L\'evy densities $\hkappa_t$. However, we need to ensure that the simulated $\hkappa_t$'s stay nonnegative at all times. Inspired by \cite{carmona2010tangentlevy}, we incorporate a scaling factor in \eqref{fo:implementation:tkappadynamics} as follows:
\begin{equation}\label{fo:implementation:tkappadynamics_mod} 
d\hkappa_t(\tau,x) = \gamma_t^2 \halpha_t(\tau,x) dt +  \gamma_t \sum_{n=1}^m \hbeta^n(\tau,x) dB_t^n,
\end{equation}   
where 
\begin{equation}\label{fo:implementation:gammat}
\gamma_t = \frac{1}{\epsilon} \bigg(\inf_{ \tau \in [0, \bar{\tau}], x\in \RR} \hkappa_t(\tau,x) \land \epsilon\bigg),
\end{equation}
with $\epsilon = 1e^{-6}$ and $\bar{\tau}=1$. Of course, this modification changes the diffusion term of $\hkappa_t$, which was estimated from historical data. However, the value of $\epsilon$ is chosen to be so small that, in the historical sample, $\gamma_t$ is always equal to one. Hence, if we use the $\hbeta^n$'s chosen in the previous subsection, the resulting dynamics are still consistent with the past observations. It is also easy to see that, since $\gamma_t$ is a scalar, the drift restriction \eqref{fo:implementation:driftcondition} is satisfied by the new drift and volatility of $\kappa$. Finally, this modification ensures that $\hkappa_t$ is almost surely nonnegative for any $t$. 

\begin{figure}[htp]
   \begin{center}
       \begin{tabular}{cc}
       \includegraphics[width=0.45\columnwidth]{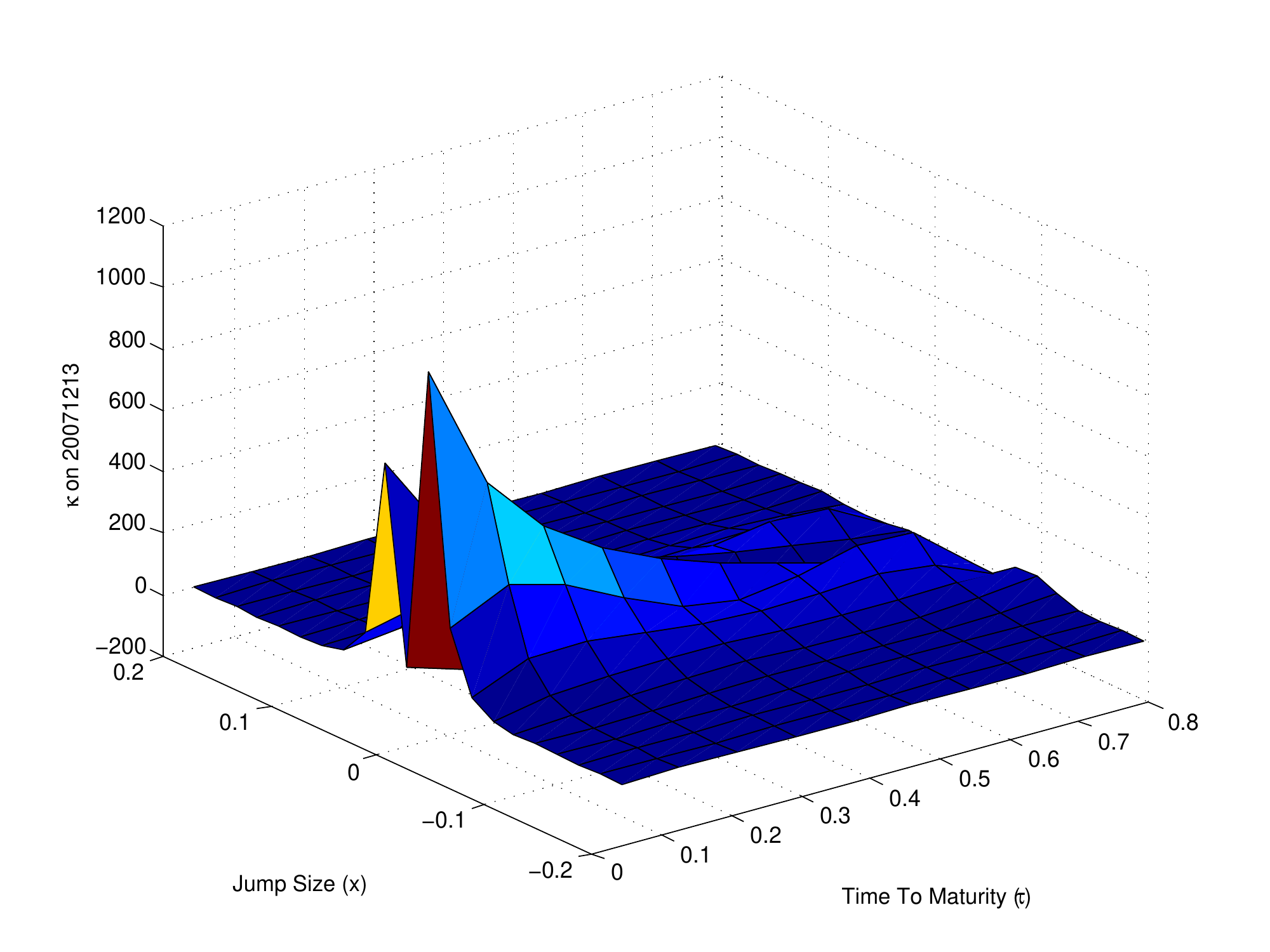}
       &
       \includegraphics[width=0.45\columnwidth]{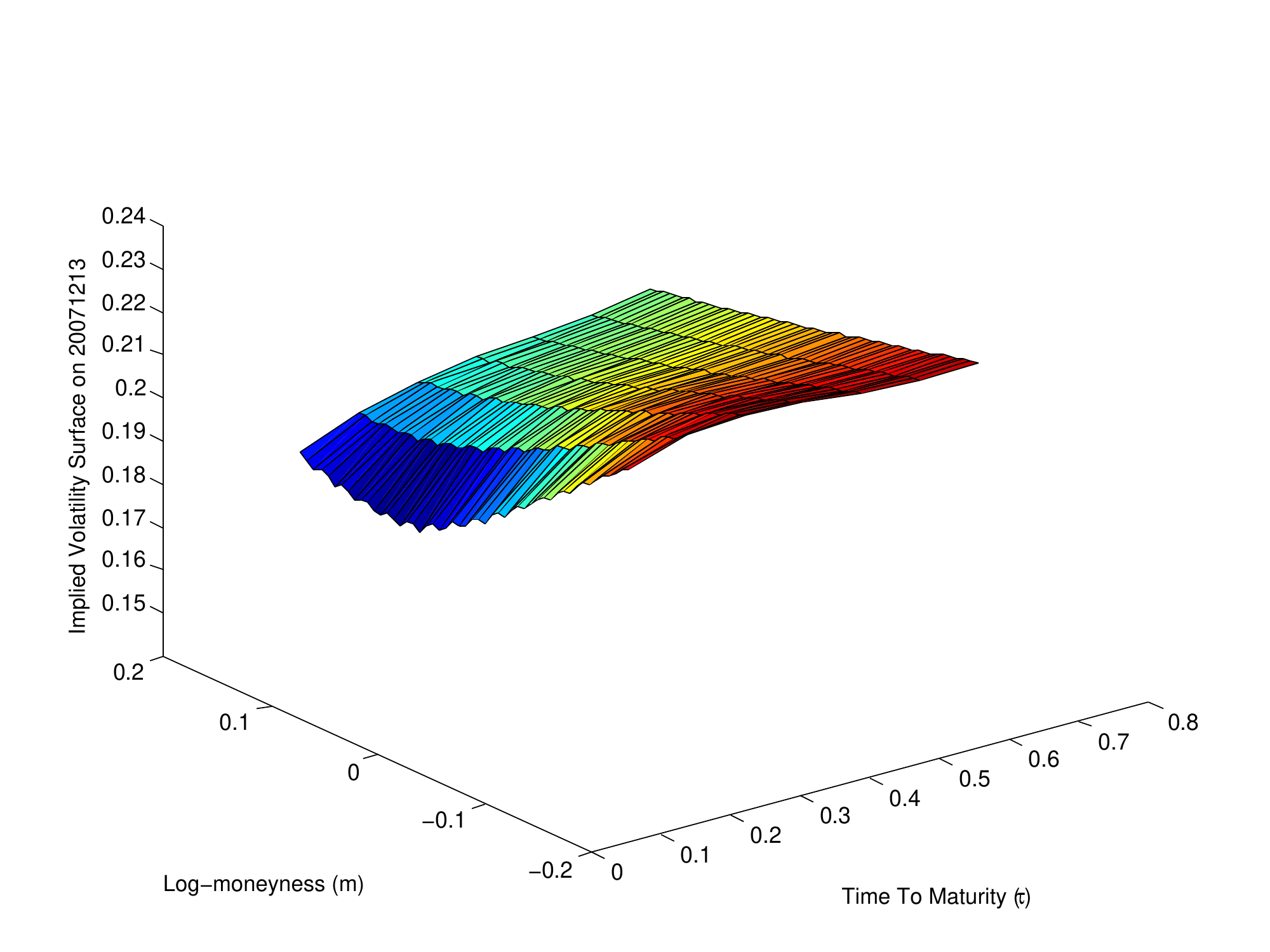}
       \\
      \fontsize{7}{12}\selectfont (a) Calibrated $\kappa$ 
     &
      \fontsize{7}{12}\selectfont (b) Calibrated Implied Volatility Surface
%      &
%      \fontsize{7}{12}\selectfont (c) $\beta^3$
      \\
      \includegraphics[width=0.45\columnwidth]{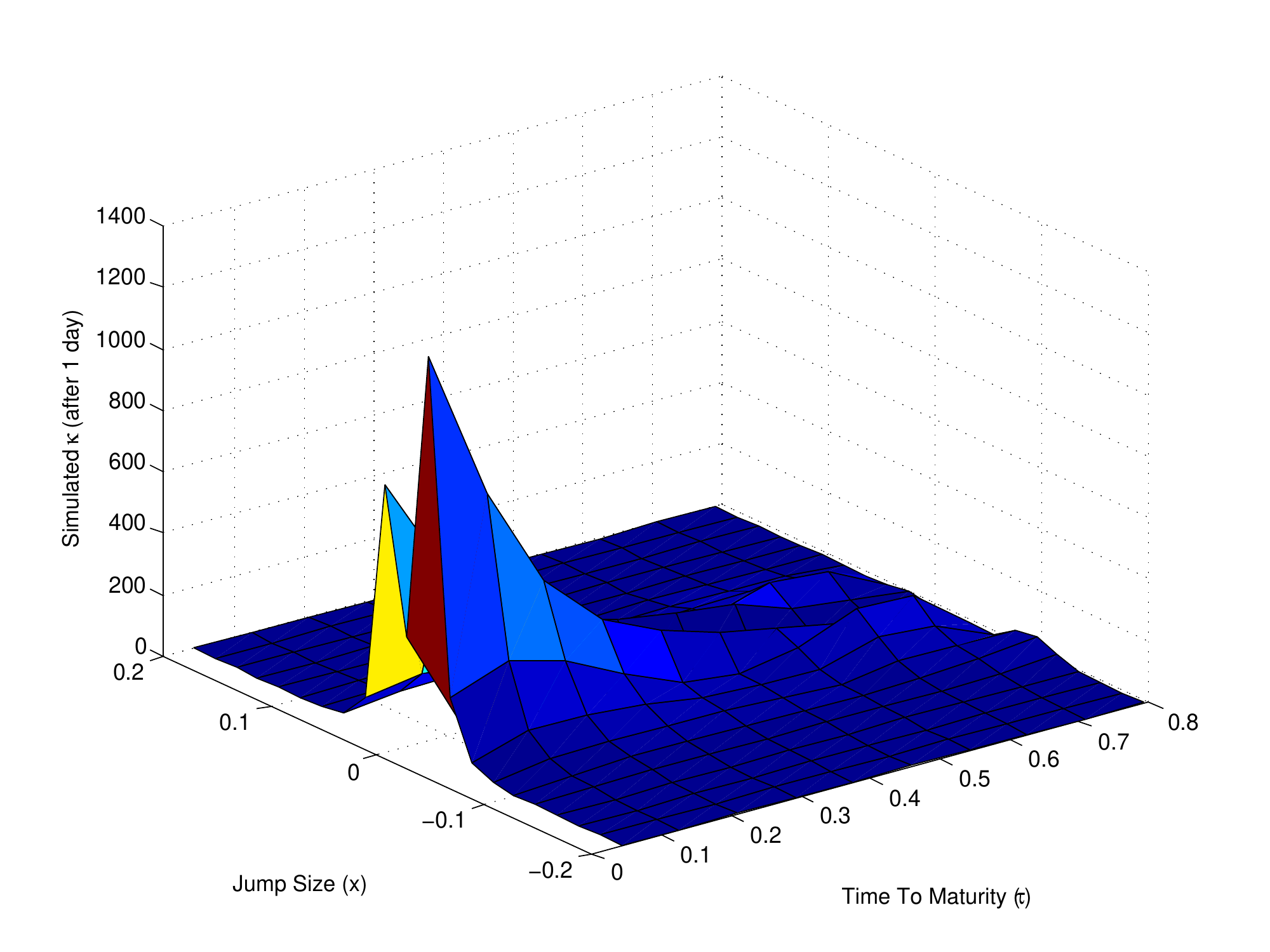}
       &
       \includegraphics[width=0.45\columnwidth]{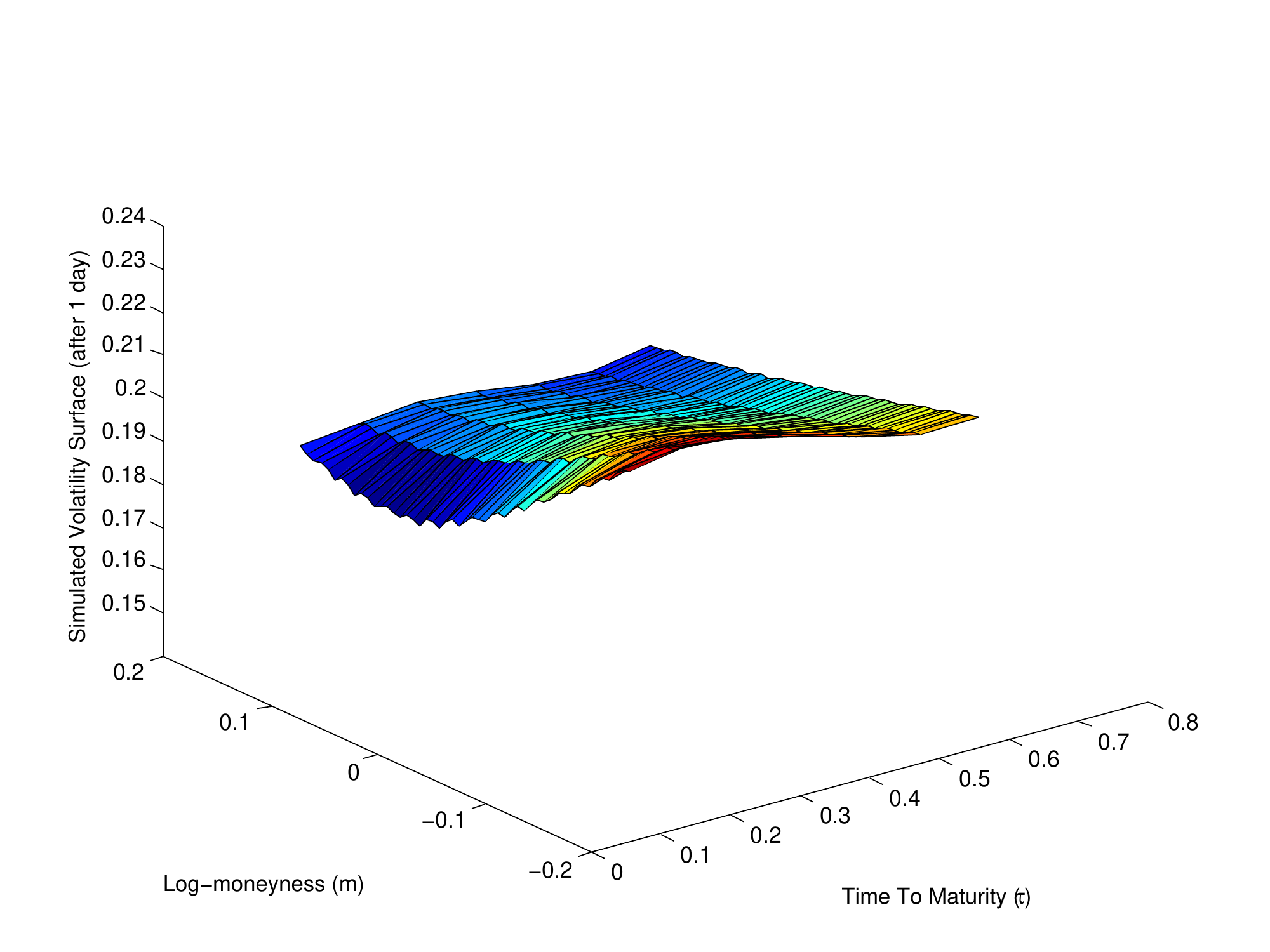}
       \\
      \fontsize{7}{12}\selectfont (c) Simulated $\kappa$ (1st day)
     &
      \fontsize{7}{12}\selectfont (d) Simulated Implied Volatility Surface (1st day)
      \\
       \includegraphics[width=0.45\columnwidth]{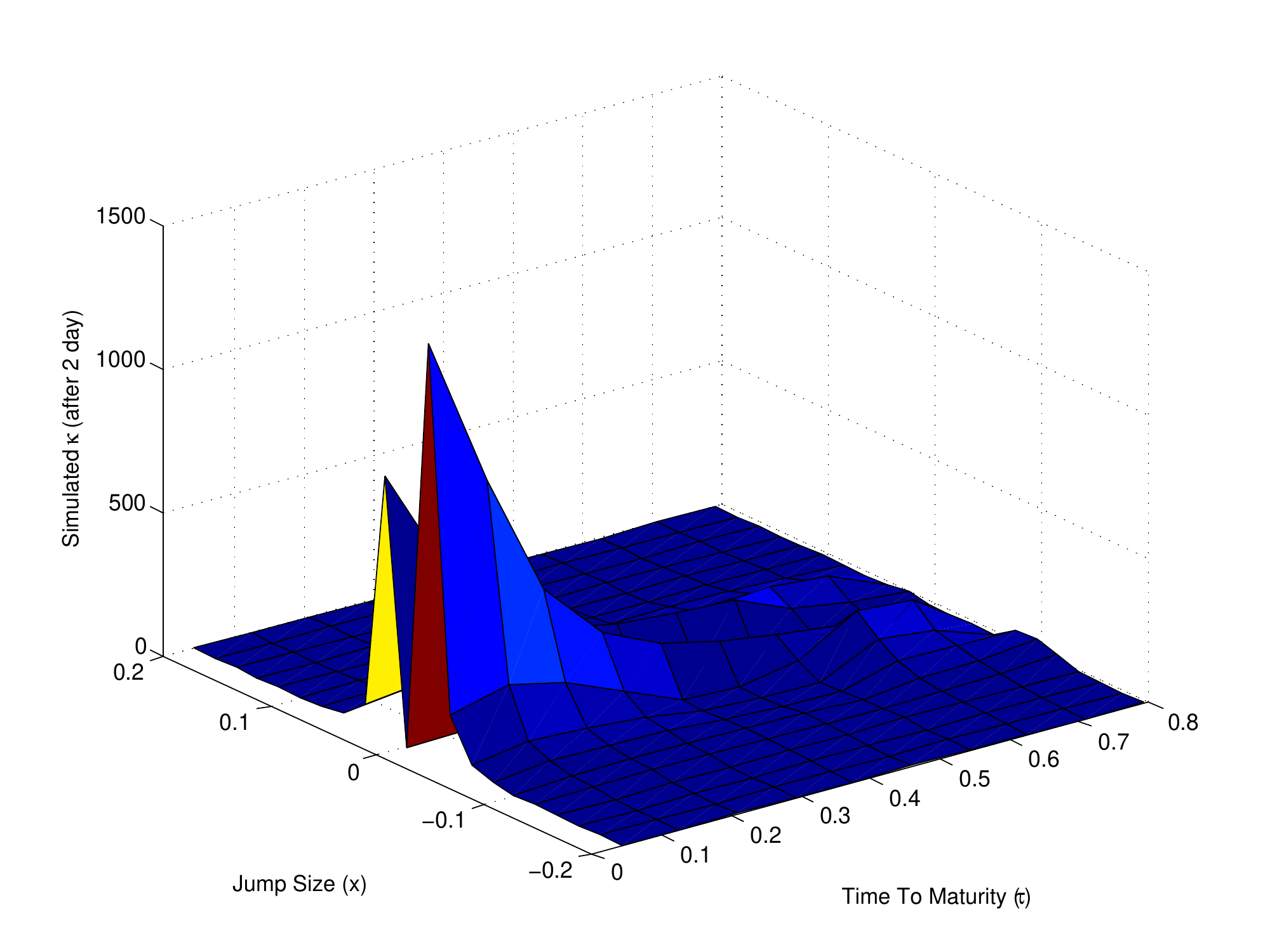}
       &
       \includegraphics[width=0.45\columnwidth]{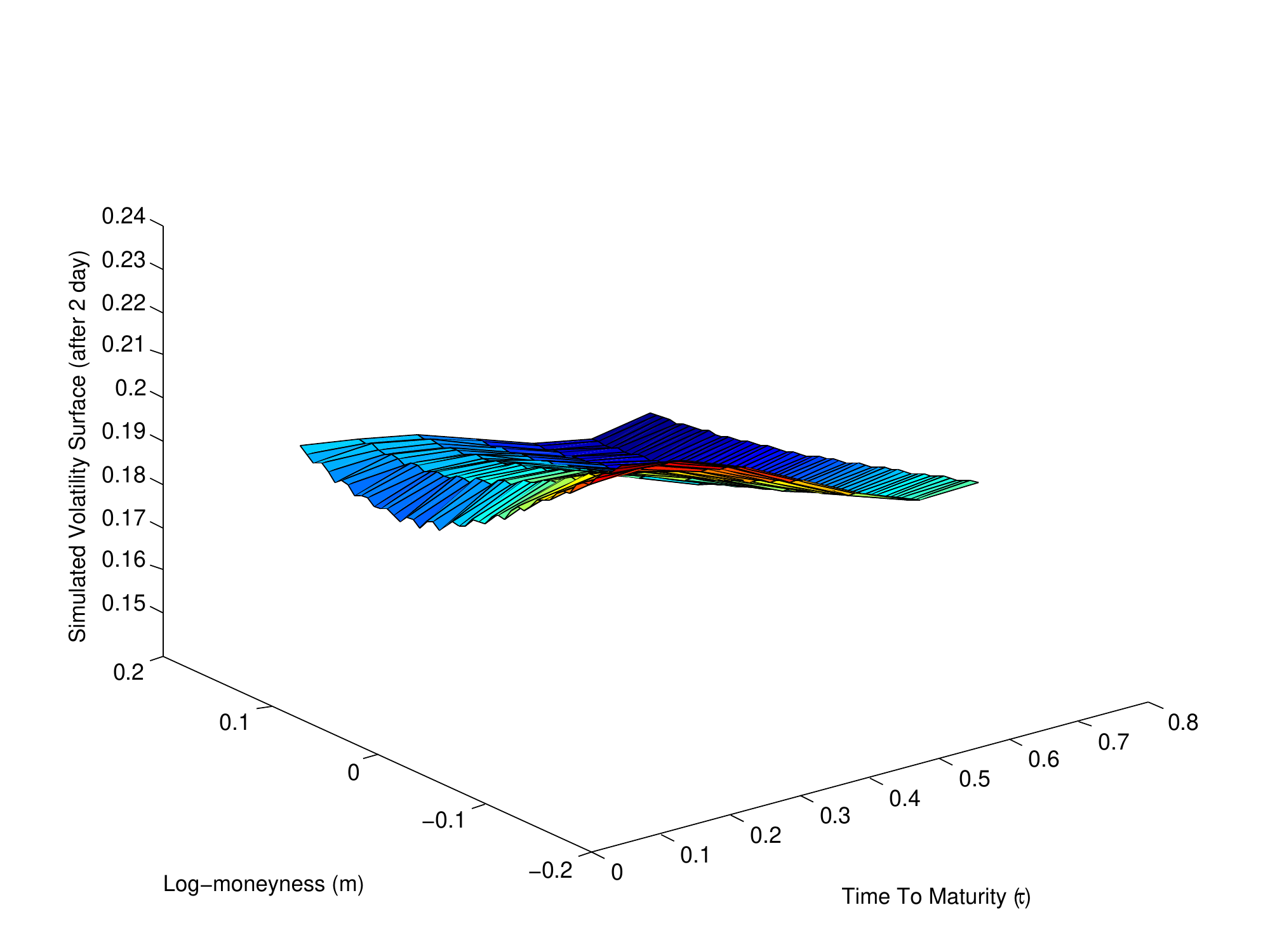}
       \\
      \fontsize{7}{12}\selectfont (e) Simulated $\kappa$ (2nd day)
     &
      \fontsize{7}{12}\selectfont (f) Simulated Implied Volatility Surface (2nd day)
\\
       \end{tabular}
   \end{center}
   \vspace{-10pt}
   \caption{Simulated $\kappa$'s and implied volatility surfaces using DETL model (1)}
   \label{fg:implementation:simulation1}
\end{figure}

\begin{figure}[htp]
   \begin{center}
       \begin{tabular}{cc}
       \includegraphics[width=0.45\columnwidth]{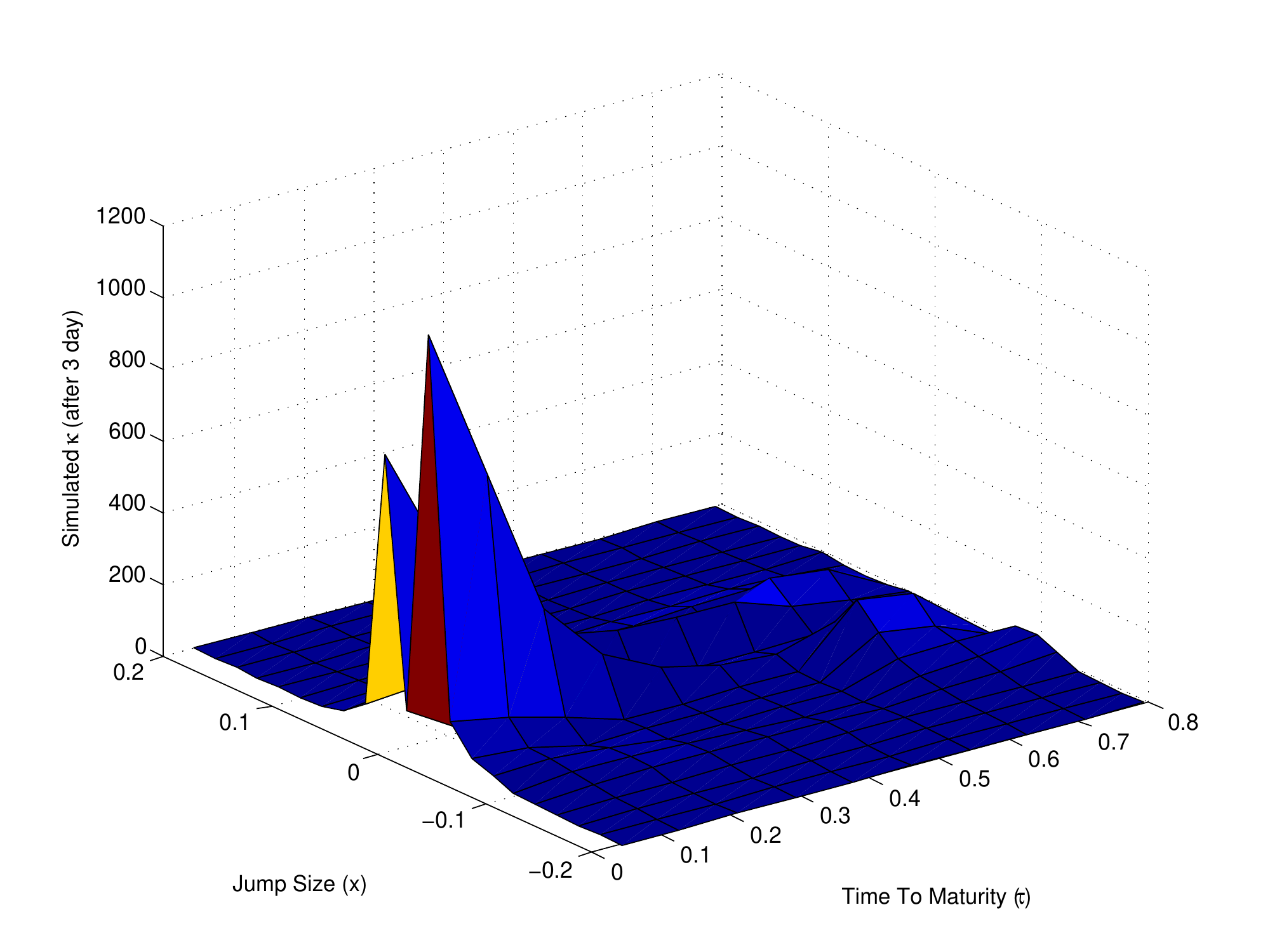}
       &
       \includegraphics[width=0.45\columnwidth]{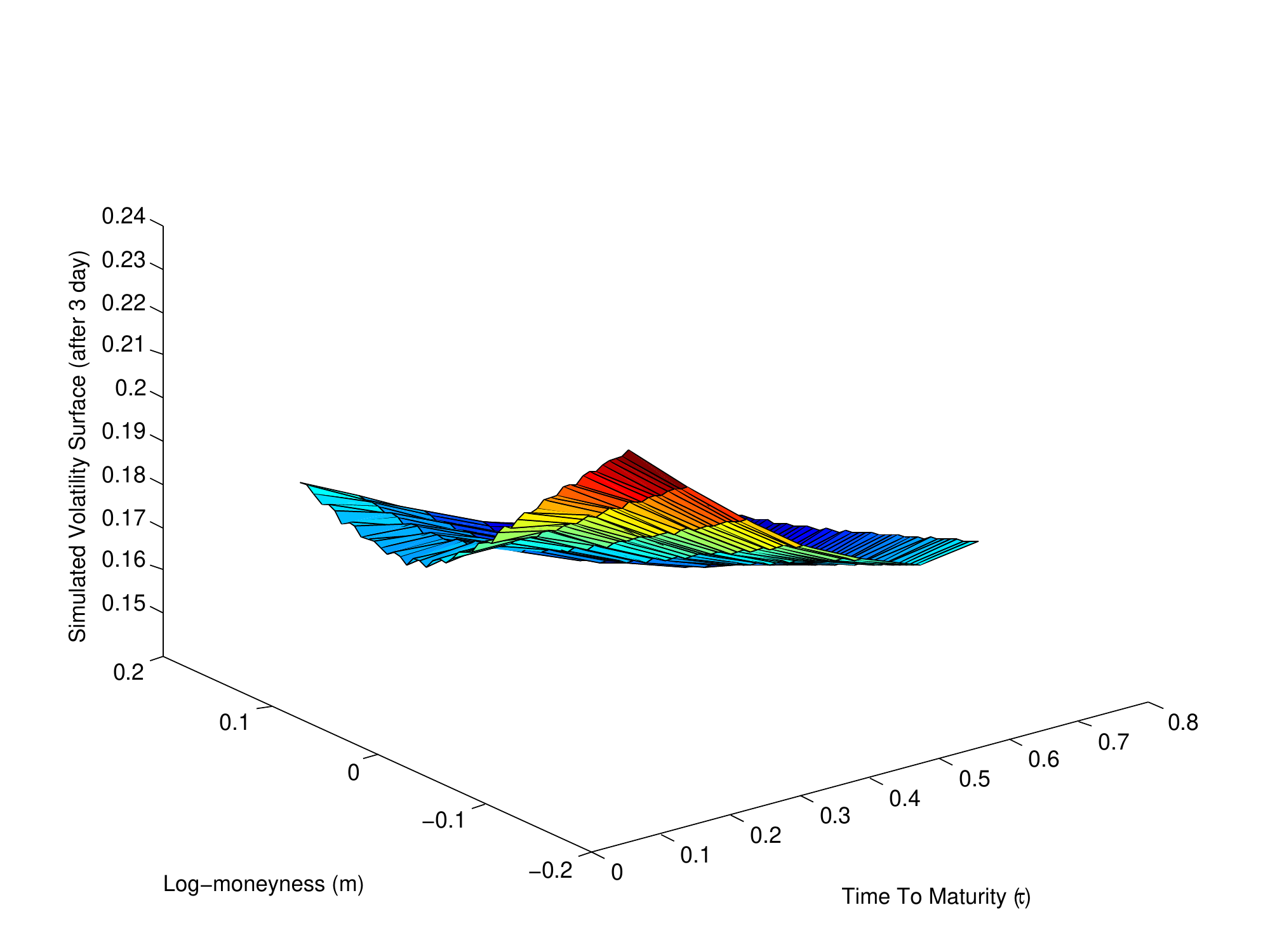}
       \\
      \fontsize{7}{12}\selectfont (a) Simulated $\kappa$ (3rd day)
     &
      \fontsize{7}{12}\selectfont (b) Simulated Implied Volatility Surface (3rd day)
%      &
%      \fontsize{7}{12}\selectfont (c) $\beta^3$
      \\
      \includegraphics[width=0.45\columnwidth]{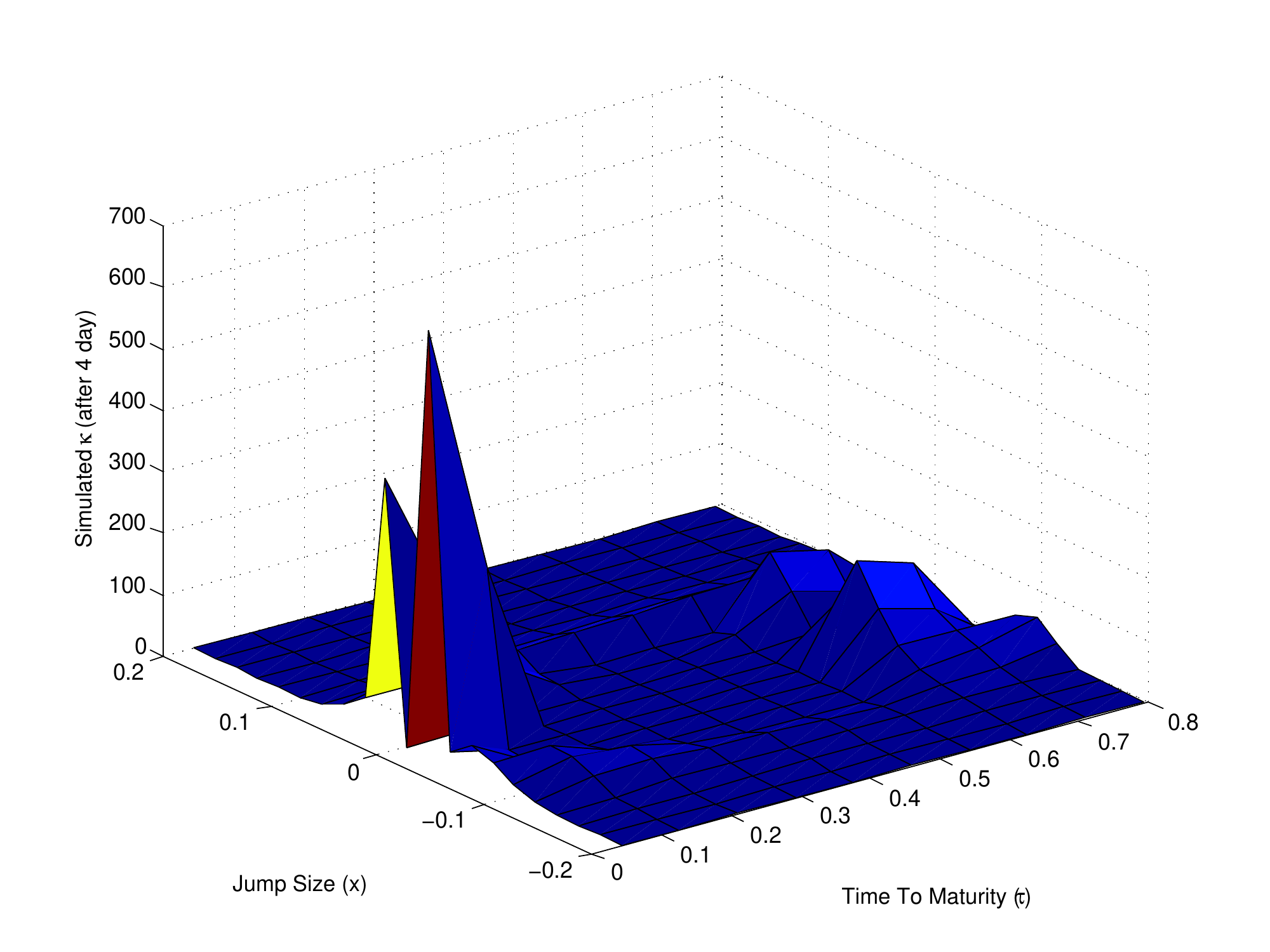}
       &
       \includegraphics[width=0.45\columnwidth]{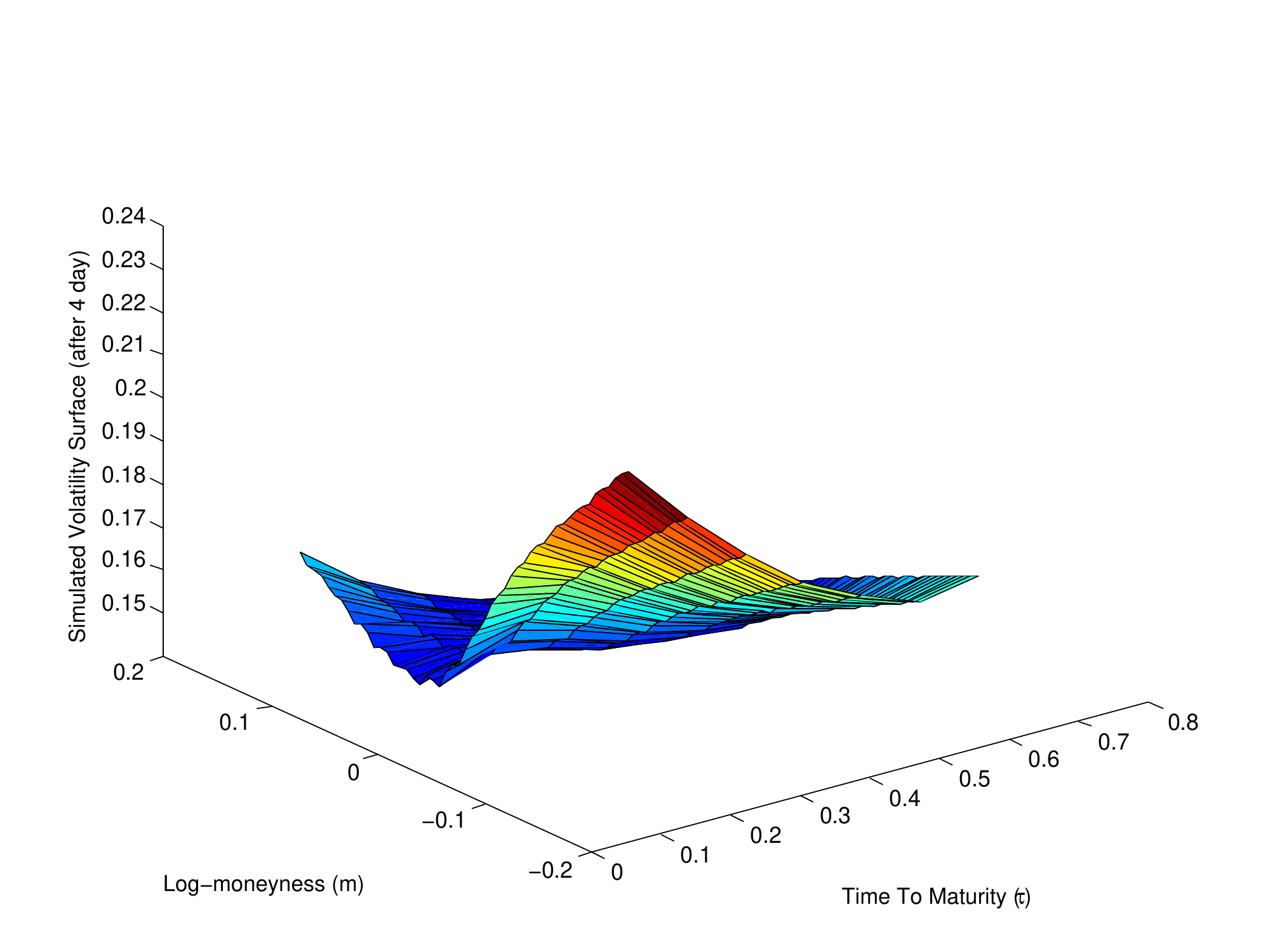}
       \\
      \fontsize{7}{12}\selectfont (c) Simulated $\kappa$ (4th day)
     &
      \fontsize{7}{12}\selectfont (d) Simulated Implied Volatility Surface (4th day)
      \\
       \includegraphics[width=0.45\columnwidth]{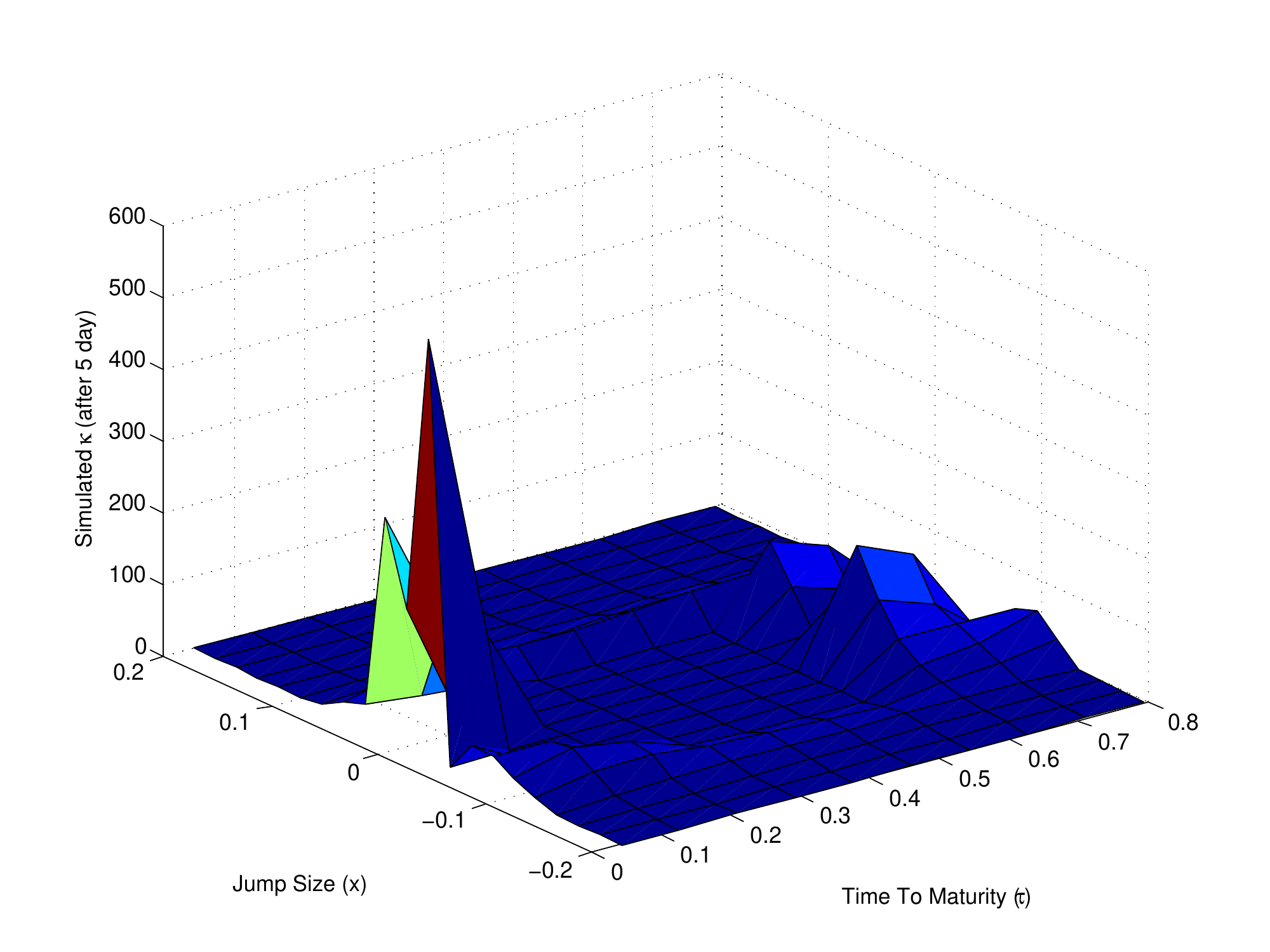}
       &
       \includegraphics[width=0.45\columnwidth]{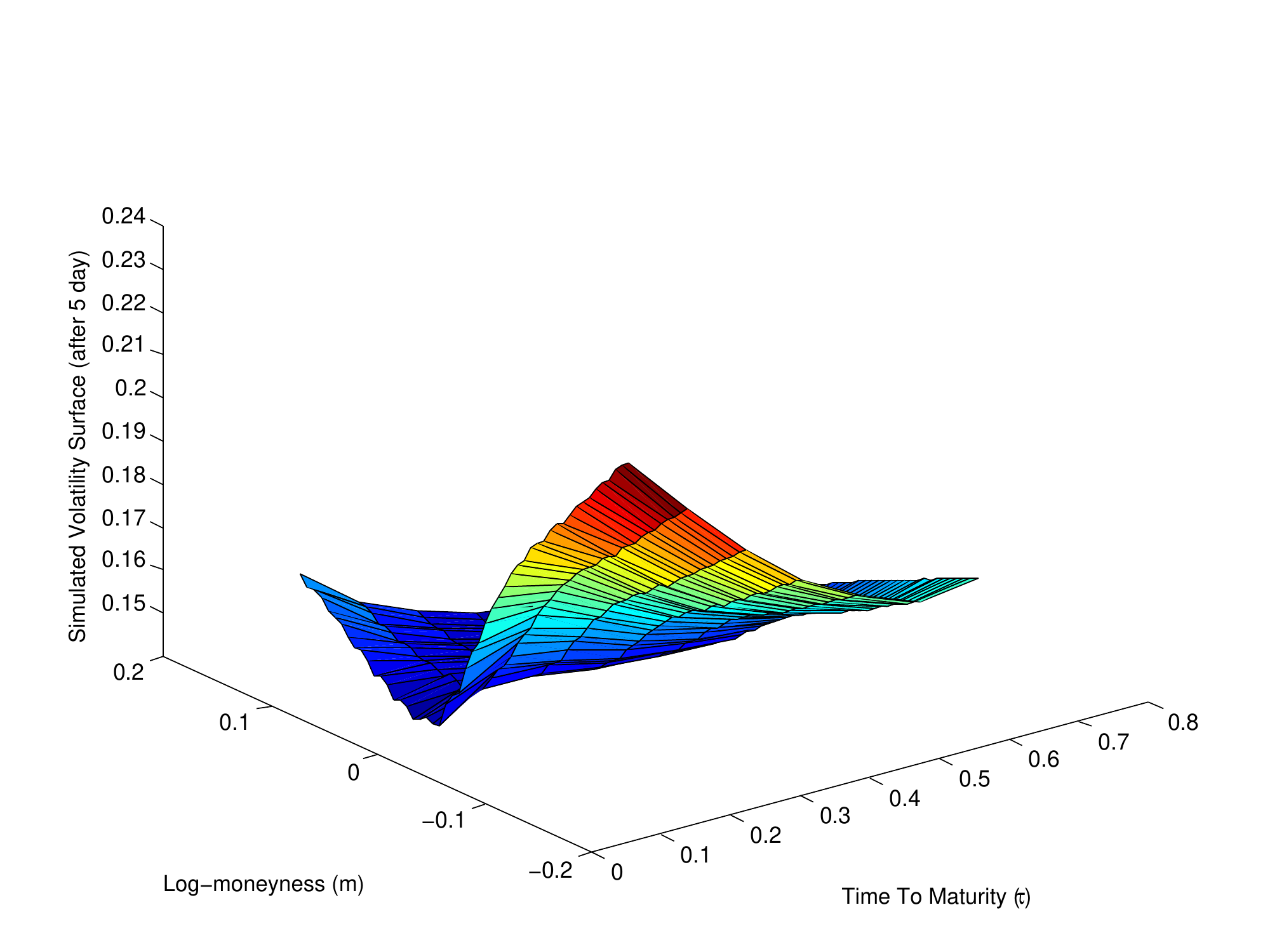}
       \\
      \fontsize{7}{12}\selectfont (e) Simulated $\kappa$ (5th day)
     &
      \fontsize{7}{12}\selectfont (f) Simulated Implied Volatility Surface (5th day)
\\
       \end{tabular}
   \end{center}
   \vspace{-10pt}
   \caption{Simulated $\kappa$'s and implied volatility surfaces using DETL model (2)}
   \label{fg:implementation:simulation2}
\end{figure}

To simulate future values of $\kappa$, we apply the explicit Euler scheme to \eqref{fo:implementation:tkappadynamics_mod}, to obtain
\begin{equation}\label{eq.detl.Euler}
\hkappa_{t+\Delta t}(\tau,x) = \hkappa_t(\tau,x) + \gamma_t^2 \halpha_t(\tau,x) \Delta t + \gamma_t\sum_{n=1}^m \hbeta^n(\tau,x) \Delta B^n_t,  
\end{equation} 
with $\Delta t$ being one day.
Having simulated $\hkappa_t$, we compute $\eta_t$ via \eqref{eq.kappa.fit.eta}. Then, for every fixed maturity $T$, the option prices in the model given by the L\'evy density $\eta_t(T,\cdot)$ can then be computed, for example, using the methods proposed in \cite{Carr99optionvaluation} or \cite{Lewis2001optionformula}. These methods are based on Fourier transform and can be implemented efficiently via numerical integration.\footnote{Please note that we cannot use \eqref{fo:implementation:Kouprice} to calculate option prices, because, even though the calibrated L\'{e}vy densities $\left\{\eta_t(T,\cdot)\right\}$ are double exponential, there is no reason to believe that the simulated $\eta$'s remain double exponential.}  
In particular, in our simulation, we use the following formula to calculate future option prices: 
\begin{equation}\label{fo:implementation:lewisopt}
C_t^{1,\kappa_t}(T,x) = 1 - \frac{e^{x/2}}{\pi} \int_0^\infty \frac{du}{u^2 + \frac{1}{4}} Re\left[\exp\left(-iu x\right) \phi_t\left(T,u-\frac{i}{2}\right)\right],
\end{equation}
where $\phi_t$ is the characteristic function of an exponential L\'{e}vy process with the L\'{e}vy density $\eta_t(T,\cdot)$, starting from one:
$$
\phi_t(T,u)  = \exp\bigg[ - iu (T-t)\int_\RR \eta_t(T,x)(e^x-1)dx  + (T-t)\int_\RR (e^{iux}-1) \eta_t(T,x)dx \bigg].  
$$
From the above option price, $C_t^{1,\kappa_t}(T,x)$, we can easily calculate the implied volatility by inverting the Black Scholes formula, assuming that $S_t=1$ and the interest and dividend rates are zero. As discussed at the very end of Subsection \ref{sec:nosym}, this value is the same as the value of implied volatility of a call option for spot level $S_t$, strike $S_t e^x$, and maturity $T$, regardless of what the level of $S_t$ is (hence, we don't need to simulate it).
Using this method, we simulate the implied volatility surfaces five days into the future starting from Dec. 13, 2007, as shown in Figures~\ref{fg:implementation:simulation1} and~\ref{fg:implementation:simulation2}. 
%From these results, we can see that the simulated $\kappa$'s, although not precisely the same, resemble the surface $\kappa$ computed on Dec. 13, 2007 in terms of its shape and magnitude. The same observation applies to the simulated implied volatility surfaces.

\section{Discrete tangent L\'{e}vy models\label{sec:dtL}}

\subsection{Model setup and consistency conditions} \label{sec:dtL:model}

In this section, we work with a different class of tangent L\'{e}vy models in order to solve the same problem -- develop a consistent Monte Carlo simulation algorithm for the future implied volatility surfaces. The new class of tangent L\'{e}vy models is called ``discrete tangent L\'{e}vy models'', as the jump sizes of the logarithm of the tangent process are restricted to finitely many values. As a result, for each fixed maturity, the corresponding L\'evy measure is purely atomic and can be represented by a finite number of parameters. There are several benefits of the new setting as opposed to the one considered in Section \ref{sec:DETLM}. On a theoretical level, the new drift restriction is simplified to a sum, as opposed to an integral which needs to be approximated numerically. In addition the mapping between option prices and L\'evy measure is simplified in this case. The latter makes the calibration problem somewhat easier, which, in turn, allows us to use non-parametric calibration procedure, which can potentially improve the quality of static fitting. The downside of this method is that, despite the existence of an explicit formula that connects the L\'evy density and options prices, solving the optimization problem associated with the non-parametric calibration is still very computationally expensive. We show how to deal with this problem further in the section. It is worth mentioning that discrete-space versions of tangent L\'evy models have been considered in \cite{Zhao2010dtl} and \cite{emmanuel2014discrete}. The results of \cite{Zhao2010dtl} are limited to a single maturity, while the present setting includes multiple maturities. The theoretical results of \cite{emmanuel2014discrete} are very similar to the ones presented in this subsection. However, our choice of a convenient subclass of discrete tangent models and its subsequent numerical implementation are different.

Similar to Section \ref{sec:DETLM}, herein, we assume that the true dynamics of $S$ under the pricing measure $\QQ$ are given by:
\begin{equation}
\label{fo:nosym:Sdynamics.disc}
S_t = S_0 + \int_0^t \int_{\RR} S_{u-}(e^x-1)[M(dx,du)-K(dx,du)],
\end{equation}
where $M$ is an integer-valued random measure with predictable compensator $K(dx,du)$. The only difference between \eqref{fo:nosym:Sdynamics.disc} and \eqref{fo:nosym:Sdynamics} is that, in the above expression, the compensator $K(dx,du)$ may not be absolutely continuous with respect to $dxdu$. In fact, herein, we restrict our analysis to the compensators $K$ of the form:
\begin{equation}\label{fo:dtL:KDef}
K(dx,du) = \sum_{j=1}^N K_u^j \delta_{x^j}(dx)du,
\end{equation}
where every $K^j$ is a nonnegative predictable process, such that
$$
\EE \int_0^{\bT} K^j_u du < \infty,
$$
and $\{x^1,\cdots,x^N\}$ is a finite subset of $\RR$ which does not change with time. Clearly, $x^j$'s correspond to the jump sizes and $K^j$'s -- to their intensities.
At any fixed time $t$, a tangent model for the underlying is given by
\begin{equation}
\label{fo:disc.tangent.proc.def}
\tilde{S}_T = S_t + \int_t^T \int_{\RR} \tilde{S}_{u-}(e^x-1)\left[N_t(dx,du)-\kappa_t(dx,du)\right],
\end{equation}
for $T\in[0,\bar{T}]$, where $N_t(dx,du)$ is a Poisson random measure associated with the jumps of $\log \tilde{S}$ whose compensator is given by a deterministic measure $\kappa_t(du,dx)$. 
Of course, to be consistent with \eqref{fo:dtL:KDef}, we only consider L\'evy measures $\kappa_t$ of the form:
\begin{equation}\label{fo:dtL:kappaDef}
\kappa_t(dx,du) = \sum_{j=1}^N\kappa_t^j(u) \delta_{x^j}(dx)du,
\end{equation}
where every $\kappa^j_t(\cdot)$ is a continuous deterministic function. With a slight abuse of notation, we denote the collection of $\kappa^j_t$, for $j=1,\ldots,N$, by $\kappa_t$. As before, we denote by $C_t^{S_t, \kappa_t}(T,x)$ the call prices produced by the above tangent model at time $t$ (cf. \eqref{fo:nosym:Cdef}). The concept of a tangent model, then, requires that \eqref{fo:nosym:C} holds for all every $t \in [0, \bar{T})$.
Next, we define the joint dynamics of $S_t$ and $\kappa_t$ by
\begin{equation}\label{fo:dtl:dyn}
\left\{
\begin{array}{c}
{ S_t = S_0 + \int_0^t \int_{\RR}S_{u-}(e^x-1)\left[M(dx,du)-\sum_{j=1}^N K_u^j \delta_{x^j}(dx)du\right],}\\
{}\\
{\kappa^j_t(T) = \kappa^j_0(T) + \int_0^{t}\alpha^j_u(T) du + \sum_{n=1}^m \int_0^{t} \beta^{j,n}_u(T) dB^n_u,
\,\,\,\,\,\,\,\,\,\,\,j=1,\ldots,N,}\\
\end{array}
\right.
\end{equation}
where $B=(B^1,\cdots,B^m)$ is an $m$-dimensional standard Brownian motion; $(\kappa_t^j)_{t\in[0,\bar{T}]}$ and $(\alpha_t^j)_{t\in[0,\bar{T}]}$ are progressively measurable stochastic processes taking values in $\mathcal{B}_d$ (as defined in \eqref{fo:review:Bdcondition}); and, for each $j\in\{1,\cdots,N\}$ and $n\in\{1,\cdots,m\}$, $(\beta^{j,n}_t)_{t\in [0,\bar{T}]}$ is a progressively measurable square integrable stochastic process with values in $\mathcal{H}_d$ (as defined in \eqref{fo:review:Hdcondition}).

As before, given the dynamics of $(S_t,\kappa_t)$, we need to ensure that they satisfy certain ``consistency conditions" in order to produce models which are, indeed, tangent to the true model almost surely at all times. 
First, we make the following assumption.
\begin{ass}\label{ass:dtl.1}
The jump sizes $\{x^1,\cdots,x^N\}$ are regularly spaced and have 0 at the center. In other words, the number of feasible jump sizes $N$ is odd, and $x^j = (j - M)d_x$ for $j = 1,\cdots,N$, with $M = (N+1)/2$ and let $d_x>0$ being the spacing between two feasible jump sizes.
\end{ass}
Note that $x^M$ corresponds to jump size $0$, therefore, the value of $\kappa^M_t(T)$ can be set arbitrarily without changing the options' prices.
For convenience, we define it as
\begin{equation} \label{eq.dtl.new.1} 
\kappa^M_t(T) = - \sum_{j\neq M} \kappa^j_t(T),
\end{equation} 
so that
\begin{equation} \label{fo:dtL:xassumption2}
\sum_{j=1}^N \kappa^j_t(T) = 0.
\end{equation}

\begin{ass}\label{ass:dtl.2}
Almost surely, for every $n=1,\ldots,m$ and almost every $t\geq0$, we have: $\beta_t^{k,n} = 0, \forall k < M/2$ or $k > 3M/2$, with $M = (N+1)/2$.
\end{ass}

Finally, we are ready to formulate the consistency conditions for discrete tangent L\'{e}vy models. The theorem presented below, essentially, follows from Proposition 9 in \cite{emmanuel2014discrete}, up to some technical differences in the definition of the code-book dynamics \eqref{fo:dtl:dyn}. However, since the results of \cite{emmanuel2014discrete} have not appeared in a publication, we present an alternative proof of this theorem in Appendix C.

\begin{theorem}\label{th:dtl.consist}
Let Assumptions \ref{ass:dtl.1} and \ref{ass:dtl.2} hold and assume, in addition, that $\left(S_t\right)_{t\in[0,\bT]}$ is a true martingale and $\kappa_t(T,x)\geq0$, almost surely for all $t\in[0,\bT)$ and almost all $(T,x)\in[t,\bT]\times\RR$.
Then the processes $\left(S_t,\kappa_t\right)_{t\in[0,\bT]}$, satisfying (\ref{fo:dtl:dyn}) and (\ref{fo:dtL:xassumption2}),  are \emph{consistent}, in the sense that \eqref{fo:nosym:C} holds, if and only if the following conditions hold almost surely for almost every
$t\in[0,\bT)$ and all $T\in(t,\bT]$:
\begin{enumerate}

\item \textbf{Drift restriction}:
\begin{equation} \label{fo:dtL:driftcondition}
\alpha_t^j(T) = -\sum_{n=1}^m \sum_{\frac{N+1}{4} \le k \le \frac{3(N+1)}{4}}\beta_t^{k,n}(T) \int_t^T \beta_t^{j+M-k,n}(u) du,\,\,\,\,\,\,\,\,\,\,\,\,\,j=1,\ldots,N.
\end{equation}

\item \textbf{Compensator specification}:
\begin{equation}\label{fo:dtL:K}
K^j_t = \kappa^j_t(t),\,\,\,\,\,\,\,\,\,\,\,\,\,\text{for all}\,\,j\neq M.
\end{equation}

\item \textbf{Symmetry condition}:
\begin{equation}\label{fo:dtL:sym}
\sum_{\frac{N+1}{4} \le j \le \frac{3(N+1)}{4}} \beta^{j,n}_t(T) (e^{x^j}-1) = 0,
\,\,\,\,\,\,\,\,\,\,\,\,\,n=1,\ldots,m.
\end{equation}
\end{enumerate}
\end{theorem}

Notice that Theorem~\ref{th:dtl.consist} and equations \eqref{fo:dtl:dyn} provide a method for constructing a market-based model for call prices, or consistent model for implied volatility surface. We start by choosing $(\beta^1_t\ldots,\beta^m)_{t\in[0,\bar{T}]}$, so that they are consistent with historical evolution, satisfy Assumption \ref{ass:dtl.2} and the following linear constraints:
\begin{equation*}
\sum_{ \frac{N+1}{4} \leq j \leq \frac{3(N+1)}{4}} \beta^{j,n}_t(T) = 0, 
\,\,\,\,\,\,\,\,\,\,\,\,\,\,\sum_{\frac{N+1}{4} \le j \le \frac{3(N+1)}{4}} \beta^{j,n}_t(T) (e^{x^j} - 1) = 0.
\end{equation*}
Then, we use the drift restriction \eqref{fo:dtL:driftcondition} and the second equation in \eqref{fo:dtl:dyn} to generate the paths of $(\kappa_t)_{t\in[0,\bar{T}]}$. Notice that (\ref{fo:dtL:xassumption2}) is satisfied with such  choice of $\alpha$ and $\beta$, which follows from
\begin{align*}
\sum_{j=1}^N \alpha^j_t(T) & = -\sum_{j=1}^N \sum_{n=1}^m \sum_{\frac{N+1}{4} \leq k \leq \frac{3(N+1)}{4}} \beta_t^{k,n}(T)\bbeta_t^{j+M-k,n}(T) \\
& = -\sum_{n=1}^m \sum_{\frac{N+1}{4} \leq k \leq \frac{3(N+1)}{4}}\beta_t^{k,n}(T) \sum_{\frac{N+1}{2}+1 \leq u \leq \frac{N+1}{2}+N}\bbeta_t^{u-k,n}(T) = 0,
\end{align*}
where the last equality holds because the range of $u$ covers $[\frac{N+1}{4}, \frac{3(N+1)}{4}]$ regardless of the value of $k$.
Having simulated the paths of $(\kappa_t)_{t\in[0,\bar{T}]}$, we compute the associated call prices $C_t^{1,\kappa_t}$. There is no need to simulate $S_t$, because, as discussed at the end of Subsection \ref{sec:nosym}, for any value of $S_t$, the implied volatility in the log-moneyness variable can be computed by the inverse Black-Scholes formula applied to $C_t^{1,\kappa_t}$.

Similar to the discussion in Subsection \ref{sec:implementation}, the implementation of discrete tangent L\'evy models consists of two stages -- estimation and simulation. The estimation, in turn, consists of two parts: static fitting and dynamic fitting. All these steps are addressed in the following subsections. The market data consists of preprocessed SPX call option prices described in Subsection~\ref{sec:implementation:data}.

\subsection{Static fitting}\label{subsec:dtL:static}

Recall that static fitting refers to the process by which we identify the additive measures $\kappa_t$. This is done ``statically'' for each day $t$.
As shown further in this subsection, in the case of discrete tangent L\'{e}vy models, the PIDE for call prices can be simplified to a finite system of ordinary differential equations (ODEs). Moreover, the discrete tangent L\'{e}vy models allow us to perform non-parametric calibration of tangent processes to market data, which helps us better understand the true structure of jump measures. By employing non-parametric fitting with more variables, we allow more flexibility in the model hence we can obtain better fitting results. However, in order to obtain reasonable and consistent jump measures, a few technical difficulties associated with non-convex optimization problems and their potential instability need to be addressed. In this subsection, we explain how we overcome these difficulties in detail.

We first describe the preliminary constructions. As specified in Section \ref{sec:dtL:model}, we assume the set of feasible jump sizes $\{x^j\}_{j=1}^N$ is equally spaced with zero at the center, so each $x^j$ can be expressed as (recall that $M = \frac{N+1}{2}$ is the center point with $x^M = 0$)
\begin{equation*}
-A + (j-1)dx \quad \text{with} \quad A = (M-1)dx = \frac{N-1}{2}dx,
\end{equation*}
and $(-A, A)$ is the truncation of the $x$-domain.
The choice of $N$ and $dx$ requires some consideration. Ideally, we would like $N$ to be as large as possible, to fully cover the structure of jumps, but a trade-off with the computation complexity and the limited data size is also important. As for the jump size spacing $dx$, it not only determines the value of $A$ (the truncation of the $x$-domain), but also has a big impact on the jump intensities $\kappa^j$'s. On the one hand, $dx$ should be large enough, so that $(-A, A)$ covers the possible range of jump sizes and option prices. On the other hand, if $dx$ is too large, the grid of jump sizes is too sparse, and the calibrated $\kappa$ might not reflect the true jump structure. A quick test on historical S\&P 500 prices in the year of 2007 shows that the magnitude of most daily jumps falls between -0.035 and 0.029. Taking these facts into account, we decided to take $N = 301$ and $dx = 0.005$ after a few trials, so the center point $M = 151$ and the space truncation $A = 0.75$.
However, the large number of variables makes the calibration (i.e. static fitting) nearly infeasible, as the procedure becomes extremely slow and the results are unstable. To address these issues, we reduce the number of parameters from $N=301$ to $Nvar = 24$, by assuming $\kappa^j = 0$ for large positive or negative jumps,  and by dividing the jump sizes into small groups and making adjacent jump sizes share the same value of $\kappa^j$. 
The reduction speeds up the procedure significantly, and the calibration results look much more reasonable as the dimension of the optimization problem is greatly reduced. At the same time, the original grid with $N = 301$ and $dx = 0.005$ is kept sufficiently large, to allow for a good quality of fit to the market call prices.

To perform the calibration, we need to develop a procedure to compute call prices from the calibration variables $\kappa^j$'s. Equation \eqref{fo:dtL:callPIDE} in Appendix C provides the following PDE for call prices: 
\begin{align} \label{fo:dtL:CPDE}
\partial_{T} C^{S_t,\kappa_t}_t(T,x)= & \underset{j:j\neq M}\sum \kappa_t^j(T) e^{x_j} C^{S_t,\kappa_t}_t(T,x-x^j) - C^{S_t,\kappa_t}_t(T,x) \cdot \underset{j:j\neq M}\sum \kappa_t^j(T) \nonumber \\
& \qquad + [\partial_x -1]C^{S_t,\kappa_t}_t(T,x) \cdot \Omega(T-t),
\end{align}
where $\Omega$ is the symmetry index defined by 
\begin{equation}\label{eq.dtl.kappasym}
\Omega(T-t) = \sum_{k\neq M} \kappa^j_t(T) \left(e^{x^j}-1\right)
\end{equation}
It is not hard to see that, if \eqref{fo:dtL:driftcondition}, \eqref{fo:dtL:sym} and Assumption \ref{ass:dtl.2} hold, then $\Omega(\cdot)$ is a deterministic function.
Notice that \eqref{fo:dtL:CPDE} is a PDE because of the presence of the derivative $\partial_x C^{S_t,\kappa_t}_t(T,x)$ on the right-hand. Let us now perform a change of variables that eliminates this derivative and reduces the equation to a (multi-dimensional) ODE. Namely, we define
\begin{equation}
\tilde{C}_t(T,x) := C^{S_t,\kappa_t}_t\left(T,x - \int_t^T \Omega(u-t) du\right)
\end{equation}
This change of variables translates \eqref{fo:dtL:CPDE} to the following ODE of $\tC_t(T,x)$:
\begin{equation}\label{fo:dtL:CODE}
\partial_T \tC_t(T,x) = \underset{j:j\neq M}\sum \kappa_t^j(T) e^{x^j} \tC_t(T,x-x^j) -  \tC_t(T,x)\underset{j:j\neq M}\sum \kappa^j_t(T) - \Omega(T-t) \cdot \tC_t(T,x),
\end{equation}
which doesn't contain any derivatives in the right hand side.
Still, we will not use \eqref{fo:dtL:CODE} directly in computing option prices. Similar to Subsection~\ref{sec:implementation:static}, we would work with option time values instead of option prices. 
In our problem, we define the modified time value function as
\begin{equation}
\tV^j_t(T) = \tC_t(T,x^j+\log S_t) - S_t(1-e^{x^j})^+.
\end{equation}
The evolution of $\tV$ at the grid-points is, then, given by
\begin{equation}\label{fo:dtL:VPIDE}
\left\{
\begin{aligned}
\partial_T \tV^i_t(T) = & \underset{j:j\neq M}\sum \kappa_t^j(T) e^{x^j} \tV_t(T,x^i-x^j) - \tV_t(T, x^i)\underset{j:j\neq M}\sum \kappa_t^j(T) \\
& \quad -S_t (1-e^{x^i})^+\underset{j:j\neq M}\sum \kappa_t^j(T) + S_t \underset{j:j\neq M}\sum \kappa_t^j(T) e^{x^j} (1-e^{x^i-x^j})^+ \\
& \quad - \Omega(T-t) \left(\tV_t(T,x^i) + S_t (1-e^{x^i})^+ \right), \quad \quad i = 1,\cdots,N\\
\tV^i_t(t) = & 0, \quad i= 1,\cdots,N.\\
\end{aligned}
\right.
\end{equation}
Let $\{T_1, T_2,\cdots,T_L\}$ denote the set of maturities available. We now describe the procedure to compute $\{\tV_t(T_l)\}_{l=1}^L$, given $\kappa_t$, using \eqref{fo:dtL:VPIDE}. Consider the time average of $\kappa^j_t(T)$ between each pair of consecutive maturities $[T_{l-1}, T_l)$ (assuming $T_0 = t$):
\begin{equation}\label{fo:dtL:barkappa}
\theta^j_t(T_l) = \frac{1}{T_l - T_{l-1}} \int_{T_{l-1}}^{T_l} \kappa^j_t(u) du, \quad 1\le l\le L.
\end{equation}
It is easy to see that the option time value doesn't change if we substitute every $\kappa^j_t(\cdot)$ to a piece-wise constant function, whose value between each $T_{l-1}$ and $T_l$ are given by $\theta^j_t(T_l)$.
With this substitution, the coefficients in the right hand side of the PDE \eqref{fo:dtL:VPIDE} become constant for $T\in(T_{l-1},T_l)$. Hence, we can solve it to obtain an iterative formula for the solution at maturities $T_l$: \begin{equation}\label{fo:dtL:PDEsol}
\tV_t(T_l) = e^{(T_l-T_{l-1})G_t(T_l)}\tV_t(T_{l-1}) + (e^{(T_l-T_{l-1})G_t(T_l)} - I)G_t^{-1}(T_l) b_t(T_l), \quad l= 1,\cdots,L,
\end{equation}
where the matrix $G_t(T_l) = \{g_{ij}(T_l)\}_{i,j= 1}^N$ has the following entries:
\begin{equation*}
g_{ij}(T_l) = \bone_{\{i \neq j, 1 \leq i-j+M \leq N\}} e^{x^{i-j+M}} \theta_t^{i-j+M}(T_l) - \bone_{\{i=j\}} \bigg(\underset{k:k\neq M}\sum \theta^k_t(T_l) + \bOmega(T_l-t) \bigg),
\end{equation*}
and $b_t(T_l) = (b_1(T_l),\cdots,b_N(T_l))^T$ is given by:
\begin{equation*}
b_i(T_l) = -S_t(1-e^{x^i})^+ \left(\underset{k:k\neq M}\sum \theta_t^k(T_l)+\bOmega(T_l-t)\right) + S_t \underset{k:k\neq M}\sum \theta_t^k(T_l) e^{x^k} (1-e^{x^i-x^k})^+,
\end{equation*}
with 
\begin{equation}\label{fo:dtL:avgsym}
\bOmega(T_l-t) = \frac{1}{T_l - T_{l-1}} \int_{T_{l-1}}^{T_l} \Omega(u-t) du. 
\end{equation}
It is easy to see that $G_t(T_l)$ and $b_t(T_l)$ only depend on $\{\theta_t^j(T_l)\}_{j=1}^N$ as does $\tV_t(T_l)$. Notice that the true time value (the value by which the option price exceeds the intrinsic value) can be recovered from the modified time value by
\begin{equation}\label{fo:dtL:trueV}
V^{S_t,\theta_t,j}_t(T) = C^{S_t,\kappa_t}_t\left(T,x^j - \int_t^T \Omega(u-t) du + \log S_t\right) - S_t\left(1-e^{x^j - \int_t^T \Omega(u-t) du}\right)^+ 
\end{equation}
\begin{equation*}
= \tV_t\left(T,x^j \right) + S_t\left(1-e^{x^j}\right)^+ - S_t\left(1-e^{x^j - \int_t^T \Omega(u-t) du}\right)^+.
\end{equation*}
Thus, \eqref{fo:dtL:PDEsol} and \eqref{fo:dtL:trueV} establish the relation between the time values $V^{S_t,\theta_t,j}_t(T_l)$'s and the average jump intensities $\theta^j_t(T_l)$'s.

Let $V_t^{mkt,j}(T_l)$ denote the market time value at time $t$ of the call option with strike $S_t e^{x^j - \int_t^T \Omega(u-t) du}$ and maturity $T_l$. If any of these strike values are not traded in the market, we use linear interpolation in log-strike to obtain the value of $V_t^{mkt,j}(T_l)$'s. Increasing $l$, we formulate the calibration problem for each $T_l$ separately: find nonnegative $\{\theta_t^j(T_l)\}_{j\neq M}$ which minimize the difference between the market time values $\{V_t^{mkt,j}(T_l)\}_{j=1}^N$ and the model time values $\{V^{S_t,\theta_t,j}_t(T_l)\}_{j=1}^N$, under the symmetry condition
\begin{equation}\label{eq.bOmega.theta}
\sum_{k\neq M} \theta^j_t(T_l) \left(e^{x^j}-1\right) = \bOmega(T_l-t),
\end{equation}
with some fixed deterministic function $\bOmega$. This symmetry condition ensures that \eqref{eq.dtl.kappasym} holds, which, in turn, implies that \eqref{fo:dtL:driftcondition}, \eqref{fo:dtL:sym} and Assumption \ref{ass:dtl.2} hold.
Notice that we formulate the calibration problem for each maturity separately because we can construct the time values $\{V^{S_t,\theta_t,j}_t(T_l)\}_{j=1}^N$ using only $\{\theta_t^j(T_l)\}_{j\neq M}$ and the already constructed (calibrated) time values for previous maturity, $\{V^{S_t,\theta_t,j}_t(T_{l-1})\}_{j=1}^N$.
The preliminary optimization problem becomes:
\begin{align}\label{fo:dtL:optimization_ori}
\min_{\theta_t(T_l) \ge 0} \quad & \sum_{j=1}^N \omega_j |V_t^{S_t,\theta_t,j}(T_l) - V_t^{mkt,j}(T_l)|^2\\
& \text{s.t.} \quad \sum_{j\neq M} \theta_t^j(T_l) (e^{x^j} - 1) = \bOmega(T_l - t), \nonumber
\end{align}
where $\omega_j = \frac{1}{|V_t^{bid,j}(T_l) - V_t^{ask,j}(T_l)|^2}$ are the weights associated with liquidities. As before, we put more weights on contracts with better liquidity, which is reflected by smaller bid-ask spread. Once the average intensities $\theta_t^j(T_l)$ are constructed for all $j$ and $l$, we recover $\left\{\kappa^j_t(\cdot)\right\}$ using \eqref{fo:dtL:barkappa} and interpolation across maturities.

\begin{figure}[htp]
   \begin{center}
       \begin{tabular}{cc}
       \includegraphics[width=0.45\columnwidth]{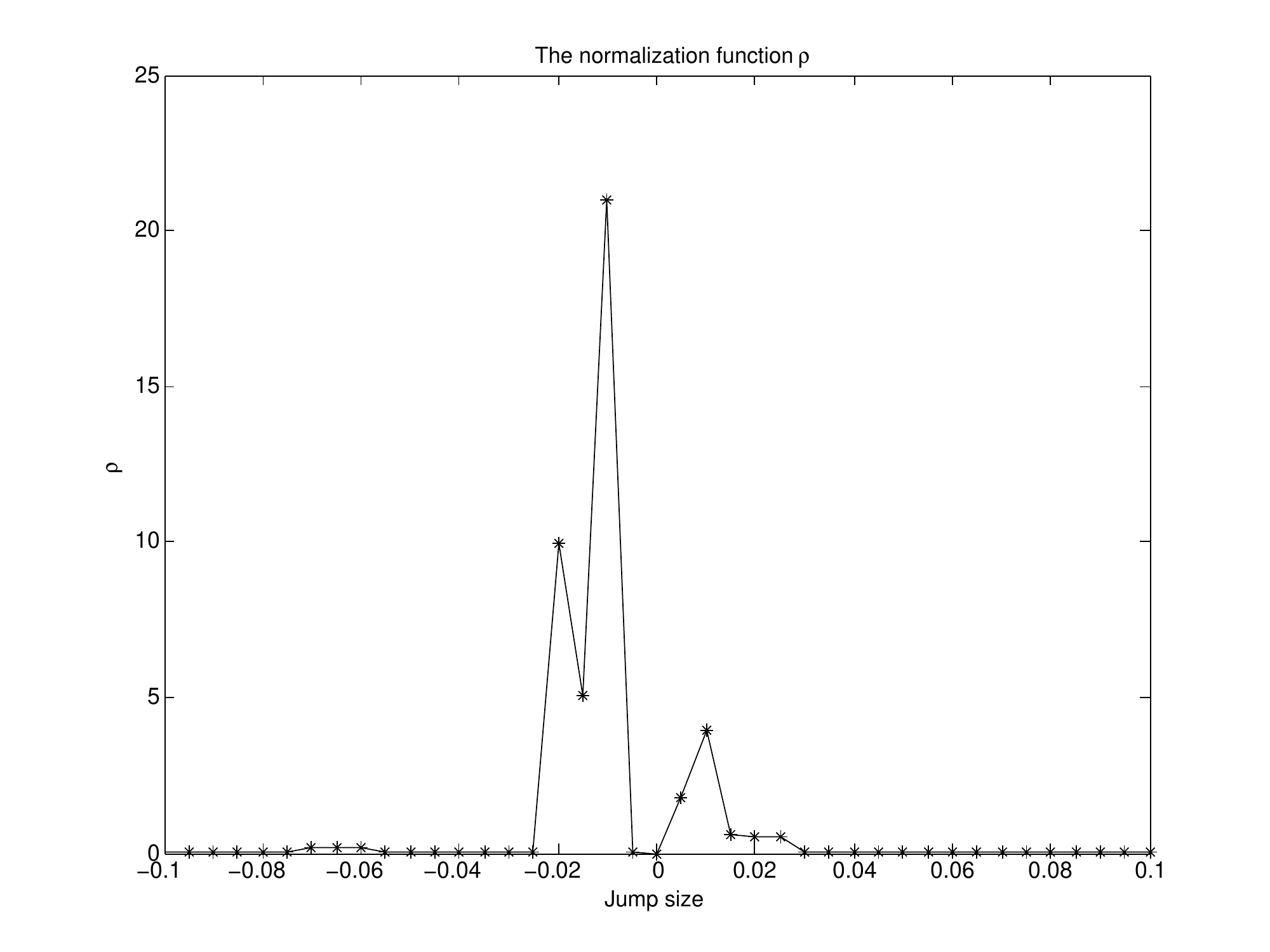}
       &
       \includegraphics[width=0.45\columnwidth]{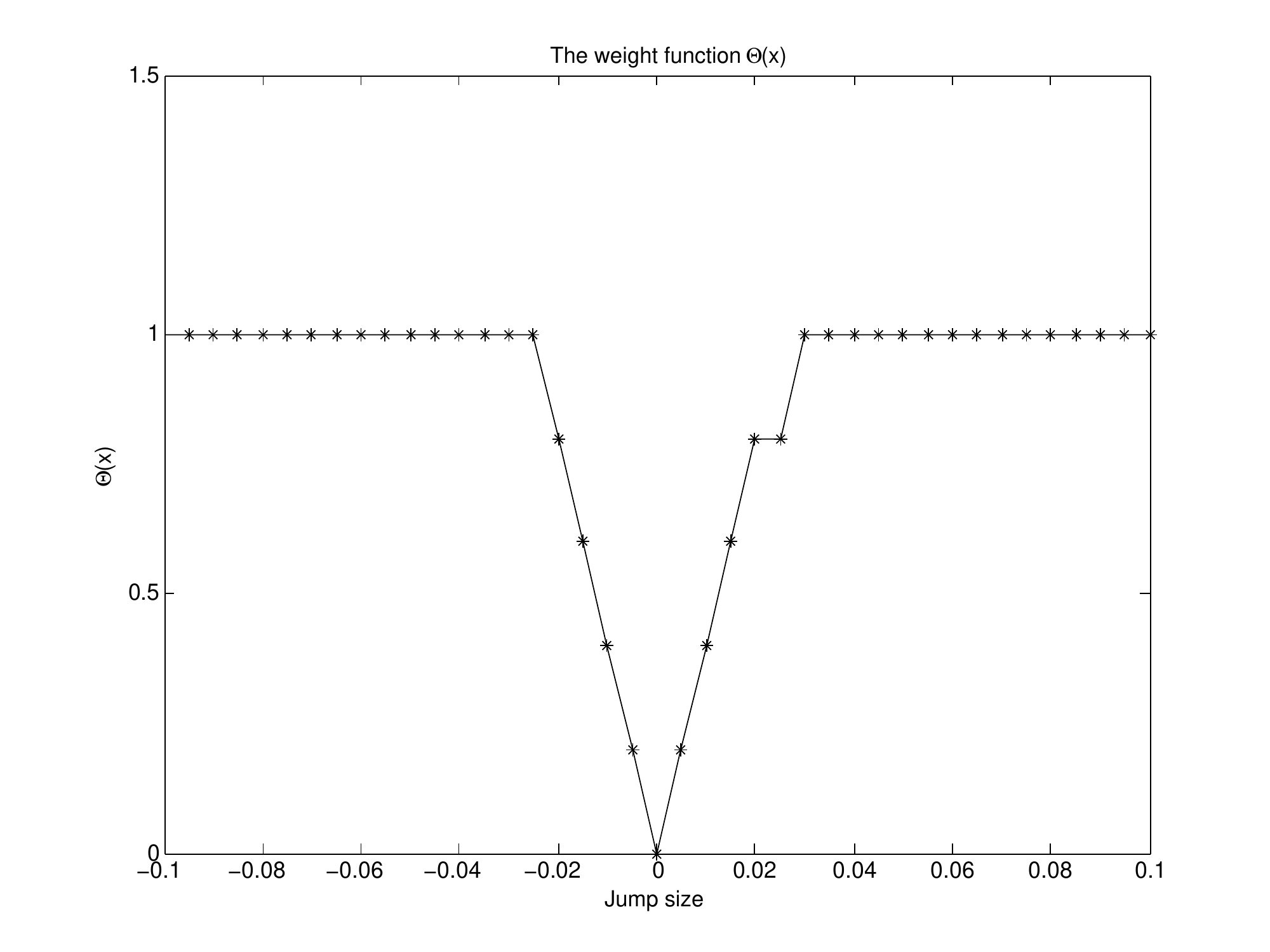}
       \\
      \fontsize{7}{12}\selectfont (a) The normalization function $\rho$ 
     &
      \fontsize{7}{12}\selectfont (b) The weight function $\Theta$
      \\
       \includegraphics[width=0.45\columnwidth]{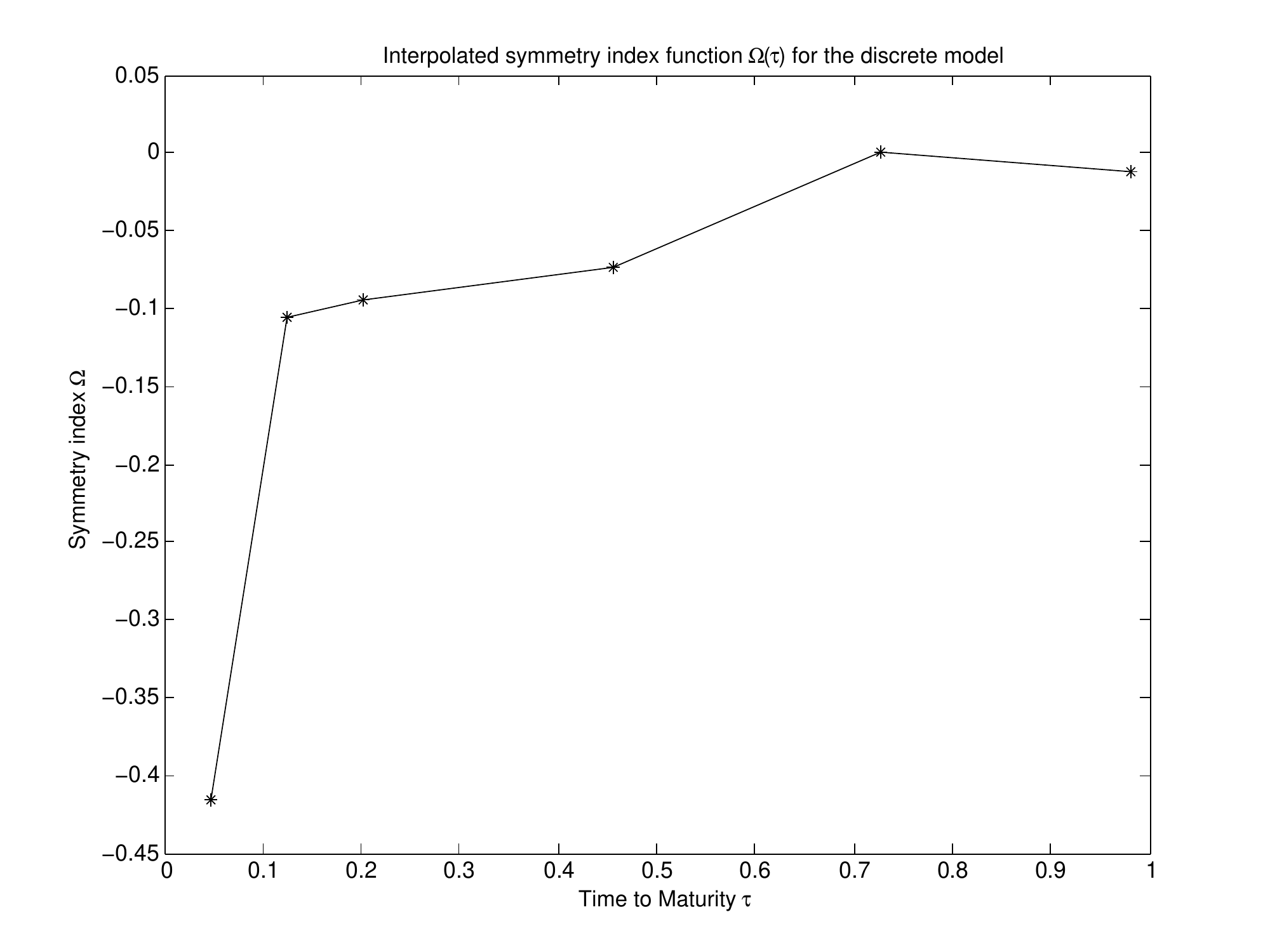}
       %&
       %\includegraphics[width=0.45\columnwidth]{sec-implementation/figures/kappa_sample.pdf}
       \\
      \fontsize{7}{12}\selectfont (c) Average symmetry index $\bOmega$ as a function of time to maturity   
     &
      \fontsize{7}{12}\selectfont       
      \\
       \end{tabular}
   \end{center}
   \vspace{-10pt}
   \caption{The normalization function $\rho$, the weight function $\Theta$, and the average symmetry index $\bOmega$}
   \label{fg:implementation.dtl.new}
\end{figure}

It is easy to see that \eqref{fo:dtL:optimization_ori} is a non-convex optimization problem, and numerical algorithm may get stuck in a local minimum. In our implementation, we took three actions: (1) choosing an appropriate normalization function and working with normalized jump intensities, (2) carefully choosing the initial point, and (3) adding several penalization terms to regularize the problem. We now describe the three actions in detail:

\begin{description}
\item[Action 1]
Notice that the average jump intensities $\theta^j_t(T)$'s can vary in size significantly across $j$ (e.g. recall that the small jumps occur much more frequently than the large ones). This, in turn, may cause numerical difficulties in the optimization.
To address this problem, we will use a normalization function denoted by $\rho(x) = \sum_{j\neq M}\rho^j \delta_{x^j}(x)$, and perform the optimization for the normalized average jump intensities 
$$
\tilde{\theta}_t^j(T_l) := \theta_t^j(T_l)/\rho^j
$$ 
We choose the normalization function $\rho$ to be given by $\theta_0(T_1)$, which is obtained by solving \eqref{fo:dtL:optimization_ori} for the first maturity and the first day in the calibration period, without a symmetry constraint. This function is shown in Figure~\ref{fg:implementation.dtl.new}.a. We then require the optimization variables $\{\tilde{\theta}_t^j(T_l)\}_{j,l}$ to stay in a reasonable scale, to avoid the overflow problems.

\item[Action 2]
Motivated by the choice of the normalization function, on each calibration day, we set the initial value for the optimization variable $\tilde{\theta}_t(T_1)$ to be a vector of 1's. We use a calibrated $\tilde{\theta}_t(T_{l-1})$ as the initial value for $\tilde{\theta}_t(T_{l})$. 

\item[Action 3]
To improve the convexity properties of the objective function and, therefore, make the optimizer be more likely to converge, we add the three penalization terms:
\begin{description}
\item[Penalization term 1]  \begin{equation}\label{fo:dtL:penalized1}
    F^l_1(\tilde{\theta}_t(T_l)) = \sum_{j\neq M} (\tilde{\theta}^j_t(T_l) - \tilde{\theta}^j_t(T_{l-1}) )^2,
    \end{equation}
with $\tilde{\theta}^j_t(T_0) = 1$. This penalization term also ensures that $\tilde{\theta}^j_t(T)$'s is not varying too much across maturities $T$. 
    
\item[Penalization term 2]
\begin{equation} \label{fo:dtL:penalized2}
F_2(\tilde{\theta}_t(T_l)) = \left( \sum_{j=1}^{M-1} (\tilde{\theta}_t^j(T) - \tilde{\theta}_t^{j-1}(T) )^2 \Theta (x^j) + \sum_{j=M+1}^{N}  (\tilde{\theta}_t^j(T) - \tilde{\theta}_t^{j+1}(T))^2 \Theta(x^j) \right)
\end{equation}
This penalization term also ensures that $\tilde{\theta}^j_t(T)$'s is not varying too much across $j$. Here, $\Theta(x)$ is a fixed weight function, with which we penalize the differences at larger jumps more heavily -- it is shown in Figure~\ref{fg:implementation.dtl.new}.b.

\item[Penalization term 3]
    \begin{equation} \label{fo:dtL:penalized3}
    F_3(\tilde{\theta}_t(T_l)) = \sum_{j=1}^N \frac{1}{\tilde{\theta}_t^j(T_i)}
\end{equation}
    This penalization term also keeps $\tilde{\theta}^j_t$'s away from zero, so that, if we simulate the future values of $\tilde{\theta}^j_{u}$, starting from $u=t$, they are less likely to touch zero.
\end{description}

Summing up the above, we formulate the resulting optimization problem. For each day $t\in[0,\bar{T}]$ and each maturity $T_l$, we run the penalized optimization
\begin{align}\label{fo:dtL:optimization_penalized}
\min_{\tilde{\theta}_t(T_l) \ge 0} \quad & \sum_{j=1}^N \omega_j \left|V_t^{S_t,\rho\tilde{\theta}_t,j}(T_l) - V_t^{mkt,j}(T_l)\right|^2 +  \epsilon_1 F^l_1(\tilde{\theta}_t(T_l)) + \epsilon_2 F_2(\tilde{\theta}_t(T_l)) + \epsilon_3 F_3(\tilde{\theta}_t(T_l)) \\
& \text{s.t.} \quad \sum_{j\neq M} \rho^j \tilde{\theta}_t^j(T_l) (e^{x^j} - 1) = \bOmega(T_l - t), \nonumber
\end{align}
The coefficients of the penalization terms $\epsilon_1$, $\epsilon_2$ and $\epsilon_3$ can be determined by a bisection method as follows.

\begin{algorithm}[H]
\label{algo:dtL:epsilon}
\caption{Algorithm for determining $\epsilon_1$, $\epsilon_2$ and $\epsilon_3$}
Run the unpenalized optimization \eqref{fo:dtL:optimization_ori} without the symmetry constraint, and record the value of the objective function as $f_0$\;
      \For{$i = 1:3$}{
      $\epsilon_i = 5$\;
      $f = 100 f_0$\;
        \While{$f > 1.05 f_0$}{
            Run the penalized optimization \eqref{fo:dtL:optimization_penalized} with the $i$-th penalization term only, and record the value of $\sum_{j=1}^N \omega_j |V_t^{S_t,\rho\tilde{\theta}_t,j}(T_l) - V_t^{mkt,j}(T_l)|^2$ as $f$\;
            $\epsilon_i = \epsilon_i/2$\;
      }
    }
\end{algorithm}

Having run the Algorithm~\ref{algo:dtL:epsilon} for different $t$ and $T_l$, we choose the following values: $\epsilon_1 = 0.3125$, $\epsilon_2 = 0.3125$, and $\epsilon_3 = 0.0012$.
\end{description}
The average symmetry index function $\bOmega(\cdot)$ (introduced in \eqref{fo:dtL:avgsym}) is then determined using $t=0$ -- the first day of the calibration horizon -- by running the penalized optimization \eqref{fo:dtL:optimization_penalized} without the symmetry constraint and setting
\begin{equation}\label{fo:dtL:symmetryinit}
\bOmega(T_l - t_0) = \sum_{j=1}^N \rho^j \tilde{\theta}_{0}^j(T_l) (e^{x^j} - 1).
\end{equation}
We interpolate linearly to obtain the values of $\bOmega(\tau)$ for all $\tau \in\RR_+$.
Before we present the results, we summarize the static fitting procedure as follows.

\begin{algorithm}[H]
\label{algo:dtL:calibration}
\caption{Algorithm for fitting $\tilde{\theta}$}
Preprocess the market data according to \eqref{fo:dtl:transform}\;
Determine $\epsilon_1$, $\epsilon_2$ and $\epsilon_3$ - the coefficients of the penalization terms - by Algorithm~\ref{algo:dtL:epsilon}\;
On $t=0$, starting from the initial point $\tilde{\theta}_{init}$, specified in Action 2, run the penalized optimization \eqref{fo:dtL:optimization_penalized}, without the symmetry constraint, and compute the average symmetry index $\bOmega$ via \eqref{fo:dtL:symmetryinit} and linear interpolation\;
For all days $t \in (0,\bT]$, run the constrained penalized optimization \eqref{fo:dtL:optimization_penalized} with $\tilde{\theta}_{init}$ and $\bOmega$. Save the time series $(\tilde{\theta}_t)_{t \in [0,\bT]}$.
\end{algorithm}

\begin{figure}[htp]
   \begin{center}
       \begin{tabular}{cc}
       \includegraphics[width=0.45\columnwidth]{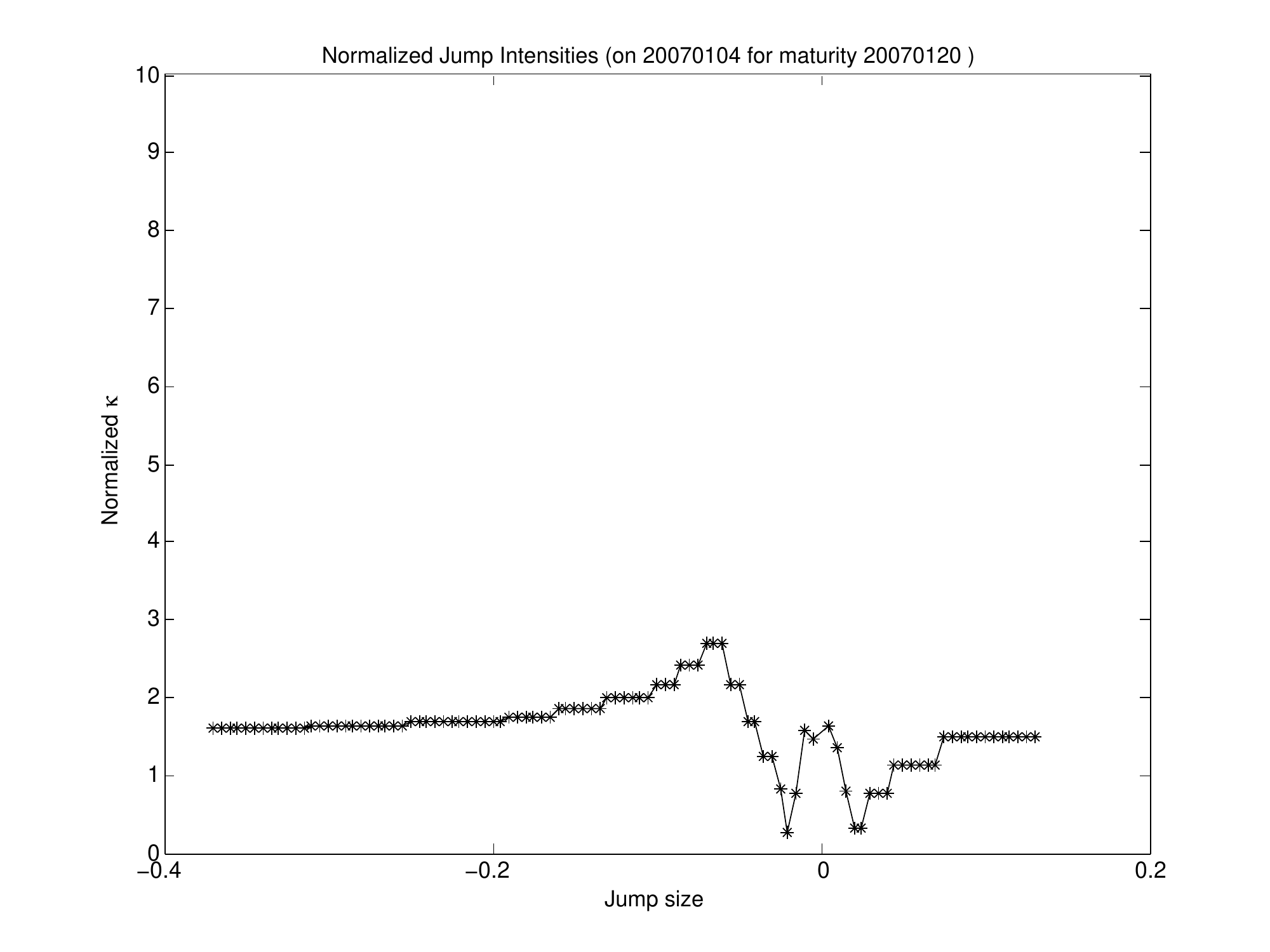}
       &
       \includegraphics[width=0.45\columnwidth]{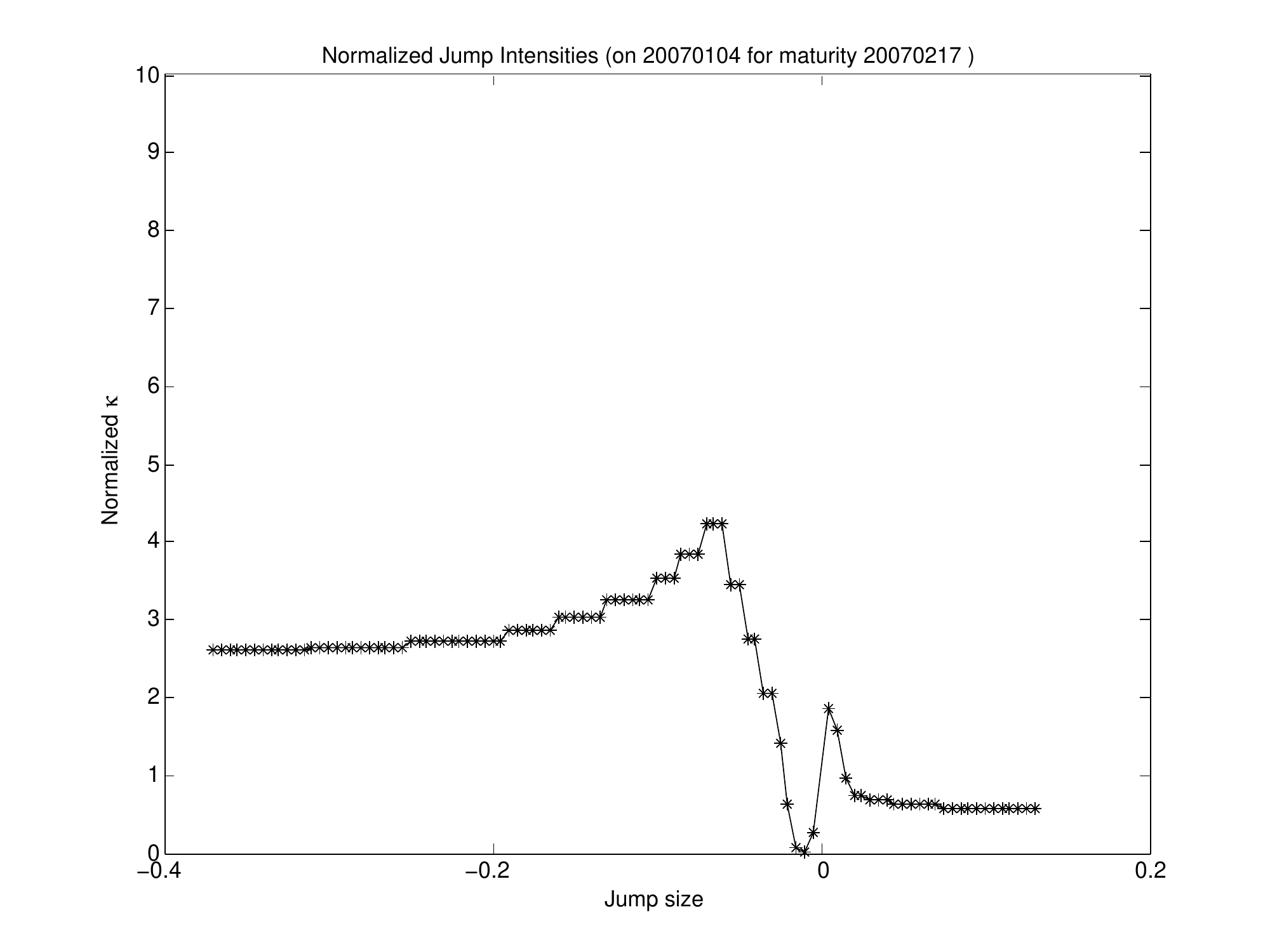}
       \\
      \fontsize{7}{12}\selectfont (a) Maturity Jan.20,2007
     &
      \fontsize{7}{12}\selectfont (b) Maturity Feb.17,2007
%      &
%      \fontsize{7}{12}\selectfont (c) $\beta^3$
      \\
      \includegraphics[width=0.45\columnwidth]{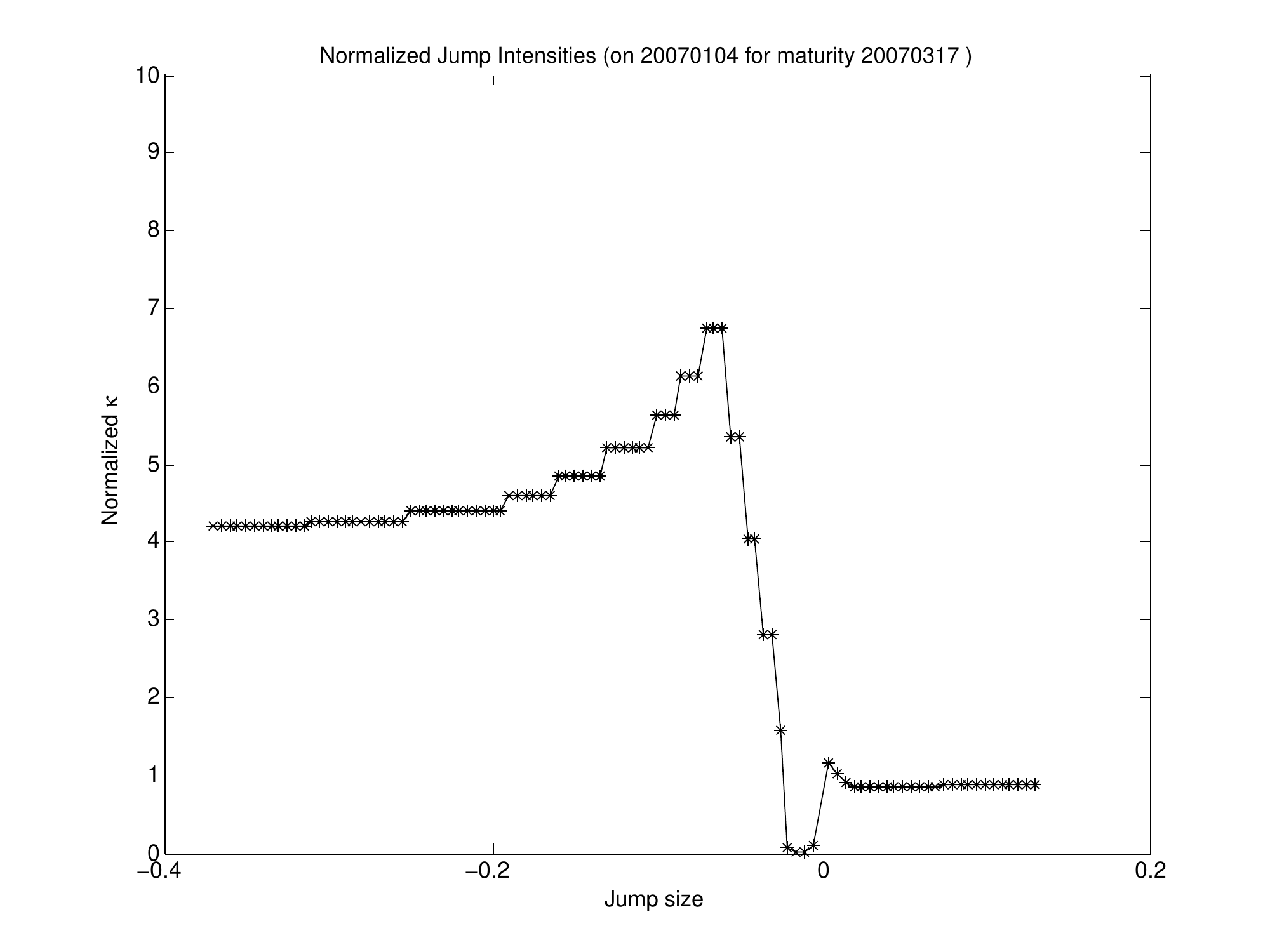}
       &
       \includegraphics[width=0.45\columnwidth]{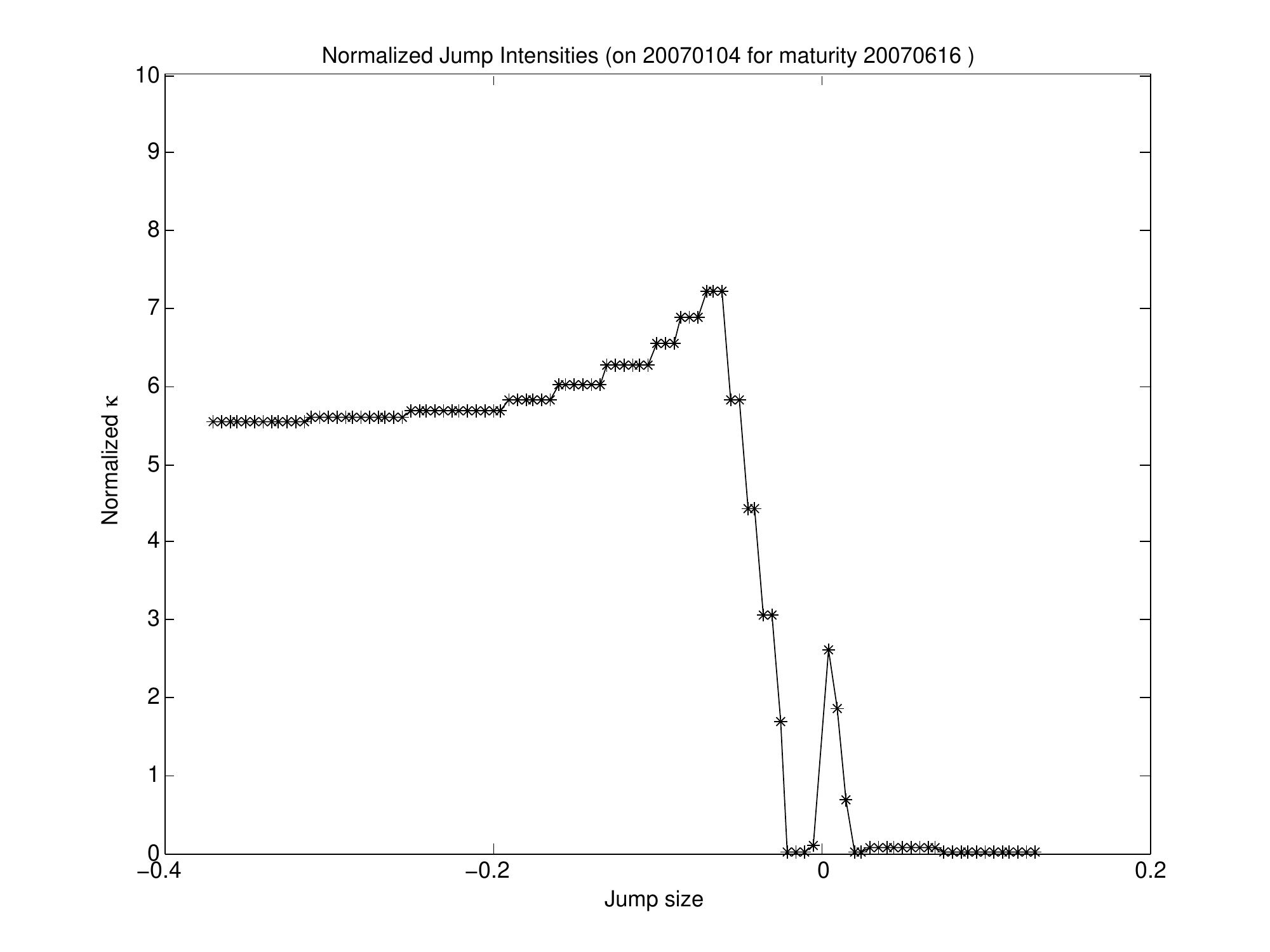}
       \\
      \fontsize{7}{12}\selectfont (c) Maturity Mar.17,2007
     &
      \fontsize{7}{12}\selectfont (d) Maturity Jun.16,2007
      \\
      \includegraphics[width=0.45\columnwidth]{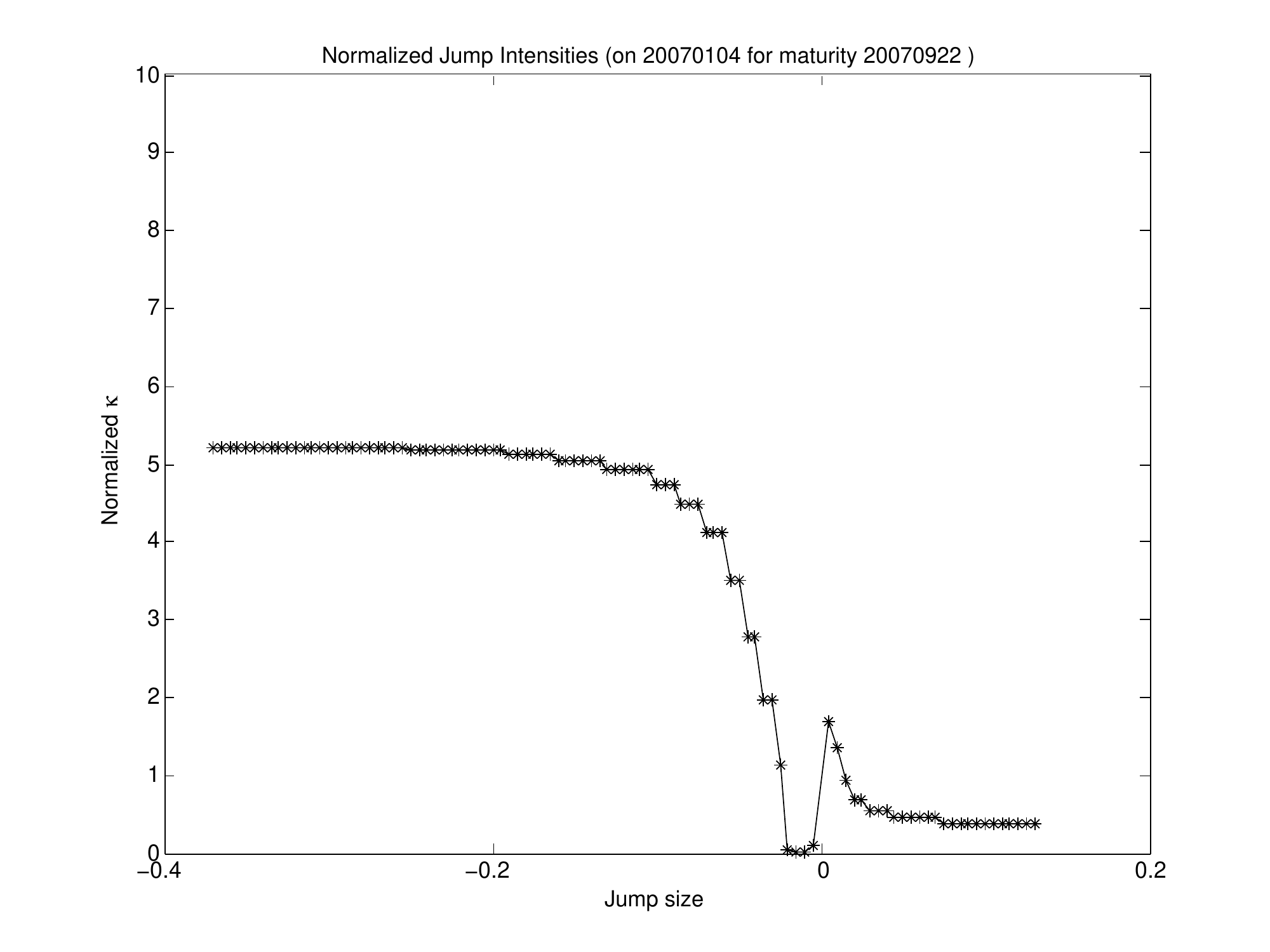}
       &
       \includegraphics[width=0.45\columnwidth]{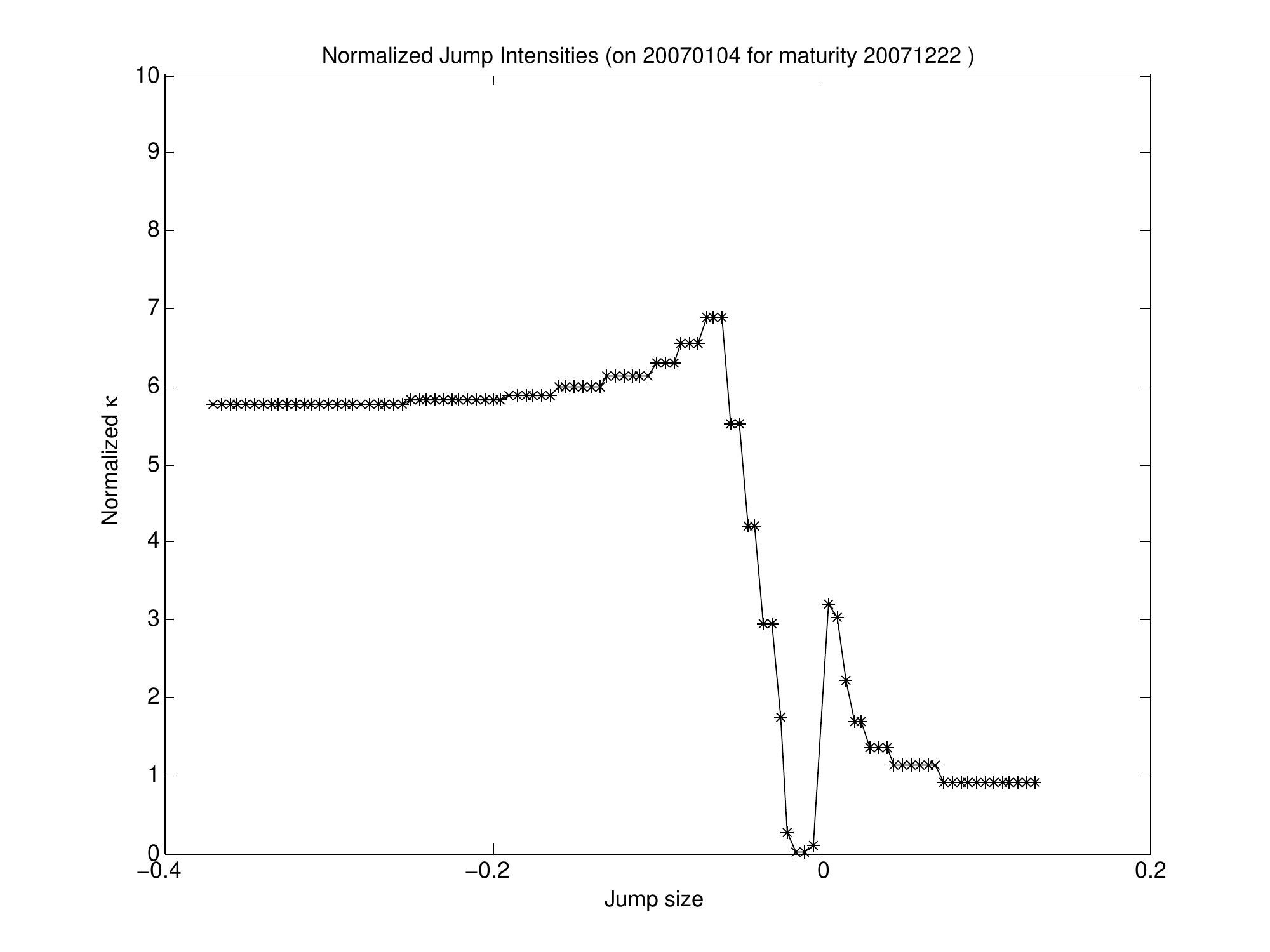}
       \\
      \fontsize{7}{12}\selectfont (e) Maturity Sep.22,2007
     &
      \fontsize{7}{12}\selectfont (f) Maturity Dec.22,2007
\\
       \end{tabular}
   \end{center}
   \vspace{-10pt}
   \caption{Calibrated average normalized jump intensities $\tilde{\theta}(T_l)$ for DTL model on the second day, Jan. 4, 2007}
   \label{fg:dtL:tkappa_day2}
\end{figure}

\begin{figure}[htp]
   \begin{center}
       \begin{tabular}{cc}
       \includegraphics[width=0.45\columnwidth]{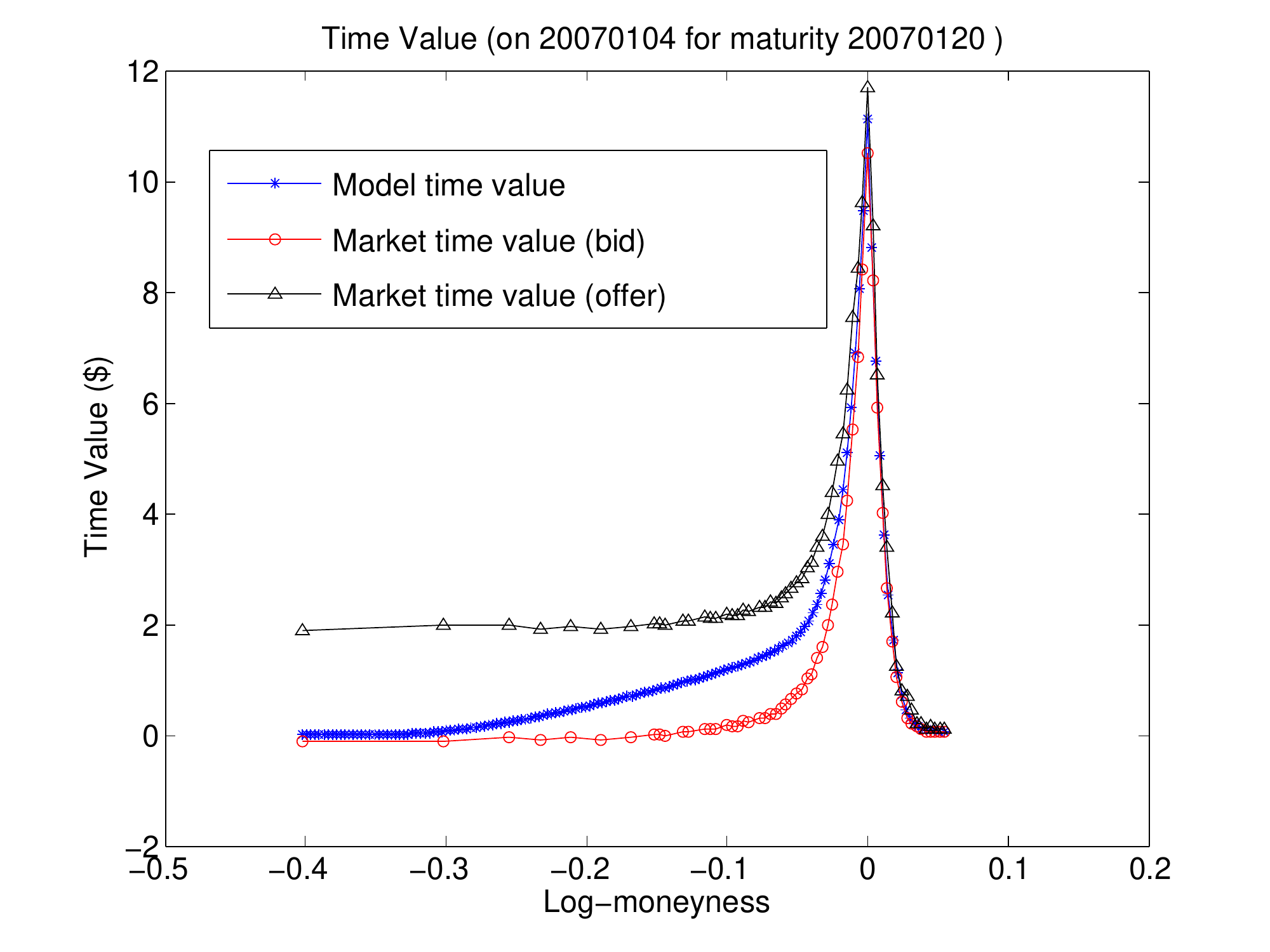}
       &
       \includegraphics[width=0.45\columnwidth]{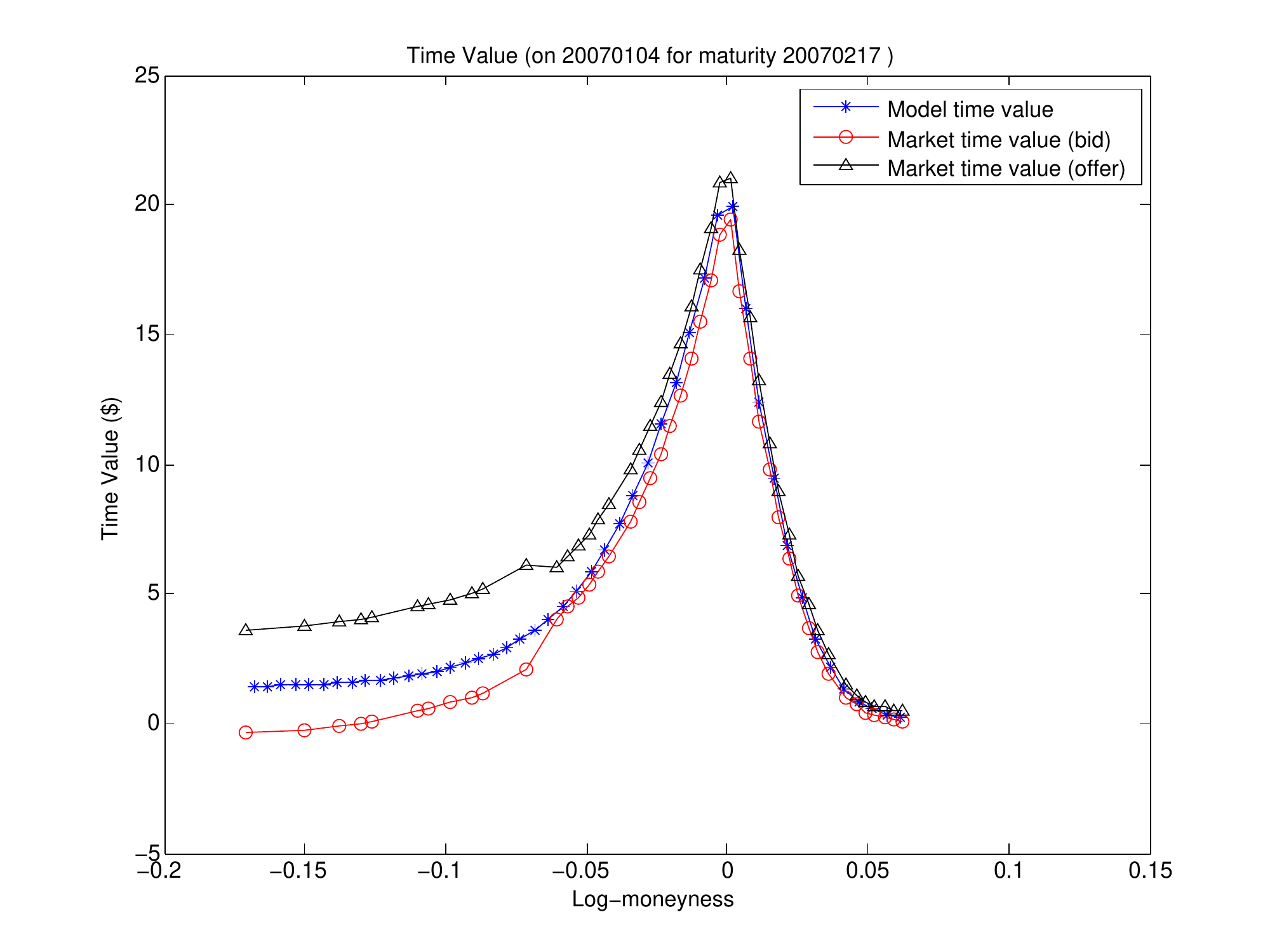}
       \\
      \fontsize{7}{12}\selectfont (a) Maturity Jan.20,2007
     &
      \fontsize{7}{12}\selectfont (b) Maturity Feb.17,2007
%      &
%      \fontsize{7}{12}\selectfont (c) $\beta^3$
      \\
      \includegraphics[width=0.45\columnwidth]{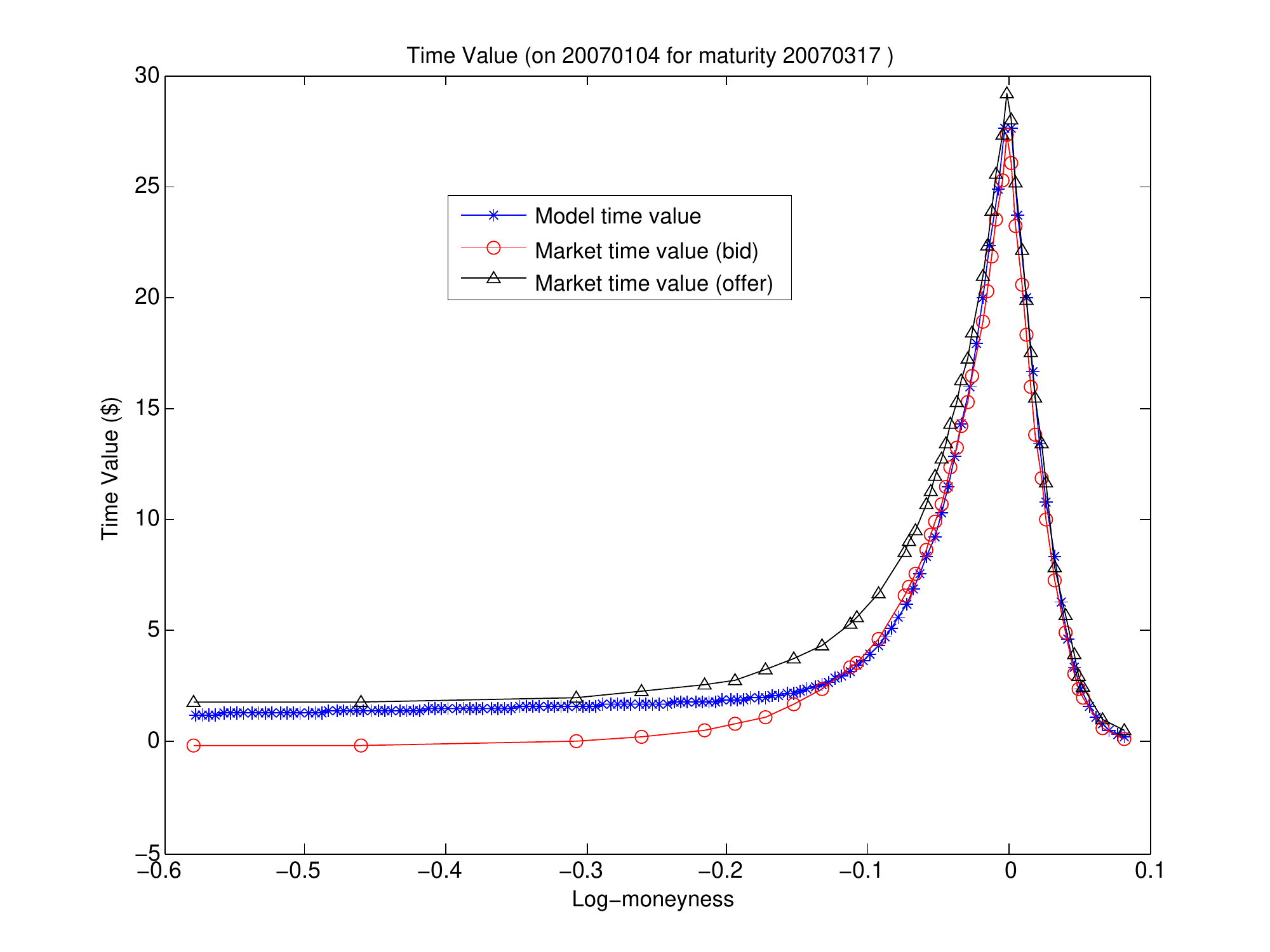}
       &
       \includegraphics[width=0.45\columnwidth]{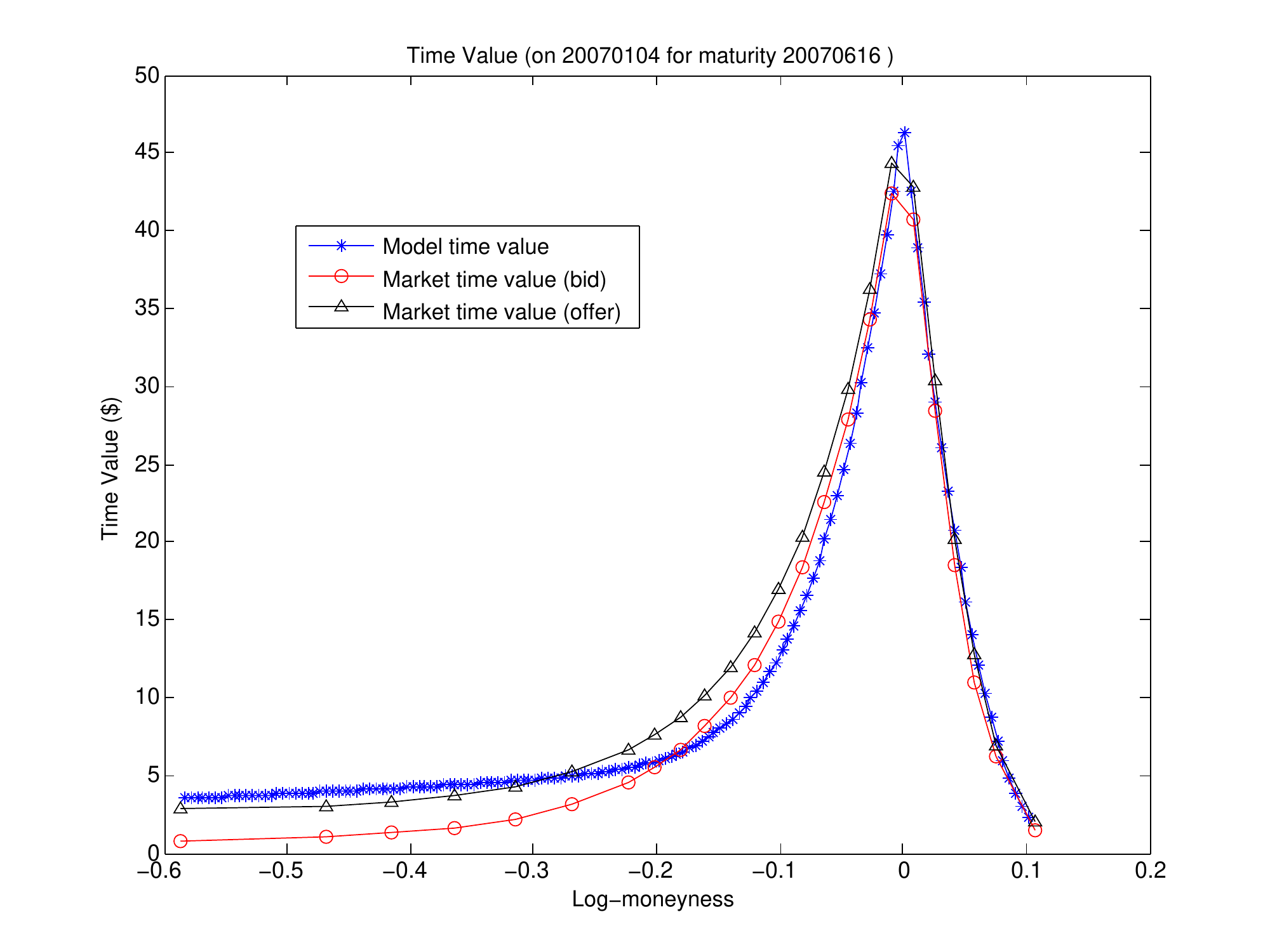}
       \\
      \fontsize{7}{12}\selectfont (c) Maturity Mar.17,2007
     &
      \fontsize{7}{12}\selectfont (d) Maturity Jun.16,2007
      \\
       \includegraphics[width=0.45\columnwidth]{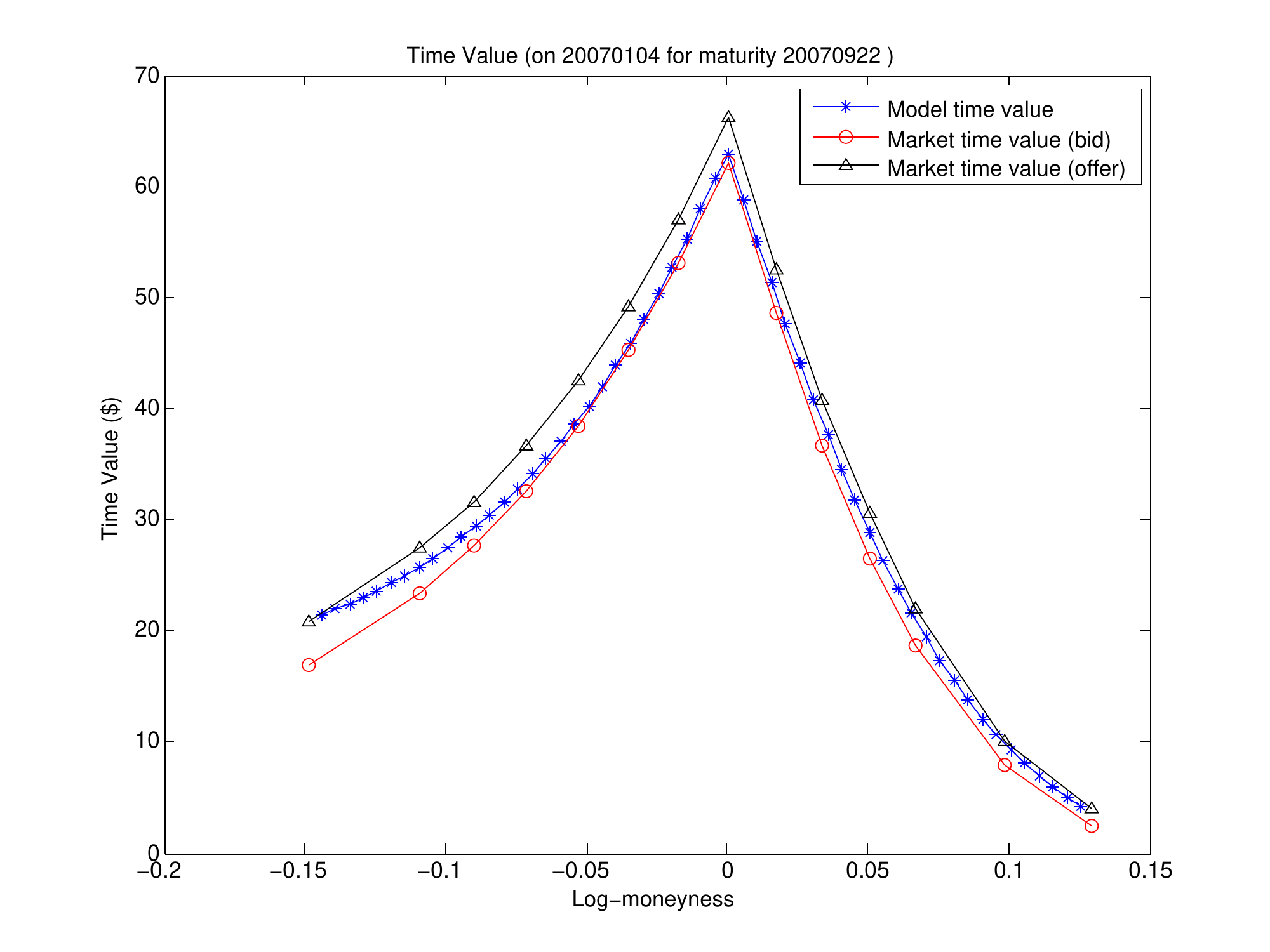}
       &
       \includegraphics[width=0.45\columnwidth]{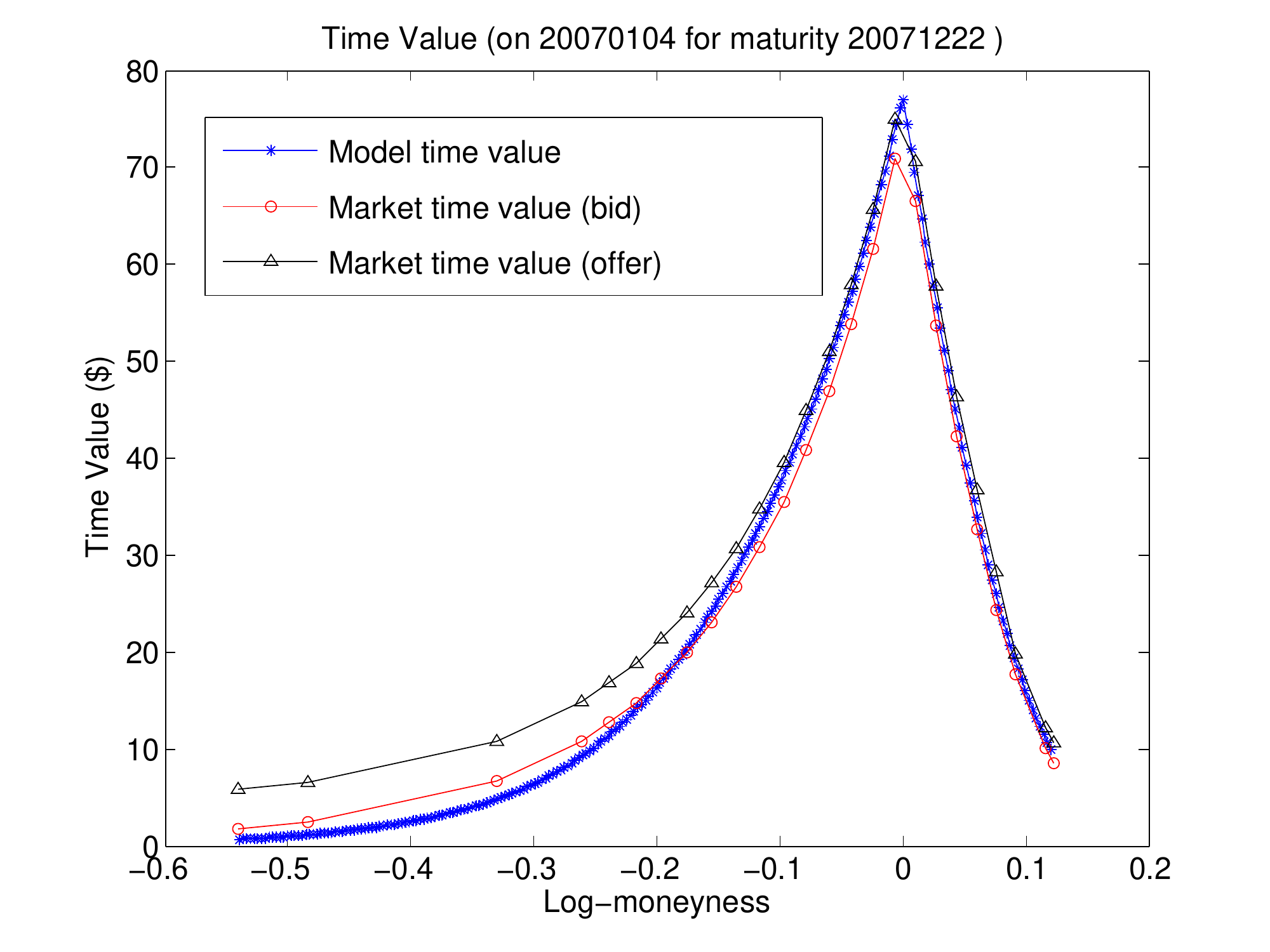}
       \\
      \fontsize{7}{12}\selectfont (e) Maturity Sep.22,2007
     &
      \fontsize{7}{12}\selectfont (f) Maturity Dec.22,2007
\\
       \end{tabular}
   \end{center}
   \vspace{-10pt}
   \caption{Calibrated time values for DTL model on the second day, Jan. 4, 2007}
   \label{fg:dtL:timevalue_day2}
\end{figure}

On Jan. 3, 2007, the first day of the calibration period, we fit $\tilde{\theta}_0(T_l)$ to the prices of call options of six different maturities $T_l$, by solving \eqref{fo:dtL:optimization_penalized} without symmetry condition. 
%The resulting $\tilde{\theta}_0(T_l)$'s are shown in Figure~\ref{fg:dtL:tkappa_day1}. 
Multiplying the results by $\rho$, we obtain the average jump intensities $\theta_0(T_l)$'s and the corresponding calibrated time values.
With the family of $\tilde{\theta}_0(T_l)$'s, calibrated on the first day, we compute the average symmetry index $\bOmega$ via \eqref{fo:dtL:symmetryinit} and linear interpolation, as shown in Figure~\ref{fg:implementation.dtl.new}.c. Using the symmetry index, we run the constrained optimization \eqref{fo:dtL:optimization_penalized} for the subsequent days. The calibration results for the next day, Jan. 4, 2007, are shown in Figures~\ref{fg:dtL:tkappa_day2} - \ref{fg:dtL:timevalue_day2}. It is clear that the fitting is very accurate in terms of matching the time values. It might be a little surprising to see the bimodal or even trimodal shape of the jump intensities; however, these were also observed in \cite{Cont2004nonparametriccalibration}. We can see that a discrete tangent L\'evy measure provides more flexibility and serves as a good nonparametric alternative to the parametric calibration performed in the previous section.

\begin{figure}[htp]
   \begin{center}
       \begin{tabular}{cc}
       \includegraphics[width=0.45\columnwidth]{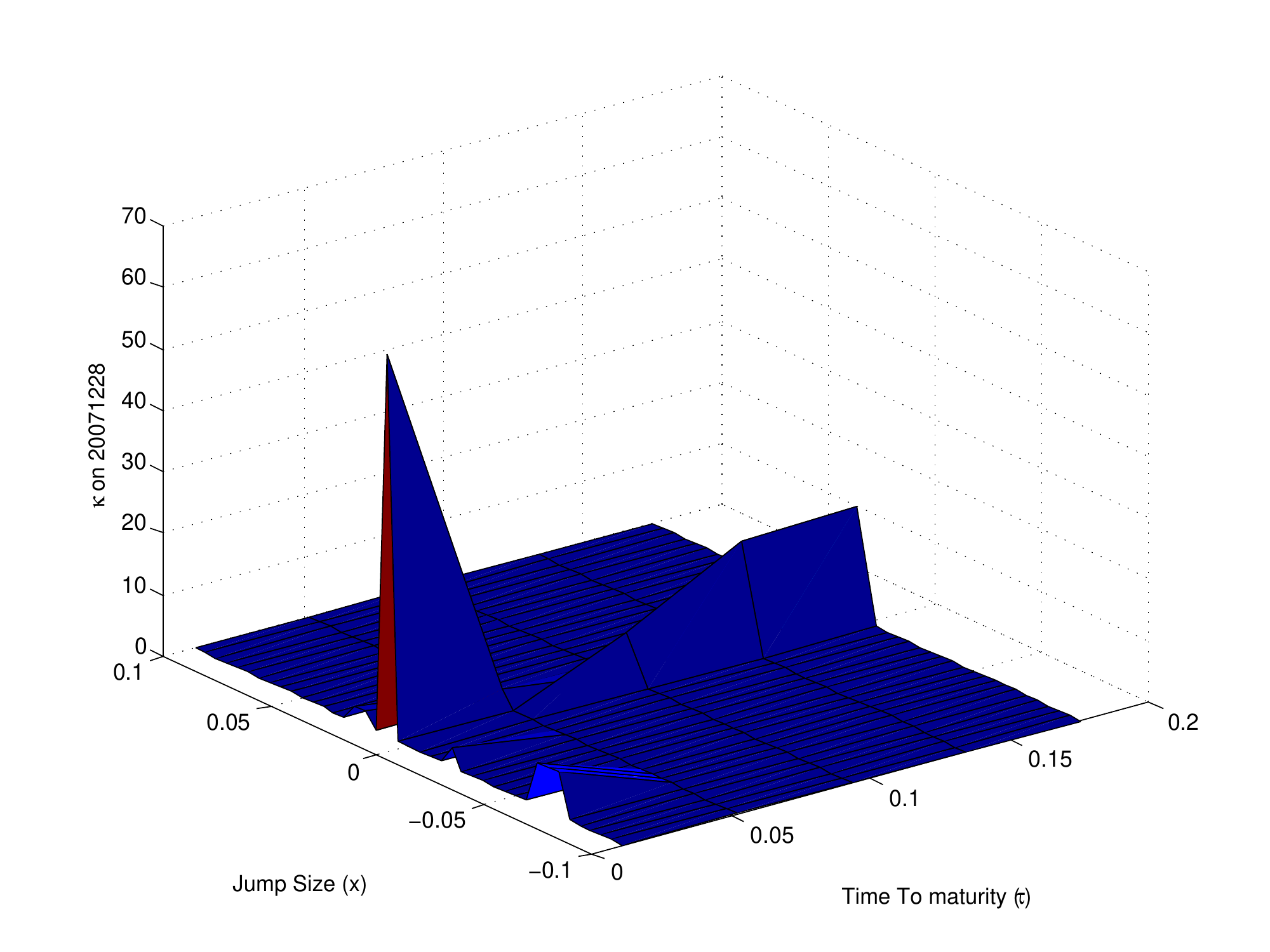}
       &
       \includegraphics[width=0.45\columnwidth]{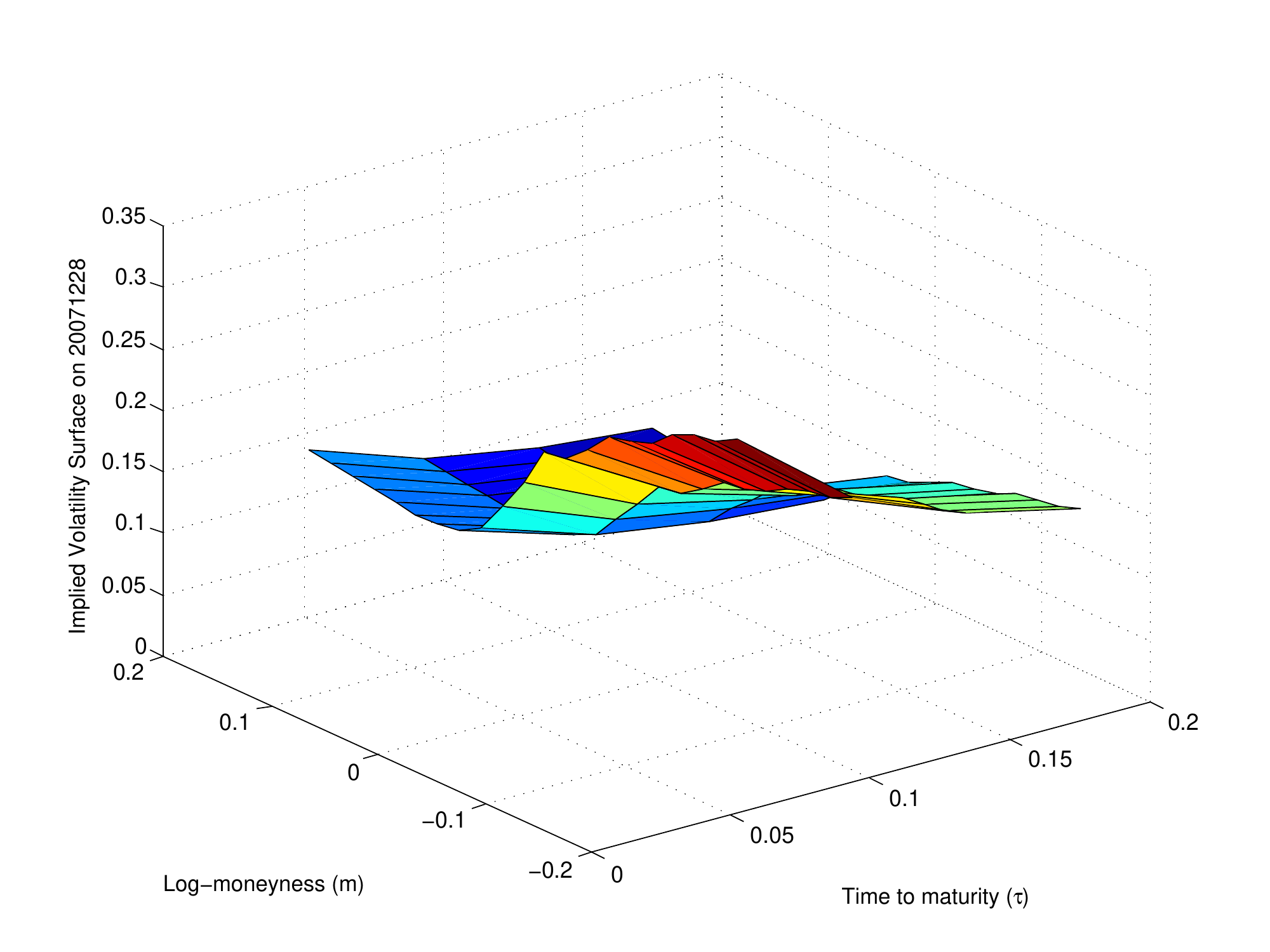}
       \\
      \fontsize{7}{12}\selectfont (a) Calibrated $\kappa$ 
     &
      \fontsize{7}{12}\selectfont (b) Calibrated Implied Volatility Surface
%      &
%      \fontsize{7}{12}\selectfont (c) $\beta^3$
      \\
      \includegraphics[width=0.45\columnwidth]{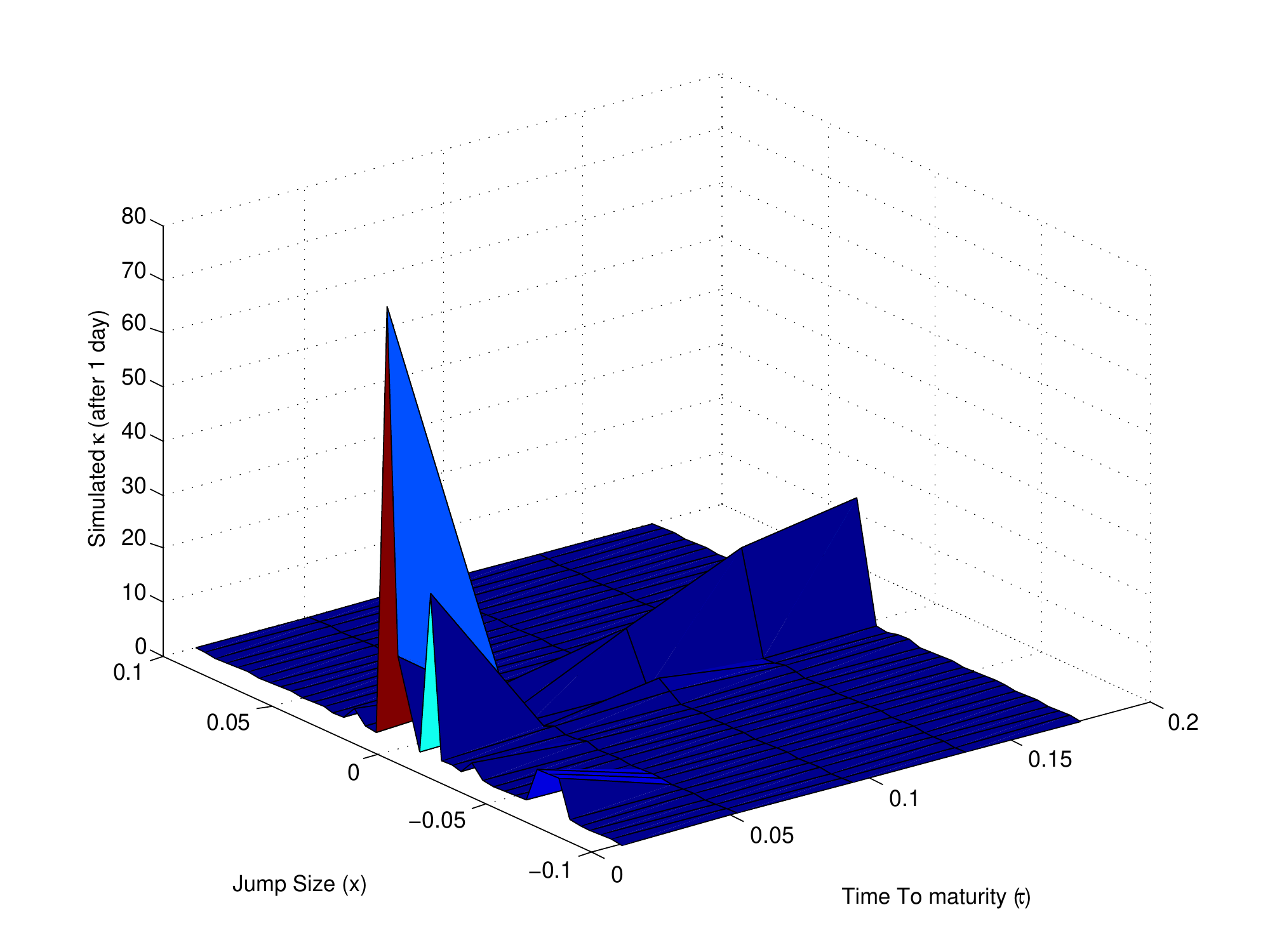}
       &
       \includegraphics[width=0.45\columnwidth]{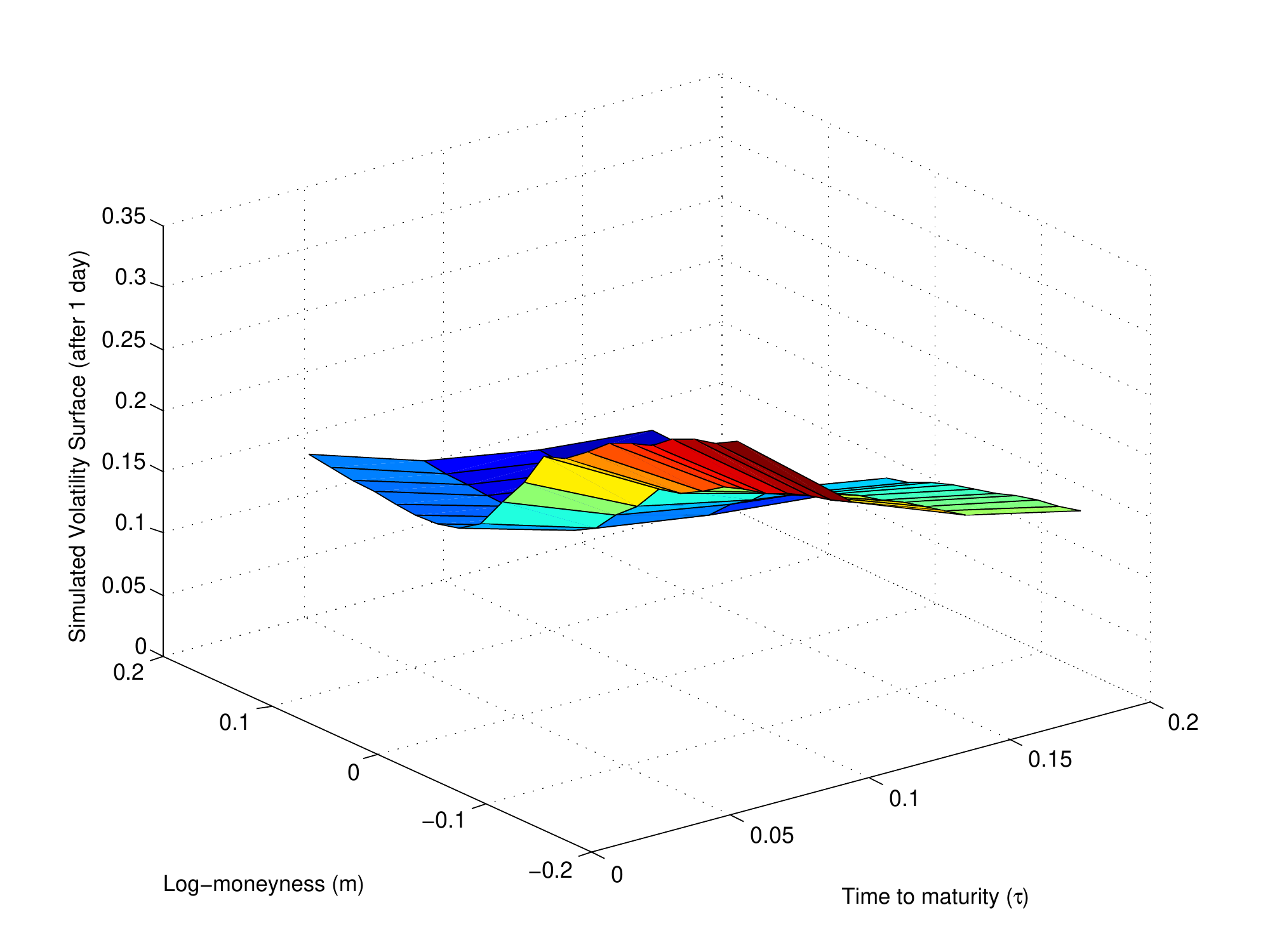}
       \\
      \fontsize{7}{12}\selectfont (c) Simulated $\kappa$ (1st day)
     &
      \fontsize{7}{12}\selectfont (d) Simulated Implied Volatility Surface (1st day)
      \\
       \includegraphics[width=0.45\columnwidth]{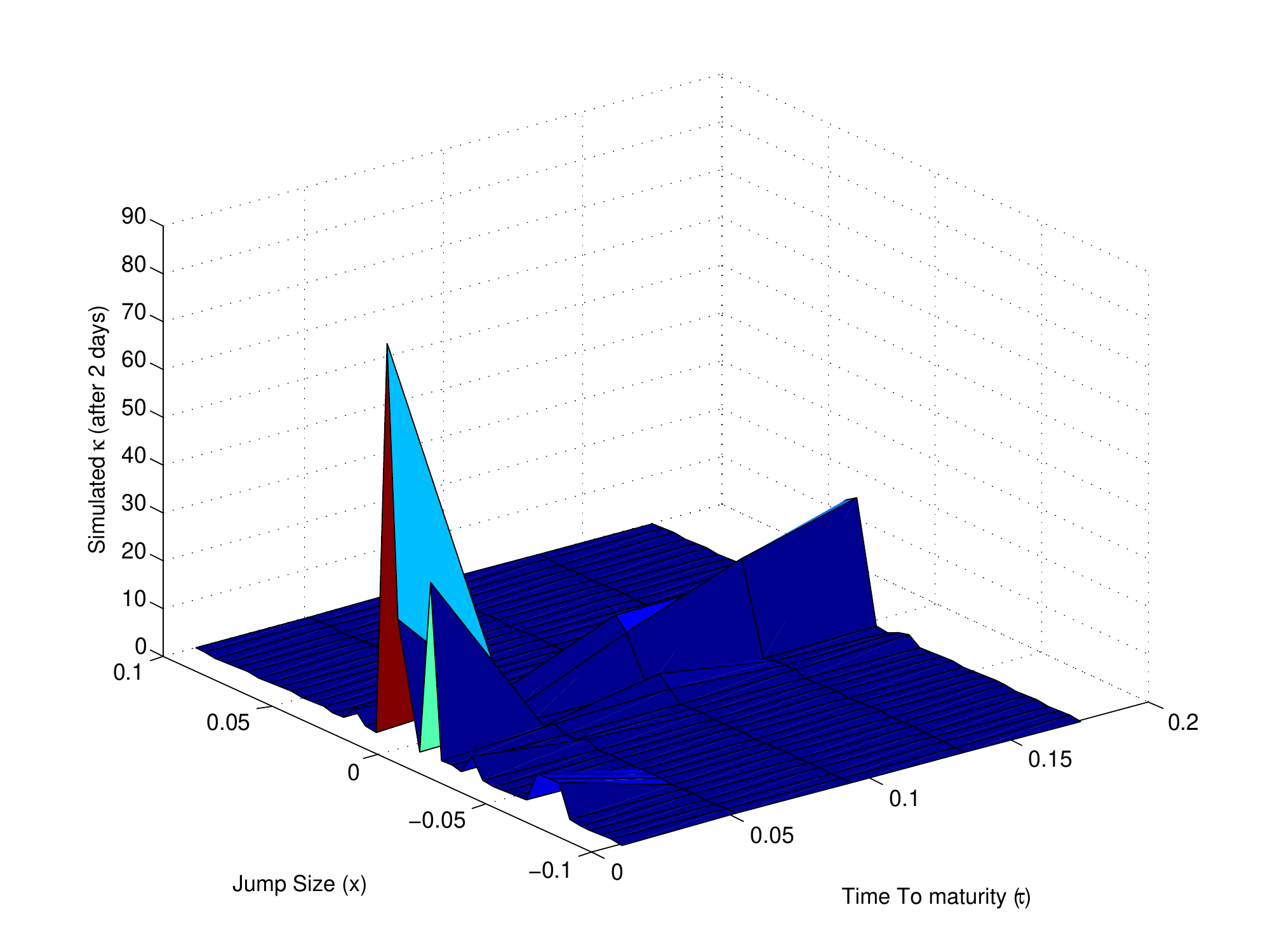}
       &
       \includegraphics[width=0.45\columnwidth]{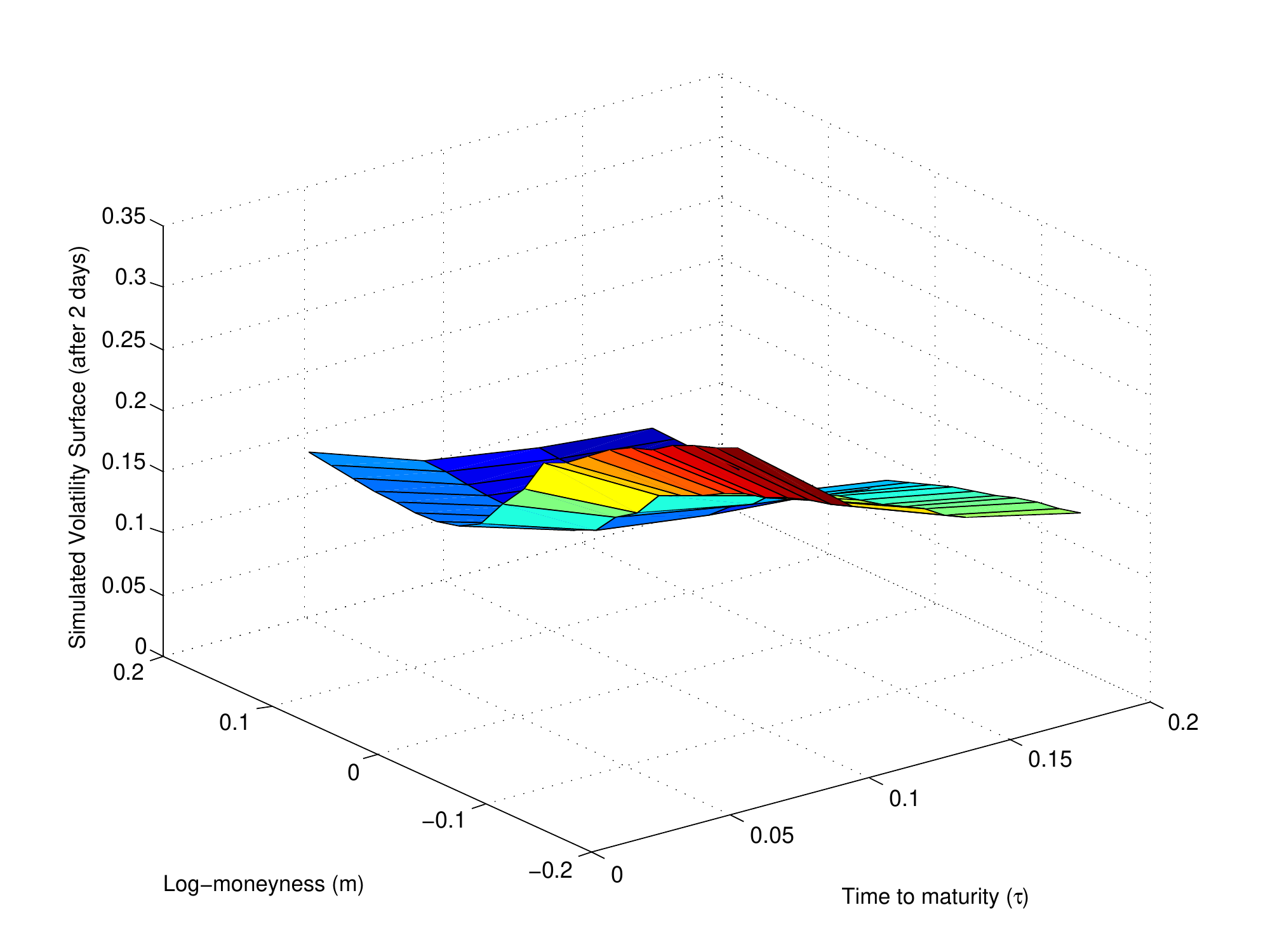}
       \\
      \fontsize{7}{12}\selectfont (e) Simulated $\kappa$ (2nd day)
     &
      \fontsize{7}{12}\selectfont (f) Simulated Implied Volatility Surface (2nd day)
\\
       \end{tabular}
   \end{center}
   \vspace{-10pt}
   \caption{Simulated $\kappa$'s and implied volatility surfaces using DTL model (1)}
   \label{fg:dtL:simulation1}
\end{figure}

Finally, we need to recover $\kappa_t$'s from the calibrated $\theta_t$'s, via \eqref{fo:dtL:barkappa}.
To do this, we assume that each 
$$
\tkappa^j_t(T)=\kappa^j_t(T)/\rho^j
$$ 
is exponential between two consecutive maturities $T_{l-1}$ and $T_l$. In other words, we search for $\tkappa_t^j(T)$ in the following form:
\begin{equation*}
\tkappa_t^j(T) = \sum_{l=1}^{L} \tkappa_t^j(T_{l-1}) e^{c^j_{l-1}(T - T_{l-1})} \bone_{(T_{l-1}, T_l]}(T),
\end{equation*}
where $c^j_0,\cdots,c^j_{L-1}$ are constants. This assumption, along with \eqref{fo:dtL:barkappa}, leads to
\begin{equation} \label{fo:dtL:kappac}
\tilde{\theta}^j_t(T_l) (T_l - T_{l-1}) = \frac{\tkappa^j_t(T_{l-1})}{c^j_{l-1}} \left(e^{c^j_{l-1}(T_l - T_{l-1})}-1 \right).
\end{equation}
The above equation suggests that, as long as we have the value of $\tkappa^j_t(T_0)$ at one point $T_0 \in [t,T_L]$, we will be able to back out $\tkappa^j_t(T)$ for all $T \in [t,T_L]$ from $\{\tilde{\theta}^j_t(T_l)\}_{l=1}^L$'s. Hence, for a fixed $j$, we will make a ``guess'' of $\tkappa^j_t(t)$ by running a single-variable optimization
\begin{align} \label{fo:dtL:yopt}
\min_{y \ge 0} \quad & \sum_{l=1}^L (\tilde{\theta}^j_t(T_l)-\tilde{\theta}^{j,y}_t(T_l))^2, \nonumber \\
& \text{s.t.} \quad -300 \leq c^j_l \leq 20, \quad l = 1,...,L, 
\end{align}
where $\tilde{\theta}^{j,y}_t(T_l)$ is the $[T_{l-1},T_l]$-average of the $\tkappa_t^j(T)$, as given by \eqref{fo:dtL:kappac}, assuming that $\tkappa^j_t(t) = y$. Once we obtain the optimal $y^*$, we set  $\tkappa^j_t(t) = y^*$ and solve for $c_l$'s via \eqref{fo:dtL:kappac}. The latter, in turn, gives us the values of $\tkappa^j_t(T)$ for all $T \in [t,\bT]$. Repeating the process for all $j \neq M$ and all $t\in[0,\bT]$, we obtain the time series of actual jump intensities $(\kappa_t)_{t\in[t_0,\bT]}$. An example of calibrated $\kappa_t$ is shown in Figure \ref{fg:dtL:simulation1}.a.

\subsection{Dynamic fitting}\label{subsec:dtL:dynamic}

Herein, we estimate the dynamics of $\kappa_t$, given the historical sample obtained from option prices in the previous subsection. Recall that the dynamics of $\kappa_t$ are given by the second equation in \eqref{fo:dtl:dyn}.
In the presence of the drift condition \eqref{fo:dtL:driftcondition}, the problem of dynamic fitting boils down to fixing the number of necessary factors $m$ and finding the factors $\{\beta^n\}_{n=1}^m$. As in Subsection~\ref{sec:implementation:dynamic}, we will use PCA to complete this task. Recall that PCA has to be applied to a stationary time series, hence we change variable from $T$ to $\tau = T-t$ and introduce 
\begin{equation*}\label{eq.dtl.hkappa.def}
\hat{\kappa}_t(\tau) = \kappa_t(t+\tau)
\end{equation*}
This gives us
\begin{equation}\label{fo:dtL:kappaDynamicsTau}
d\hat{\kappa}_{t}^j(\tau) = \hat{\alpha}^j_t(\tau) dt + \sum_{n=1}^m \hat{\beta}^{j,n}_t(\tau) dB_t^n,
\end{equation} 
with
\begin{equation}\label{eq.dtl.halpha.hbeta.def}
\hat{\alpha}^j_t(\tau) = \alpha^j_t(t+\tau) + \frac{\partial \kappa^j_t(t+\tau,x)}{\partial T}, 
\,\,\,\,\,\,\,\,\,\,\,\,\,\,\,\hat{\beta}^{j,n}_t(\tau) =  \beta^{j,n}_t(t+\tau).
\end{equation}

\begin{figure}[htp]
  \begin{center}
       \begin{tabular}{cc}
       \includegraphics[width=0.45\columnwidth]{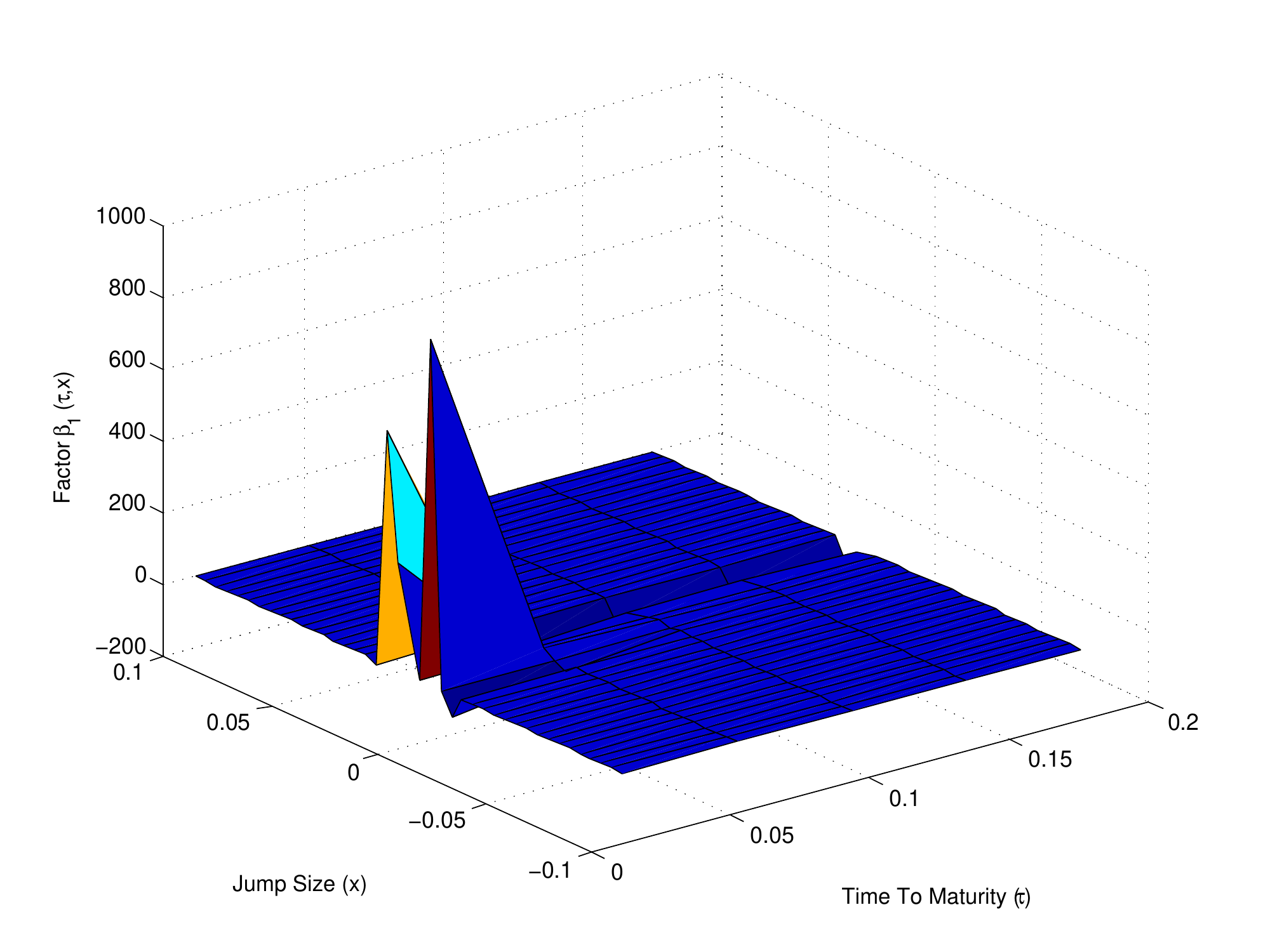}
       &
       \includegraphics[width=0.45\columnwidth]{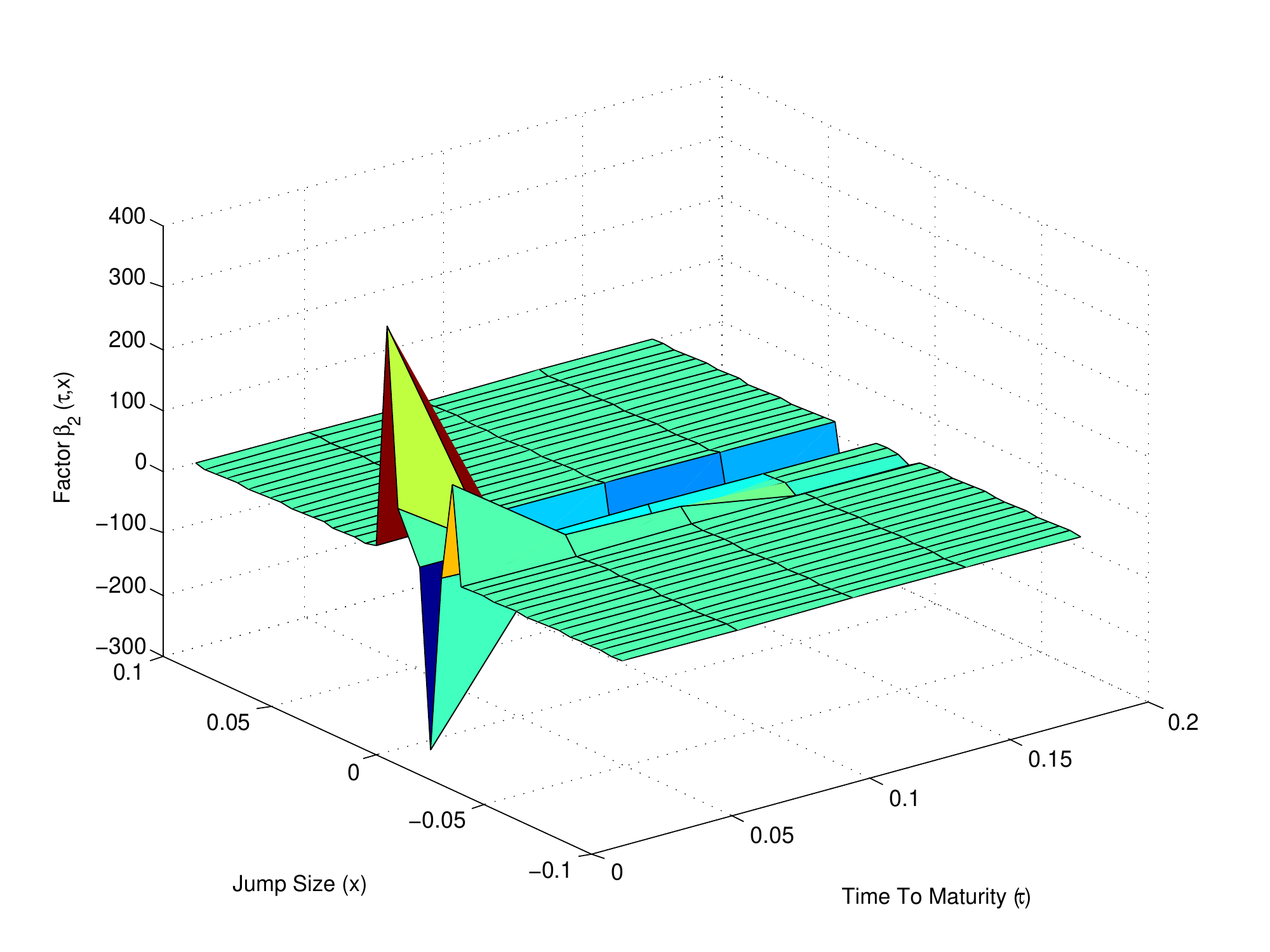}
       \\
      \fontsize{7}{12}\selectfont (a) The first eigenmode scaled by $\sqrt{\lambda_1}$
      &
      \fontsize{7}{12}\selectfont (b) The second eigenmode scaled by $\sqrt{\lambda_2}$
      \\
      \includegraphics[width=0.45\columnwidth]{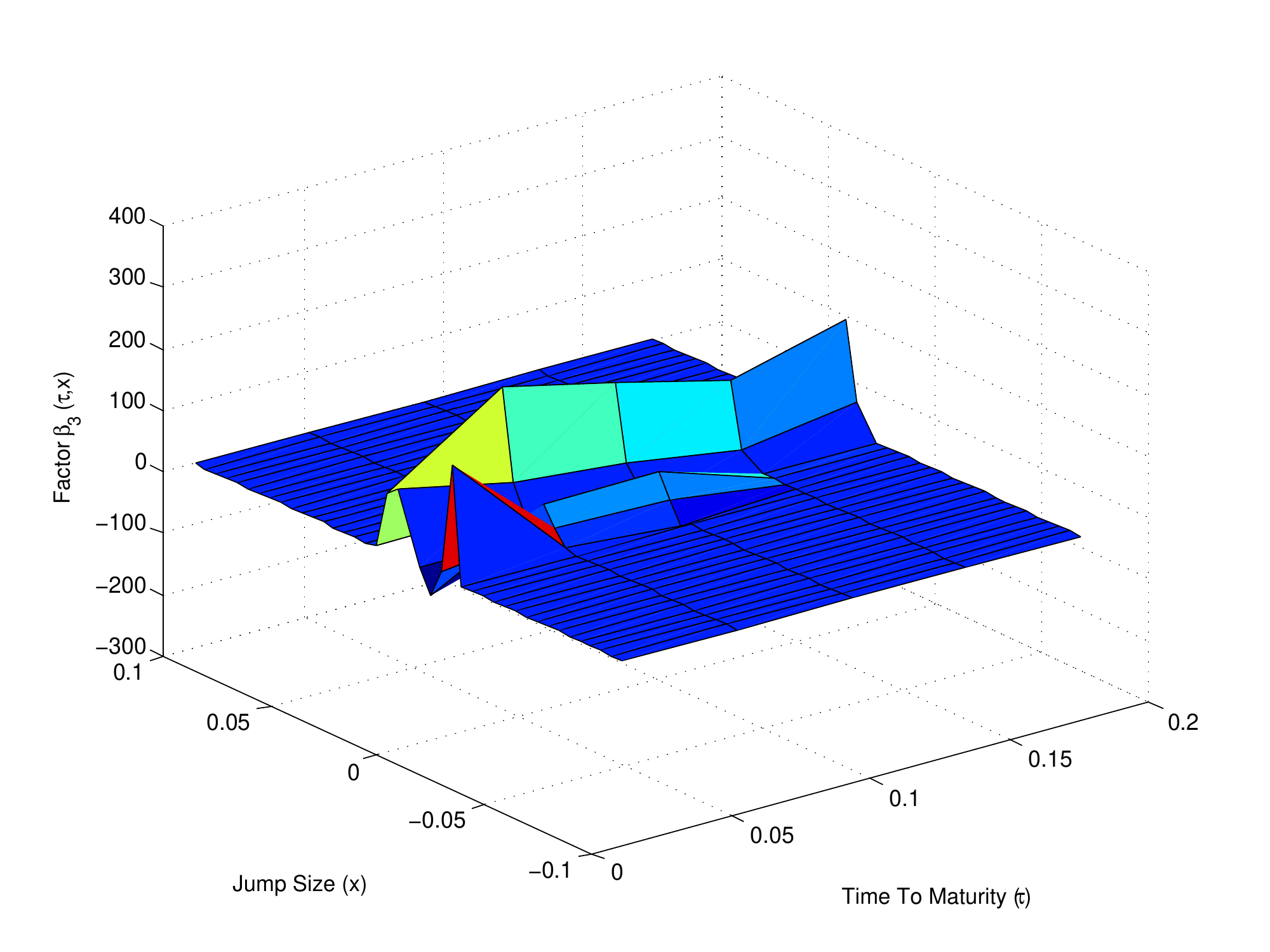}
       &
       \includegraphics[width=0.45\columnwidth]{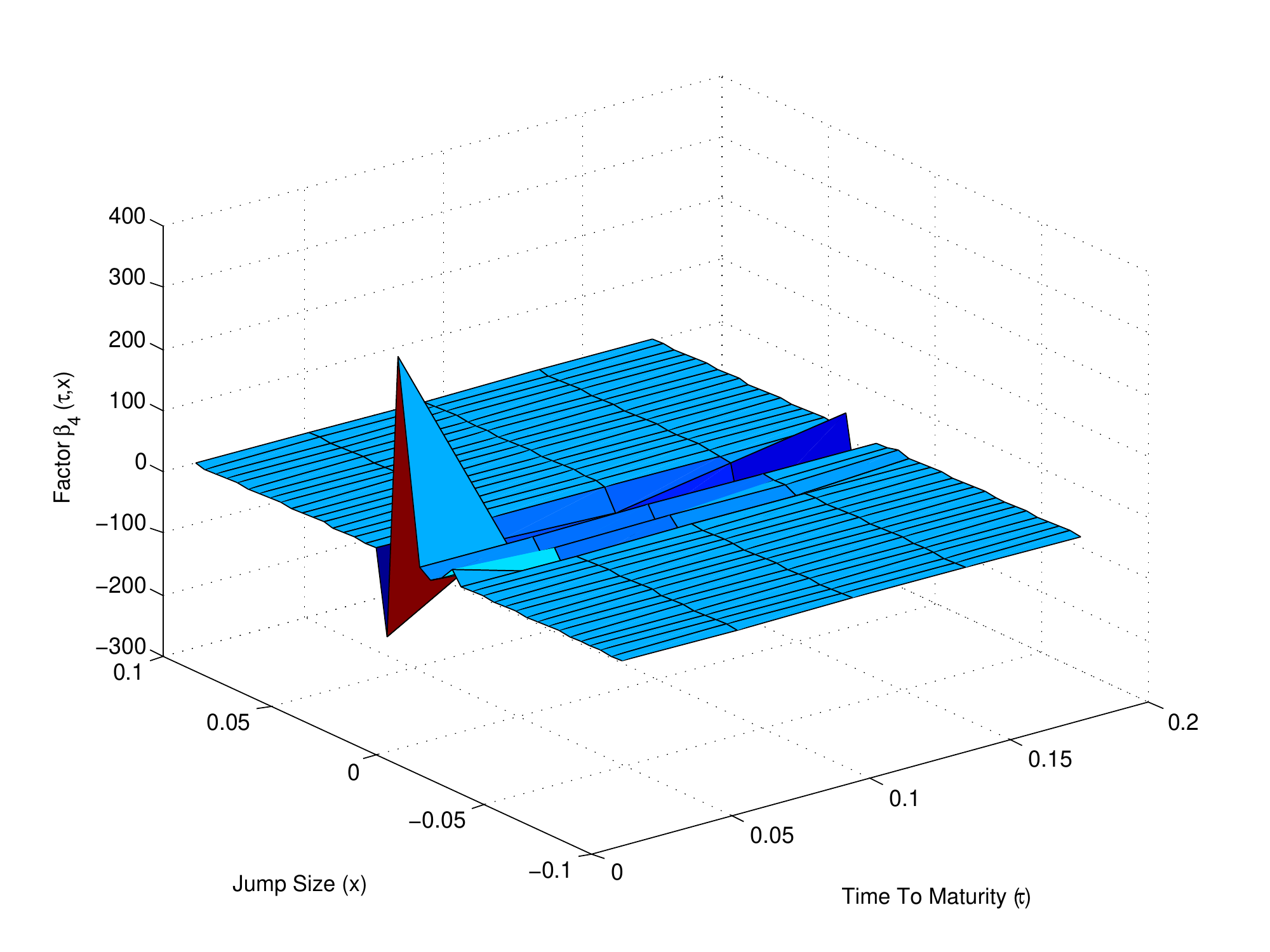}
       \\
      \fontsize{7}{12}\selectfont (c) The third eigenmode scaled by $\sqrt{\lambda_3}$
      &
      \fontsize{7}{12}\selectfont (d) The fourth eigenmode scaled by $\sqrt{\lambda_4}$
      \\
       \end{tabular}
   \end{center}
   \vspace{-10pt}
   \caption{Eigenvalues and eigenmodes of $\Delta\hkappa$ for DTL model}
   \label{fg:dtL:eigenmodes}
\end{figure}

\begin{figure}[htp]
  \begin{center}
       \begin{tabular}{cc}
       \includegraphics[width=0.45\columnwidth]{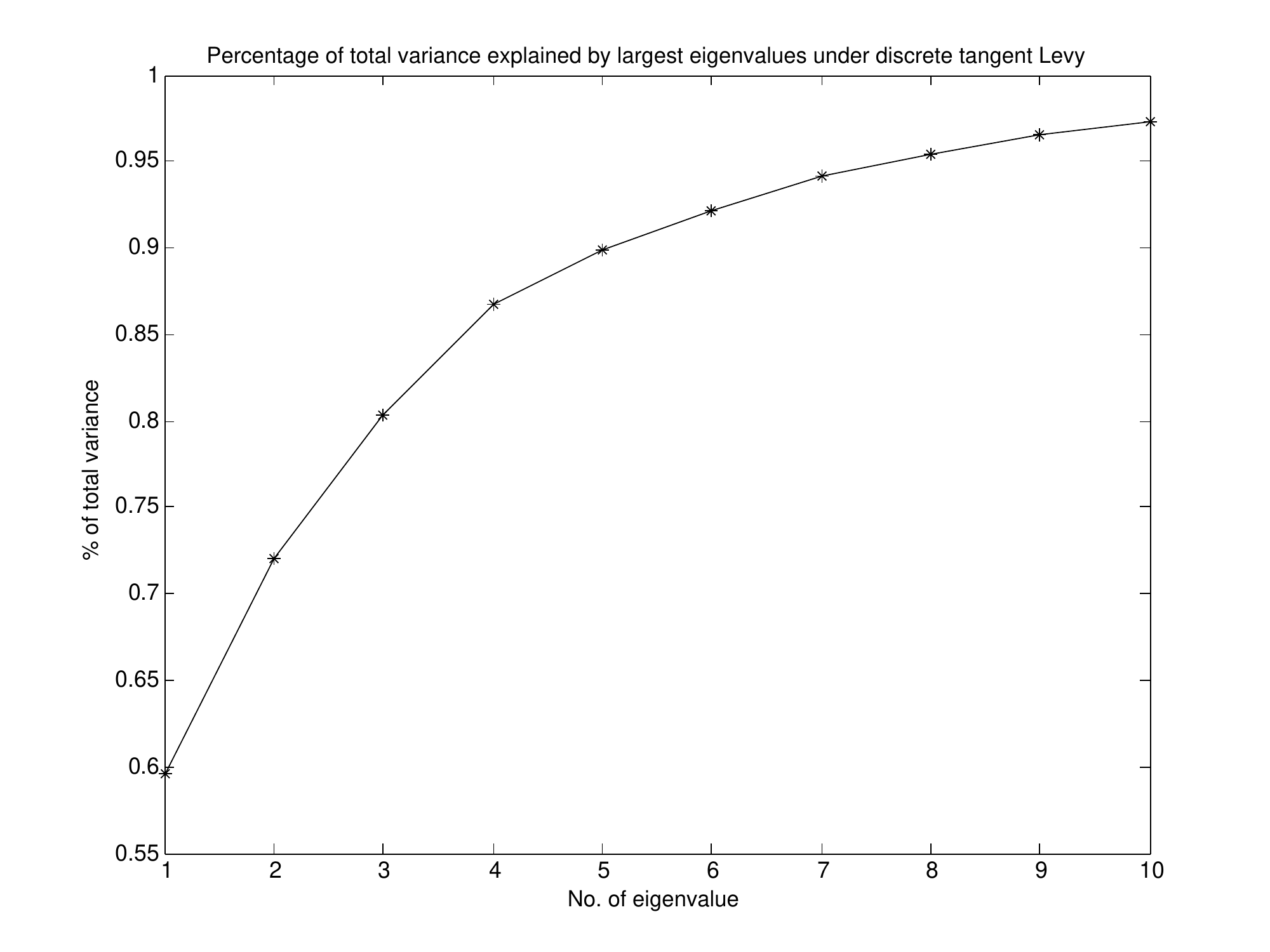}
       &
       \includegraphics[width=0.45\columnwidth]{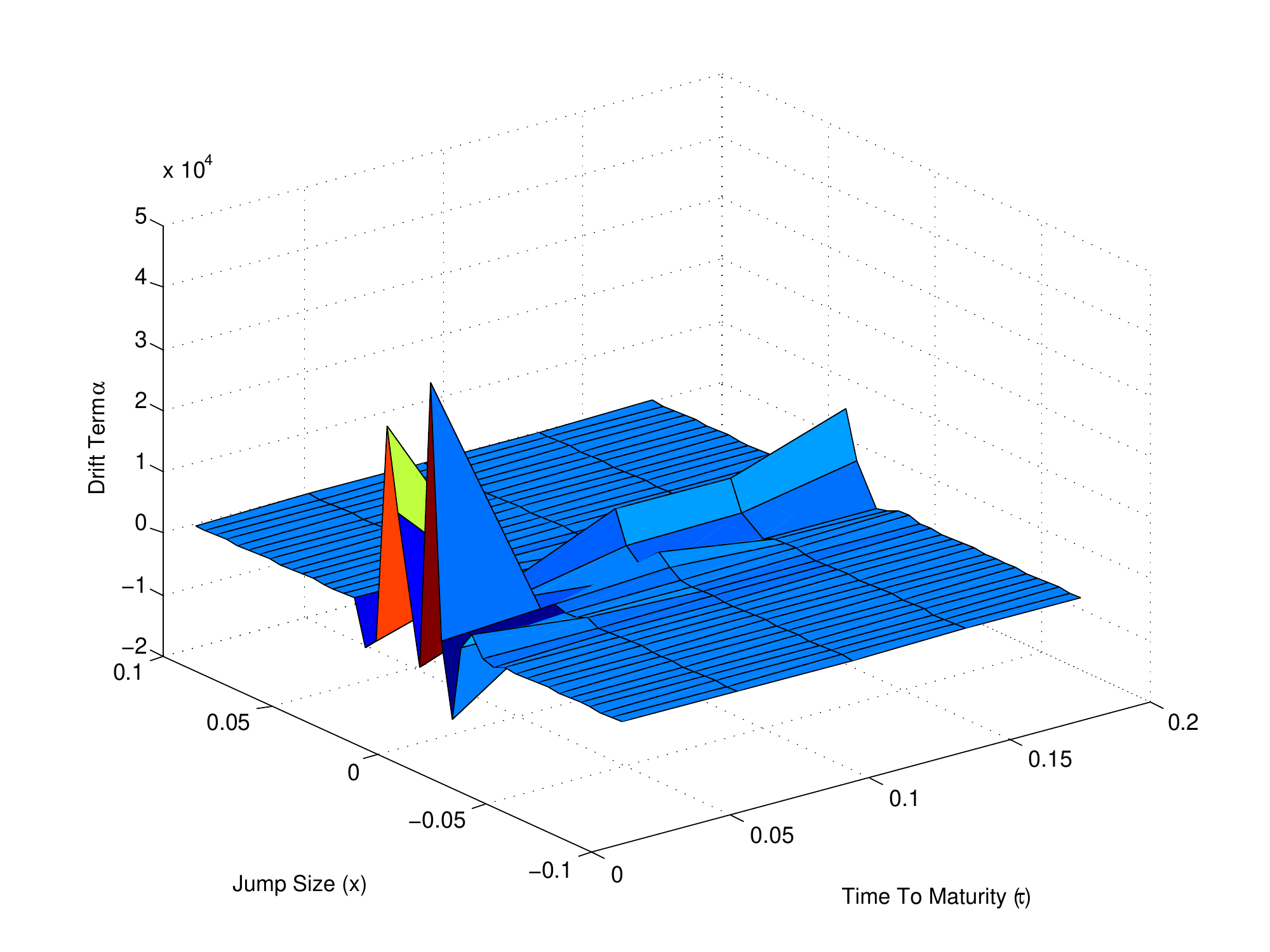}
       \\
      \fontsize{7}{12}\selectfont (a) Percentage of variance explained by the eigenmodes
      &
      \fontsize{7}{12}\selectfont (b) The drift term $\alpha$ for DTL model
      \\
      \end{tabular}
   \end{center}
   \vspace{-10pt}
   \caption{Eigenvalues and eigenmodes of $\Delta\hkappa$ for DTL model}
   \label{fg:dtL:eigenvalue.and.drift}
\end{figure}

For each day $t$, we choose 4 maturities, and for each maturity $T$, we have 24 jump intensities $\hat{\kappa}^j_t(T)$'s. So there are $4\times 24 = 96$ points on the random surface $\Delta \hat{\kappa}^j_t(T) = \hat{\kappa}^{j}_{t}(T) - \hat{\kappa}^{j}_{t-1}(T)$, which is quite large compared to the number of observations, 251. This could lead to inaccuracy in estimating the covariance matrix, thus hampering the estimation of $\hat{\beta}^n$'s. To reduce the number of points, we discard the jump sizes whose jump intensities $\hat{\kappa}^j$ are consistently very small. This gives us a much smaller set of 7 jump sizes, or 28 points on the surface of $\Delta\hat{\kappa}_t$. We perform PCA for the resulting time series $\{\Delta\hat{\kappa}_t\}$, to obtain the values of each $\hbeta^{j,n}(T)$ at the chosen points $(j,T)$. To obtain its values at other points, we simply interpolate $\hbeta^{\cdot,n}(\cdot)$ linearly.
The results of PCA are shown in Figure~\ref{fg:dtL:eigenvalue.and.drift}. We can see that the first $4$ eigenvalues account for over $86\%$ of the total variance. Hence, we set
\begin{equation*}
m = 4  \quad \text{and} \quad \hat{\beta}^n =  \sqrt{\lambda_n} f_n, \quad n = 1,2,3,4,
\end{equation*}
where $\lambda_n$ is the $n$-th largest eigenvalue, and $f_n$ is the corresponding eigenfunction.   
The so-obtained $\hat{\beta}^n$'s are also shown in Figure~\ref{fg:dtL:eigenmodes} as functions of $x$ and $\tau$. After transforming $\hbeta$ to $\beta$ via \eqref{eq.dtl.halpha.hbeta.def}, we apply the drift condition \eqref{fo:dtL:driftcondition} to compute the drift term $\alpha$ that guarantees the absence of dynamic arbitrage. A plot of $\alpha$ is shown in Figure~\ref{fg:dtL:eigenvalue.and.drift}. Finally, we use \eqref{eq.dtl.halpha.hbeta.def} once more to obtain $\halpha$ from $\alpha$.

\subsection{Monte Carlo simulation of implied volatility surfaces}\label{sec:dtL:simulation}

\begin{figure}[htp]
   \begin{center}
       \begin{tabular}{cc}
       \includegraphics[width=0.45\columnwidth]{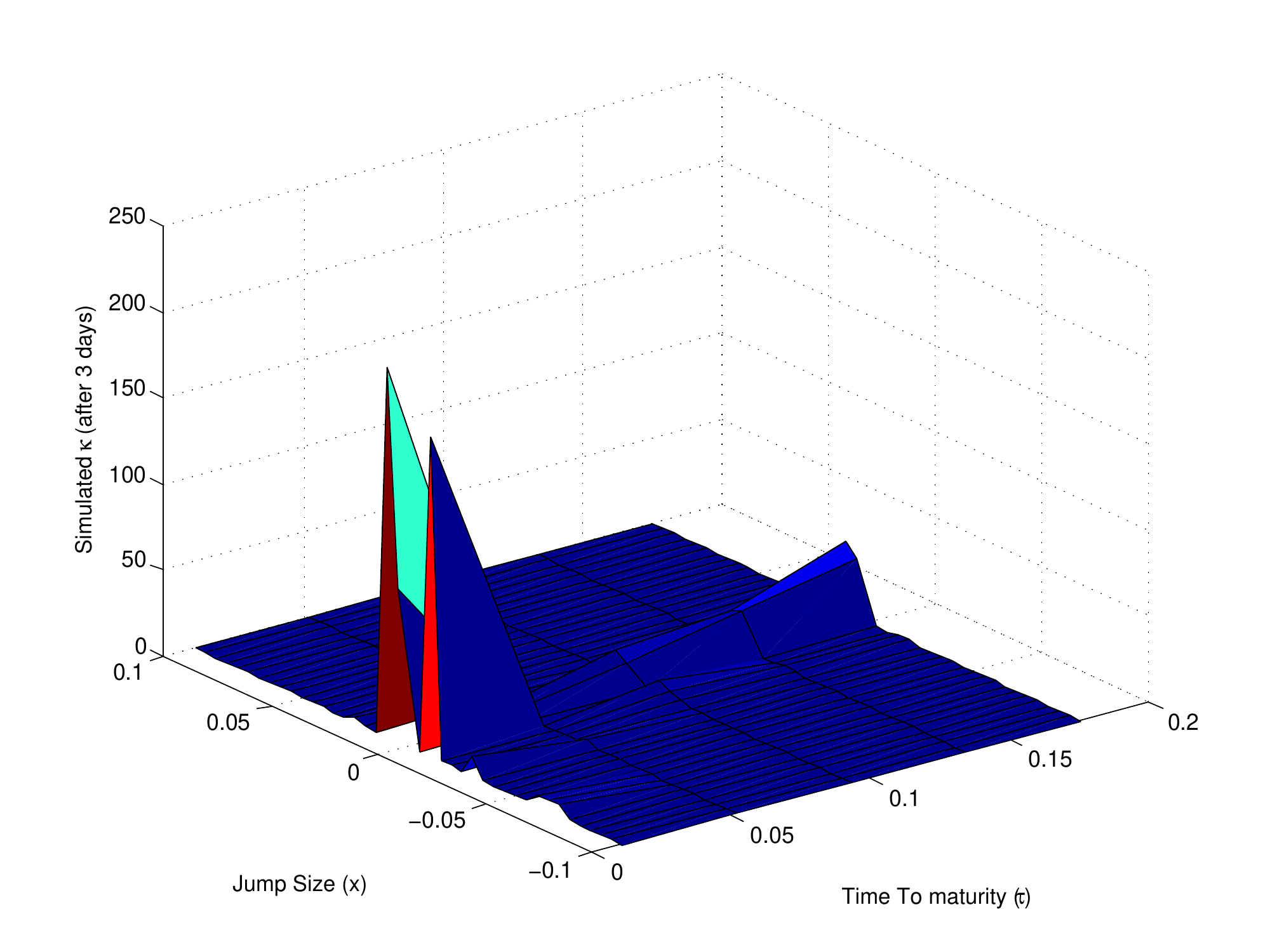}
       &
       \includegraphics[width=0.45\columnwidth]{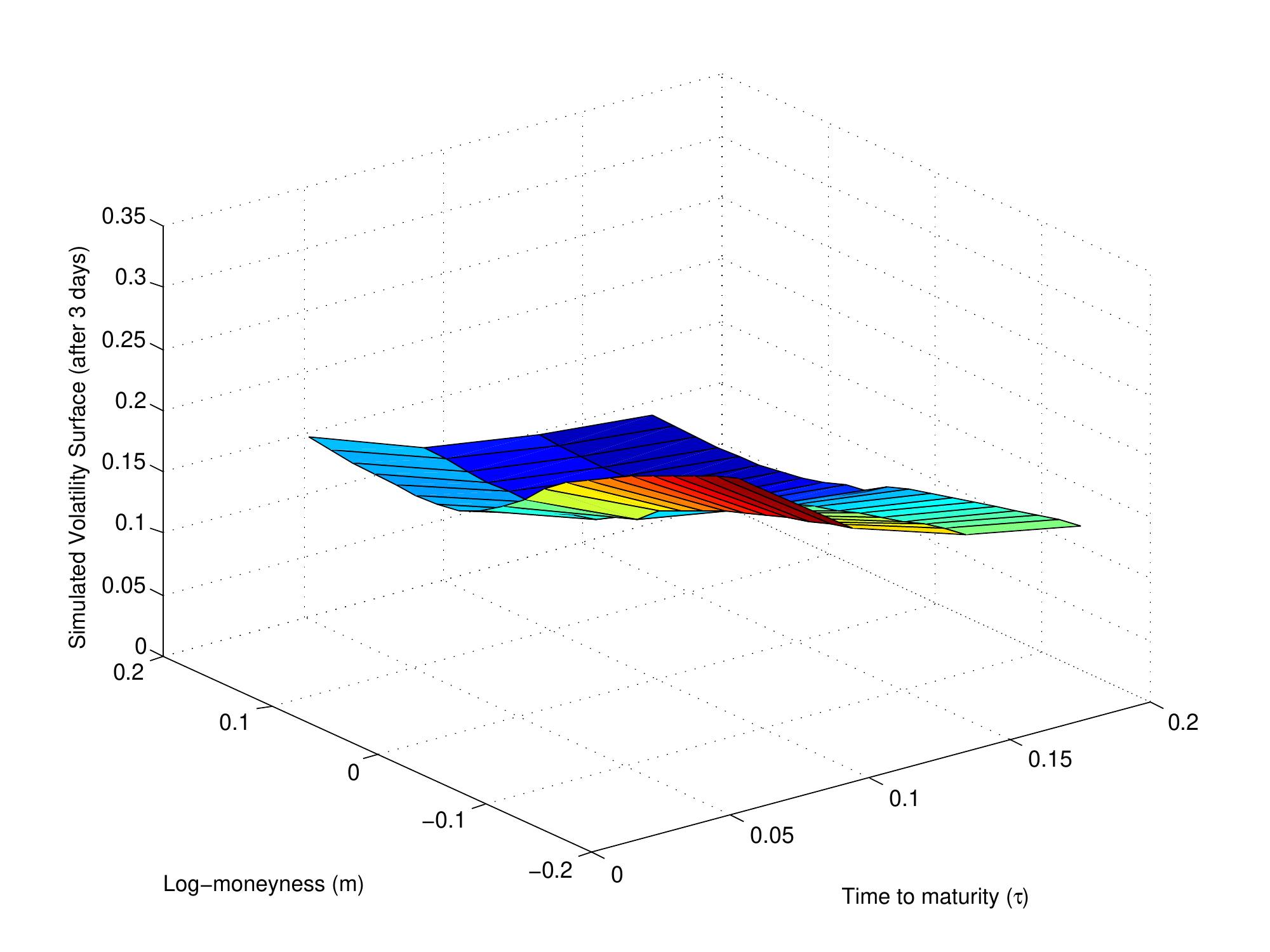}
       \\
      \fontsize{7}{12}\selectfont (a) Simulated $\kappa$ (3rd day)
     &
      \fontsize{7}{12}\selectfont (b) Simulated Implied Volatility Surface (3rd day)
%      &
%      \fontsize{7}{12}\selectfont (c) $\beta^3$
      \\
      \includegraphics[width=0.45\columnwidth]{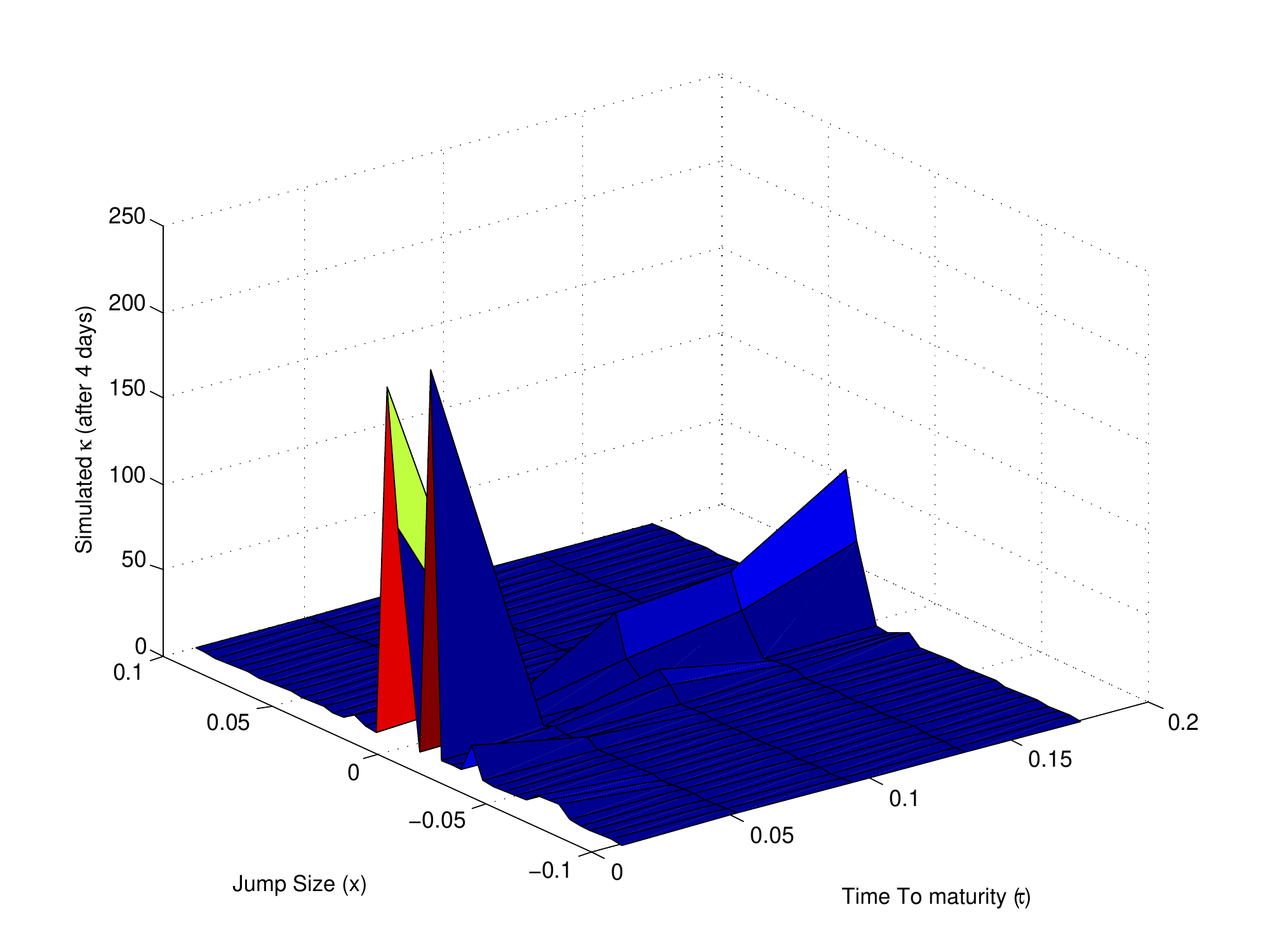}
       &
       \includegraphics[width=0.45\columnwidth]{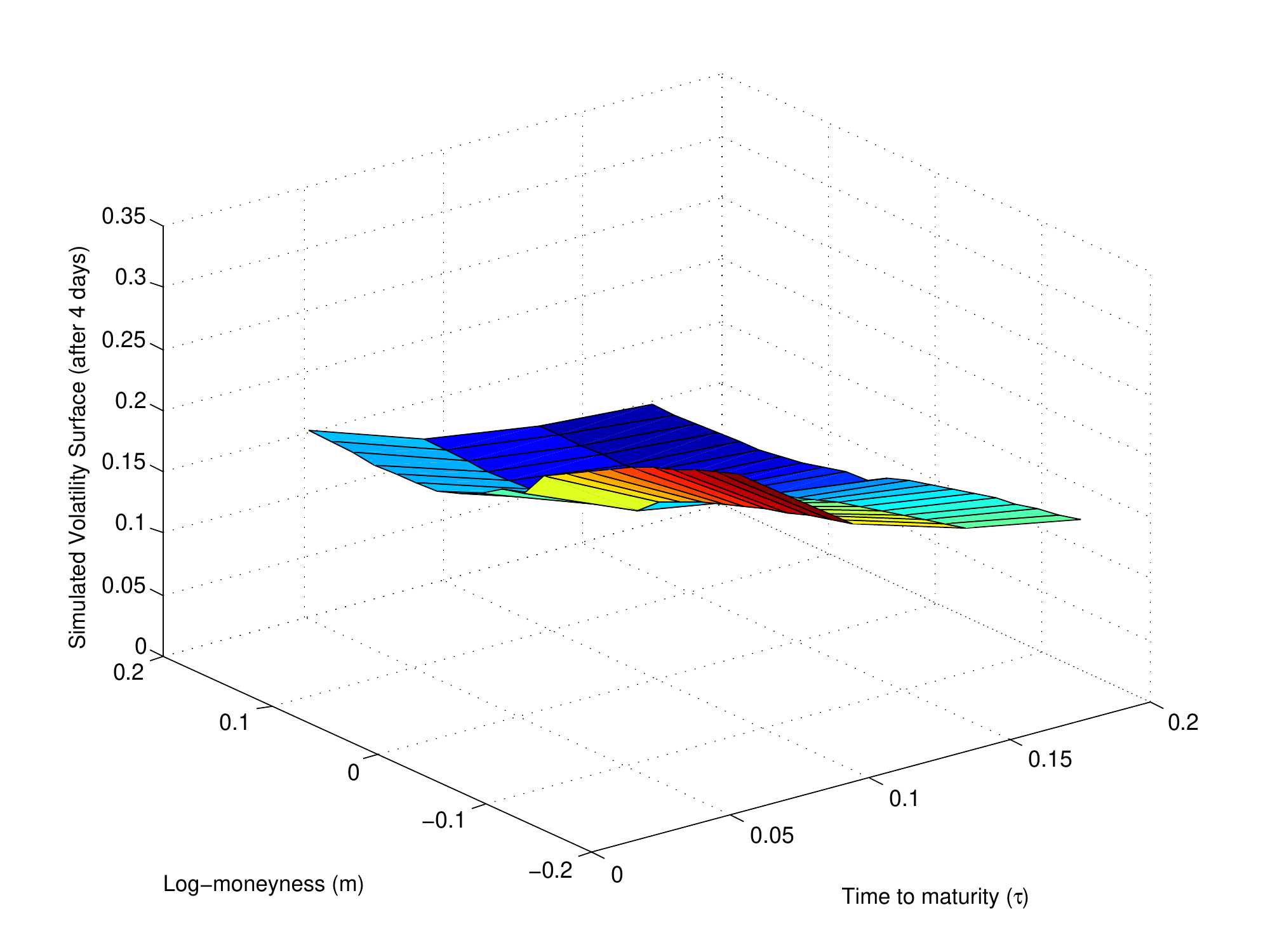}
       \\
      \fontsize{7}{12}\selectfont (c) Simulated $\kappa$ (4th day)
     &
      \fontsize{7}{12}\selectfont (d) Simulated Implied Volatility Surface (4th day)
      \\
       \includegraphics[width=0.45\columnwidth]{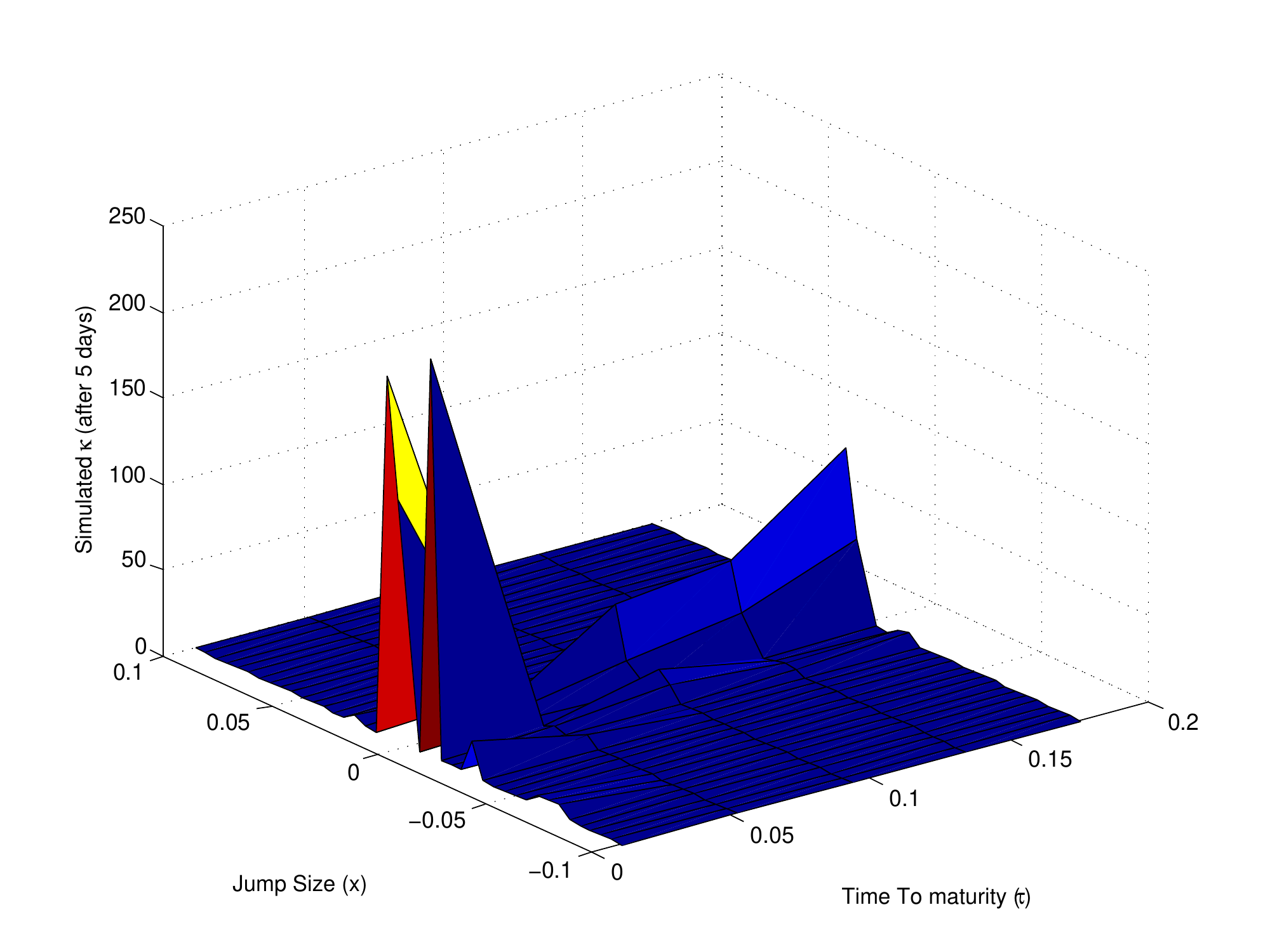}
       &
       \includegraphics[width=0.45\columnwidth]{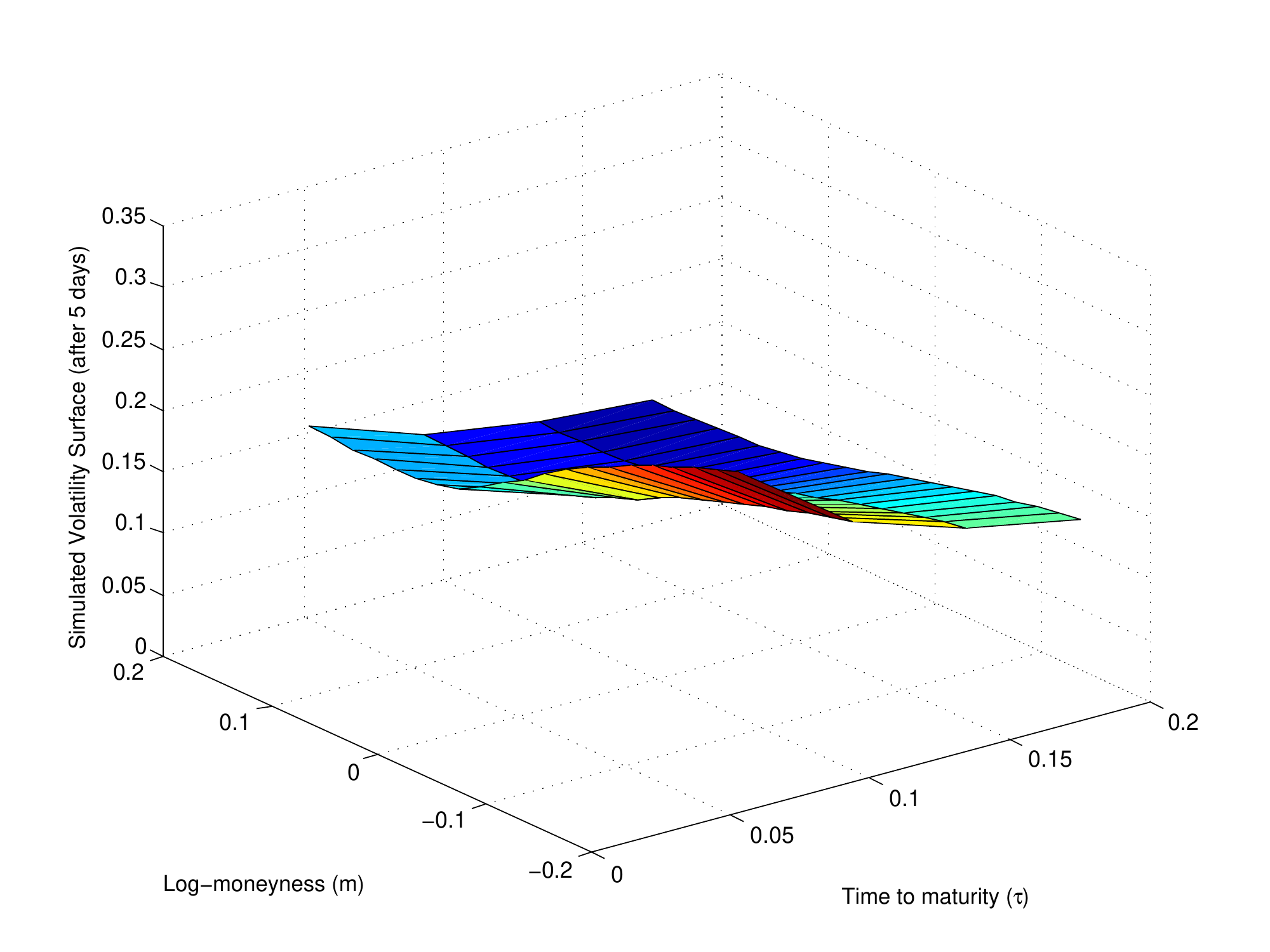}
       \\
      \fontsize{7}{12}\selectfont (e) Simulated $\kappa$ (5th day)
     &
      \fontsize{7}{12}\selectfont (f) Simulated Implied Volatility Surface (5th day)
\\
       \end{tabular}
   \end{center}
   \vspace{-10pt}
   \caption{Simulated $\kappa$'s and implied volatility surfaces using DTL model (2)}
   \label{fg:dtL:simulation2}
\end{figure}

Using the results of static and dynamic fitting, herein, we simulate the future paths of implied volatility surfaces that are free of arbitrage and are consistent with the present and historical observations. 
Similar to Section~\ref{sec:implementation:simulation}, we incorporate a scaling factor $\gamma_t$ in \eqref{fo:dtL:kappaDynamicsTau}, to ensure the non-negativity of $\hkappa_t^j$'s:
\begin{equation}\label{fo:dtL:kappaDynamicsTau_mod}
d\hat{\kappa}_{t}^j(\tau) = \gamma_t^2 \hat{\alpha}^j(\tau) dt + \gamma_t \sum_{n=1}^m \hat{\beta}^{j,n}(\tau) dB_t^n,
\end{equation}
where 
\begin{equation}\label{fo:dtL:gammat}
\gamma_t = \frac{1}{\epsilon} \bigg(\min_{ j\neq M, \tau \in [0, \bar{\tau}]} \hkappa_t^j(\tau) \land \epsilon\bigg),
\end{equation}
with $\epsilon = 1e^{-6}$ and $\bar{\tau}=0.5$. 
Our strategy is now similar to the previous section. Namely, we use the explicit Euler scheme for \eqref{fo:dtL:kappaDynamicsTau_mod} to simulate future $\hat{\kappa}$'s: 
\begin{equation}\label{eq.dtl.Euler}
\hkappa_{t+\Delta t}^j(\tau) = \hkappa_t^j(\tau) + \gamma_t^2\halpha^j(\tau) \Delta t+ \gamma_t \sum_{n=1}^m \hbeta^{j,n}(\tau) \Delta B^n_t, 
\end{equation} 
with $\Delta t$ being one day.
For each sample path, we convert simulated $\hat{\kappa}$ to $\kappa_t(T) = \hat{\kappa}_t(T-t)$ and compute the average intensity $\theta$ defined in \eqref{fo:dtL:barkappa} by numerical integration. We then apply the iterative formula \eqref{fo:dtL:PDEsol} to the average intensity $\theta$ to compute the time values and, in turn, option prices $C^{1,\kappa_t}_t(T,x)$ for different strikes $e^x$ and maturities $T$ on any given day $t$. Finally, we invert the Black-Scholes formula to obtain implied volatilities. 
We perform the simulation starting from Dec. 28, 2007. Moving five days forward,
the simulated $\kappa_t$'s and the corresponding implied volatility surfaces are shown in Figures~\ref{fg:dtL:simulation1} and~\ref{fg:dtL:simulation2}. We see that the simulated $\kappa_t$'s maintain similar shape as the $\kappa$  on Dec. 28, 2007, and the implied volatility surfaces are consistent with what we normally observe in the market. Using this approach, we can generate as many samples as needed, and use them as scenarios of future implied volatility surfaces in various applications, such as pricing forward-starting options or solving problems of risk management. We illustrate this idea in Section~\ref{sec:comparison}.

\section{Empirical analysis of the performance of tangent L\'{e}vy models\label{sec:comparison}}

In this section, we discuss the importance of consistency in modeling derivatives prices. 
As we know, an investment manager's portfolio or a trader's trading book often contains multiple financial derivatives written on the same underlying. As a simple example, an equity trader might hold a calendar spread and a butterfly spread at the same time. To properly manage the risk, one needs to understand the joint dynamics of these derivatives, for which a consistent modeling framework is crucial.
Tangent L\'{e}vy models (as any market-based model) are built to achieve this goal precisely. This is due to the fact that not only present but also historical information contained in the time series of options' prices is used in the estimation of model dynamics. Classical stochastic volatility models cannot capture the historical evolution of options' prices, henc,e there is a reason to believe that market-based models would lead to better performance in portfolio management. 
To show that tangent L\'{e}vy models do indeed work better, here, we test the two tangent L\'{e}vy models implemented in Sections~\ref{sec:DETLM} and \ref{sec:dtL} using the following portfolio choice problem. The results are, then, compared against one of the most popular volatility models in the industry -- the Stochastic Alpha Beta Rho (SABR) model.

\subsection{The variance-minimizing portfolio choice problem}
This example is a simplified Markowitz-type portfolio optimization problem. Consider a portfolio manager who needs to decide how he/she should balance a portfolio of SPX options so that its risk is minimized. Among the many definitions of portfolio risk, we adopt the one used in the classic Markowitz problem (for example, see Section 6.6 of \cite{luenberger1997investment}) -- namely, the standard deviation of the portfolio return over a given (future) time period. 
Notice that this is not a typical Markowitz portfolio problem, given we are not considering the trade-off between return and risk as a typical Markowitz problem would. As a matter of fact, we would assume that the portfolio manager lives in a risk-neutral world, so that the expected return is normalized. We admit that lacking excess return might make the example less exciting, but it helps us compare the model performance in an apples-to-apples fashion. With the normalized return, there is no need to worry about the impact of different market views portfolio managers might build into the investment decisions. Of course, without such a trade-off, there is a trivial solution to the portfolio choice problem -- do not invest at all, reducing the risk to zero. To make the problem non-trivial, we require that the value of the portfolio at the time when it is constructed must be equal to a fixed number $M$. Such a restriction is relevant if the manager makes profits off the commission, proportional to the size of the investment portfolio he/she manages. For example, an option market maker might want to know the optimal inventory so that he/she can adjust the quoting strategy accordingly to reach the portfolio composition with minimal inventory risk. Or, a broker dealer might need to know her optimal position in options over the next several days to meet the risk and capital requirements. 
%In all cases, the consistency in modeling the product dynamics that tangent L\'{e}vy models possess, as we will illustrate in the following sections, is very desirable.

We now formulate this problem mathematically. Let us assume that there are $n$ options with the same maturity $T$ but with different strikes $K_1,...,K_n$ in the portfolio. Let $C_u(K_i)$ be the time-$u$ price of the $K_i$-struck option, and let $\omega_i$ be the quantity of this option in the portfolio, with a negative $\omega_i$ representing to short-selling. The weights $\omega_i$ have to be determined at the initial time $d$. The portfolio value at any future time $t$ is simply $V_t = \sum_{i=1}^n \omega_i C_t(K_i)$, and the return over a $u$-day period is $R_u = V_{d+u}/V_d$. For simplicity, we assume that the risk-free rate and the dividend yield are both zero, so the expectation of $R_u$ is simply $1$. For a given $u\in(0,T)$, to determine the portfolio weights, we need to solve the following convex optimization problem: 
%\begin{equation}\label{fo:compare:optR}
%\min_{\omega \in \mathbb{R}^n} \quad \EE(R_u - 1)^2
%\end{equation}
%Here, $C_0 = [C_0(K_1),...,C_0(K_n)]^T$ is the initial option price vector and $\omega = [\omega_1,...,\omega_n]^T$ is the option quantity vector. This optimization problem can be easily converted into the following convex optimization problem 
\begin{align*}
\min_{\omega \in \mathbb{R}^n} \quad \EE(R_u - 1)^2 
&= \frac{1}{M^2} \min_{\omega \in \mathbb{R}^n} \quad \EE(V_u - M)^2 \\
& \text{s.t.} \quad V_d = \omega^T C_d = M, 
\end{align*}
where $M \in \RR, M > 0$ is the initial value of the portfolio. This is equivalent to 
\begin{align}\label{fo:compare:opt}
\min_{\omega \in \mathbb{R}^n} \quad & \omega^T \Lambda_u \omega\\
& \text{s.t.} \quad V_d = \omega^T C_d = M, \nonumber
\end{align}
where $\Lambda_u = \EE[(C_{d+u} - C_d)^2]$ is the covariance matrix of the time-$d+u$ options' prices.  It is easy to see that the closed-form solution to the quadratic optimization \eqref{fo:compare:opt} is
\begin{equation}\label{fo:compare:weights}
\omega = \frac{ M\Lambda_u^{-1} C_d}{C_d^T \Lambda_u^{-1}C_d}.
\end{equation}
Thus, as it is well-known, the key to solving this optimization problem is to estimate the covariance matrix $\Lambda_u$. To do this, we compute the sample covariance matrix using the time-$d+u$ option prices simulated under each model. Note that, to obtain a fair comparison, the parameters of each model are only estimated using the options data prior to day $d$. Then, given $N$ samples of the time-$d+u$ options' prices, 
$$
C^{(j)} = \left[C^{(j)}_{d+u}(K_1),...,C^{(j)}_{d+u}(K_n)\right]^T,\,\,\, j = 1,...,N,
$$ 
the sample covariance matrix is estimated as 
\begin{equation}\label{fo:compare:cov}
\Lambda_u = \frac{1}{N-1} \sum_{j=1}^{N} (C^{(j)}-C_d) (C^{(j)}-C_d)^T.
\end{equation} 
Different models generate different simulated paths of options' prices, which then lead to different optimal weights. Naturally, how these optimal weights perform in the real world serves as an indicator of the model consistency. To be more specific, a consistent model should be able to generate portfolios with smaller standard deviation in the returns. 
To estimate the standard deviation of portfolio returns, we define the figure of merit $Q$ as the \textit{average realized deviation} of the portfolio return in the testing period, i.e. 
\begin{equation}\label{fo:compare:Q}
Q = \sqrt{\frac{1}{N_{test}} \sum_{k=1}^{N_{test}} (R^k_u - 1)^2},
\end{equation}
where $N_{test}$ is the number of trials, and $R^k_u$ is the actual portfolio return (given by market data) over a $u$-day period, with initial day $d_k$ and with the optimal weights $\omega^k$, obtained by \eqref{fo:compare:weights} on day $d_k$. Different trials correspond to different initial days $d_k$, i.e. 
\begin{equation}
R^k_u 
%= \frac{\sum_{i=1}^n \omega^k_i C_{D_k+u}(K_i)}{\sum_{i=1}^n \omega^k_i C_{D_k}(K_i)} 
= \frac{1}{M}\sum_{i=1}^n \omega^k_i C_{d_k+u}(K_i).  
\end{equation}
Recall that, by assumption, the mean of $R^k_u$ should always be 1. To make this assumption be consistent with the data, we choose a relatively small time horizon $u$.

\subsection{Simulation algorithms}\label{sec:compare:algo}
As mentioned in the previous subsection, to find the optimal portfolio, we need to estimate the covariance matrix using simulated option prices. In this section, we describe the simulation algorithms for each model. 
\begin{itemize}
\item {\bf Double exponential tangent L\'{e}vy model}.
For this experiment, we need to simulate both the underlying process $S$ and the non-homogeneous L\'evy density $\kappa$. For the double exponential tangent L\'{e}vy model, in particular, we need to complete the following two steps to move one step ahead from $t$ to $t+\Delta t$: 
\begin{itemize}
\item Step 1: Simulate the underlying process by
\begin{equation*}
S_{t+\Delta t} = S_t \exp\{ -\int_{\RR}(e^x-1)\kappa_t(t,x)dxdt + \sum_{k=1}^{N_t} J_k\}.
\end{equation*}  
Here $\kappa_t(t,x)$ is the additive density for immediate maturity $T=t$, $N_t$ is the number of jumps during the $(t,t+\Delta t]$ period, which has a Poisson distribution with parameter $\lambda \Delta t$, where $\lambda = \int_{\RR} \kappa_t(t,x) dx$, and $J_k$'s are the jump sizes having the disturbution $\frac{1}{\lambda}\kappa_t(t,x)dx$. Notice that we approximate the jump component of $\log S$ with a compound Poisson process, which is reasonable given that the jump activity is finite in our setting.   

\item Step 2: Simulate the L\'evy density $\kappa_{t+\Delta t}$ via \eqref{eq.detl.Euler}.
\end{itemize}
Simulating $u$ days ahead requires repeating the two steps $u$ times. We can then use the Fourier transform methods, as described in Subsection~\ref{sec:implementation:simulation}, to calculate time-$u$ option prices, and estimate the covariance matrix to obtain optimal weights. 

\item {\bf Discrete tangent L\'{e}vy model}. 
In this case, we need to complete two similar steps to move from $t$ to $t+\Delta t$: 
\begin{itemize}
\item Step 1: Simulate the underlying process by
\begin{equation*}
S_{t+\Delta t} = S_t \exp\left( -\sum_{j=1}^N (e^{x^j}-1)\kappa_t^j(t)\Delta t + \sum_{k=1}^{N_t} J_k\right).
\end{equation*}  
In the above, $\kappa_t^j(t)$'s are the jump intensities for immediate maturity $T=t$, $N_t$ is the number of jumps during the $(t,t+\Delta t]$ period, which has a Poisson distribution with parameter $\lambda \Delta t$, where $\lambda = \sum_{j\neq M} \kappa_t^j(t)$, and $J_k$'s are the jump sizes having the distribution $\frac{1}{\lambda}\sum_{j\neq M} \kappa_t^j(t) \delta_{x^j}(dx)$. 

\item Step 2: Simulate the intensity $\kappa_{t+\Delta t}$ via \eqref{eq.dtl.Euler}.
\end{itemize}
Repeating the two steps $u$ times, we simulate the L\'evy measure $\kappa$ and the underlying $S$ in $u$ days. We then use the iterative formula \eqref{fo:dtL:PDEsol} to calculate time-$u$ time values, from which we deduce the time-$u$ option prices and estimate the covariance matrix, to obtain the optimal portfolio weights. 

\item {\bf SABR model}. 
The simulation based on SABR model is slightly easier. SABR model, as proposed by Hagan et al. in \cite{hagan2002managing},  describes the dynamics of the forward price $F$ and the volatility $\alpha$ as follows:
\begin{align}\label{fo:compare:sabr}
dF_t & = \alpha_t F_t^\beta dB_t^1, \nonumber \\
d\alpha_t & = \nu \alpha_t dB_t^2,
\end{align}
where $F$ and $\alpha$ are correlated through $dB_t^1 dB_t^2 = \rho dt$. \cite{hagan2002managing} provides the following asymptotic formula for the time-$t$ implied volatility under the SABR model:
\begin{align}\label{fo:compare:sabrimpvol}
\sigma_t(K,T,F_t,\alpha_t) & \approx \frac{\alpha_t}{(F_t K)^{(1-\beta)/2}\big\{1+\frac{(1-\beta)^2}{24}\log^2 F_t/K + \frac{(1-\beta)^4}{1920} \log^4 F_t/K \big\}} \cdot \bigg(\frac{z}{x(z)}\bigg) \cdot \nonumber \\
& \quad \left\{1 + \left[\frac{(1-\beta)^2}{24} \frac{\alpha_t^2}{(F_t K)^{(1-\beta)}} + \frac{1}{4} \frac{\rho\beta\nu\alpha_t}{(F_tK)^{(1-\beta)/2}} + \frac{2-3\rho^2}{24} \nu^2 \right] (T-t) \right\}, 
\end{align}
where $K$ is the strike value, $T$ is the maturity, $F_t$ is the current spot level, and $z$ and $x(z)$ are defined as
\begin{align}
z &= \frac{\nu}{\alpha_t} (F_t K)^{(1-\beta)/2} \log \frac{F_t}{K}, \nonumber \\
x(z) &= \log \bigg\{ \frac{\sqrt{1-2\rho z + z^2} + z - \rho}{1-\rho} \bigg\}. 
\end{align}
As for the parameters' values, \cite{hagan2002managing} suggests that $\beta$ can be fixed in advance and \cite{west2005calibration} verifies empirically that this is a reasonable assumption. In our example, we will use two values of $\beta$: $\beta = 1$ and $\beta = 0.7$. $\beta = 1$ is probably the most natural choice for equity market as it mimics a log-nomal model most closely, and $\beta = 0.7$ is widely used on trading desks as it provides better results for risk management. The other parameters -- the current volatility $\alpha_t$, the volatility of volatility $\nu$ and the correlation $\rho$ -- will be calibrated to market prices by minimizing the sum of squared differences between the market call prices and those produced by the model, calculated with \eqref{fo:compare:sabrimpvol}. With the parameters calibrated on the initial day $d_k$, the forward price and the volatility can be simulated as follows: 
\begin{align}
F_{t+\Delta t} & = F_t e^{-0.5 \alpha_t^2 \Delta t + \alpha_t \Delta B_t^1}, \nonumber\\
\alpha_{t+\Delta t} & = \alpha_t e^{-0.5 \nu^2 \Delta t + \nu(\rho \Delta B_t^1 + \sqrt{1-\rho^2} \Delta B_t^2)},
\end{align} 
where $B_t^1$ and $B_t^2$ are independent. The time-$u$ implied volatilities and option prices can then be computed via \eqref{fo:compare:sabrimpvol}, with the simulated spot $F_{t+u}$ and volatility $\alpha_{t+u}$. 
\end{itemize}

\subsection{Results of empirical analysis}
In this section, we will go through the test procedure in detail and present the test results for the following models: 
\begin{itemize}
\item Double exponential tangent L\'{e}vy model (DETL).
\item Discrete tangent L\'{e}vy model (DTL).
\item SABR model with $\beta = 1$.
\item SABR model with $\beta = 0.7$.
\end{itemize}

Each model will be run in two periods: (I) Jan. 2007- Aug. 2008 and (II) Jan. 2011 - Dec. 2012. For each period, we use the first year's data as a training sample, to estimate the parameters of the tangent L\'{e}vy models, and we use the rest of the data as the testing sample, to compute the figure of merit $Q$ defined in  \eqref{fo:compare:Q}. The division between training and testing samples is shown in Table~\ref{tb:compare:periods}. Please note that we cut off the first period at August 2008 to reduce the impact of the financial crisis.  
The tests will be run on a portfolio of call options and underlying -- referred to as a ``(C + S) portfolio" -- with three, four and five strikes. In each case, we pick every other strike starting from the strike closest to the underlying spot value (in other words, closest to at-the-money) at the moment when the portfolio is constructed. We pick these options because their market prices are most accurate. Assuming the set of available strikes is $K_1 < K_2 < ... < K_n$ and the spot $S$ satisfies $K_{i-1} < S < K_i$, Table~\ref{tb:compare:strikes} illustrates the strikes used in each case.  

For all portfolios, we use a simulation horizon of $u=8$ days, and, at the time $d_k$ when the portfolio is constructed, the options have maturity of $T=d_k+u+30$, so that their time-to-maturity becomes $30$ days when the given simulation period ends. We also assume the budget constraint $M=1$. In addition to the figure of merit $Q$, we also check the \textit{average predicted deviation} defined as
\begin{equation}\label{fo:compare:P}
P = \sqrt{\frac{1}{N_{test}} \sum_{k=1}^{N_{test}} (\omega^{k*})^T \Lambda_{d_k} \omega^{k*}},
\end{equation}
where $\omega^{k*}$ is the set of optimal weights obtained via \eqref{fo:compare:weights} on day $d_k$. The difference between $Q$ and $P$ is another measure of the accuracy of a model's prediction. 
Besides the predicted and realized deviation, one may be interested in how much the optimal portfolio weights fluctuate across the initial days $d_k$. To measure this fluctuation, we define the \emph{average quantity oscillation} index $K$:
\begin{equation}\label{fo:compare:K}
K = \frac{1}{n}\sum_{i=1}^{n}\bigg(\frac{1}{N_{test}} \sum_{k=1}^{N_{test}-1} \left|\omega_i^{(k+1)*} - \omega_i^{k*}\right|\bigg), 
\end{equation}
where $\omega_i^{k*}$ is the quantity of the $K_i$-struck option in the optimal portfolio constructed on day $d_k$.

\subsubsection{Period I}
\label{subse:part1}

For Period I, the parameters estimation for the two tangent L\'{e}vy models is described in Sections~\ref{sec:DETLM} and \ref{sec:dtL} respectively. Following the simulation algorithms outlined in Subsection~\ref{sec:compare:algo}, for every initial day $d_k$ in the testing sample, we simulating 500 sample paths for the underlying and the option prices, using each model, and starting with the actual prices observed on day $d_k$. In the simulation with tangent L\'evy models, we use the drift $\alpha$ and volatility $\beta$ estimated from the training sample (SABR model does not allow for any use of past prices). Using the simulated prices, for each model, we calculate the average predicted deviation $P$, according to \eqref{fo:compare:P}, and estimate the optimal portfolio weights $\omega^{k*}$ via \eqref{fo:compare:weights}. Using these weights, we construct the corresponding and follow its value using the \emph{actual market prices} in the time period $(d_k,d_k+u]$, to compute the average realized deviation $Q$ via  \eqref{fo:compare:Q}.

\begin{figure}[htp]
   \begin{center}
       \begin{tabular}{cc}
       \includegraphics[width=0.45\columnwidth]{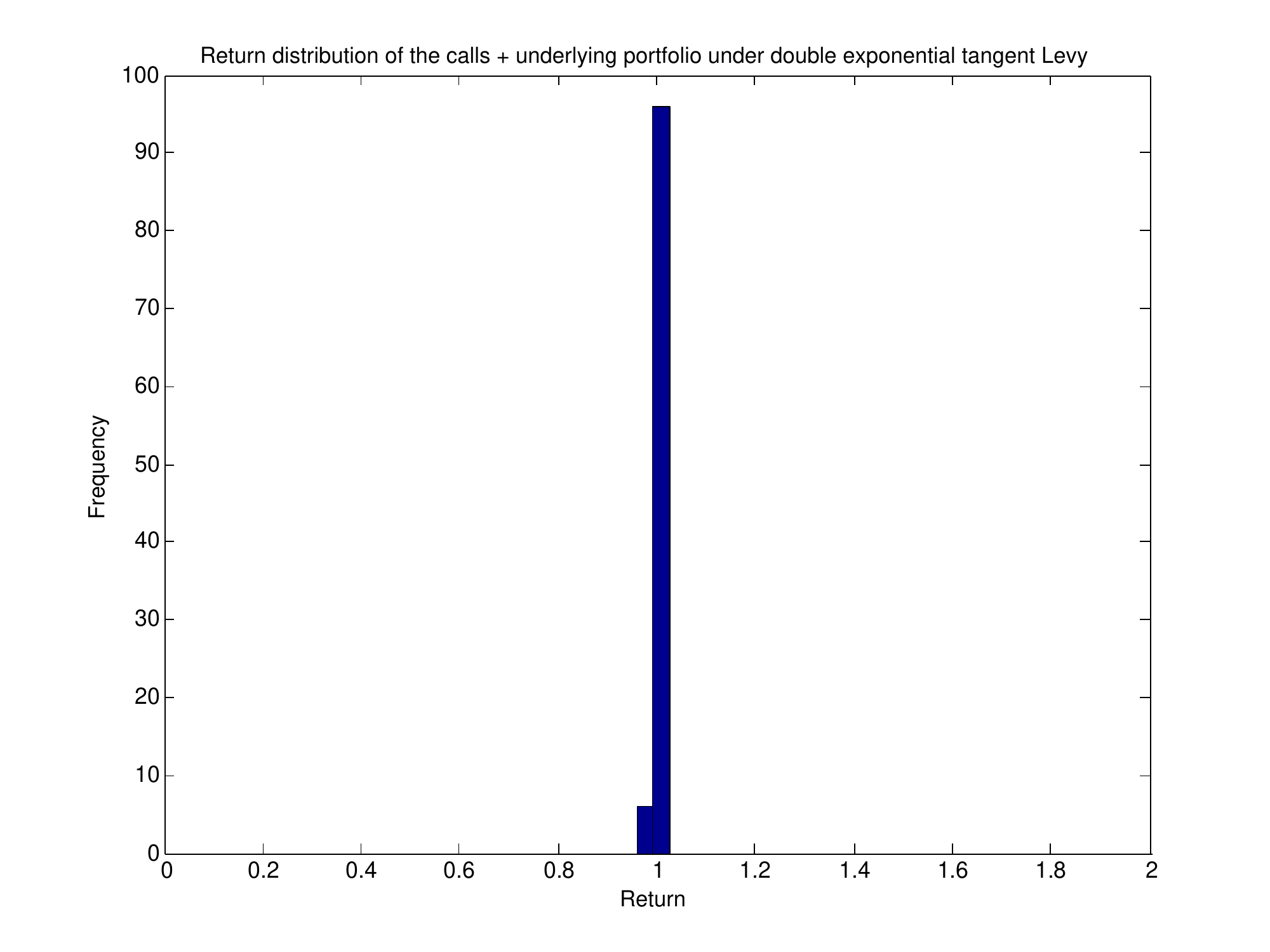}
       &
       \includegraphics[width=0.45\columnwidth]{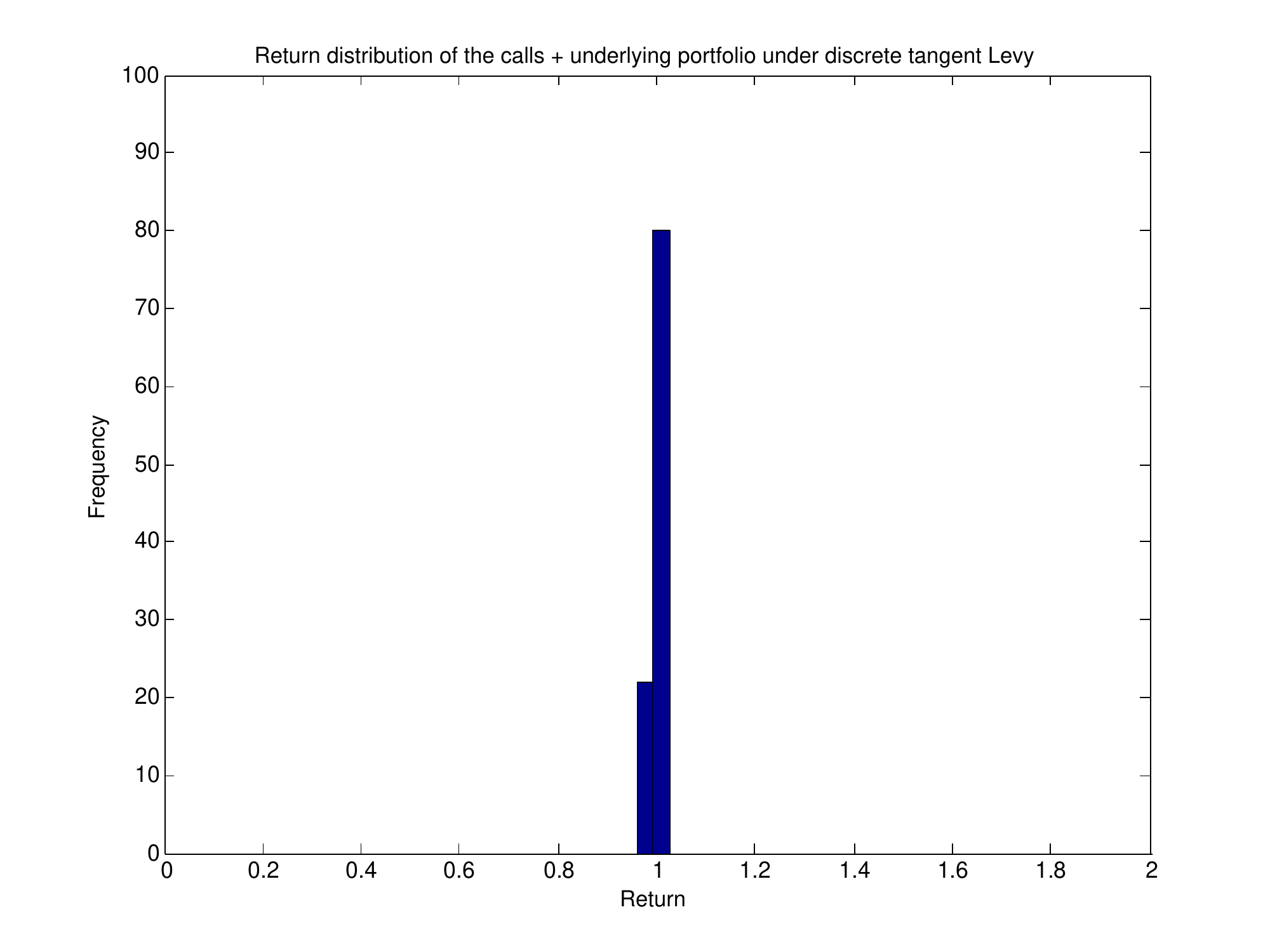}
       \\
       \fontsize{7}{12}\selectfont (a) Under double exponential tangent L\'{e}vy model
      &
       \fontsize{7}{12}\selectfont (b) Under discrete tangent L\'{e}vy model
       \\
       \includegraphics[width=0.45\columnwidth]{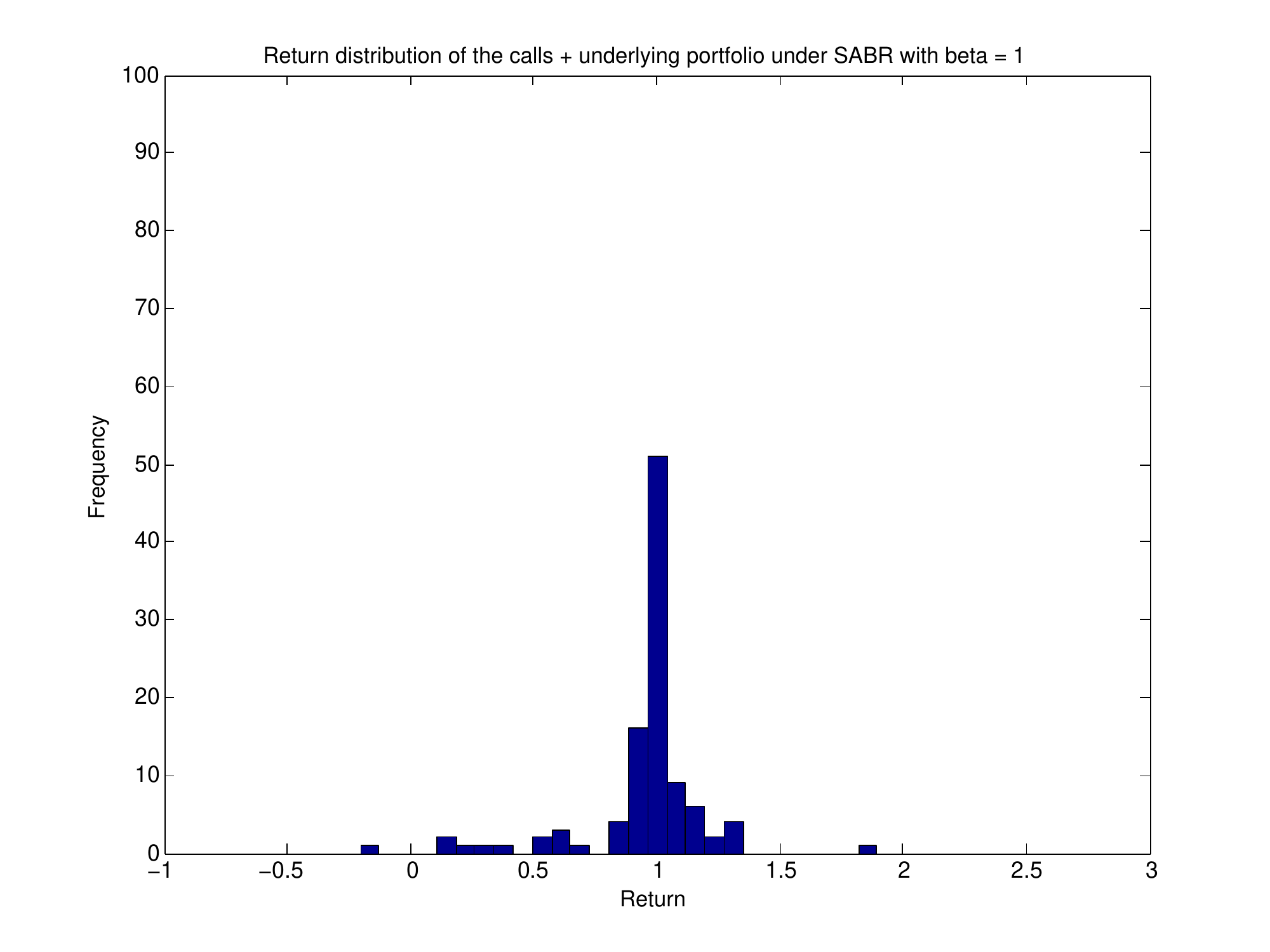}
       &
       \includegraphics[width=0.45\columnwidth]{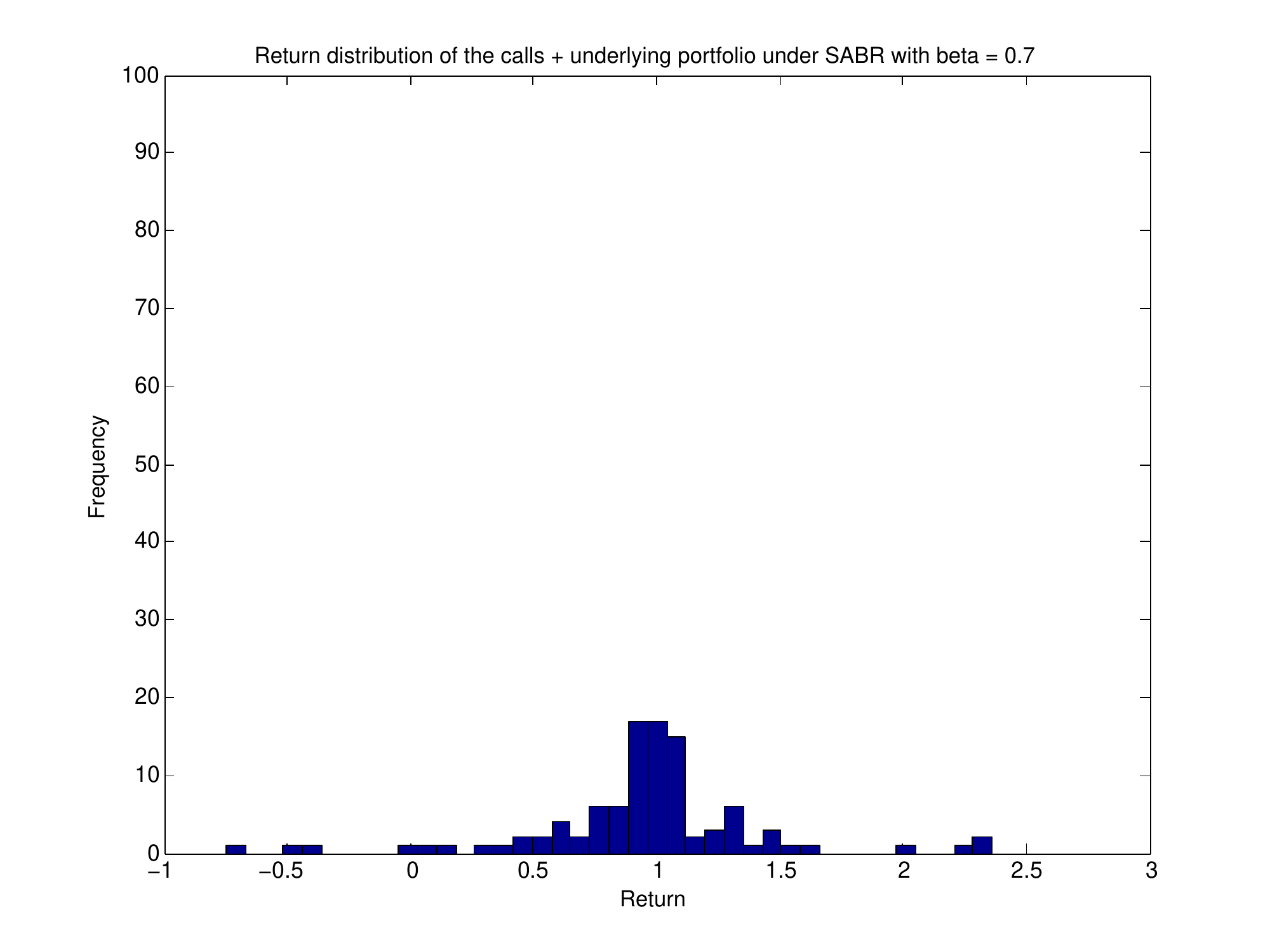} 
        \\
        \fontsize{7}{12}\selectfont (c) Under SABR model with $\beta = 1$
      &
       \fontsize{7}{12}\selectfont (d) Under SABR model with $\beta = 0.7$
       \\
       \end{tabular}
   \end{center}
   \vspace{-10pt}
   \caption{Distribution of the 8-day returns of (C + S) portfolio with 5 strikes in Period I. Different scales are used to show more details.}
   \label{fg:compare:retdistA1}
\end{figure}

\begin{figure}[htp]
   \begin{center}
       \begin{tabular}{cc}
       \includegraphics[width=0.45\columnwidth]{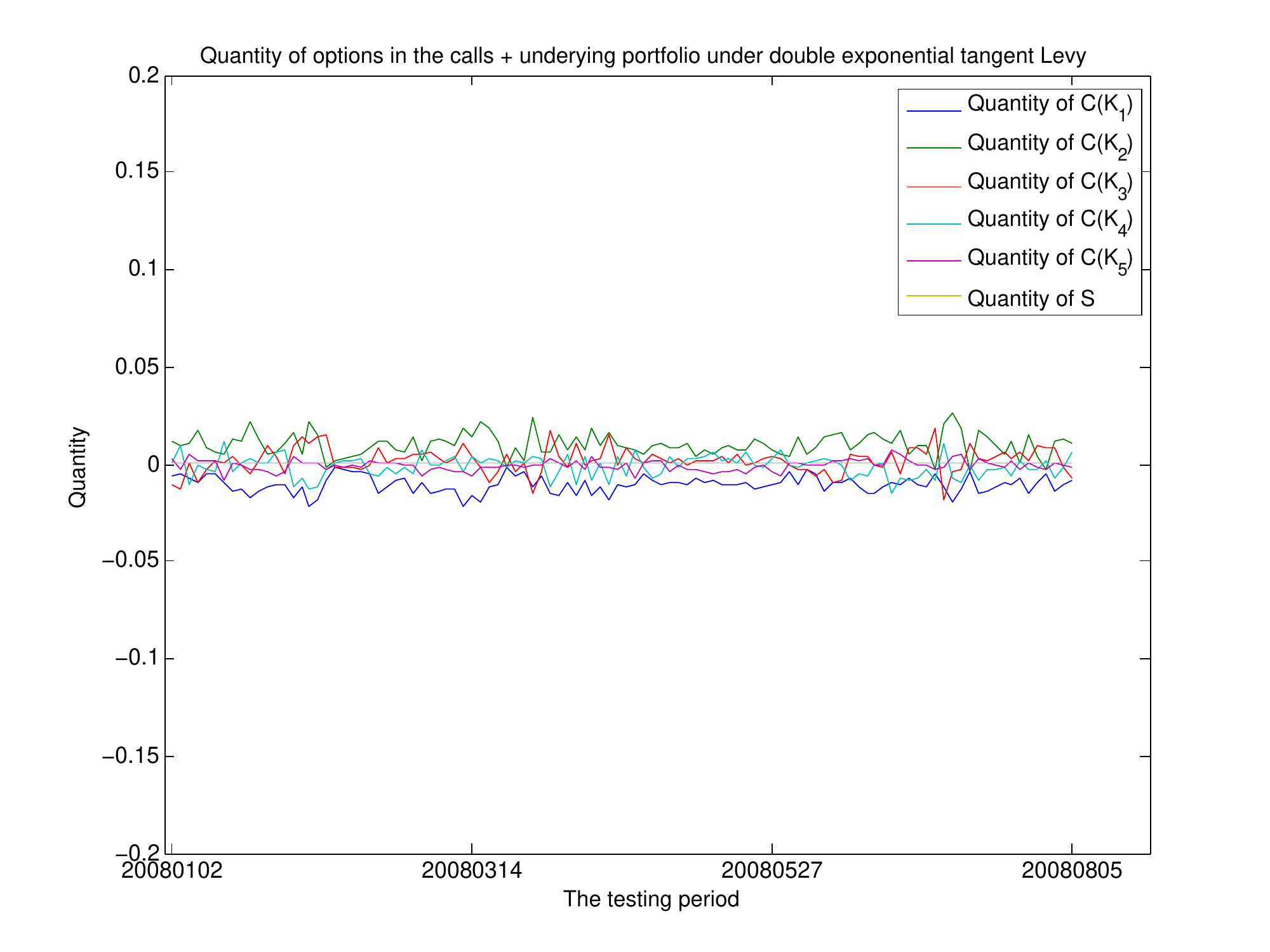}
       &
       \includegraphics[width=0.45\columnwidth]{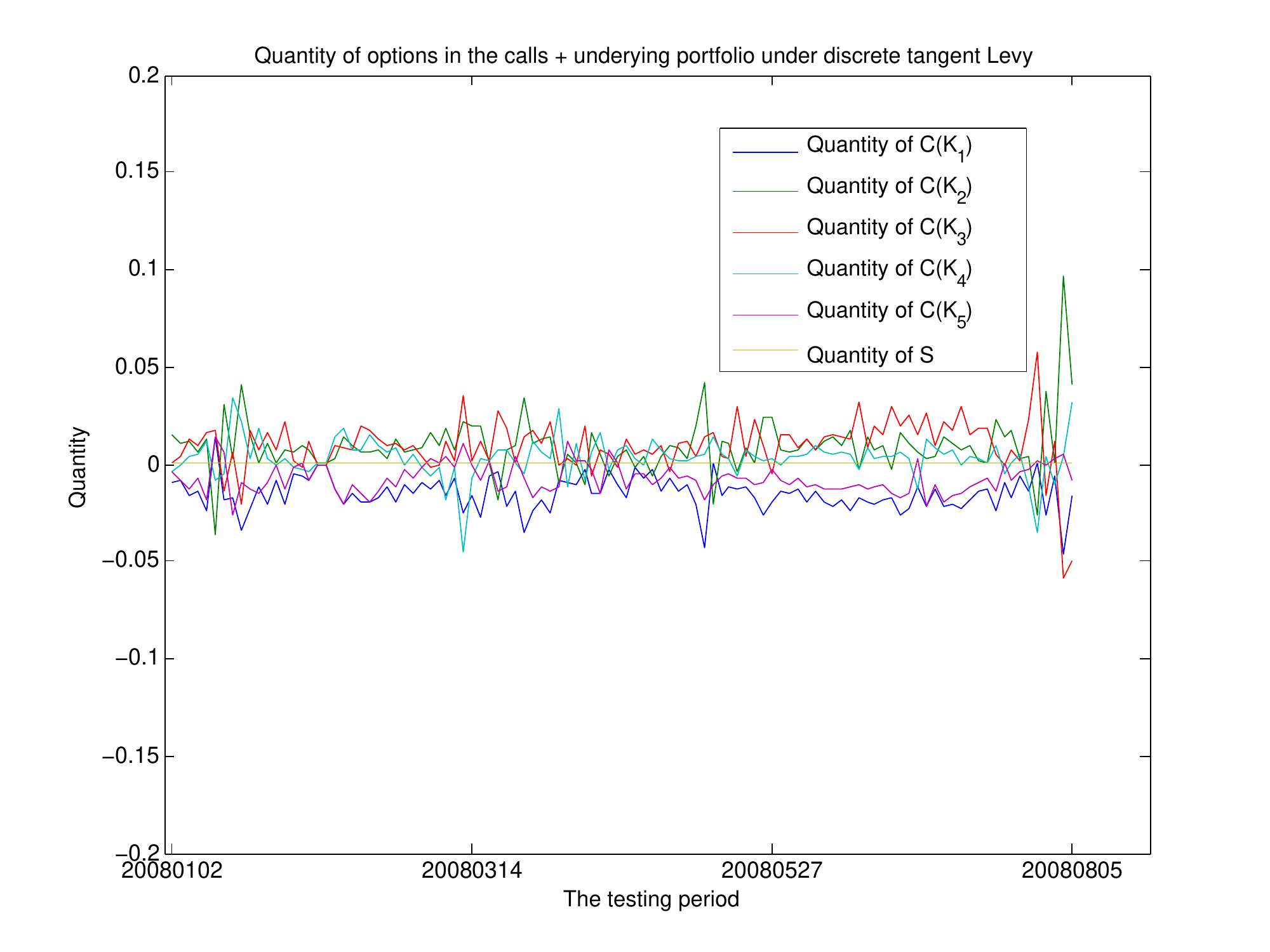}
       \\
       \fontsize{7}{12}\selectfont (a) Under double exponential tangent L\'{e}vy model
      &
       \fontsize{7}{12}\selectfont (b) Under discrete tangent L\'{e}vy model
       \\
       \includegraphics[width=0.45\columnwidth]{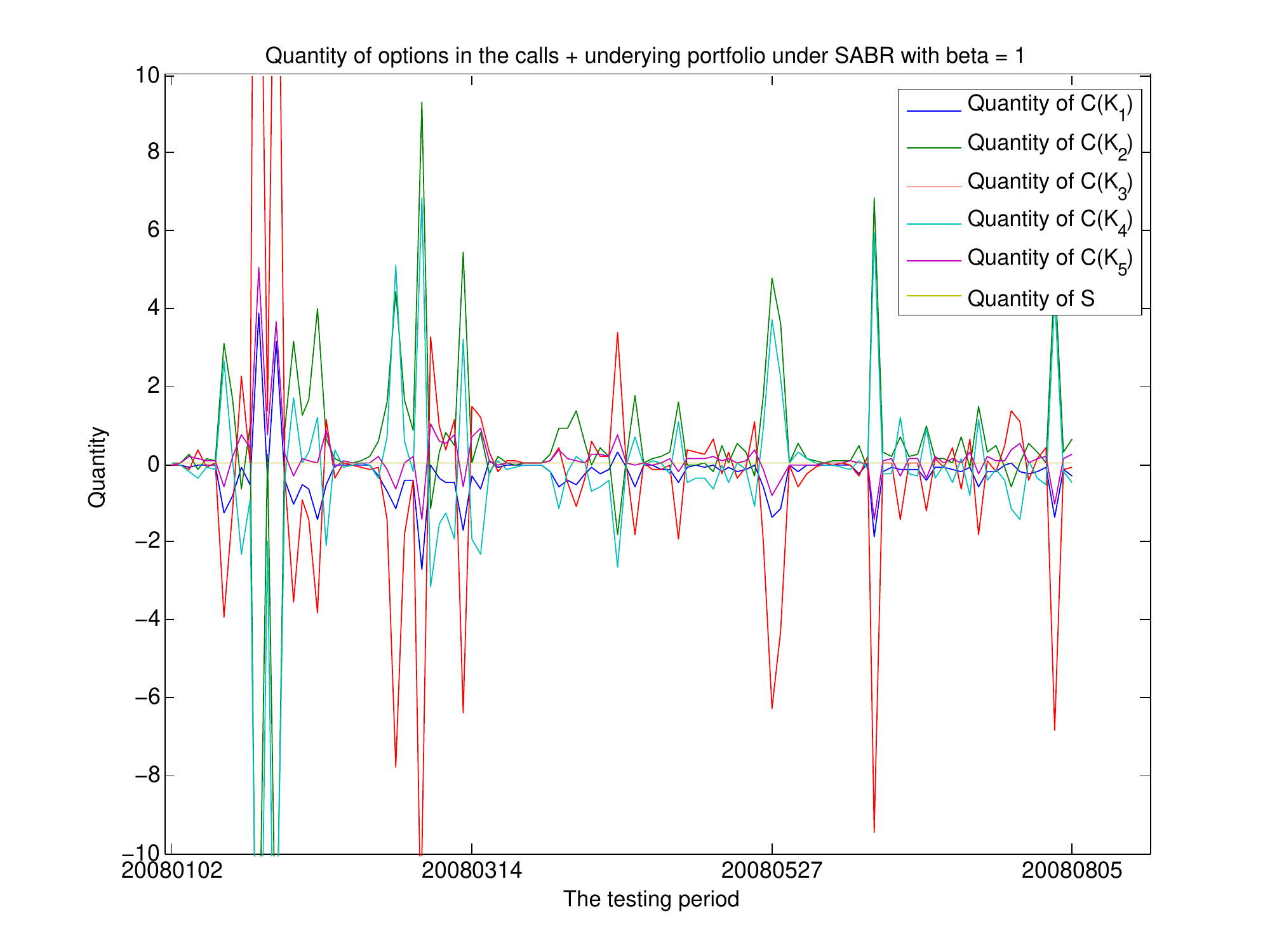}
       &
       \includegraphics[width=0.45\columnwidth]{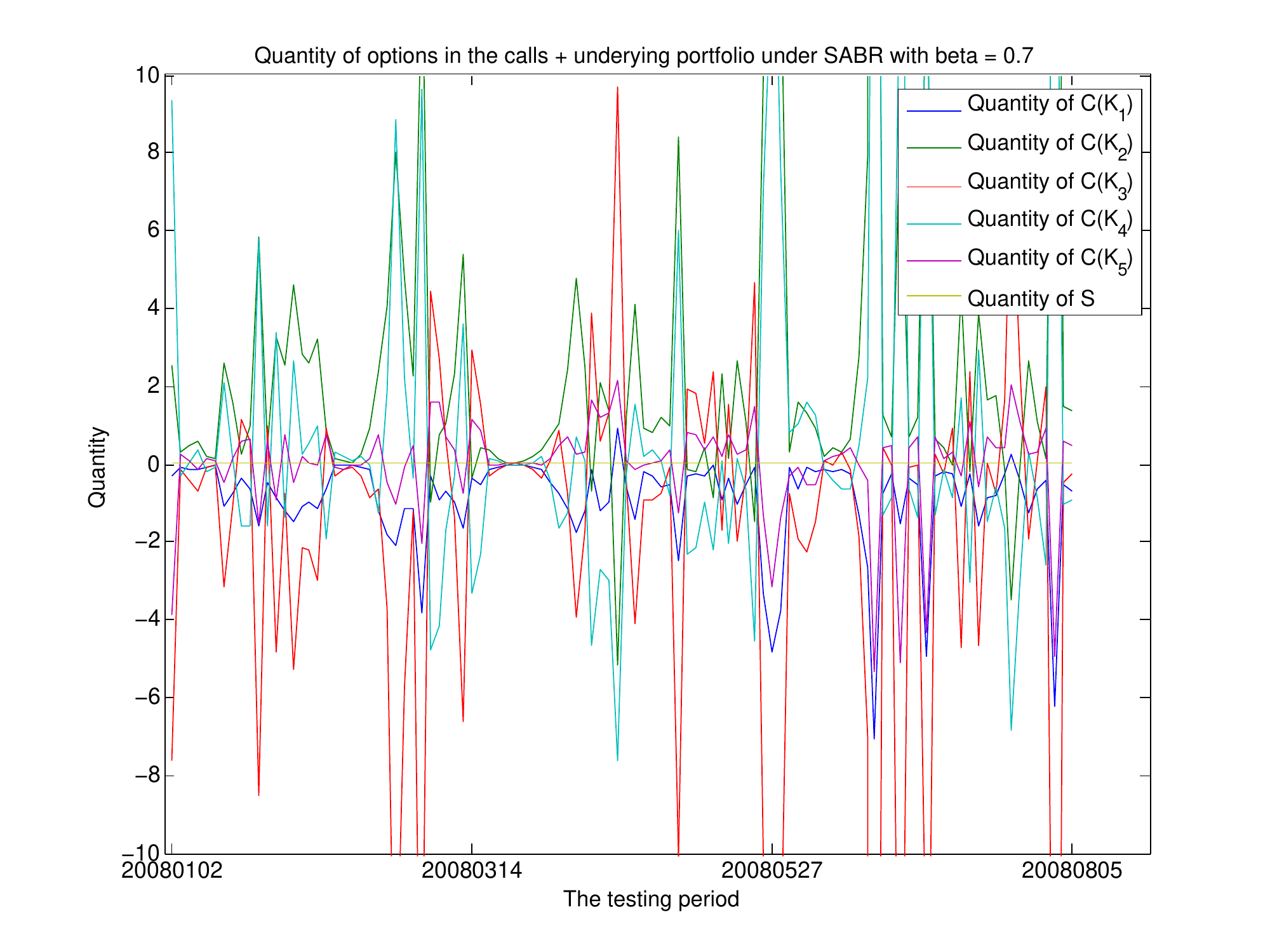} 
        \\
        \fontsize{7}{12}\selectfont (c) Under SABR model with $\beta = 1$
      &
       \fontsize{7}{12}\selectfont (d) Under SABR model with $\beta = 0.7$
       \\
       \end{tabular}
   \end{center}
   \vspace{-10pt}
   \caption{Option quantities in (C + S) portfolio with 5 strikes in Period I. Different scales are used to show more details}
   \label{fg:compare:quanA1}
\end{figure}

The results are shown in Table~\ref{tb:compare:pA1}. It is easy to see that, for a portfolio with 5 strikes, the tangent L\'{e}vy models produce much smaller values of $Q$ than the one produced by a SABR model, indicating that the tangent L\'{e}vy models do a much better job at finding the minimal-variance portfolio. This can also be seen in Figure~\ref{fg:compare:retdistA1}, which shows that the distribution of realized returns is much more concentrated around 1 under the tangent L\'{e}vy models than under the SABR model. 
Furthermore, if we look at the difference between $Q$ and $P$, we can see that it is much smaller for tangent L\'{e}vy models than for the SABR model. This suggests that the tangent models produce a more reliable prediction of the risk of an option portfolio (as measured by the standard deviation of its return) than the SABR model. Besides a small return fluctuation, another nice feature of tangent L\'{e}vy models is the stability of optimal option quantities across the initial days $d_k$. Figure~\ref{fg:compare:quanA1} shows the optimal quantities of options and underlying index, in the portfolio with 5 strikes, across all initial days in the testing period, for every model. Similarly, Table~\ref{tb:compare:qA1} shows the average quantity oscillation $K$ (defined in \eqref{fo:compare:K}) for all portfolios and all models. 
It is easy to see that the portfolio weights constructed via the tangent L\'{e}vy models are much more stable than those constructed using the SABR model. This can be explained by the fact that the parameters of tangent L\'evy models are estimated from both the present and historical option prices, while the classical stochastic volatility models, such as SABR, can only be calibrated to the option prices available on day $d_k$. It is well known (and obvious intuitively) that an estimate based on a larger sample is more robust. Thus, the ability of tangent L\'evy models to be fitted to the historical options prices makes their output (in this case, the optimal portfolio weights) more stable.

\begin{figure}[htp]
   \begin{center}
       \begin{tabular}{cc}
       \includegraphics[width=0.45\columnwidth]{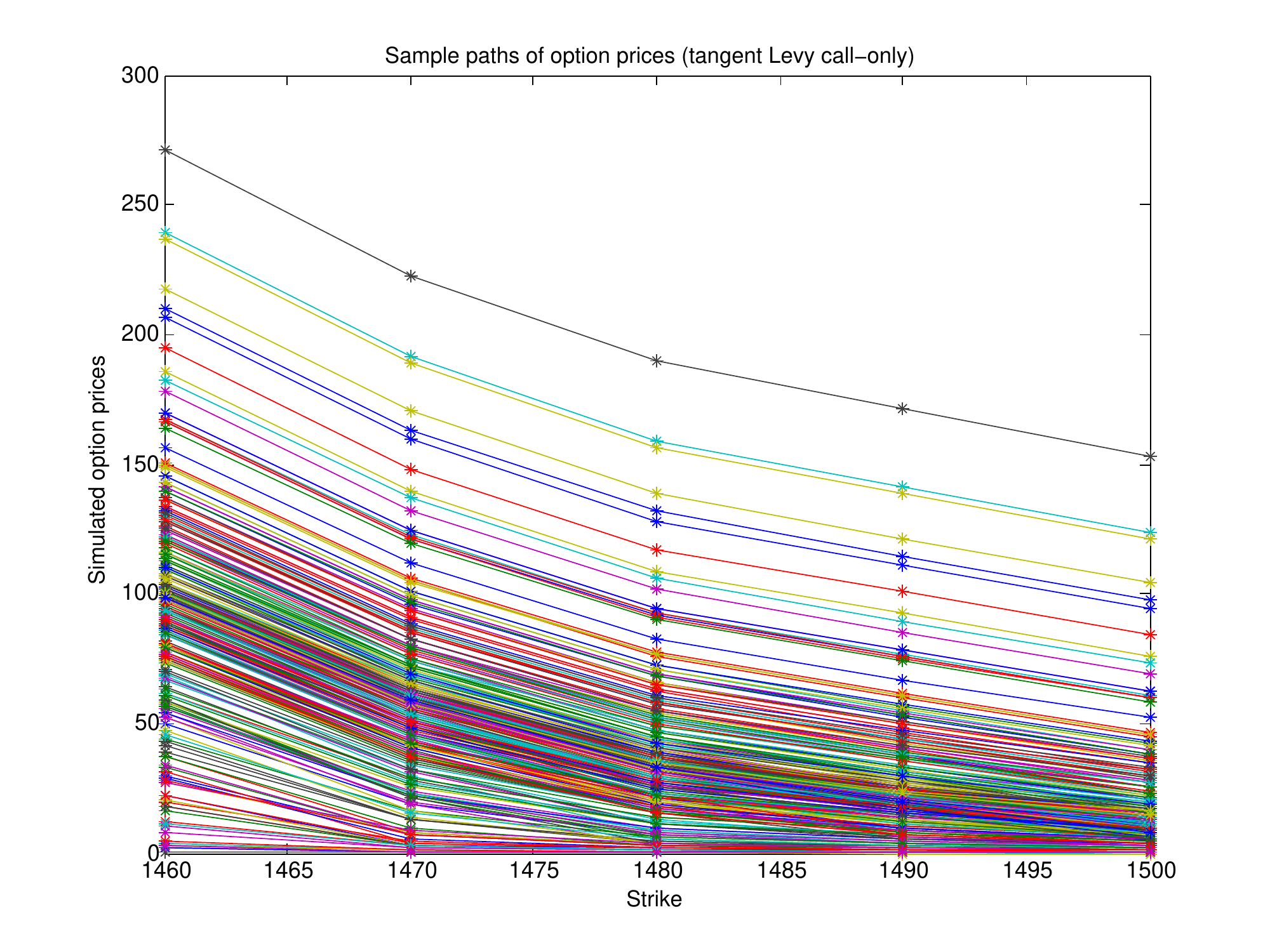}
       &
       \includegraphics[width=0.45\columnwidth]{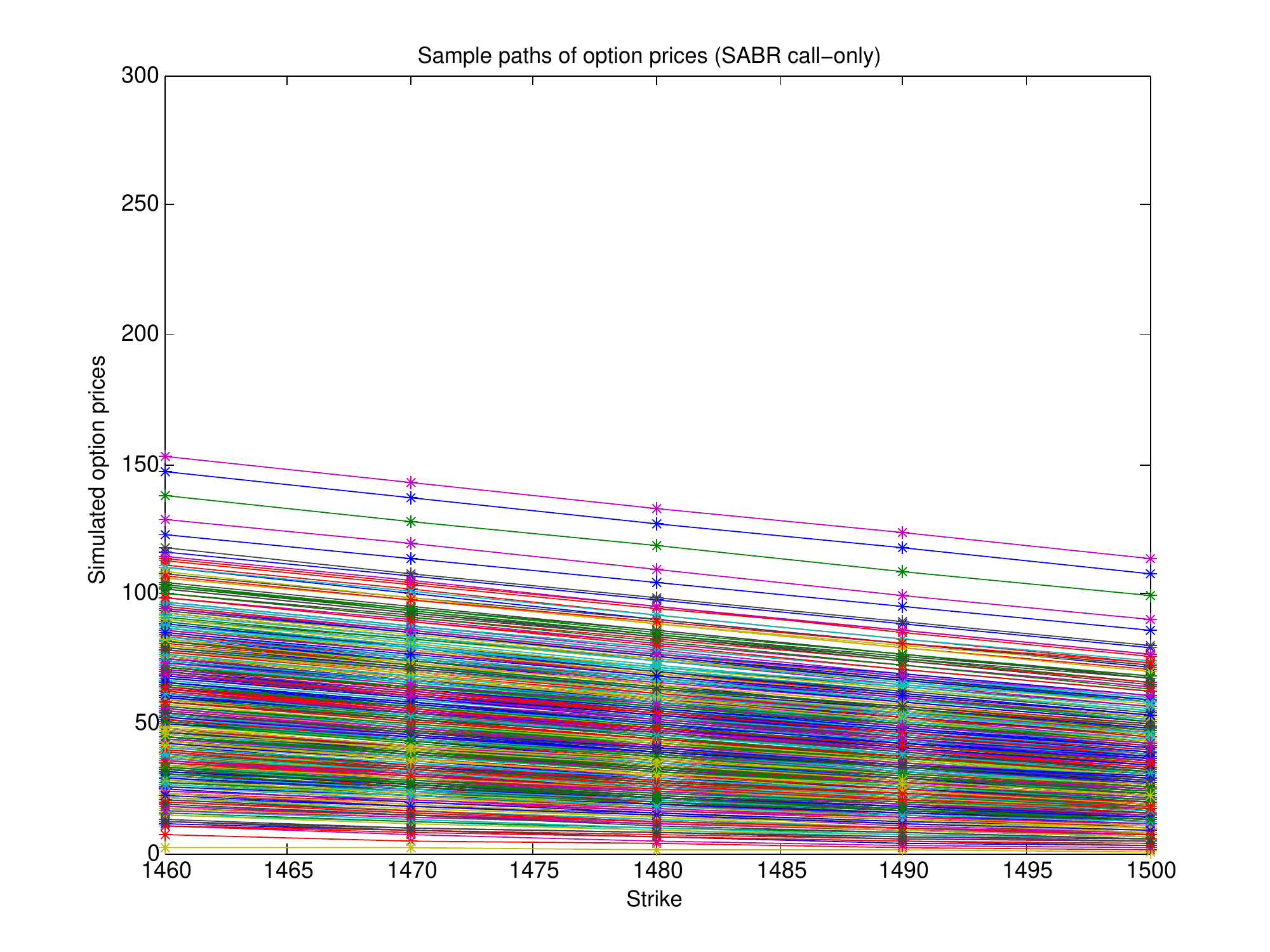}
       \\
          \fontsize{7}{12}\selectfont (a) Under double exponential tangent L\'{e}vy model
      &
       \fontsize{7}{12}\selectfont (b) Under SABR model with $\beta = 1$
          \end{tabular}
   \end{center}
   \vspace{-10pt}
   \caption{Terminal option prices in (C + S) portfolio, as functions of strike, simulated using 500 sample paths}
   \label{fg:compare:callpaths}
\end{figure}

Tables~\ref{tb:compare:pA1} and \ref{tb:compare:qA1} also show that the difference between the performance of tangent L\'evy models and the performance of SABR model shrinks as the number of strikes in the portfolio decreases. This is not a surprise: as the number of strikes decreases, the number of degrees of freedom in the dynamics of option prices, which have to be captured by the model, decreases as well. Eventually, for a very small number of strikes, the SABR models are doing relatively well. However, even in the case of 3 strikes, the tangent L\'{e}vy models do at least as good as SABR (although at a higher computational cost). Thus, the real benefit of using tangent L\'evy models is, of course, only visible when the number of options in the portfolio is relatively large.
Figure~\ref{fg:compare:callpaths} provides a visual explanation for tangent L\'{e}vy models' outperformance. It shows the simulated call option prices, as functions of strike, at the end of the simulation period in 500 sample paths under the double exponential tangent L\'{e}vy model and under the SABR model with $\beta=1$. It is easy to see that the SABR model allows for very limited shapes of call price curves, while the tangent L\'{e}vy model is able to generate a lot more various shapes. It is the lack of variety of different scenarios for the joint evolution of call prices (not simply the lack of parameters in the model) that prohibits the classical stochastic volatility models, such as the SABR model, from capturing the true dynamics of option prices (or implied volatility surface) contained in the historical data.

It is worth noting that the discrete tangent L\'{e}vy model doesn't perform as well as the double exponential tangent L\'{e}vy model, even though it is still superior to SABR model (assuming sufficiently many strikes). 
This can be explained by the fact that the results of the nonparametric static fitting of a tangent L\'{e}vy density, discussed in Subsection~\ref{subsec:dtL:static}, often lack stability, introducing additional noise into the model and, hence, damaging its performance. It could be an interesting area for future work to improve the robustness of static fitting of discrete tangent L\'{e}vy models.

%Therefore, if the actual future prices become very different from the projected ones (this happens from time to time), the resulting portfolio weights might be relatively stable under tangent L\'{e}vy models, but would vary significantly under SABR, leading to the instability of the optimal weights. From a technical perspective, this is because the lack of variety under SABR model makes the covariance matrix $\Lambda_u$ ill-conditioned. Tangent L\'{e}vy models work much better on this front, and are able to produce simulated option prices that are consistent with real-world dynamics. 

\subsubsection{Period II}
\label{subse:period2}

Herein, we repeat the same analysis for Period II. The main purpose of this analysis is to show that the outperformance of tangent L\'evy models is not due to our choice of a testing period, but that it is a persistent property. First, we need to estimate the parameters of the two tangent L\'{e}vy models using the data of year 2011. The calibration procedures are exactly the same as the ones described in Sections~\ref{sec:DETLM} and \ref{sec:dtL}, so we only present the main results here.

\begin{figure}[h]
   \begin{center}
       \begin{tabular}{cc}
       \includegraphics[width=0.45\columnwidth]{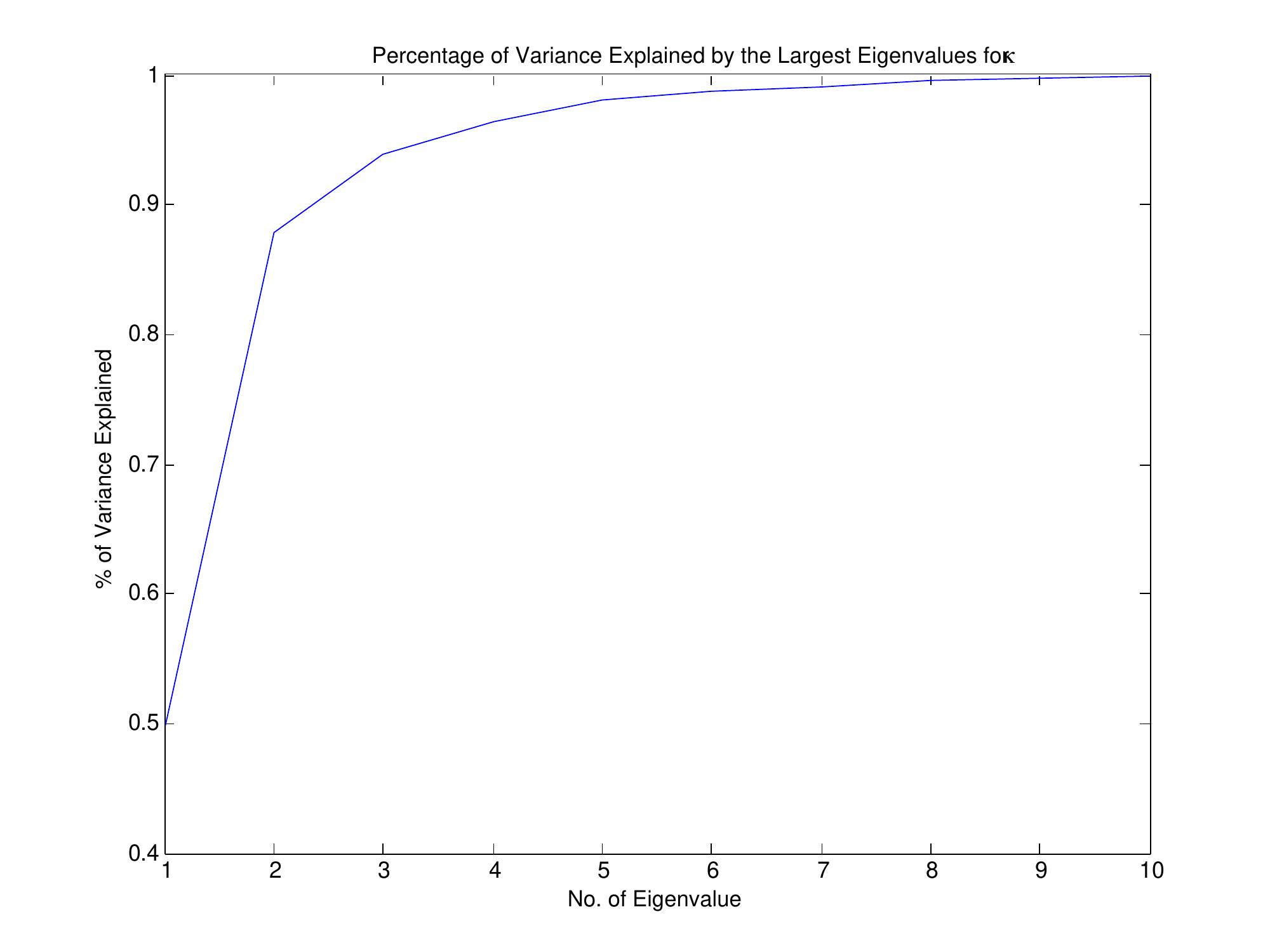}
       &
       \includegraphics[width=0.45\columnwidth]{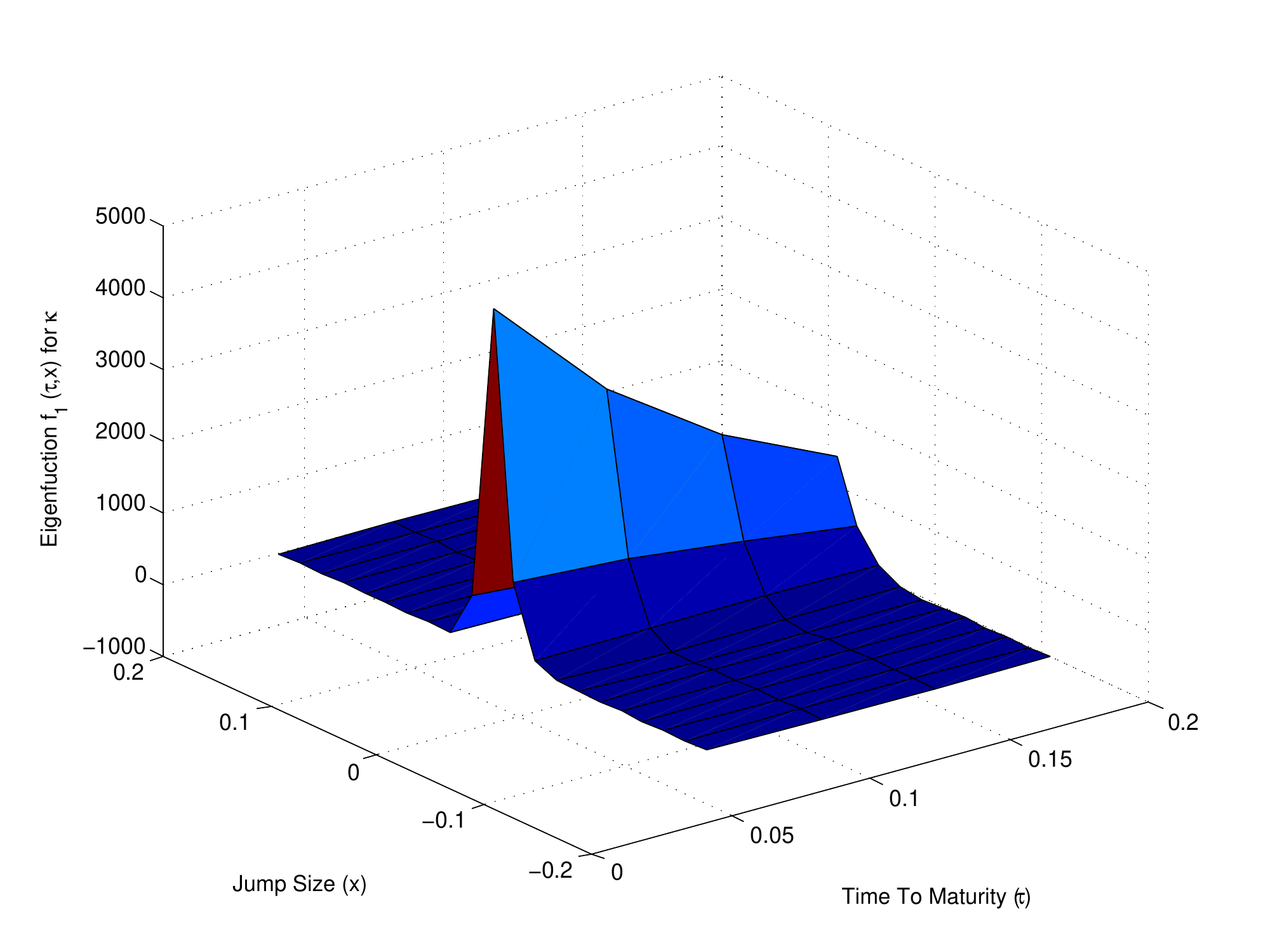}
       \\
      \fontsize{7}{12}\selectfont (a) Percentage of variance explained by the first eigenmodes
     &
      \fontsize{7}{12}\selectfont (b) The first eigenmode scaled by $\sqrt{\lambda_1}$
      \\
       \includegraphics[width=0.45\columnwidth]{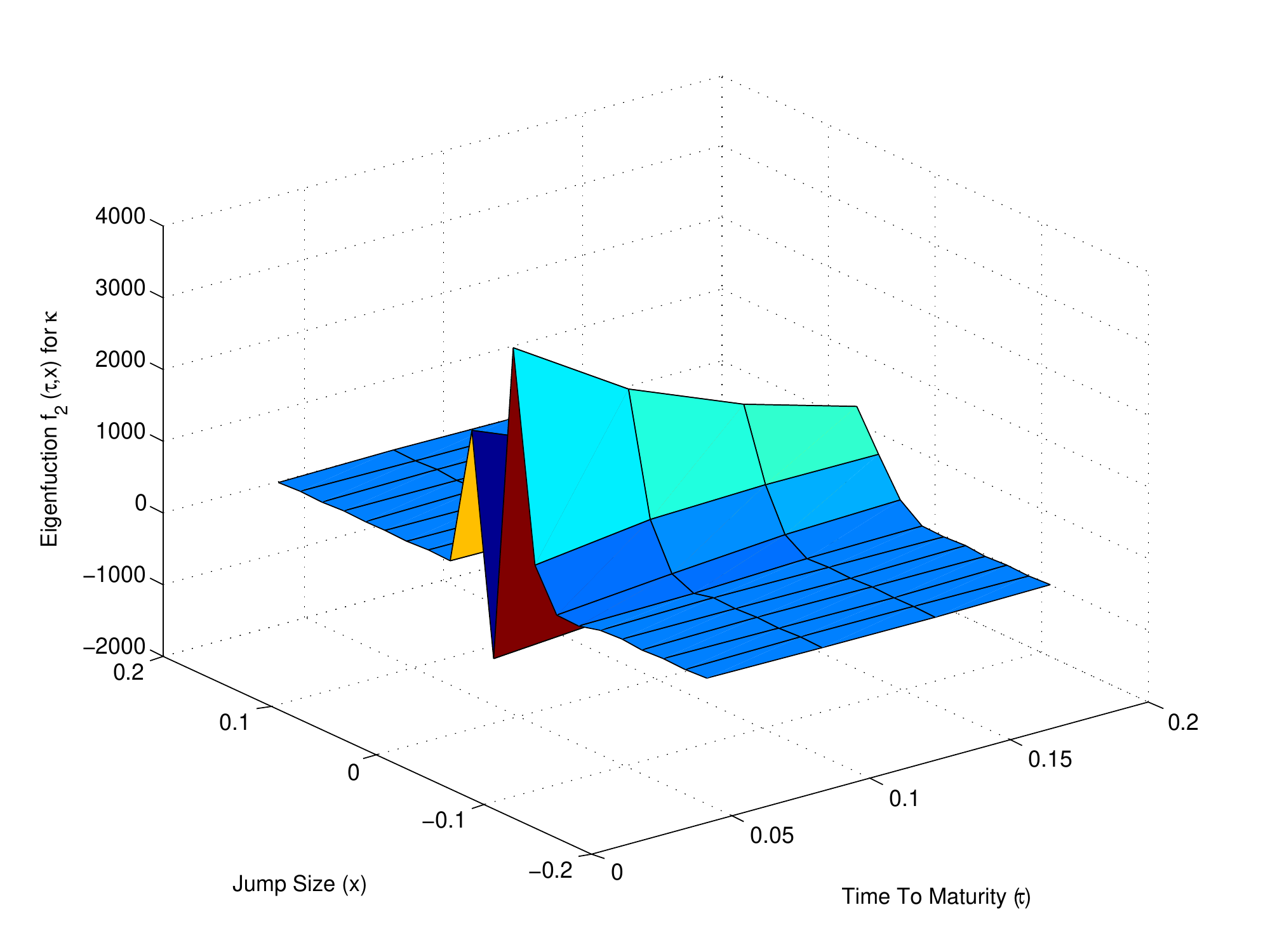}
       &
       \includegraphics[width=0.45\columnwidth]{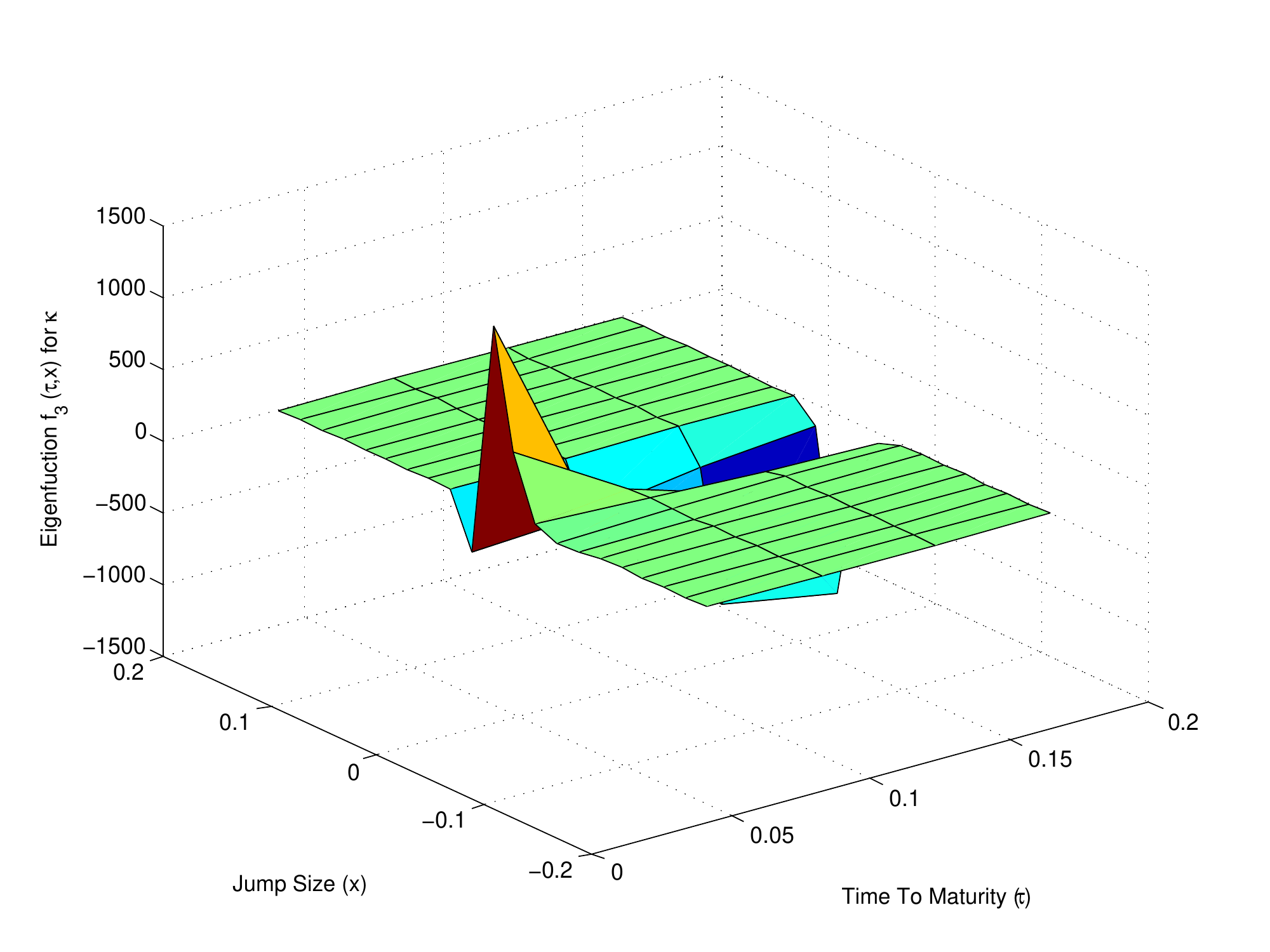}
       \\
      \fontsize{7}{12}\selectfont (c) The second eigenmode scaled by  $\sqrt{\lambda_2}$
     &
      \fontsize{7}{12}\selectfont (d) The third eigenmode  scaled by $\sqrt{\lambda_3}$    
      \\
       \end{tabular}
   \end{center}
   \vspace{-10pt}
   \caption{Eigenvalues and eigenmodes of $\Delta\hkappa$ under DETL, estimated using 2011 data}
   \label{fg:compare:eigen_kou}
\end{figure}

\begin{figure}[htp]
   \begin{center}
       \begin{tabular}{cc}
       \includegraphics[width=0.45\columnwidth]{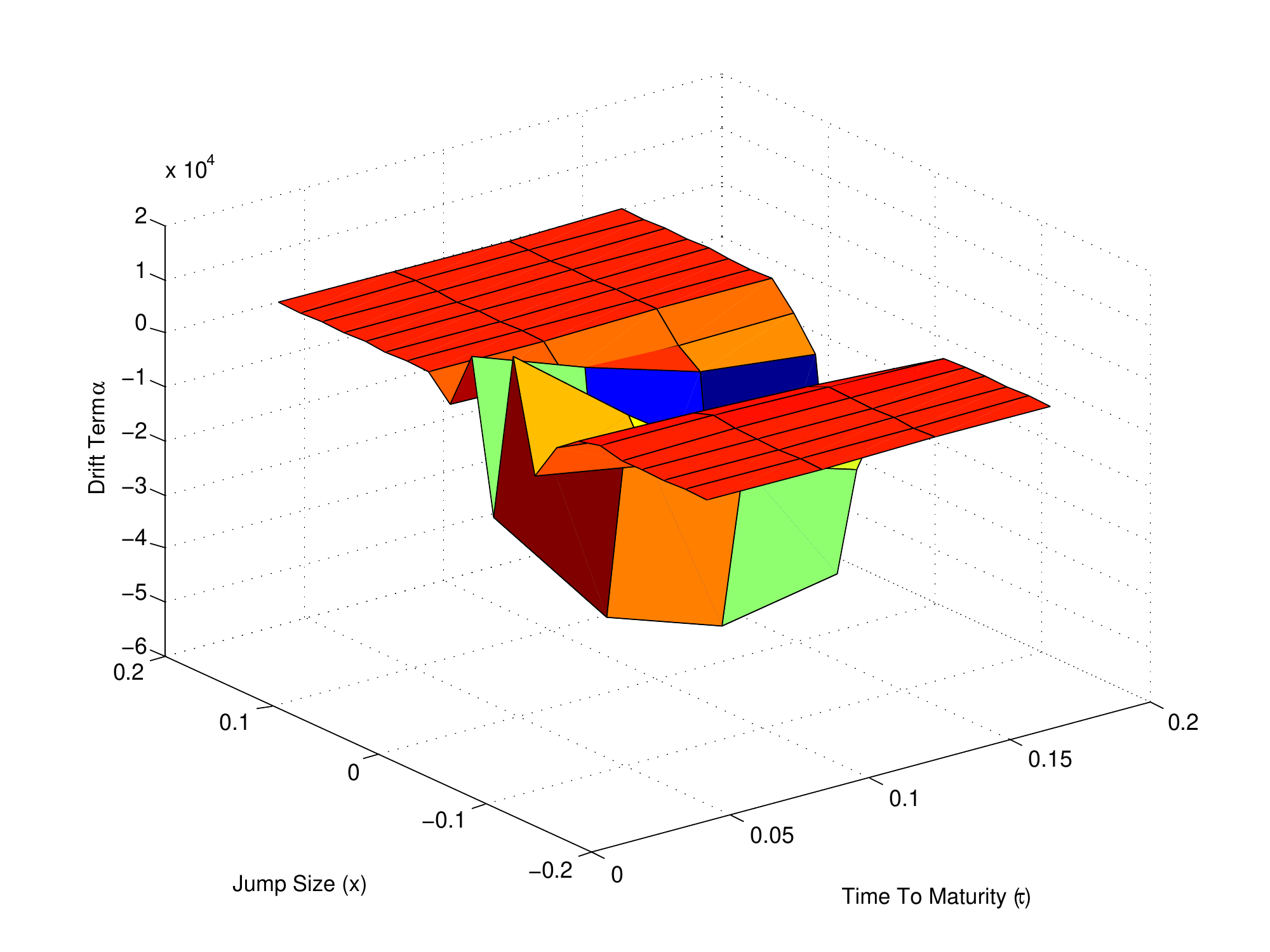}
       &
       \includegraphics[width=0.45\columnwidth]{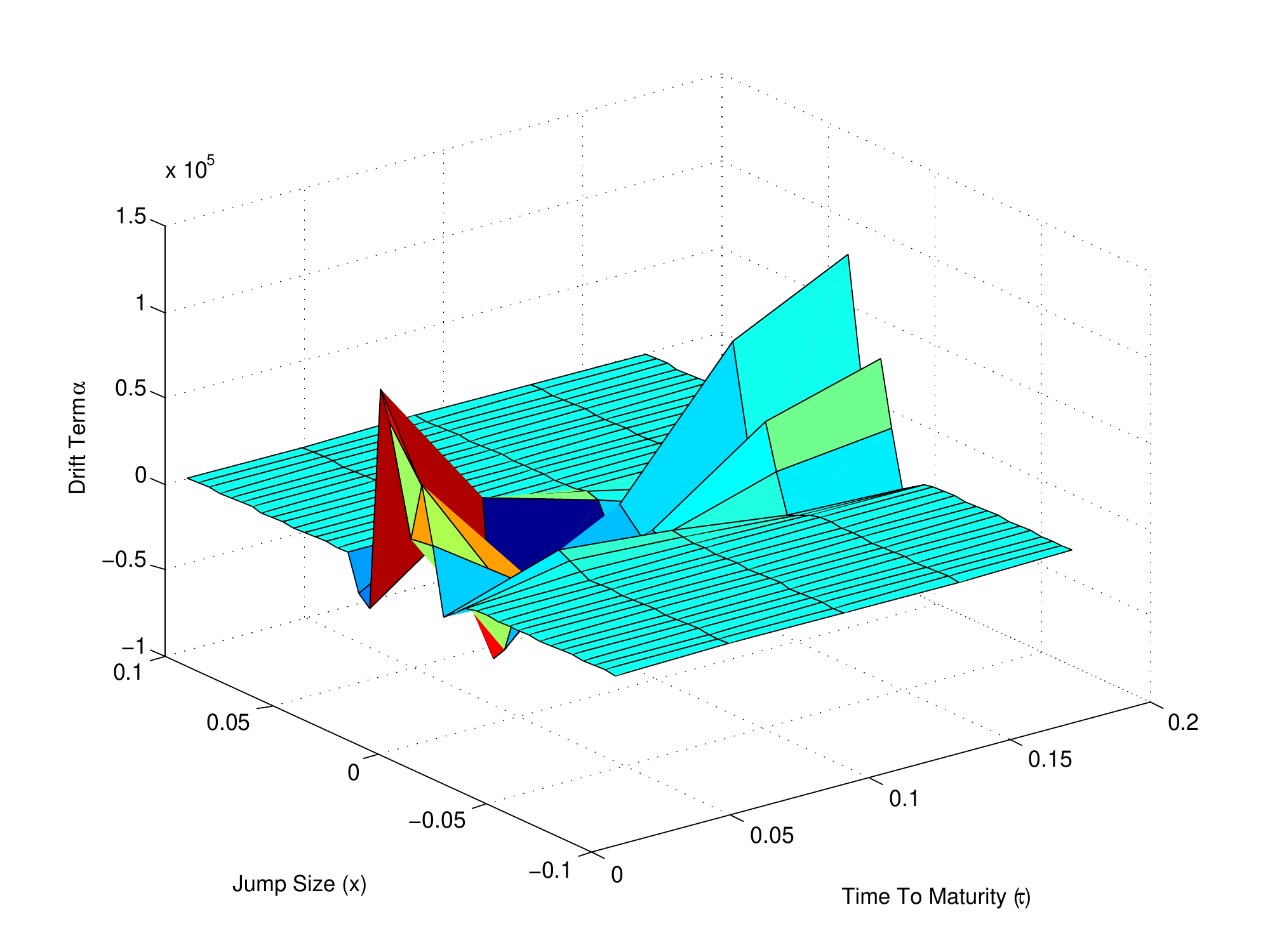}
       \\
       \fontsize{7}{12}\selectfont (a) Under double exponential tangent L\'{e}vy model
      &
       \fontsize{7}{12}\selectfont (b) Under discrete tangent L\'{e}vy model
        \end{tabular}
   \end{center}
   \vspace{-10pt}
   \caption{The drift terms $\alpha$ under DETL and DTL, estimated using 2011 data}
   \label{fg:compare:drift}
\end{figure}

\begin{figure}[h]
   \begin{center}
       \begin{tabular}{cc}
       \includegraphics[width=0.45\columnwidth]{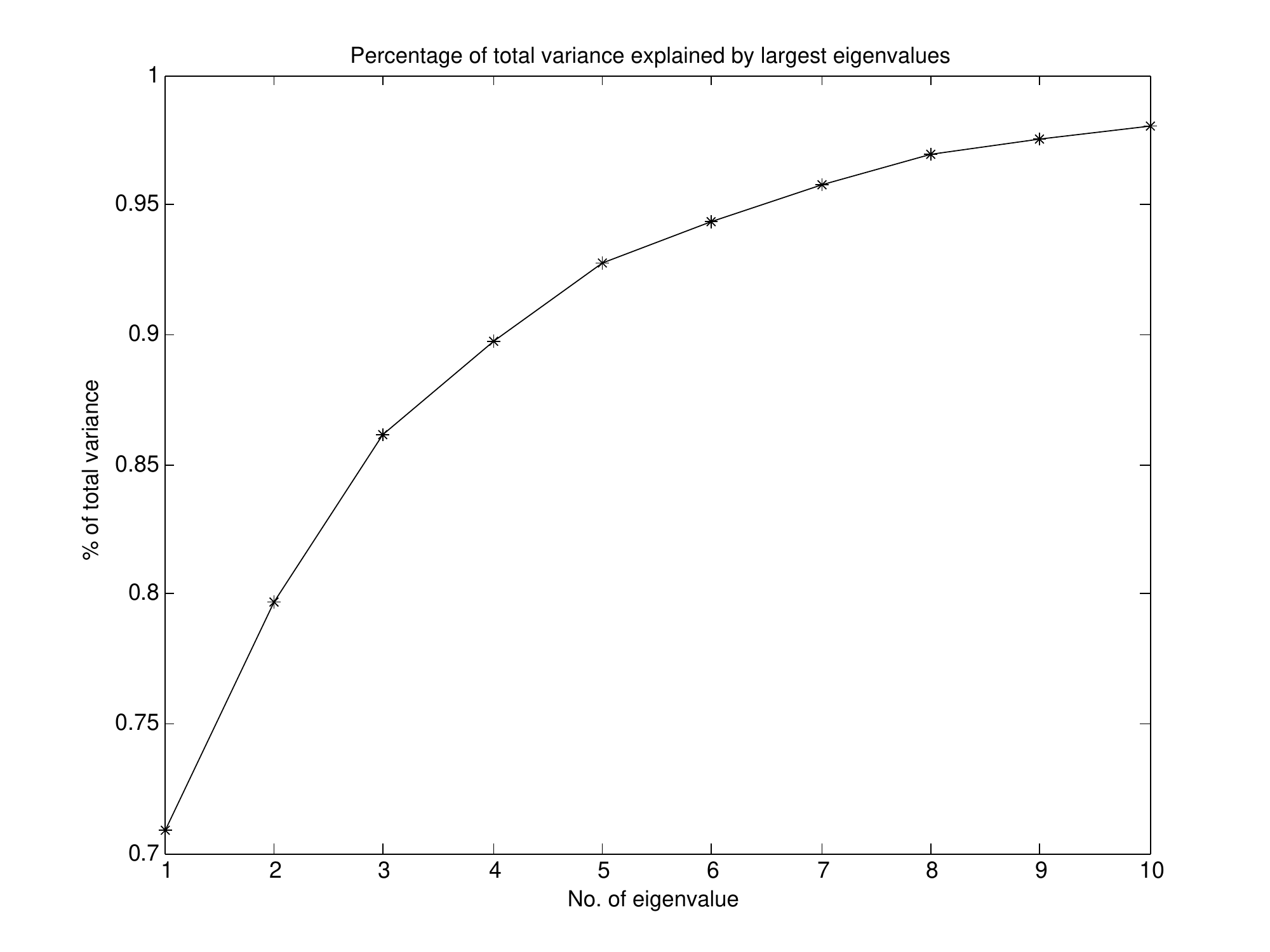}
       &
       \includegraphics[width=0.45\columnwidth]{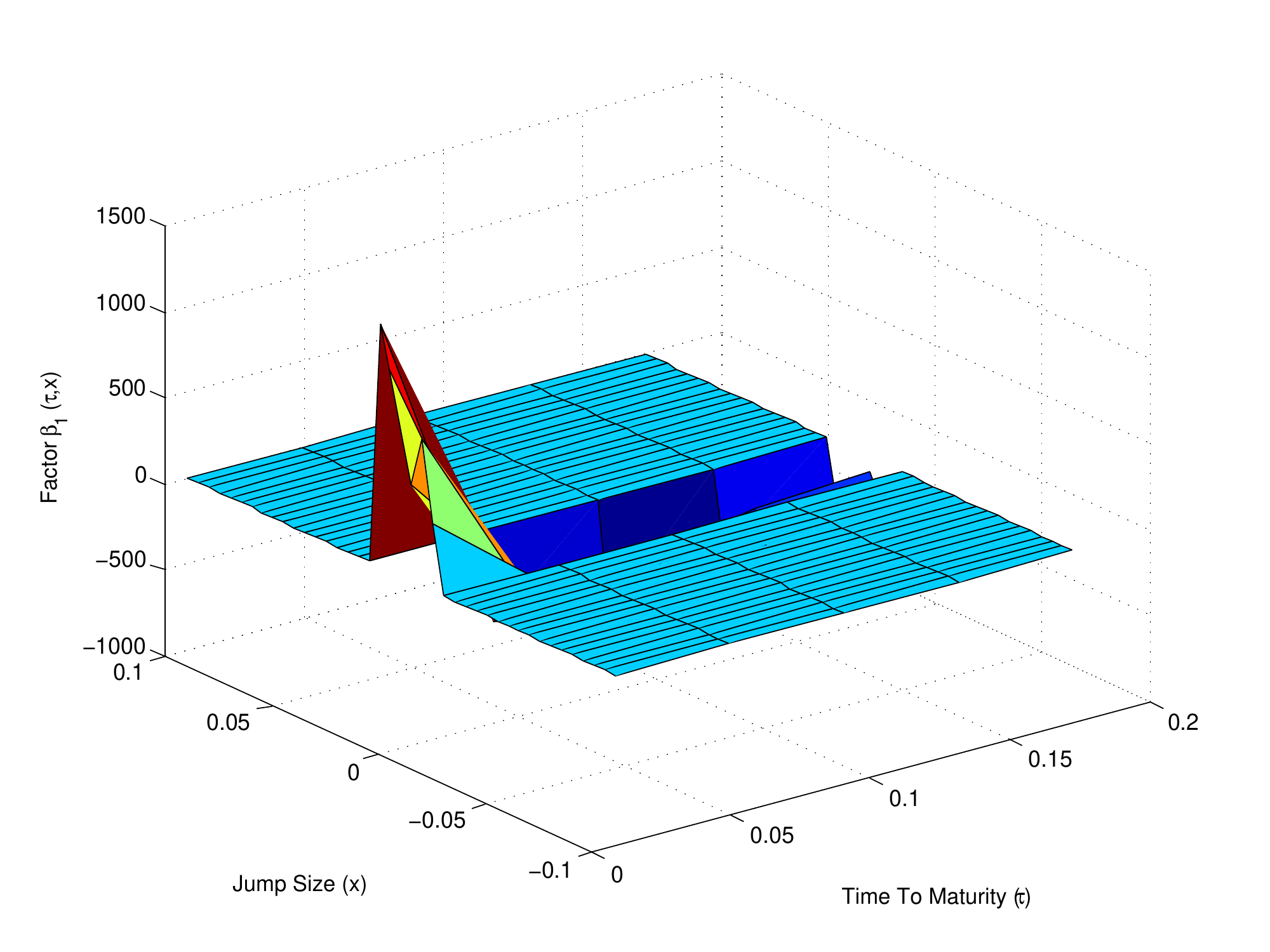}
       \\
      \fontsize{7}{12}\selectfont (a) Percentage of variance explained by the first eigenmodes
     &
      \fontsize{7}{12}\selectfont (b) The first eigenmode scaled by $\sqrt{\lambda_1}$
      \\
       \includegraphics[width=0.45\columnwidth]{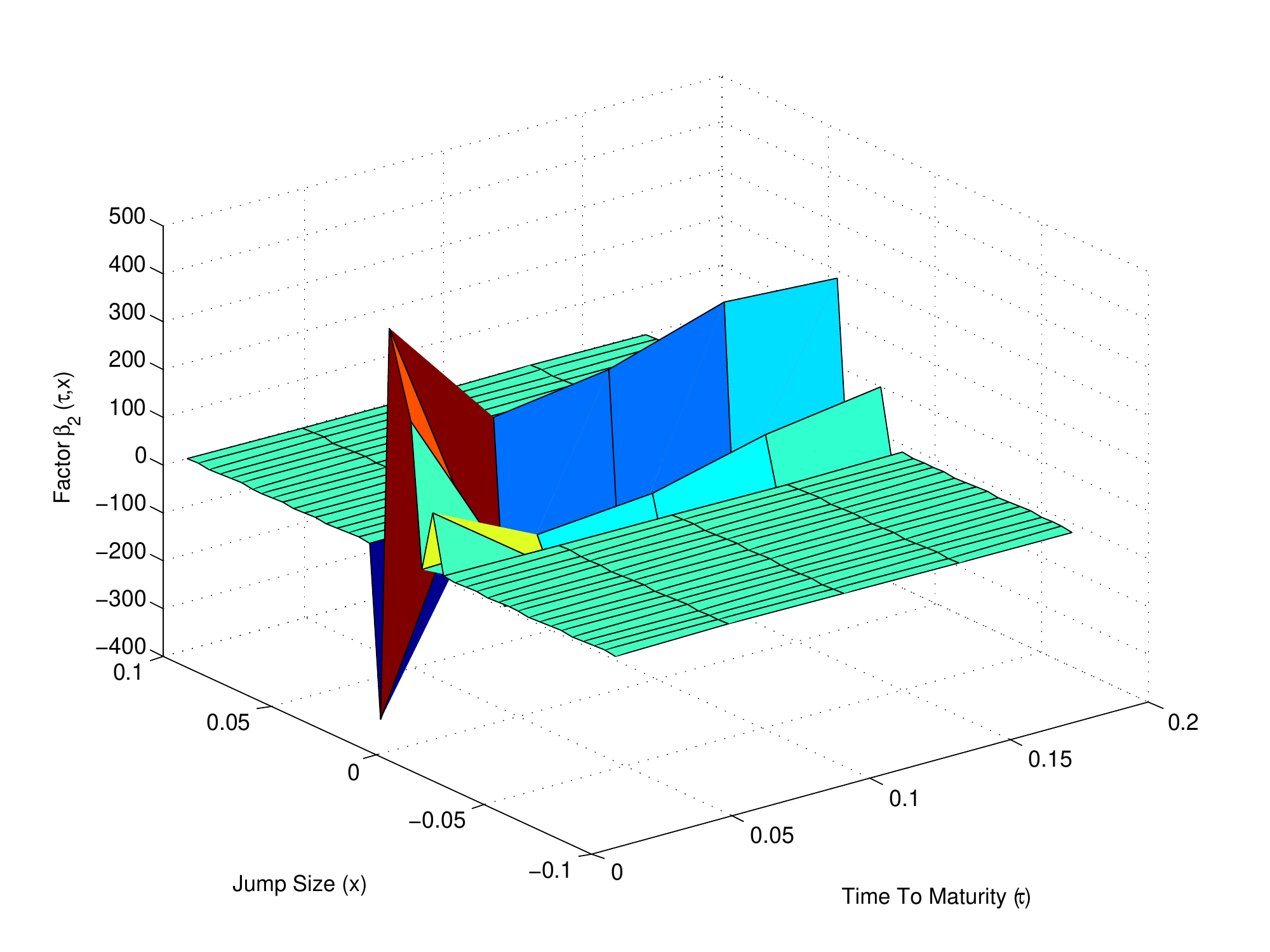}
       &
       \includegraphics[width=0.45\columnwidth]{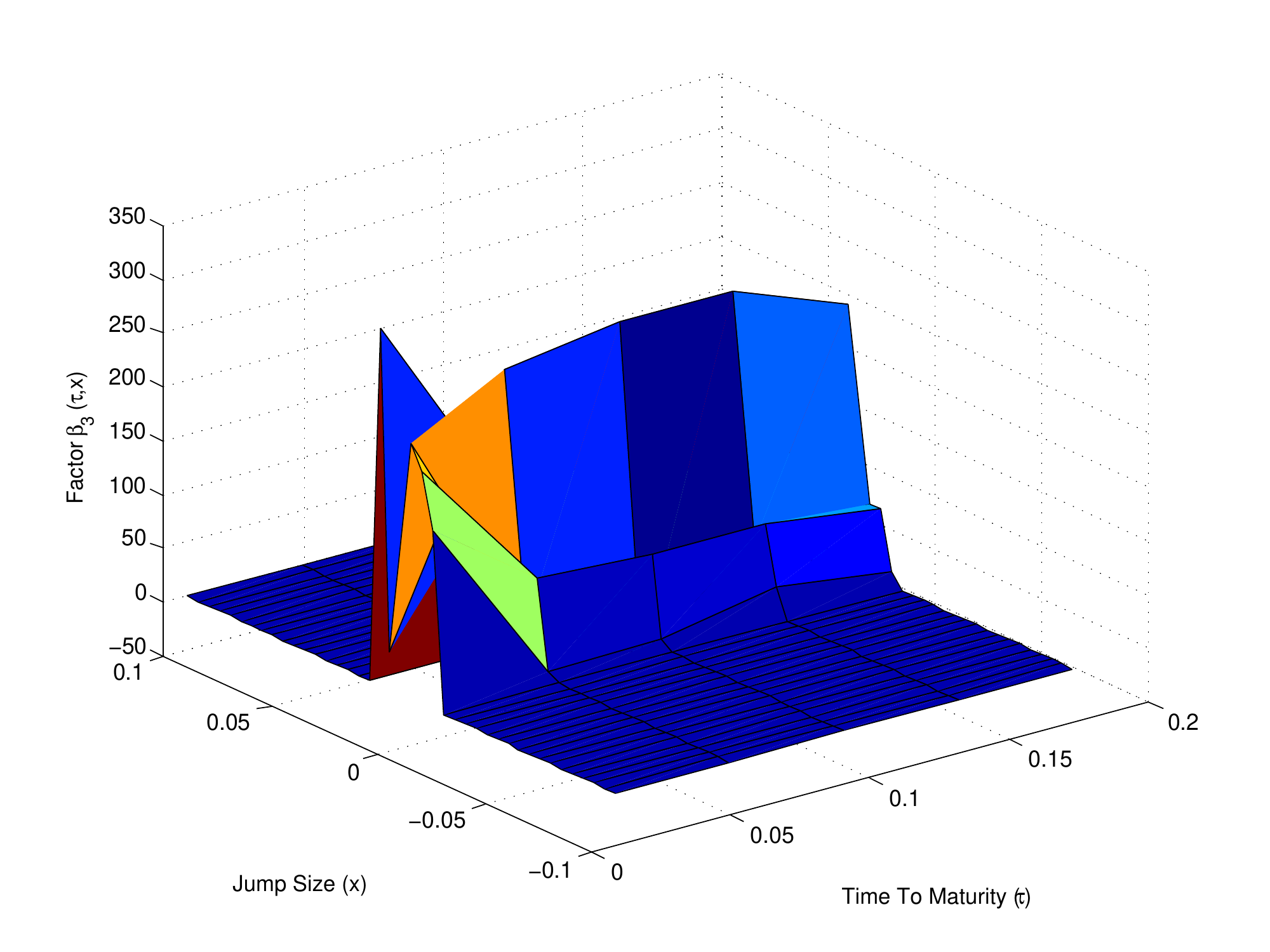}
       \\
      \fontsize{7}{12}\selectfont (c) The second eigenmode scaled by  $\sqrt{\lambda_2}$
     &
      \fontsize{7}{12}\selectfont (d) The third eigenmode  scaled by $\sqrt{\lambda_3}$    
      \\
       \end{tabular}
   \end{center}
   \vspace{-10pt}
   \caption{Eigenvalues and eigenmodes under DTL, estimated using 2011 data}
   \label{fg:compare:eigen_dtl}
\end{figure}

\begin{itemize}
\item In the case of double exponential tangent L\'{e}vy model, the first three eigenmodes explain over 93\% of the variance. The eigenvalues and the eigenmodes are shown in Figure~\ref{fg:compare:eigen_kou}, and the corresponding drift term $\alpha$ is shown in Figure~\ref{fg:compare:drift}.a. Comparing to Figures~\ref{fg:implementation:eigen} and \ref{fg:implementation:drift}, we see that these results are almost the same as for the year 2007, suggesting that this model is very robust. 

\item In the case of discrete tangent L\'{e}vy model, the first three eigenmodes explain over 86\%  of the variance. The eigenvalues, the eigenmodes and the drift term are shown in Figures~\ref{fg:compare:eigen_dtl} and \ref{fg:compare:drift}.b. Note that these graphs are slightly different from those we obtained for year 2007 (Figures~\ref{fg:dtL:eigenmodes} and \ref{fg:dtL:alpha}), although having similar patterns. The lack of robustness can be explained by the fact, in the case of DTL models, we use a non-parametric fit to the data, which may offer a better fit quality, but, at the same time, is known to be less robust than the parametric estimation. 
\end{itemize}

\begin{figure}[h]
   \begin{center}
       \begin{tabular}{cc}
       \includegraphics[width=0.45\columnwidth]{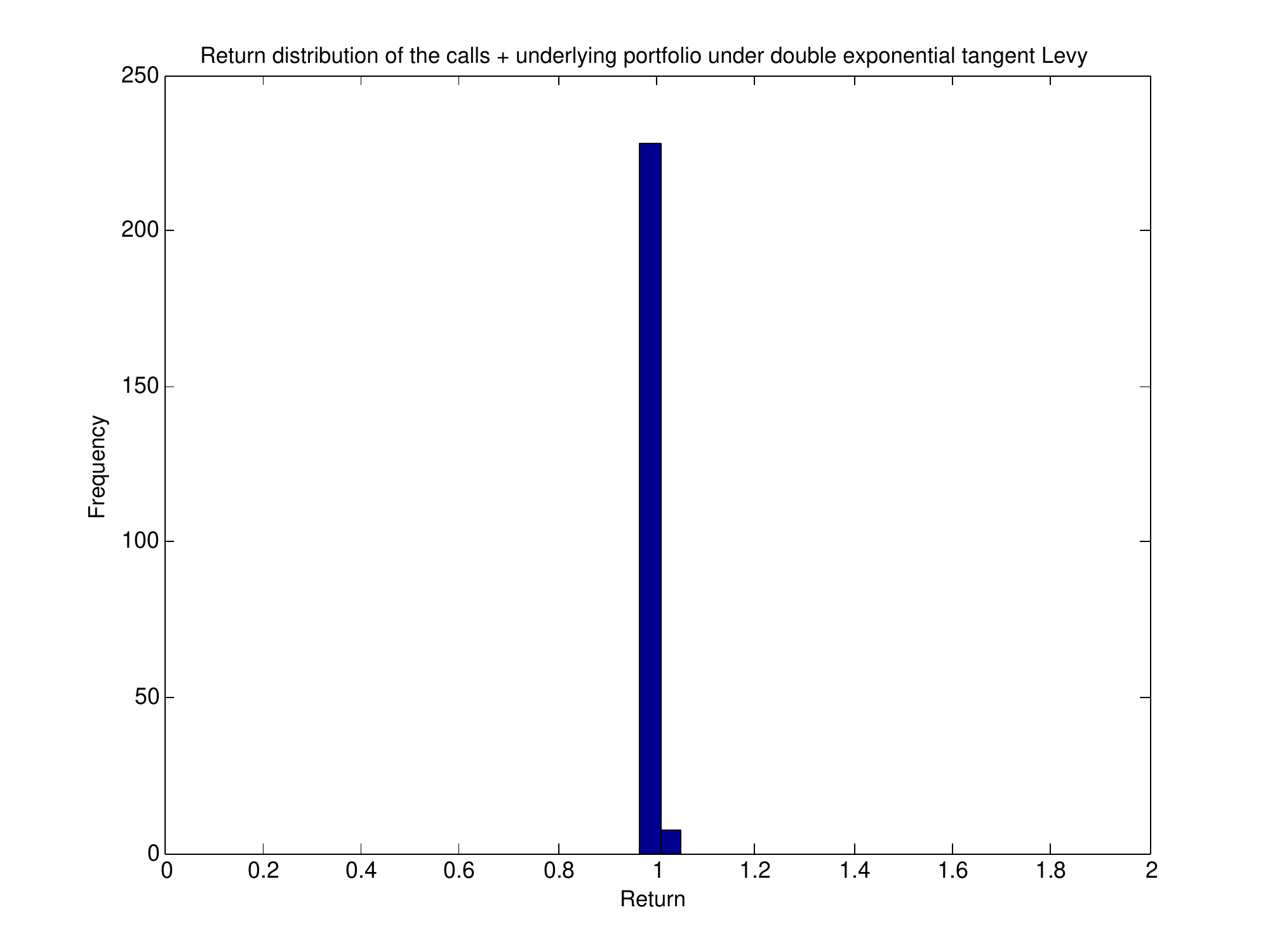}
       &
       \includegraphics[width=0.45\columnwidth]{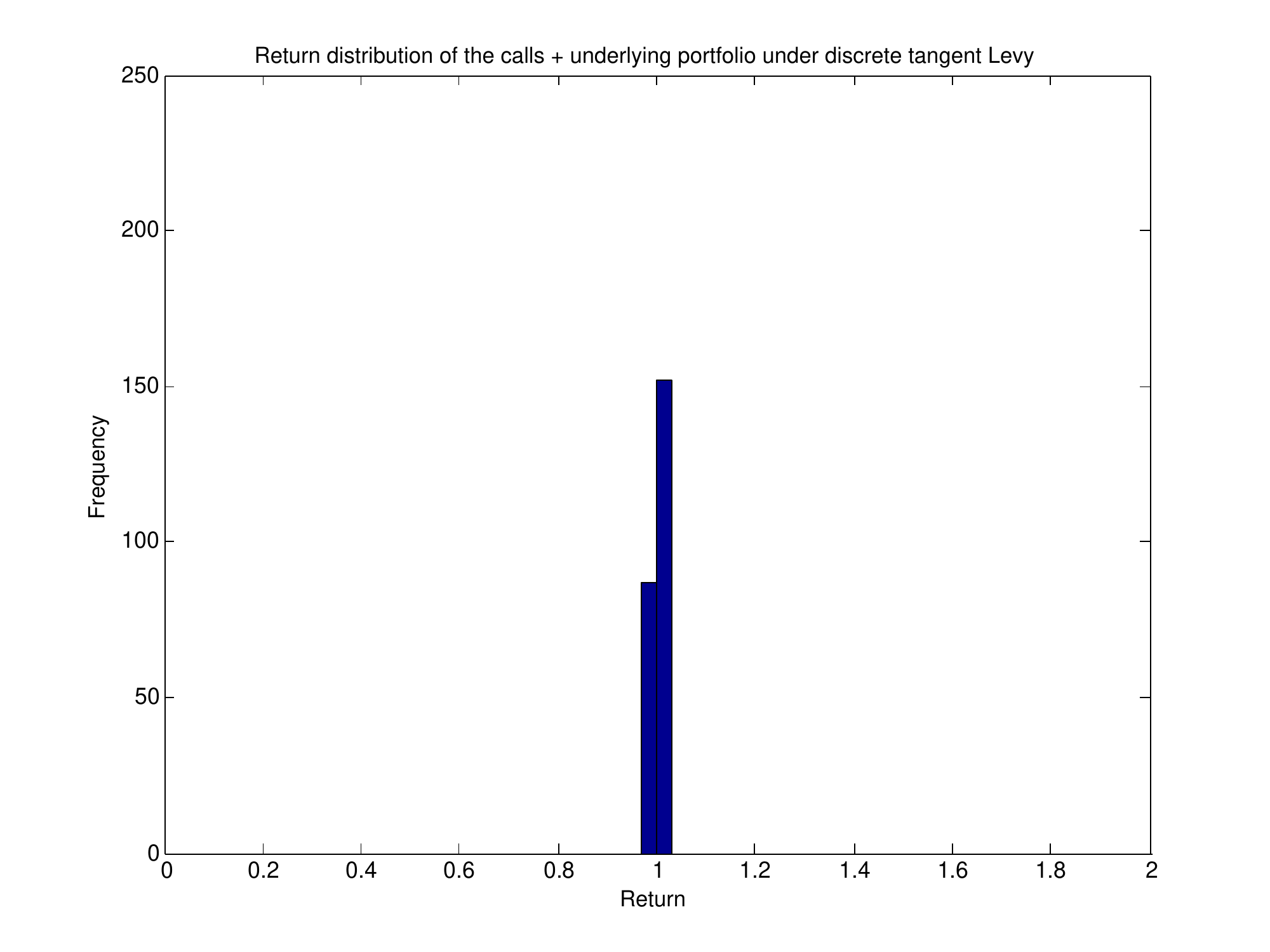}
       \\
       \fontsize{7}{12}\selectfont (a) Under double exponential tangent L\'{e}vy model
      &
       \fontsize{7}{12}\selectfont (b) Under discrete tangent L\'{e}vy model
       \\
       \includegraphics[width=0.45\columnwidth]{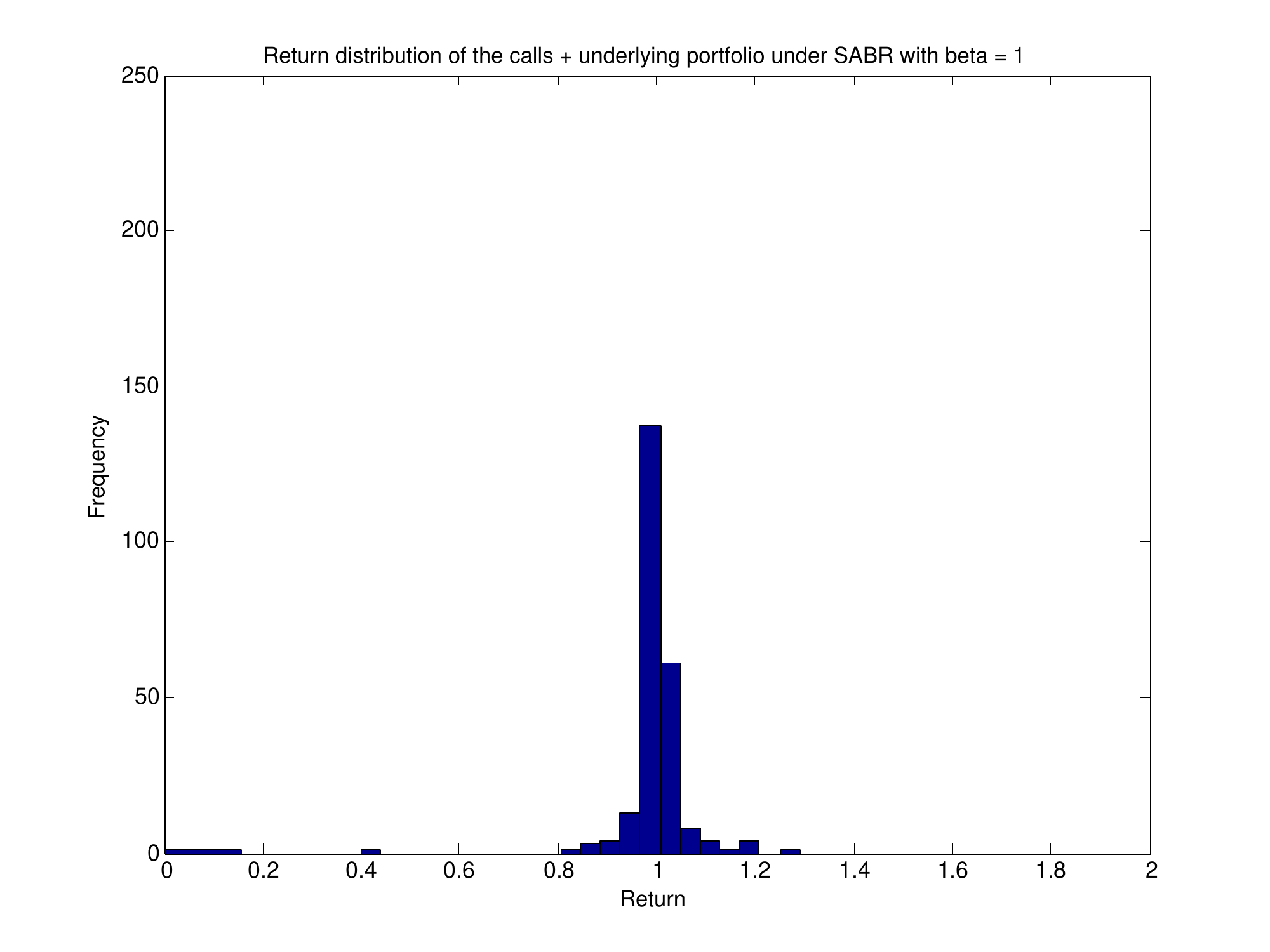}
       &
       \includegraphics[width=0.45\columnwidth]{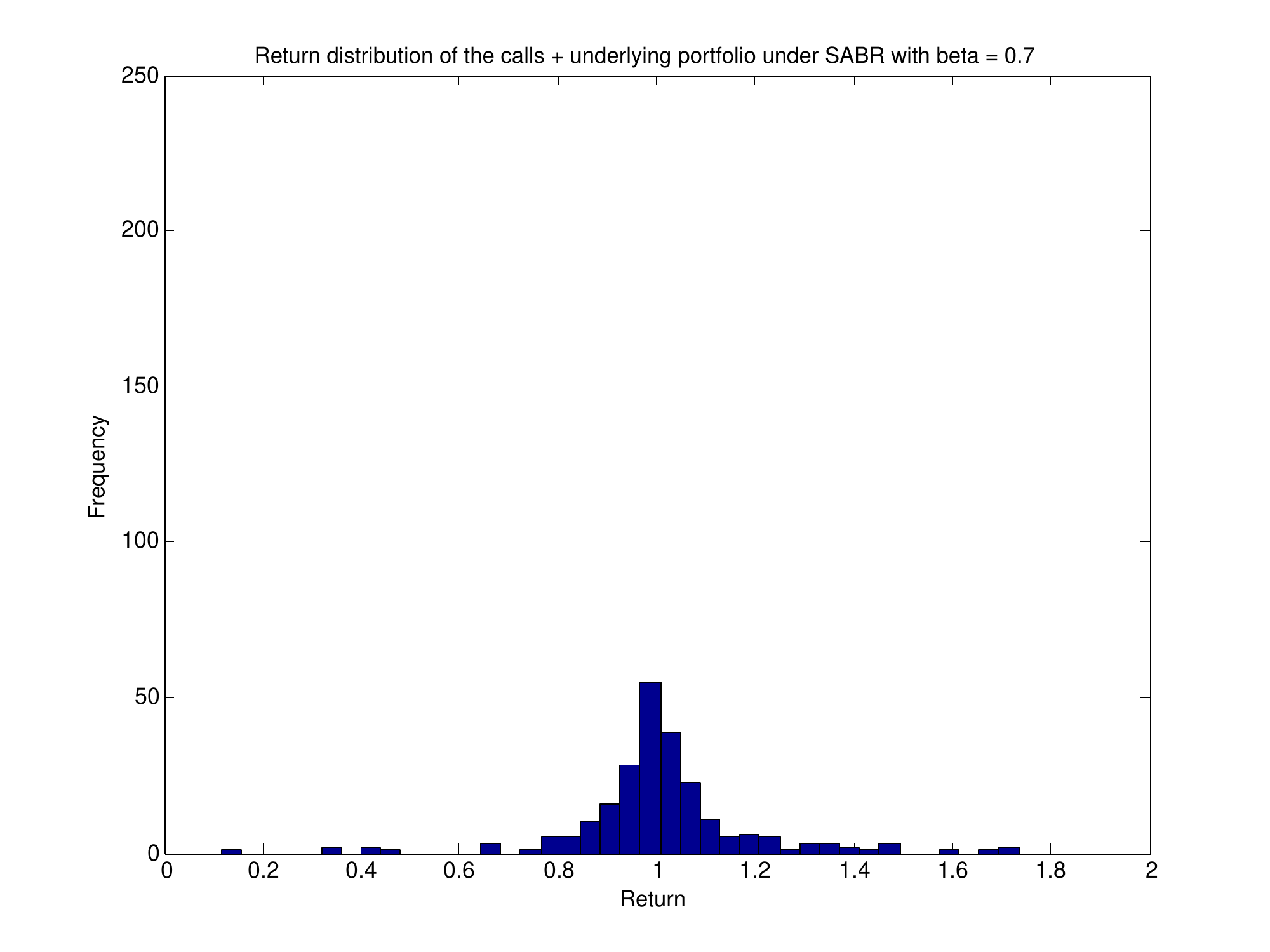} 
        \\
        \fontsize{7}{12}\selectfont (c) Under SABR model with $\beta = 1$
      &
       \fontsize{7}{12}\selectfont (d) Under SABR model with $\beta = 0.7$
       \\
       \end{tabular}
   \end{center}
   \vspace{-10pt}
   \caption{Distribution of the 8-day returns of (C + S) portfolio with 5 strikes in Period II}
   \label{fg:compare:retdistA2}
\end{figure}

\begin{figure}[h]
   \begin{center}
       \begin{tabular}{cc}
       \includegraphics[width=0.45\columnwidth]{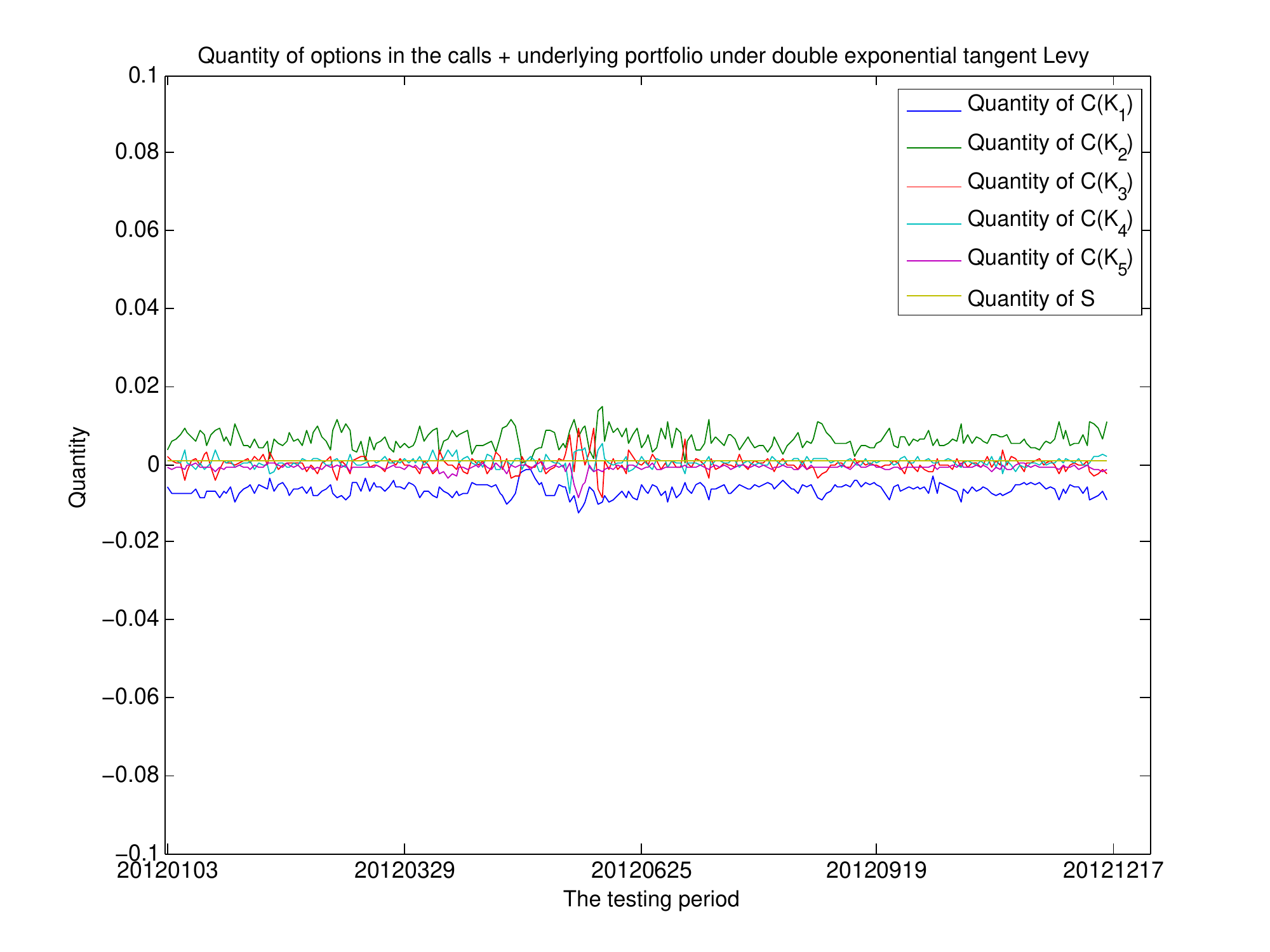}
       &
       \includegraphics[width=0.45\columnwidth]{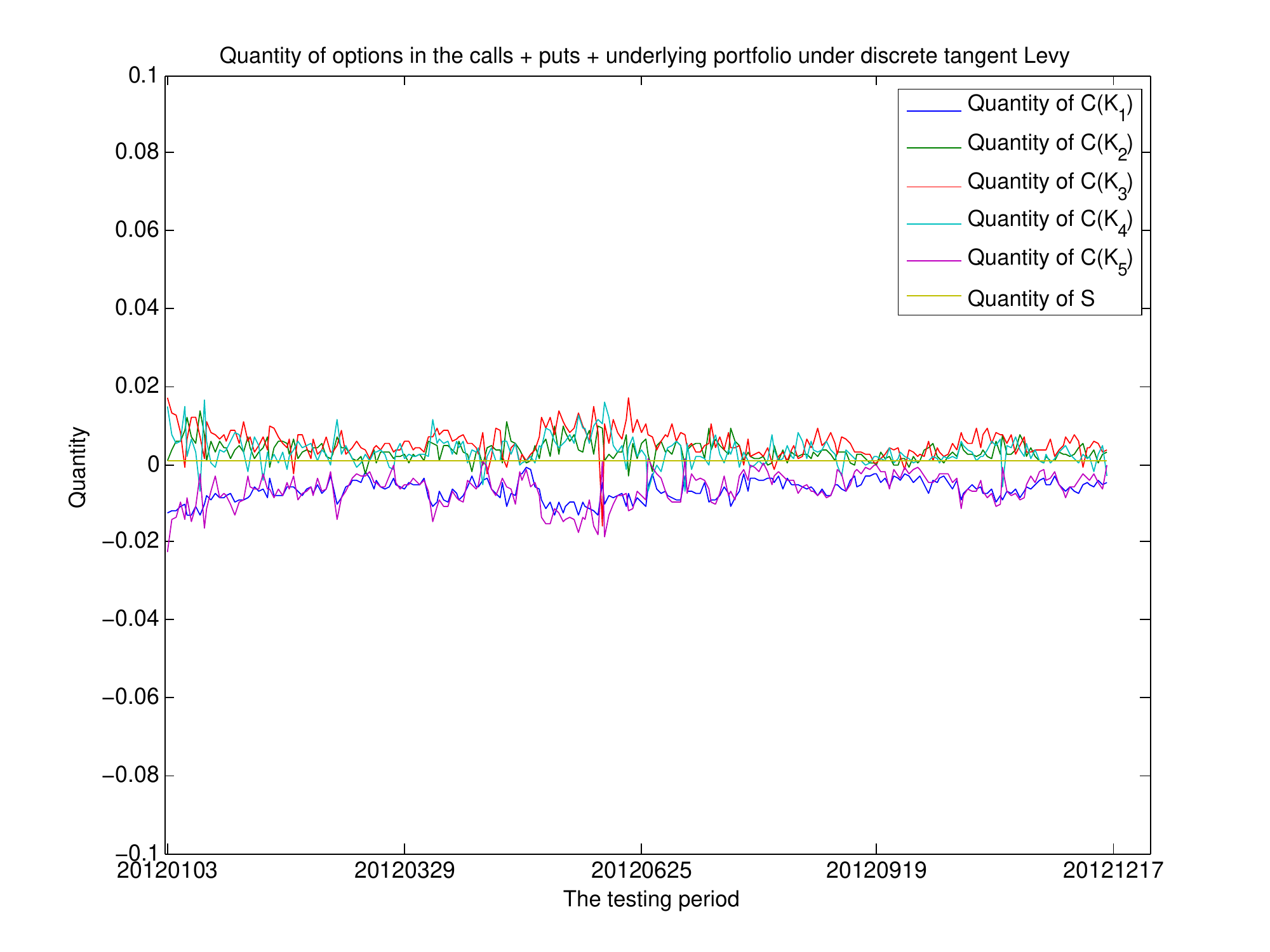}
       \\
       \fontsize{7}{12}\selectfont (a) Under double exponential tangent L\'{e}vy model
      &
       \fontsize{7}{12}\selectfont (b) Under discrete tangent L\'{e}vy model
       \\
       \includegraphics[width=0.45\columnwidth]{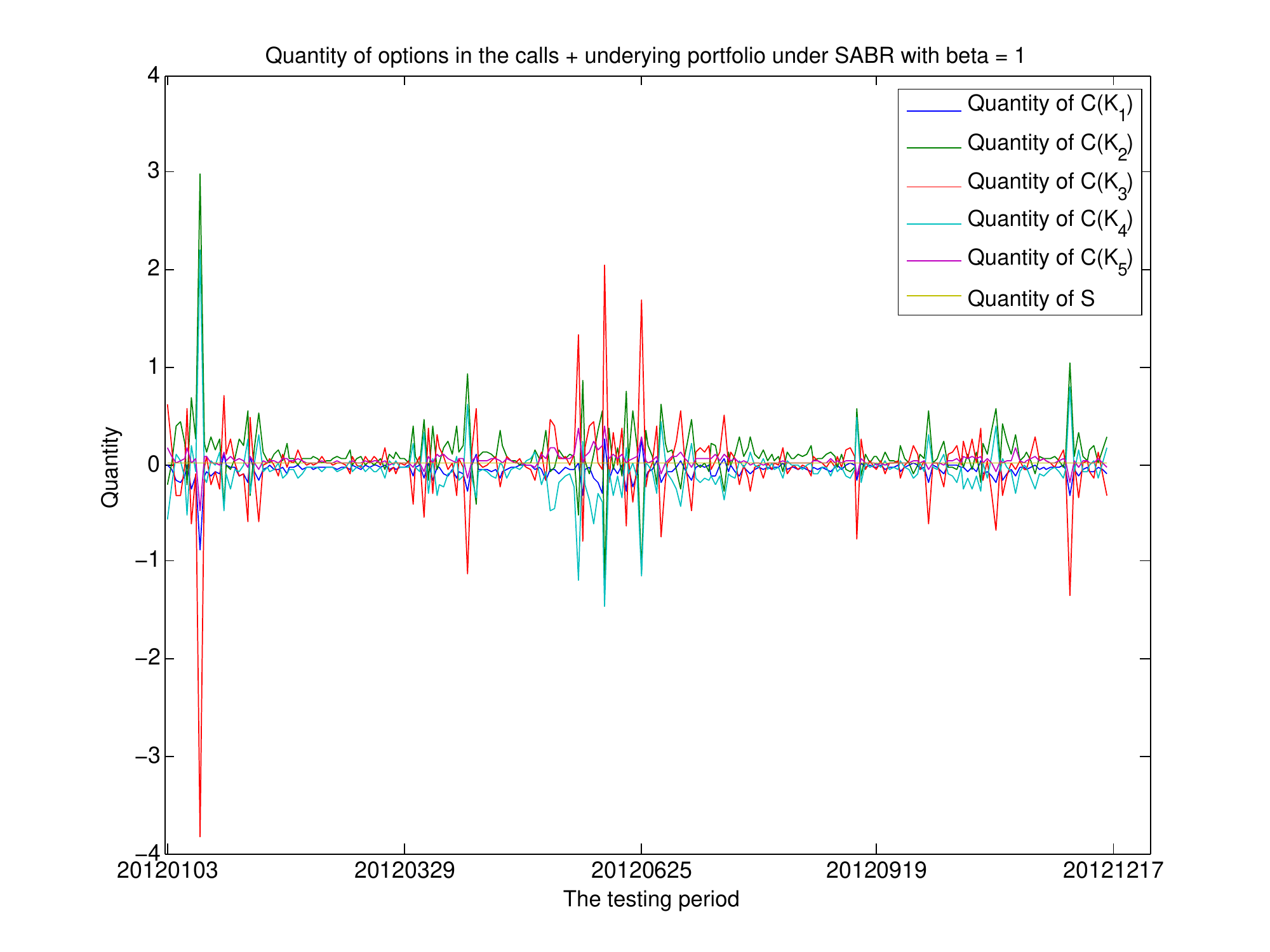}
       &
       \includegraphics[width=0.45\columnwidth]{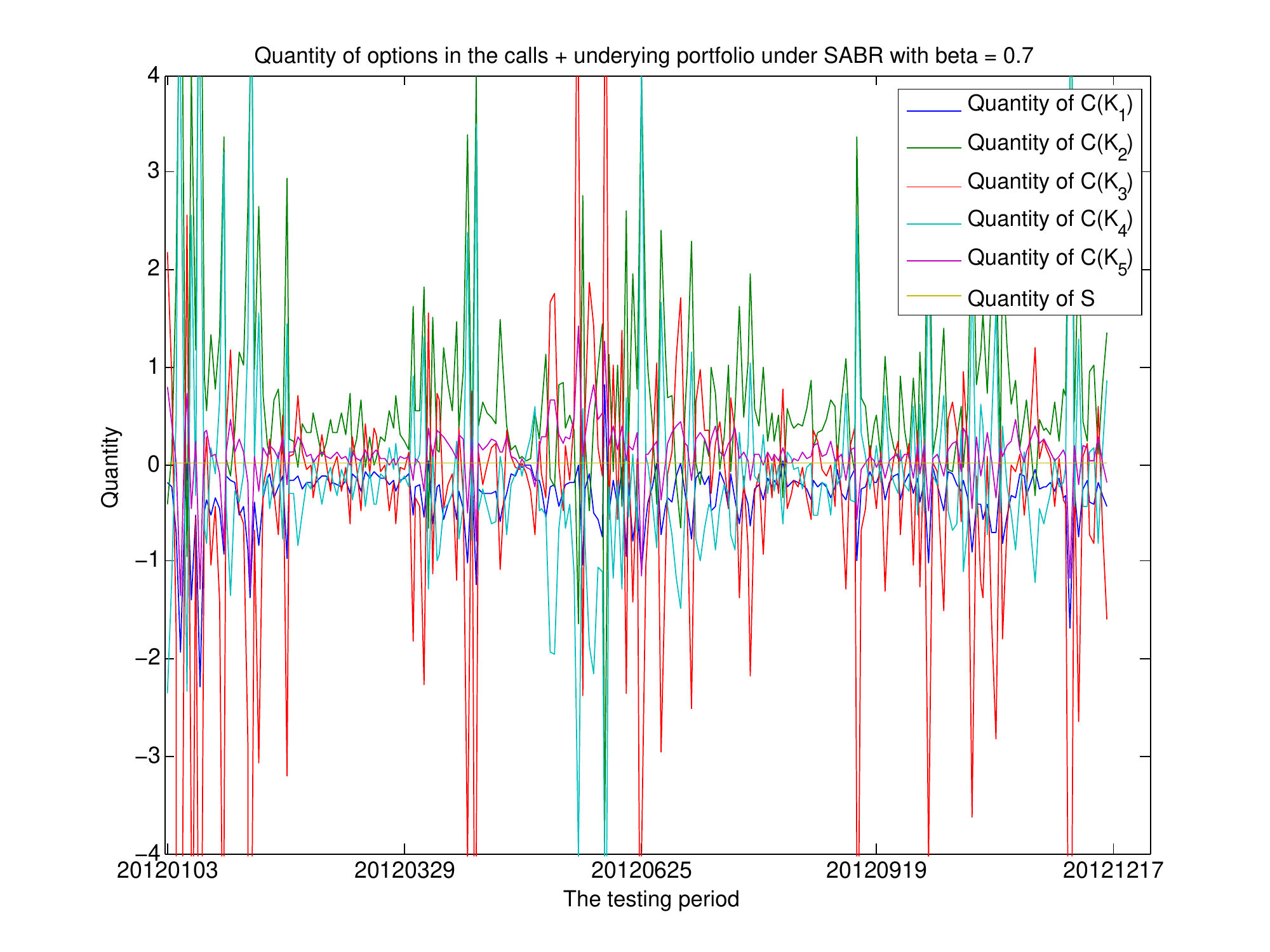} 
        \\
        \fontsize{7}{12}\selectfont (c) Under SABR model with $\beta = 1$
      &
       \fontsize{7}{12}\selectfont (d) Under SABR model with $\beta = 0.7$
       \\
       \end{tabular}
   \end{center}
   \vspace{-10pt}
   \caption{Option quantities in (C + S) portfolio with 5 strikes in Period II. Different scales are used to show more details}
   \label{fg:compare:quanA2}
\end{figure}

Once the estimation is completed, we can repeat the same simulation and testing procedures as in Subsection \ref{subse:part1}, to obtain the results shown in Tables~\ref{tb:compare:pA2} and \ref{tb:compare:qA2}, as well as in Figures~\ref{fg:compare:retdistA2} and \ref{fg:compare:quanA2}. These results confirm the finding of Subsection \ref{subse:part1}: for sufficiently many strikes in the portfolio, the tangent L\'{e}vy models do a much better job at finding a portfolio with smallest variance, their predictions for the variance are more reliable, and the portfolio weights are more stable.

\section{Conclusion}\label{se:conclusion}
In this paper, we implement and test two types of market-based models for European-type options. These types of models can be viewed as numerically tractable specifications of the tangent L\'evy models proposed in \cite{carmona2010tangentlevy} and \cite{carmona2011tangentlevy}. In particular, they provide a method for generating Monte Carlo samples of future implied volatility surfaces, in a way that is consistent with their past and present values.
We estimate the parameters of these models using real market data, for two periods: 2007-2008 and 2011-2012. The estimation procedure is described in a lot of detail, so that it can be reproduced by any interested reader. 

In addition, we use the estimated models and the real market data to conduct an empirical study using, whose goal is to compare the performance of market-based models with the performance of classical stochastic volatility models. We choose the problem of minimal-variance portfolio choice as a measure of model performance and compare the two tangent L\'evy models to SABR model. Our study demonstrates that the tangent L\'{e}vy models do a much better job at finding a portfolio with smallest variance, their predictions for the variance are more reliable, and the portfolio weights are more stable. To the best of our knowledge, this is the first example of empirical analysis that provides a convincing evidence of the outperformance of the market-based models for European options using real market data.

Our work is also subject to certain limitations, which suggest directions of future research. 
One of the most serious limitations of our work is the lack of numerical stability at the stage of static fitting, discussed at the end of Subsection \ref{subse:part1}. To address this issue, one has to come up with a family of tangent L\'evy densities that is rich enough -- so that it can approximate well the option prices observed in the market -- and, at the same time, not too large -- so that the calibration procedure is numerically tractable and more stable. This is a balance that seems to be hard to find. Another extension is to search for other families of tangent models -- not necessarily based on L\'evy processes. For example, it is not hard to find a combination of arbitrage-free prices of two call options, with the same maturity and different strikes, which cannot be approximated with an arbitrary precision (i.e. both prices at the same time) by the exponential L\'evy models. This means that, in principle, the market prices of call options (even with the bid-ask spread) may be such that there is no tangent L\'evy model that can match them (provided they contain more than one strikes). This, in turn, motivates the search for other families of models, which can always fit an arbitrary family of arbitrage-free option prices. An example of such family is provided in \cite{LVG}, but the existence and description of consistent dynamics within this family of models remains an open question.

\section{Appendix A}\label{se:spaces}
Here, we define the Banach spaces associated with tangent L\'{e}vy processes. 
\begin{itemize}
\item $\mathcal{B}_0$ is a Banach space of Borel measurable functions satisfying
\begin{equation}\label{fo:review:B0}
\|f\|_{\mathcal{B}_0} := \int_\mathbb{R}(|x| \wedge 1)|x|(1+e^x)|f(x)|dx < \infty.
\end{equation}

\item $\mathcal{B}$ is a Banach space of absolutely continuous functions $f: [0, \bar{T}] \rightarrow \mathcal{B}_0$ satisfying
\begin{equation}\label{fo:review:B}
\|f\|_\mathcal{B} := \|f(0)\|_{\mathcal{B}_0} + \int_0^{\bar{T}} \| \frac{d}{du} f(u) \|_{\mathcal{B}_0} du < \infty.
\end{equation}

\item $\mathcal{H}_0$ is a Hilbert space of Borel measurable functions $f: \mathbb{R} \rightarrow \mathbb{R}$ satisfying
\begin{equation}
\|f\|^2_{\mathcal{H}_0} := \int_\mathbb{R}|x|^4(1+e^x)^2|f(x)|^2dx < \infty.
\end{equation}

\item $\mathcal{H}$ is a Hilbert space of absolutely continuous functions $f: [0, \bar{T}] \rightarrow \mathcal{H}_0$ satisfying
\begin{equation}\label{fo:review:H}
\|f\|^2_\mathcal{H} := \|f(0)\|^2_{\mathcal{H}_0} + \int_0^{\bar{T}} \| \frac{d}{du} f(u) \|^2_{\mathcal{H}_0} du < \infty.
\end{equation}

\item $C([0,\bar{T}])$ is a Banach space of continuous functions $f: [0,\bar{T}] \rightarrow \RR$ satisfying
\begin{equation}\label{fo:review:CoT}
\sup_{x\in \RR} |f(x)| < \infty.
\end{equation}

\item $W^{1,2}([0,\bar{T}])$ is a Hilbert space of absolutely continuous functions $f: [0, \bar{T}] \rightarrow \RR$ satisfying
\begin{equation}\label{fo:review:W12}
|f(0)|^2 + \int_0^{\bar{T}} | \frac{d}{du} f(u) |^2 < \infty.
\end{equation}

\item $\mathcal{B}_d$ is a Banach space of absolutely continuous functions $f: [0,\bar{T}] \rightarrow \RR$ satisfying
\begin{equation} \label{fo:review:Bdcondition}
\|f\|_{\mathcal{B}_d} := |f(0)| + \int_0^{\bar{T}} | \frac{d}{du} f(u) | du < \infty.
\end{equation}
Here the subscript $_d$ is used to indicate the ``discrete'' models. 

\item $\mathcal{H}_d$ is the Hilbert space of absolutely continuous functions $f: [0,\bar{T}] \rightarrow \RR$ satisfying
\begin{equation} \label{fo:review:Hdcondition}
\|f\|^2_{\mathcal{H}_d} := |f(0)|^2 + \int_0^{\bar{T}} | \frac{d}{du} f(u)|^2 du < \infty.
\end{equation}
\end{itemize}

We know that $\mathcal{H}_0 \subset \mathcal{B}_0$, $\mathcal{H} \subset \mathcal{B}$, $W^{1,2}([0,\bar{T}]) \subset C([0,\bar{T}])$ and $\mathcal{H}_d \subset \mathcal{B}_d$. In addition, it is not hard to see that the completion of $\mathcal{H}_0$ is $\mathcal{B}_0$ with respect to the norm $\| \cdot \|_{\mathcal{B}_0}$. Similarly, the completion of $\mathcal{H}$ is $\mathcal{B}$ with respect to $\| \cdot \|_{\mathcal{B}}$,  the completion of $W^{1,2}([0,\bar{T}])$ is $C([0,\bar{T}])$ with respect to the ``sup'' norm, and the completion of $\mathcal{H}_d$ is $\mathcal{B}_d$ with respect to the $\|\cdot\|_{\mathcal{B}_d}$ norm. Hence, we conclude that the couples $\left(\mathcal{H}, \mathcal{B}\right)$, $\left(W^{1,2}([0,\bar{T}]), C([0,\bar{T}]) \right)$, and $\left(\mathcal{H}_d, \mathcal{B}_d\right)$ are all conditional Banach spaces (see III 5.3 in \cite{Kuo1975GaussianMeasure} for definition).

\section{Appendix B}
Proof of Lemma~\ref{cor:implementation:koupricing}.
The proof is similar to the one given in \cite{Kou2002model} except that $Z(T) = \mu T + \sum_{i=1}^{N_T} Y_i$ now follows a gamma distribution in the absence of the diffusion term.  The tail probability is given by 
$$
\PP\{Z(T) \geq a\} = \Psi(\mu, \lambda, p, \lambda_1, \lambda_2; a, T),
$$
with $\Psi$ given in \eqref{fo:implementation:psi}.
If we set $V_i = \exp(Y_i)$ for $i = 1,\cdots,N$, the drift term has to satisfy $\mu = - \lambda\EE[V_i-1]$ for $S_t$ to be a martingale, so the dynamics become
\begin{equation*}
d S_t = - \lambda\EE[V_i-1] S_{t-} dt + S_{t-} \cdot d\left[\sum_{i=1}^{N_t}(V_i-1)\right].
\end{equation*}
Let $\zeta = \EE[V_i-1] = \frac{p\lambda_1}{\lambda_1 - 1} + \frac{(1-p) \lambda_2}{\lambda_2+1} - 1$.  Using results on equivalence of measures for compound Poisson processes (see Proposition 9.6 in \cite{Cont2004financialmodeling} for example), we can see the time-$t$ price of a call option with maturity $T$ and strike $K$ is
\begin{align*}
C_t(T,K) & = \EE[(S_T - K)^+|\mathcal{F}_t] = \EE[S_T \bone_{S_T>K}|\mathcal{F}_t] - \EE[K \bone_{S_T>K}|\mathcal{F}_t], \\
& = S_t \cdot \Psi \left( - \lambda \zeta, \lambda^*, p^*, \lambda^*_1, \lambda^*_2; \log(\frac{K}{S_t}), T-t \right) \\
& \qquad \qquad - K \cdot \Psi \left( - \lambda \zeta, \lambda, p, \lambda_1, \lambda_2; \log(\frac{K}{S_t}), T-t \right),
\end{align*}
where $\lambda^* = \lambda(\zeta+1), \ p^* = \frac{p}{1+\zeta}\cdot\frac{\lambda_1}{\lambda_1-1}, \ \lambda_1^* = \lambda_1-1$ and $\lambda_2^* = \lambda_2+1$.

\section{Appendix C}

Proof of Theorem \ref{th:dtl.consist}.
Let us introduce the double exponential tail function $\psi$, for any L\'evy measure $\nu$ (defined on $\RR\setminus\left\{ 0\right\}$), with finite exponential moments, and all $x\in\RR\setminus\left\{ 0\right\}$:
\begin{equation}\label{eq.dtl.psi.gen}
\psi(\nu,x) := \left\{
\begin{array}{cc}
{\int_{-\infty}^{x} (e^x-e^z) \nu(dz)} & {x<0}\\
{} & {}\\
{\int_{x}^{\infty} (e^z-e^x) \nu(dz)} & {x>0.}\\
\end{array}
\right.
\end{equation}
Notice that, since $\nu$ is a L\'evy measure, $\psi(\nu,\cdot)\in\mathbb{L}^1(\RR)$. In addition, $\psi(\nu,\cdot)$ determines $\nu$ uniquely.
%\begin{equation}
%\label{fo:dtL:newpsi}
%\psi(\kappa;x)= \left\{
%\begin{array}{cc}
%\sum_{x^j<x}\kappa^j(T)(e^x-e^{x^j}) & {x<0}\\
%-\sum_{x^j>x}\kappa^j(T)(e^x-e^{x^j})  & {x>0}.
%\end{array}
%\right.
%\end{equation}
Using the above definition of $\psi$, we repeat the derivations in \cite{Carr2004locallevy}, to derive the associated partial integro-differential equation (PIDE) for call prices $C^{S_t,\kappa_t}(T,x)$ in the $(T,x)$ variables: 
\begin{equation}\label{eq.dtl.calls.psi}
\left\{
\begin{array}{c}
{\partial_{T} C^{S_t,\kappa_t}_t(T,x)=\int_{\RR} \psi(\kappa_t(T),x-y) \left(\partial^2_{xx} - \partial_x\right) C^{S_t,\kappa_t}_t(T,y) dy}\\
{}\\
{C^{S_t,\kappa_t}_t(t,x) = (S_t-e^x)^+,}\\
\end{array}
\right.
\end{equation}
where we treat $\kappa_t(T)$ as a measure: $\kappa_t(T) = \sum_{j\neq M} \kappa^j_t(T) \delta_{x^j}$.
Integrating by parts in \eqref{eq.dtl.calls.psi}, we make use of the discrete structure of $\kappa_t(T)$ to replace the integral with a summation and obtain the following PDE:
\begin{equation}\label{fo:dtL:callPIDE}
\left\{
\begin{array}{rl}
\partial_{T} C_t^{S_t,\kappa_t}(T,x)= & \underset{j:j\neq M}\sum \kappa_t^j(T) e^{x^j} C_t^{S_t, \kappa_t}(T,x-x^j) - C_t^{S_t,\kappa_t}(T,x) \cdot \underset{j:j\neq M}\sum \kappa_t^j(T) \\
& \qquad + [\partial_x -1]C_t^{S_t,\kappa_t}(T,x) \cdot \underset{j:j\neq M} \sum \kappa_t^j(T)(e^{x^j}-1) \\
C_t^{S_t,\kappa_t}(t,x) = & (S_t-e^x)^+.
\end{array}
\right.
\end{equation}
Similar to \cite{carmona2010tangentlevy}, we introduce $\Delta_t^{S_t,\kappa_t}(T,x) = - \partial_x C_t^{S_t,\kappa_t}(T,x)$. Differentiating and taking Fourier transform in $x$ on both sides of \eqref{fo:dtL:callPIDE}, we obtain
\begin{equation*}
\left\{
\begin{array}{rl}
\partial_{T} \hD_t^{S_t,\kappa_t}(T,\xi)= & \hD_t^{S_t,\kappa_t}(T,\xi) \left( \underset{j:j\neq M}\sum \kappa_t^j(T) \left( e^{(1-2\pi i\xi) x^j} - 1 + (2\pi i\xi - 1) (e^{x^j}-1)\right) \right)  \\
\hD_t^{S_t,\kappa_t}(t,\xi) = & \frac{e^{(1-2\pi i\xi) \log S_t}} {1-2\pi i\xi},
\end{array}
\right.
\end{equation*}
whose solution is given by
\begin{align}
\hD_t^{S_t,\kappa_t}(T,\xi) = \frac{e^{(1-2\pi i\xi)\log S_t}}{1-2\pi i\xi} \exp \left(\underset{j:j\neq M}\sum \int_t^T \kappa_t^j(u)du \cdot \left( e^{(1-2\pi i\xi) x^j} - (1 - 2\pi i\xi)e^{x^j}- 2\pi i\xi \right) \right).
\end{align}
Repeating the first part of the proof of Theorem 4.7 in \cite{carmona2010tangentlevy}, we see that the dynamics of $(S_t,\kappa_t)$ are consistent with the true underlying dynamics if and only if
\begin{equation*}
\left(\langle \hD_t^{S_t,\kappa_t}(T,\cdot), \phi \rangle := \int_{\RR}\hD_t^{S_t,\kappa_t}(T,\xi)\phi(\xi)d\xi\right)_{t\in[0,T)}
\end{equation*}
is a local martingale for any $\phi\in \mathcal{S}$, where $\mathcal{S}$ is the Schwartz space of fast-decaying functions (cf. \cite{carmona2010tangentlevy}).
This condition, in turn, is equivalent to the drift term in the semimartingale decomposition of $\hD_t^{S_t,\kappa_t}(T,\xi)$ being zero. An application of the generalized It\^{o}'s formula (cf. Theorem III.5.4 in \cite{Kuo}) to $\hD_t^{S_t,\kappa_t}(T,\xi)$ gives the following expression for the drift:
\begin{align*}
\Gamma_t(T,\xi) := & \hD_t^{S_t,\kappa_t}(T,\xi) \cdot \bigg[ -\underset{j:j\neq M}\sum \kappa_t^j(t) \left( e^{(1-2\pi i\xi) x^j} - (1 - 2\pi i\xi)e^{x^j} - 2\pi i\xi \right) \\
& \qquad \quad \quad + \underset{j:j\neq M}\sum \int_t^T \alpha_t^j(u)du \cdot \left( e^{(1-2\pi i\xi) x^j} - (1 - 2\pi i\xi)e^{x^j}- 2\pi i\xi \right) \\
& \qquad \quad \quad + \half \sum_{n=1}^m \left( \underset{j:j\neq M}\sum  \int_t^T \beta_t^{j,n}(u)du \cdot \left( e^{(1-2\pi i\xi) x^j} - (1 - 2\pi i\xi)e^{x^j}- 2\pi i\xi \right)\right)^2  \\
&\qquad \quad \quad +  \sum_{j\neq M} \int_\RR \left(e^{(1-2 \pi i\xi)x} - e^x(1-2\pi i \xi) - 2\pi i\xi\right)K^j_t \delta_{x^j}(dx) \bigg].
\end{align*}
The normalized drift term $\Gamma_t(T,\xi)/\hD_t^{S_t,\kappa_t}(T,\xi)$ is absolutely continuous as a function of $T \in [t,\bar{T}]$, therefore the condition $\Gamma_t(T,\xi)\equiv0$ is equivalent to
\begin{equation}\label{fo:dtL:cond1}
\frac{\lim_{T\downarrow t}\Gamma_t(T,\xi)}{\lim_{T\downarrow t} \hD_t^{S_t,\kappa_t}(T,\xi)} \equiv 0,
\end{equation}
and
\begin{equation}\label{fo:dtL:cond2}
\partial_T \frac{\Gamma_t(T,\xi)}{\hD_t^{S_t,\kappa_t}(T,\xi)} \equiv 0.
\end{equation}
A direct calculation shows that
\begin{align*}
\sum_{j\neq M}\int_{\RR}(e^{(1-2 \pi i\xi)x} - e^x(1-2\pi i \xi) - 2\pi i\xi) K^j_t \delta_{x^j}(dx) & = -2\pi (2\pi \xi^2+i\xi) \int_{\RR} e^{-2\pi i x \xi} \psi\left(\sum_{j\neq M}K^j_t \delta_{x^j}; x\right) dx, \\
\underset{j:j\neq M}\sum \kappa_t^j(t) \left( e^{(1-2\pi i\xi) x^j} - (1 - 2\pi i\xi)e^{x^j} - 2\pi i\xi \right) & = -2\pi (2\pi \xi^2+i\xi) \int_{\RR} e^{-2\pi i x \xi} \psi\left(\sum_{j\neq M} \kappa^j_t(t)\delta_{x^j}; x\right) dx.
 \end{align*}
Along with the uniqueness of Fourier transform and its inverse, the above equation shows that \eqref{fo:dtL:cond1} is equivalent to the compensator condition \eqref{fo:dtL:K}.
Next, \eqref{fo:dtL:cond2} gives us:
\begin{align}\label{fo:dtL:driftab}
& \underset{j:j\neq M}\sum \alpha_t^j(T) \cdot \left( e^{(1-2\pi i\xi) x^j} - (1 - 2\pi i\xi)e^{x^j}- 2\pi i\xi \right)  \nonumber \\
= & -\sum_{n=1}^m \left(\underset{k:k\neq M}\sum \beta_t^{k,n}(T)(e^{(1-2\pi i\xi) x^k} - (1 - 2\pi i\xi)e^{x^k}- 2\pi i\xi) \right) \nonumber \\
&\qquad \qquad \qquad \qquad \cdot \left( \underset{l:l\neq M}\sum \bbeta_t^{l,n}(T) ( e^{(1-2\pi i\xi) x^l} - (1 - 2\pi i\xi)e^{x^l}- 2\pi i\xi)\right),
\end{align}
with $\bbeta_t^{j,n}(T) = \int_t^T \beta_t^{j,n}(u) du$. The equation holds for all $\xi \in \RR$. Dividing both sides by $\xi^2$ and letting $\xi \rightarrow \infty$, we obtain
\begin{equation*}
0 = \sum_{n=1}^m \left(\underset{k:k\neq M}\sum \beta_t^{k,n}(T) (e^{x^k} - 1)\right) \left(\underset{l:l\neq M}\sum \bbeta_t^{l,n}(T)(e^{x^l} - 1)\right) 
\end{equation*}
\begin{equation*}
\phantom{?????????????????????}= \frac{1}{2} \partial_T \sum_{n=1}^m \left(\underset{k:k\neq M}\sum \bbeta_t^{k,n}(T)(e^{x^k} - 1)\right)^2 
\qquad \forall 0\leq t \leq T \leq \bT.
\end{equation*}
from which we conclude that $\beta$ has to satisfy the symmetry condition \eqref{fo:dtL:sym}.
%\begin{equation}\label{fo:dtL:symbeta}
%\underset{k:k\neq M}\sum \beta_t^{k,n}(T) (e^{x^k} - 1) = \sum_{k=1}^N \beta_t^{k,n}(T) (e^{x^k} - 1) = 0.
%\end{equation}
This symmetry condition also allows us to simplify \eqref{fo:dtL:driftab} to
\begin{align*}
& \underset{j:j\neq M}\sum \alpha_t^j(T) \cdot \left( e^{(1-2\pi i\xi) x^j} - (1 - 2\pi i\xi)e^{x^j}- 2\pi i\xi \right) \\
= & -\sum_{n=1}^m \left(\underset{k:k\neq M}\sum \beta_t^{k,n}(T)(e^{(1-2\pi i\xi) x^k} - e^{x^k}) \right)\cdot \left( \underset{l:l\neq M}\sum \bbeta_t^{l,n}(T) ( e^{(1-2\pi i\xi) x^l} - e^{x^l})\right).
\end{align*}
Now, dividing both sides of the above equation by $\xi$, and let $\xi \rightarrow \infty$, we see that $\alpha$ also has to satisfy the symmetry condition
\begin{equation}\label{fo:dtL:symalpha}
\underset{j:j\neq M}\sum \alpha_t^{j}(T) (e^{x^j} - 1) = \sum_{k=1}^N \alpha_t^{j}(T) (e^{x^j} - 1) = 0.
\end{equation}
Recall that we defined $\kappa^M$ so that \eqref{fo:dtL:xassumption2} holds: $\sum_{j=1}^N \kappa^j_t(T) = 0$. This relation is preserved for all $t\in[0,\bT]$ if and only if
\begin{equation}\label{fo:dtL:zerosumbeta}
\sum_{j=1}^N \alpha^j_t(T) = 0, \quad \quad \sum_{j=1}^N \beta^{j,n}_t(T)  = 0, \quad n= 1, \cdots, m.
\end{equation}
Substituting \eqref{fo:dtL:symalpha}, \eqref{fo:dtL:sym} and \eqref{fo:dtL:zerosumbeta} into \eqref{fo:dtL:driftab}, we obtain
\begin{equation} \label{fo:dtL:driftab2}
\sum_{j=1}^N \alpha_t^j(T) \cdot e^{(1-2\pi i\xi) x^j} =-\sum_{n=1}^m \left(\sum_{k=1}^N \beta_t^{k,n}(T)\cdot e^{(1-2\pi i\xi) x^k} \right) \left( \sum_{l=1}^N \bbeta_t^{l,n}(T) \cdot e^{(1-2\pi i\xi) x^l}\right).
\end{equation}
Both right and left hand sides of the above equation can be expressed as linear combinations of $\{e^{(1-2\pi i \xi)x^j}\}_{j = 1,\cdots,N}$. The latter functions are linearly independent, therefore, \eqref{fo:dtL:driftab2} is equivalent to a system of equations, in which we equate the coefficients in front of every basis function. This, in combination with Assumption \ref{ass:dtl.2}, yields \eqref{fo:dtL:driftcondition}.
%And the symmetry condition \eqref{fo:dtL:sym} is implied by \eqref{fo:dtL:symalpha} and \eqref{fo:dtL:symbeta}. 
Finally, we notice that \eqref{fo:dtL:sym} and \eqref{fo:dtL:driftcondition} imply \eqref{fo:dtL:symalpha}. As \eqref{fo:dtL:zerosumbeta} always holds (by the definition of $\kappa^M$), we can reverse the above derivations to show that \eqref{fo:dtL:sym} and \eqref{fo:dtL:driftcondition} imply \eqref{fo:dtL:driftab}. Thus, we have shown that \eqref{fo:dtL:cond2} is equivalent to \eqref{fo:dtL:sym} and \eqref{fo:dtL:driftcondition}, which completes the proof of the theorem.

\section{Appendix D}

\begin{table}[b] 
\centering
\caption{Time periods}\label{tb:implementation:SPX}
\begin{tabular}{|c|c|c|} 
 \hline
 & Jan. 2007 - Aug. 2008 & Jan. 2011 - Dec. 2012\\
\hline
\# of days &  419 & 502 \\
\hline
Range of SPX spot price & \$1214.9 - \$1565.2  & \$1099.2 - \$1465.8 \\
 \hline
\end{tabular}
\end{table}

\begin{table}[b] 
\centering
\caption{Testing periods}\label{tb:compare:periods}
\begin{tabular}{|c|c|c|} 
 \hline
Period & Training period & Testing period\\
\hline
I & Jan. 2007 - Dec. 2007 & Jan. 2008 - Aug. 2008 \\
II & Jan. 2011 - Dec. 2011 & Jan. 2012 - Dec. 2012\\
 \hline
\end{tabular}
\end{table}

\begin{table}[tb]
\centering
 \caption{Strikes used in each portfolio}\label{tb:compare:strikes}
\begin{tabular}{|c|l|} 
\hline
\# of strikes &  \multicolumn{1}{c|}{Strikes used}\\ 
\hline
5 & $K_{i-3}$(call), $K_{i-1}$(call), $K_{i+1}$(call), $K_{i+3}$(call), $K_{i+5}$(call) \\ \hline
4 & $K_{i-3}$(call), $K_{i-1}$(call), $K_{i+1}$(call), $K_{i+3}$(call)\\ \hline
3 & $K_{i-1}$(call), $K_{i+1}$(call), $K_{i+3}$(call)\\
\hline
%\multirow{3}{*}{C + P + S}  &  5 & $K_{i-2}$(put), $K_{i}$(put), $K_{i+1}$(call), $K_{i+3}$(call), $K_{i+5}$(call)\\ \cline{2-3}
%& 4 & $K_{i-2}$(put), $K_{i}$(put), $K_{i+1}$(call), $K_{i+3}$(call)\\ \cline{2-3}
%& 3 & $K_{i}$(put), $K_{i+1}$(call), $K_{i+3}$(call)\\ 
%\hline
\end{tabular}
\end{table}

\begin{table}[hb]
\centering
 \caption{Average deviation of (C + S) portfolio in period I}\label{tb:compare:pA1}
\begin{tabular}{|c|c|r|r|r|r|} 
\cline{2-6}
\multicolumn{1}{r|}{}&  \# of strikes & DETL &  DTL & SABR ($\beta = 1$) & SABR ($\beta = 0.7$) \\ 
\hline \hline
\multirow{3}{*}{$\begin{array}{c}  \text{Averaged realized}\\\text{deviation } Q  \end{array}$} &  5 & 0.55\% & 1.13\% & 84.97\% & 111.42\%\\ \cline{2-6}
& 4 & 0.54\% & 1.07\% & 4.69\% & 24.43\%\\ \cline{2-6}
& 3 & 0.64\% & 0.91\% & 2.18\% & 10.50\%\\
\hline \hline
\multirow{3}{*}{$\begin{array}{c}  \text{Averaged predicted}\\\text{deviation } P \end{array}$}  &  5 & 0.87\% & 0.88\% & 0.19\% & 9.33\% \\ \cline{2-6}
& 4 & 0.88\% & 0.92\% & 0.30\% & 9.66\%\\ \cline{2-6}
& 3 & 1.05\% & 1.03\%& 0.53\% & 10.29\% \\ 
\hline
\end{tabular}
\end{table}

\begin{table}[hb]
\centering
 \caption{Average quantity oscillation $K$ (as defined in \eqref{fo:compare:K}) in (C + S) portfolio in Period I}\label{tb:compare:qA1}
\begin{tabular}{|c|r|r|r|r|} 
\hline
\# of strikes & DETL &  DTL & SABR ($\beta = 1$) & SABR ($\beta = 0.7$) \\ \hline
5 & 0.0039 & 0.0082 & 1.1747 & 2.3846 \\ \hline
4 & 0.0038 & 0.0081 & 0.1339 & 0.4629 \\ \hline
3 & 0.0027 & 0.0075 & 0.0263 & 0.0807 \\ 
\hline
\end{tabular}
\end{table}

\begin{table}[h]
\centering
 \caption{Average deviation of (C + S) portfolio in Period II}\label{tb:compare:pA2}
\begin{tabular}{|c|c|r|r|r|r|} 
\cline{2-6}
\multicolumn{1}{r|}{}&  \# of strikes & DETL &  DTL & SABR ($\beta = 1$) & SABR ($\beta = 0.7$) \\ 
\hline \hline
\multirow{3}{*}{$\begin{array}{c}  \text{Average realized}\\ \text{deviation } Q  \end{array}$} &  5 & 0.41\% & 0.63\% & 9.07\% & 33.22\%\\ \cline{2-6}
& 4 & 0.42\% & 0.57\% & 3.51\% & 17.61\%\\ \cline{2-6}
& 3 & 0.42\% & 0.60\% & 0.90\% & 5.22\%\\
\hline \hline
\multirow{3}{*}{$\begin{array}{c}  \text{Average predicted}\\ \text{deviation } P  \end{array}$}  &  5 & 0.79\% & 0.48\% & 0.36\% & 7.98\% \\ \cline{2-6}
& 4 & 0.79\% & 0.55\% & 0.43\% & 8.11\%\\ \cline{2-6}
& 3 & 0.94\% & 0.65\%& 0.62\% & 8.46\% \\ 
\hline
\end{tabular}
\end{table}

\begin{table}[hb]
\centering
 \caption{Average quantity oscillation $K$ (as defined in \eqref{fo:compare:K}) of (C + S) portfolio with 5 strikes in Period II}\label{tb:compare:qA2}
\begin{tabular}{|c|r|r|r|r|} 
\hline
\# of strikes & DETL &  DTL & SABR ($\beta = 1$) & SABR ($\beta = 0.7$) \\ \hline
5 & 0.0011 & 0.0020 & 0.1410 & 0.6642 \\ \hline
4 & 0.0012 & 0.0023 & 0.0537 & 0.2736 \\ \hline
3 & 0.0011 & 0.0021 & 0.0145 & 0.0474 \\ 
\hline
\end{tabular}
\end{table}

%\clearpage

%\section{Appendix E}

\clearpage

%\begin{figure}[htp]
%  \begin{center}
%       \includegraphics[width=0.5\columnwidth]{sec-dtL/figures/eigenvalue_dtl.pdf}
%   \end{center}
%   \vspace{-10pt}
%   \caption{Percentage of variance explained by the eigenmodes}
%   \label{fg:dtL:eigenvalue}
%\end{figure}

%\clearpage

%\clearpage

\clearpage

\bibliographystyle{plain}
\bibliography{SimTL}

\end{document}